\documentclass[12pt]{book}

\advance \topmargin by -\headheight
\advance \topmargin by -\headsep

\evensidemargin \oddsidemargin
\usepackage{amsmath,amsfonts,amssymb,amsthm,latexsym,epsf}

 \advance\hoffset by -3mm  
 \advance\voffset by  8mm  

\def\CC{{\mathchoice
{\rm C\mkern-8mu\vrule height1.45ex depth-.05ex 
width.05em\mkern9mu\kern-.05em}
{\rm C\mkern-8mu\vrule height1.45ex depth-.05ex 
width.05em\mkern9mu\kern-.05em}
{\rm C\mkern-8mu\vrule height1ex depth-.07ex 
width.035em\mkern9mu\kern-.035em}
{\rm C\mkern-8mu\vrule height.65ex depth-.1ex 
width.025em\mkern8mu\kern-.025em}}}

\def\RR{{\rm I\kern-1.6pt {\rm R}}}

\def\ZZ{{\rm Z}\kern-3.8pt {\rm Z} \kern2pt}

\newcommand{\beq}{\begin{equation}}
\newcommand{\eeq}{\end{equation}}

\newcommand{\bear}{\begin{eqnarray}}
\newcommand{\eear}{\end{eqnarray}}

\def\to{\rightarrow}

\def\dal{\dot\alpha_}

\def\to{\rightarrow}

\def\sqr#1#2{{\vcenter{\vbox{\hrule height.#2pt
         \hbox{\vrule width.#2pt height#1pt \kern#1pt
            \vrule width.#2pt}
         \hrule height.#2pt}}}}
\def\dal{\mathop{\mathchoice\sqr75\sqr75\sqr{3.75}4\sqr34}\nolimits}
\def\bOmega{\boldsymbol{\Omega}}

\newfont{\namefont}{cmr10}
\newfont{\addfont}{cmti7 scaled 1440}
\newfont{\boldmathfont}{cmbx10}
\newfont{\headfontb}{cmbx10 scaled 1728}
\renewcommand{\theequation}{{\rm\thesection.\arabic{equation}}}

\def\br{\begin{eqnarray}}
\def\er{\end{eqnarray}}
\def\brs{\begin{eqnarray*}}
\def\ers{\end{eqnarray*}}
\def\be{\begin{equation}}
\def\ee{\end{equation}}
\def\bd{\begin{description}}
\def\ed{\end{description}}

\newtheorem{prop}{Proposici\'on}[section]
\newtheorem{defi}{Definici\'on}[section]
\newtheorem{theorem}{Teorema}[section]
\newtheorem{lema}{Lema}[section]
\newtheorem{corolario}{Corolario}[section]

\def\bedef{\begin{defi}}
\def\endef{\end{defi}}

\def\bethe{\begin{theorem} }
\def\enthe{ \end{theorem} }

\def\beprop{\begin{prop} }
\def\enprop{ \end{prop} }

\def\belem{\begin{lema}}
\def\enlem{\end{lema}}

\def\becor{\begin{corolario}}
\def\encor{\end{corolario}}

\def\pa{\partial}
\def\comm#1#2{\left[#1\, ,\,#2\right]}
\def\anticomm#1#2{\left\{#1\, ,\,#2\right\}}

\def\a{\alpha}
\def\b{\beta}
\def\d{\delta}
\def\D{\Delta}
\def\eps{\varepsilon}

\def\g{\gamma}
\def\G{\Gamma}

\def\l{\lambda}
\def\L{\Lambda}

\def\m{\mu}
\def\n{\nu}

\def\O{\Omega}

\def\P{\Phi}
\def\pa{\partial}
\def\r{\rho}

\def\r{\rho}
\def\s{\sigma}
\def\S{\Sigma}

\def\th{\theta}

\def\bnabla{\boldsymbol{\nabla}}
\def\bfBox{\boldsymbol{\Box}}
\def\bW{\boldsymbol{W}}
\def\bG{\boldsymbol{\Gamma}}

\def\sqr#1#2{{\vcenter{\vbox{\hrule height.#2pt
         \hbox{\vrule width.#2pt height#1pt \kern#1pt
            \vrule width.#2pt}
         \hrule height.#2pt}}}}
\def\dal{\mathop{\mathchoice\sqr64\sqr64\sqr{3.75}4\sqr34}\nolimits}

\begin{document}


\begin{titlepage}

\begin{center} 
\large \sf  UNIVERSIDADE DE SANTIAGO DE COMPOSTELA

\vspace{.3cm}

\large Departamento de F\'\i sica de Part\'\i culas 
\end{center}

\vspace{5cm}

\begin{center} 
\LARGE  \bf The beta function of gauge theories \\at two loops in differential renormalization
\end{center}

\vspace{3cm}

\begin{center} 
Tesis presentada para optar al grado de Doctor en F\'isica.
\end{center}

\vspace{4cm}

\begin{center} 
{\sf\bf \large C\'esar Seijas Naya}

\sf Santiago de Compostela, febrero 2007.
\end{center} 

\end{titlepage}

\frontmatter
\chapter{Agradecimientos}

\begin{flushright}
\begin{it}
Si sale, sale. Y si no sale \\
hay que volver a empezar. \\
Lo dem\'as son fantas\'ias \\
\end{it}
{\bf{Edouard Manet}}
\end{flushright}

En un trabajo que ha llevado tanto tiempo, es normal el tener un mont\'on de gente a la que agradecerle el apoyo prestado para terminarlo. Espero que no se me olvide nadie, ya que como dijo Quevedo: ``el agradecimiento es la parte principal de un hombre de bien''.

En primer lugar, tengo que agradecer al profesor Javier Mas la oportunidad que me ha brindado de poder realizar esta tesis doctoral. Entre otras muchas cosas, y dejando a parte toda la f\'isica que he aprendido con \'el, me gustar\'ia destacar su apoyo en las diferentes circunstancias por las que he pasado durante todo este tiempo, as\'i como el haber intentado ense\~narme a tener la visi\'on cr\'itica necesaria para llevar a cabo una investigaci\'on. Tambi\'en me gustar\'ia agradecer la ayuda que me han brindado los profesores Manuel P\'erez-Victoria y Jose Ignacio Latorre. A Manuel tengo que darle las gracias por haber compartido conmigo su amplio conocimiento de renormalizaci\'on diferencial y simetr\'ias, y por su disponibilidad en todo momento para revisar y responder mis dudas de estos temas. En cuanto a Jose Ignacio, aunque no llegamos a trabajar juntos directamente, le estoy muy agradecido por haberme proporcionado informaci\'on muy valiosa de su trabajo en el desarrollo del m\'etodo de  renormalizaci\'on diferencial.  

Tambi\'en tengo que darles las gracias a mis compa\~neros de despacho y departamento de la facultad. Tanto en los buenos momentos (que fueron la mayor\'ia) como en los malos (realmente muy pocos) me disteis todo vuestro apoyo. Much\'isimas gracias. 

En cuanto a ``la banda'', qu\'e os puedo decir que no sepais. Siempre hab\'eis estado ah\'i, y soy muy consciente de lo afortunado que he sido por eso. Nunca os agradecer\'e lo suficiente el haberme aguantado y apoyado durante todo este tiempo.

Uno de los mejores recuerdos que tengo en mi vida es el paso por Santiago. Y esto es as\'i, gracias a la gente excepcional que conoc\'i all\'i. Lo m\'as seguro es que poco a poco nos vayamos alejando cada vez m\'as, pero el compa\~nerismo y las horas de estudio y ocio compartidas con vosotros son algo que nunca olvidar\'e. 

Tambi\'en tengo muchas cosas que agradecer a la gente que encontr\'e en mi etapa madrile\~na. En un momento lleno de cambios e incertidumbres en mi vida, fuisteis el apoyo que necesitaba. Adem\'as, si tuve el empuje de retomar los c\'alculos cuando estaba en lo que parec\'ia un callej\'on sin salida, fue en gran parte ayudado por vosotros. Gracias de todo coraz\'on.

Tampoco quiero dejar olvidados a todos los compa\~neros (bueno, y en muchos casos ya amigos) con los que he trabajado en Softgal Gesti\'on. Realmente, vosotros me habeis demostrado que el capital humano es la mayor riqueza que tiene una empresa. 

Finalmente, si el hombre es uno mismo y sus circunstancias, una gran parte de las m\'ias son mi familia. Si he llegado a poder escribir este trabajo, ha sido por  vosotros. Os quiero.

\tableofcontents

\chapter{Introduction}
At the beginning of the past century, two basic blocks of the modern physics were established: Quantum Theory and the Theory of Relativity. In the following years, Quantum Field Theory was developed in order to made both theories compatible (or, to be more precise, Quantum Mechanics and the Special Theory of Relativity, as the complete connection with General Relativity is still an open problem). One of the surprising points of this theory is that some of the calculations involved divergent quantities. At first this was thought to be a problem, but soon it was found that it was inherent to any Quantum Field Theory calculation, reflecting only the infinite degrees of freedom of these processes. Finally, the standard way of treating these divergences was established to be a two-step work:

\begin{itemize}
\item First of all, we have to {\bf{regularize}} the divergence. The idea is to introduce an extra parameter (the regulator) in terms of which we can rewrite the divergent expression as a finite function, being the infinite result a certain limit of this parameter.

\item Once we have parametrized the divergence in terms of this regulator, the next step is to drop off the divergent part of each diagram, retaining only the finite part. This is what we call {\bf{renormalization}}. The easiest way of performing this is to modify the coefficients of the terms of the action, making them regulator-dependent, so that we have new terms in the calculation (counter-terms) that can be adjusted to cancel the divergent parts. The mass scale at which this procedure is applied is called {\em{renormalization scale}}.
\end{itemize}

Renormalizable theories are those theories where the previous procedure can be applied and only the values of a few parameters are affected. The origin of the harmlessness of the quantum fluctuations in these theories can be easily understood with a method developed by Wilson \cite{Wilson:1973jj}. Here, using a functional approach, we begin by considering a theory which has an ultraviolet cutoff scale $\L$. Then, integrating over a momentum shell, we define the theory e have a new (infinitesimally) lower cutoff scale $\L^{\prime}$. Although at first this redefinition implies that an infinite set of new different terms can appear in the lagrangian, it can be shown that in a renormalizable theory only a finite subset of them tend to grow if we iterate this procedure, whereas the rest vanishes. The differential equations that govern the flow of the coefficients of the lagrangian are called {\em{renormalization group}} equations. Another approach to the {\em{renormalization group}} is the Callan-Symanzik equation \cite{Callan:1970yg,Symanzik:1970rt}, which is obtained from the arbitrariness in the election of the scale that we use to impose the renormalization conditions when obtaining the parameters of a renormalized field theory. So, with the Callan-Symanzik equation, the renormalization group flows are obtained by looking at how the parameters of the theory depend on the renormalization scale. The shifts of the coupling constants are reflected in one special parameter of the equation, which is called the {\em{beta function}}. Hence, by studying this parameter, we can obtain relevant physical information, as the validity of perturbative approach to obtain the short- or large-distance behaviour of the theory.

Central to the physics is the idea of symmetry, which is the invariance of a physical system under some kind of transformation. When Quantum Field Theory was developed, two different types of symmetries were found: the space-time symmetries, generated by the Poincare group, and on the other hand internal symmetries. It was shown that both types of symmetries can not be non-trivially mixed \cite{Coleman:1967ad}, unless we consider fermionic symmetry generators \cite{Haag:1974qh}(ie., operators that interchange fermions with bosons and vice versa). These generators allow us to obtain an extended Poincare algebra, which is called {\em{supersymmetry algebra}}. Since its discovery, and although it has not yet been experimentally verified, supersymmetry has become one of the basic elements of modern theoretical physics. Among the different reasons for that, we can stand out that is a key ingredient of the theoretical efforts for the unification of gravity and the other forces of nature (e.g., supergravity and superstring theories), and it provides models that are simpler to study and quantize, as the symmetry between fermions and bosons implies that some ``miraculous'' cancellations occur in the calculations.   

As a renormalized quantum theory should have (if possible) the same symmetries as the classical one, among the relevant features that we have to maintain in a renormalization procedure, we have the invariance with respect to local symmetry transformations, which is called {\em{gauge invariance}}. However, the quantization procedures force us to loose gauge invariance in the intermediate results (for example, we have to pick up only one representative gauge field of each gauge orbit in a functional quantization approach). To maintain explicit gauge invariance in every step of the calculations, the {\em{background field method}} \cite{DeWitt:1967ub} was developed. Here, the gauge field is split into two parts: quantum and background. We quantize the first one, which implies that we have to break the gauge invariance on it. At the same time, the second field is treated as a classical one, and therefore gauge invariance is retained in terms of it. This has relevant consequences: for example, it imposes a relation between the gauge coupling and background field renormalizations and allows us to obtain the beta function from a calculation of only the background field two-point function.

When quantizing a gauge theory, it was found that only some renormalization procedures can preserve explicitly the symmetries, except for some exceptional cases called {\em{anomalies}}, being the most successful one dimensional renormalization. The key point of this method is to rewrite the original divergent integrals in four dimensions as integrals in $D$ dimensions, being the regulator the $\eps$ parameter of this continuous dimension $D = 4 - 2 \eps$. As we have stated before, it preserves explicitly gauge invariance, making also the calculations easy to perform (even in the higher-loop cases). However, this procedure has some drawbacks. In concrete, due to the fact of changing the space-time dimension, some incongruities are expected to appear when applying this method to a dimension-dependent theory as can be a supersymmetric one.

To solve this problem, and offer an alternative renormalization procedure that works only in four dimensions, {\bf{differential renormalization}} (DiffR) was developed \cite{Freedman:1991tk}. The basics of the method are to work in coordinate space rather in momentum space, rewriting expressions that are too singular to have a well defined Fourier transform in terms of derivatives of less singular ones. With this prescription, it can be shown that the coefficients of the renormalization group equations that are satisfied by the correlators of the theory are easily obtained. At the same time, we stay all the time in four dimensions, making this a suitable renormalization procedure when dealing with supersymmetric theories. However, one important practical difficult arises. Although gauge invariance is not broken, to recover the explicit form of this invariance in the final results we have to fix the ambiguities generated by the method. So, we have to impose {\em{a posteriori}} the Ward identities. 

This important point was solved (at one loop) by the introduction of {\bf{Constrained Differential Renormalization}} (CDR) \cite{delAguila:1997kw}. The basic idea here is to give a minimal set of rules to manipulate singular expressions, so that all the ambiguities of the calculations are fixed {\em{a priori}}. At the same time, all of these manipulations are required to be compatible with the symmetries that have to be maintained. With this prescription, it can be seen that the renormalized expressions directly fulfil the Ward identities without any adjustment.

The objective of this work is to show that differential renormalization can be easily and applied to the renormalization of gauge theories at the two-loop level. In concrete, we will show that with this method we can obtain with little effort the two-loop coefficient of the expansion of the beta function of these theories. We have to point out that, although it is only fully developed for the one-loop case, to perform some of these calculations we will use CDR prescriptions. This is due to the fact that when imposing CDR at the one-loop level, the coefficients of the logarithms of the mass-scales of the two-loop renormalized expression get fixed {\em{a priori}}. No Ward identities are needed to be used. Also, we will show that differential renormalization clearly distinguishes between ultraviolet and infrared divergences as both are renormalized with different and independent mass-scales. This is not the case for dimensional regularization, where both types of divergences get mixed in the results, as they are renormalized with the same dimensional parameter $\eps$. Hence, this feature allows us to revisit one controversial point: the origin (ultraviolet of infrared) of the higher-order perturbative contributions to the beta function in supersymmetric gauge theories. Originally, Novikov, Shifman, Vainshtein and Zakharov obtained the so-called ``exact beta function'' of $N=1$ SYM ($\b_{NSVZ}$) by means of instanton analysis \cite{Novikov:1983uc}, where the origin of the higher-order contributions was clearly infrared. However, this was questioned by Arkani-Hamed and Murayama \cite{Arkani-Hamed:1997ut,Arkani-Hamed:1997mj}, as they were able to obtain $\b_{NSVZ}$ in a purely wilsonian framework, which only depends on the ultraviolet properties of the theory. With our approach, we will obtain perturbatively the two-loop coefficient of $\b_{NSVZ}$ with the advantage of having the UV and IR divergences clearly separated.   

The structure of the work is as follows: In the first chapter, we made a brief presentation of DiffR and CDR, showing also how the results of the latter can be used in two-loop calculations. In the second chapter, we give a complete treatment of the calculation of the beta function of two of the most relevant abelian gauge theories: QED and SuperQED. Although these two theories were yet renormalized in the literature with standard DiffR, we will re-obtain their two-loop beta functions without imposing Ward identities. The third chapter is devoted to the renormalization of non-abelian gauge theories, studying the concrete models of Yang-Mills and SuperYang-Mills. Finally, we present our conclusions. In appendices \ref{ap_SUSY} and \ref{ap_BFM} we made a brief presentation of our supersymmetic conventions and the background field method respectively. In appendix \ref{ap_Gauge}, in order to obtain the function that takes into account the running of the gauge parameter in the RG equations, we evaluate the one-loop RG equations for the quantum gauge field two-point functions of each theory that we treat. Finally, in appendix \ref{ap_calc} we list some identities and calculations that are used in this work.

\mainmatter

\chapter{Differential Renormalization and CDR}
\label{chap1}

\section{Differential Renormalization}

Differential Renormalization (DiffR) \cite{Freedman:1991tk} is a renormalization method in real space that consists in replacing coordinate-space expressions that are too singular by derivatives of less singular ones. This method does not need cutoff nor explicit counterterms, although they are implicitly used when performing formal integration by parts. The basic idea is that divergent expressions are well defined for non-coincident points, but at short distances the amplitude is too singular and does not have a Fourier transform. Hence, to renormalize we are instructed by the method to replace the divergent expression with the derivative of a less singular one that has the same values as the original outside the origin, but with a well defined Fourier transform (if formal integration by parts is used with the derivatives). This method is especially well suited for dimensional dependent theories (such as supersymmetric theories), because all the time we stay in four dimensions, which is not the case for dimensional regularization or dimensional reduction.

As an example consider the one-loop contribution of $\l \phi^4$ theory. The bare expression is
\br
\Gamma (x_1 , x_2 , x_3, x_4 ) &=& \frac{\l^2}{2} \left[ \d^{(4)} (x_1 - x_2 )\d^{(4)} (x_3 - x_4) [ \D(x_1 - x_4)]^2 + (2\; perms) \right] \;, \nonumber \\
\er
where $\D(x-y)$ is the massless propagator
\br
\D(x-y) \equiv \D_{xy} &=& \frac{1}{(4 \pi^2)} \frac{1}{(x-y)^2} \;.
\er
At short distance $\frac{1}{x^4}$ does not have a well defined Fourier transform, and DiffR proposes to replace it for the solution of
\br
\frac{1}{x^4} = \Box G(x^2) \; \; \; \; \; ~ x \ne 0 \;,   \label{eq_dif}
\er
which is
\br
\frac{1}{x^4} \rightarrow \left[\frac{1}{x^4} \right]_R &=& - \frac{1}{4} \Box \frac{ \ln x^2 M^2}{x^2} \;. \label{ren_D2} 
\er
Both expressions coincide for $x\neq 0$, but the new one has a well defined Fourier transform if we neglect the divergent surface terms that appears upon integrating by parts the d'alembertian. It is in these surface terms where the counterterms hide, and by applying formal integration by parts \cite{Freedman:1991tk} we are implicitly taking them into account, as we will detail later. Thus, with the renormalized expression we obtain 
\br
\int d^4 x \; e^{i p \cdot x} \left[\frac{1}{x^4} \right]_R &=& - \frac{1}{4} 
\int d^4 x e^{i p \cdot x} \Box \frac{ \ln x^2 M^2}{x^2} = \frac{p^2}{4}
 \int d^4 e^{i p \cdot x} \frac{\ln x^2 M^2}{x^2}
\nonumber \\ &=& - \pi^2 \ln \left( \frac{p^2}{{\bar{M}}^2}\right) \;.
\er
A constant with mass dimension $M$ has been introduced for dimensional reasons. It parametrizes the {\em local ambiguity}
\br
 \Box \frac{\ln x^2 {M'}^2}{x^2} =  \Box \frac{\ln x^2 M^2}{x^2}  + 2\ln\frac{M'}{M} ~\delta(x) \;.
 \label{ambigu}
\er
A crucial observation is that this shift $M\to M'$  can be absorbed in a rescaling of the coupling constant $\l$ \cite{Freedman:1991tk}. This is a hint that renormalized amplitudes satisfy renormalization group equations, with M playing the r\^ole of the renormalization group scale.

Let us take a closer look to the implicit counterterms that we are using in all of this procedure (this is discussed in \cite{Freedman:1991tk} and with more detail in \cite{Freedman:1992gr}). As we have stated previously, along with the substitution of the divergent expression with the solution of the differential equation, we also have to use the following formal integration by parts prescription
\br
\int d^4 x \; \frac{1}{x^4} T (x) \equiv - \frac{1}{4} \int d^4 x \; ( \Box \frac{ \ln x^2 M^2}{x^2} ) T(x)\equiv - \frac{1}{4} \int d^4 x \;  \frac{ \ln x^2 M^2}{x^2} \Box T(x) \;, \nonumber \\
\er

i.e., we have neglected divergent surface terms. If we made this calculation again, but excluding a ball $ {\cal{B}}_\eps$ of radius $\eps$ around the origin and keeping surface terms we have
\br
\int_{R^4 / {\cal{B}}_\eps} d^4 x \;  T (x) \Box \frac{ \ln x^2 M^2}{x^2}  &=& \int_{S_{\eps}} d \sigma_{\m} \; T(x) \pa_{\m} \frac{ \ln x^2 M^2}{x^2}  \nonumber \\
& & - \int_{R^4 / {\cal{B}}_\eps} d^4 x \; \pa_{\m} T (x) \pa_{\m} \frac{ \ln x^2 M^2}{x^2} \;.
\er

The contribution of the surface integral can be found to be
\br
\int_{S_{\eps}} d \sigma_{\m} \; T(x) \pa_{\m} \frac{ \ln x^2 M^2}{x^2} = 4 \pi^2 T(0) ( 1 - \ln \eps^2 M^2) + {\cal{O}}(\eps) \;.
\er

This is divergent as $\eps \rightarrow 0$. However, this singular contribution in the 4-point function can be cancelled if we add to the action a local counterterm proportional to $\int d^4 x \; \phi^4 (x) ( 1 - \ln \eps^2 M^2)$. Hence, as we have seen, the formal integration by parts rule is valid because we are implicitly using these counterterms. At the same time, we have to remark that the regularization method does not require us to make explicit use of them in any calculation.

\subsection{Higher Loops}
\label{Higher_Loops}
Differential renormalization can be applied not only to one-loop diagrams, but to multi-loop expressions. In general, new scales appear corresponding to the renormalization of the different subdiagrams that form the total expression. From the various types of subdivergences that can occur in a typical higher-loop diagram, we will take a closer look to one of them, where independent scales are neatly seen to appear at each stage: nested divergences. As an example we consider the following amplitude which, in principle, could form part of a bigger one: $\D(x-y) I^1(x-y)$, where $I^1(x)$ is  
\be
I^1(x-y) = \int d^4 u \D_{xu} \D^2_{yu} \;. \label{def_I1}
\ee
It corresponds to a diagram that looks as follows
\begin{figure}[ht]
\centerline{\epsfbox{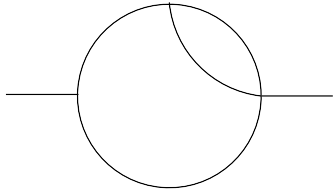}}
\caption{Two-loop diagram with nested divergences.}
\end{figure}

As can be seen, divergences occur whenever 2 points come together. We can renormalize them starting from the most inner one, and proceeding recursively
\begin{eqnarray}
\left[ \D_{xy} \int d^4 u \D_{xu} [\D^2_{yu}]_{R}\right]_{R}
 &=&
\left[- \frac{1}{4 (4 \pi^2)^4} \frac{1}{(x-y)^2} \int d^4 u \; \frac{1}{(x-u)^2}
 \Box \frac{\ln (y-u)^2 M^2_1 }{(y-u)^2}  \right]_{R}
\nonumber\\
&=&
\left[\frac{1}{4 (4 \pi^2)^3} \frac{\ln (x-y)^2 M_1^2 }{(x-y)^4}  \right]_{R}
\nonumber\\
&=& 
 - \frac{1}{32(4 \pi^2)^3} \Box \frac{ \ln^2 (x-y)^2 M_1^2 + 2 \ln (x-y)^2 M_2^{2}}{(x-y)^2} \label{High_loop_nested}
\end{eqnarray}
where, in going to the second line, we have integrated by parts the d'alembertian and  made use of $\Box \frac{1}{(x-u)^2} = \delta(x-u)$. We observe the appearance of an independent scale associate with each renormalization step. 

A systematic implementation of differential renormalization to all orders in perturbation theory was presented in \cite{Latorre:1993xh}. The basics of the method are the separation of the divergences in two groups: one corresponds to divergences derived from two points collapsing, and the other to three or more points simultaneously closing up. For the first one the singularity is replaced with the renormalized form (once the derivatives are pulled in front), whereas the other one can be shown to be recursively written as two-point function problems of the first type. This procedure follows the BPHZ renormalization program and guarantees that differential renormalization maintains unitarity and it can be applied consistently (fulfilling locality and Lorentz invariance) to all orders \cite{Latorre:1993xh}.
 
\subsection{Massive theories}

Differential renormalization of massive theories has been studied in \cite{Freedman:1991tk,Haagensen:1992am}. The appearance of a bare mass does not interfere with the method, since DiffR is related to short-distance singularities and masses only change the long-distance behaviour of the correlators. Although in this work we will only deal with massless theories, we will give briefly as an example how this procedure works with massive $\l \phi^4$. The propagator of a particle of mass $m$ is
\br
\D_{m} (x) &=& \frac{1}{4 \pi^2} \sqrt{ \frac{m^2}{x^2}} K_1 ( \sqrt{m^2 x^2}) 
\er
where $K_1$ is a modified Bessel function. Let us now consider again the 4-point function contribution; in this case is clear that the expression we have to renormalize is
\br
\left[ \sqrt{ \frac{m^2}{x^2}} K_1 ( \sqrt{m^2 x^2}) \right]^2 \;.
\er

We have to solve the massive generalization of the differential equation (\ref{eq_dif}), which has a solution of the form of 
\br
\left[ \sqrt{ \frac{m^2}{x^2}} K_1 ( \sqrt{m^2 x^2}) \right]^2_R &=& \frac{1}{2} ( \Box - 4 m^2) \sqrt{\frac{m^2}{x^2}} K_0 ( \sqrt{m^2 x^2}) K_1 ( \sqrt{m^2 x^2})  \nonumber \\
& & + \pi^2 \ln \frac{ \bar{M}^2}{m^2} \d (x) \;, \label{ren_mass_D2}
\er
where $ \bar{M} = 2 M / \gamma$ and $\gamma$ is the Euler constant. The general solution has a contact term which depends on a new mass parameter $M$. This guarantees that in the limit where $m \rightarrow 0$ the renormalized expressions for (\ref{ren_D2}) and (\ref{ren_mass_D2}) coincide.

\subsection{IR divergences}
\label{IR_divergences}
DiffR can be also applied to expression with IR divergences \cite{Mas:2002xh}, {\em i.e.}
expressions that exhibit a divergence for $p^\mu \to 0$.  
The idea is to  apply a dual version of Differential Renormalization to such quantities
\be
\left[ \frac{1}{p^4} \right]_{\tilde{R}} = - \frac{1}{4}{\dal}_p
\frac{\ln p^2/\bar{M}_{IR}^2}{p^2} + a_{IR} \delta(p)
\label{basicIRidentity} \; .
\end{equation}
We have defined for convenience $\bar{M}_{IR}=2M_{IR}/\gamma_E$,
where $\gamma_E$ is Euler's constant, and distinguished the IR scale
from the UV one. As DiffR is an implementation of Bogoliubov's $R$ operation (an operation that yields directly renormalized correlation functions satisfying renormalization group equations), in momentum space this is an explicit realization of the so-called $\tilde{R}$ operation that subtracts IR divergences. Again, diagrams with IR subdivergences are treated according to a recursion formula~\cite{Chetyrkin:nn,Popov:1984xm} analogous to the UV one.

\begin{figure}[ht]
\centerline{\epsfbox{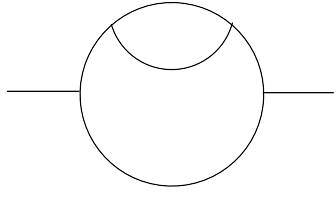}}
\caption{Two-loop diagram with UV and IR divergences.}
\label{Fig_IR_div}
\end{figure}

Now, when going to higher loops a new effect will show up, namely the coexistence of UV and IR scales. Let us start by examining a prototypical diagram where such divergences arise in a $\lambda \phi^3$ type theory (figure \ref{Fig_IR_div}). The associated amplitude that we have in this case  is of the form $G (x-y) = \Delta(x-y) I^0(x-y)$, where $I^0$ is
\br
I^0(x-y) &=& \int d^4 u d^4 v \; \D_{xu} \D_{yv} \D^2_{uv} \;.
\er
We begin by renormalizing the inner UV divergence as
\br
I^0_R (x-y) &=& - \frac{1}{4 (4 \pi^2)^2} \int d^4 u d^4 v \; \Box \frac{\ln (u-v)^2 M^2}{(u-v)^2} \D_{xu} \D_{vy} \nonumber \\
&=& \frac{1}{4 (4 \pi^2)^2} \int d^4 u \; \frac{ \ln (u-y)^2 M^2}{(u-y)^2} \D_{xu} \;.
\er
In order to renormalize the IR divergence we have to pass to momentum space
\br
I^0_R (x-y) &=& - \frac{1}{4(4 \pi^2)^5} \int d^4 u d^4 p d^4 q \; \frac{ \ln p^2/M^2}{p^2} \frac{1}{q^2} e^{-i p(u-y)} e^{-i q (x-u)} \nonumber \\
&=& - \frac{1}{4 (4 \pi^2)^3} \int d^4 p \; \frac{\ln p^2/M^2}{p^4} e^{-i p(x-y)} \;.
\er

We explicitly observe that the IR singularity at $p \to 0$ involves an UV scale $M$.

Since UV and IR overall divergences are local in coordinate and momentum space, respectively, the $R$ and $\tilde{R}$ operations commute, and one can define an operation $R^*=\tilde{R}R$ to renormalize both UV and IR divergences~\cite{Chetyrkin:nn,Popov:1984xm}. The fact that the UV and IR renormalizations decouple means that the UV and IR renormalization scales should be independent. This is a non-trivial point that in DiffR can be achieved by a careful adjustment of the local terms involving both scales\footnote{IR DiffR was investigated in~\cite{Avdeev:jp} where it was concluded that the combination of UV and IR DiffR was inconsistent, as the results depended on the order in which integrations were performed. According to \cite{Smirnov:1994km}, however, this corresponds to the natural arbitrariness of the IR renormalization, and this author has actually proposed in~\cite{Smirnov:1996yi} a consistent version of DiffR that deals with both UV and IR divergences. Our approach here will be closer to the original version of DiffR}.  

As we have to guarantee that the IR renormalization commute with a rescaling of $M$, we have to fulfill the relation
\be
M \frac{\d}{\d M} \left[\frac{\ln
p^2/\bar{M}^2}{p^4}\right]_{\tilde{R}} =
\left[M \frac{\d}{\d M} \frac{\ln
p^2/\bar{M}^2}{p^4}\right]_{\tilde{R}} \;. \label{IR_relation}
\end{equation}
If we consider the usual expression for the renormalization of  $(\ln p^2 / \bar{M}^2)/p^4$ 
\br
\left[ \frac{\ln p^2 / \bar{M}^2}{p^4} \right]_{\tilde{R}} = -\frac{1}{8} \Box_p \frac{ \ln^2 p^2 / \bar{M}^2 + 2 \ln p^2 / \bar{M_{IR}}^2 }{p^2} \;,
\er
we find that the left hand side of (\ref{IR_relation}) is
\br
M \frac{\d}{\d M} \left[\frac{\ln p^2/\bar{M}^2}{p^4}\right]_{\tilde{R}} &=& \frac{1}{2} \Box_p \frac{ \ln p^2 / \bar{M}_{IR}^2 + \ln {\bar{M}}^2 / \bar{M}_{IR}^2}{p^2} \;,
\er
whereas the right and side has the form of 
\br
\left[M \frac{\d}{\d M} \frac{\ln p^2/\bar{M}^2}{p^4}\right]_{\tilde{R}} &=& - 2 \left[ \frac{1}{p^4} \right]_R \nonumber \\
&=& \frac{1}{2} \Box_p \frac{ \ln p^2 / \bar{M}_{IR}^2}{p^2} \;,
\er
So, the second expression differs from the first one in a local term in momentum space, that will be a non local one in position space. Hence, in order to fulfill (\ref{IR_relation}), we propose the following minimal solution that does the job \cite{Mas:2002xh}
\be
\left[\frac{\ln p^2/\bar{M}^2}{p^4}\right]_{\tilde{R}} =
-\frac{1}{8} {\dal}_p \frac{-\ln^2 p^2/\bar{M}_{IR}^2 + 2 \ln
p^2/\bar{M}_{IR}^2 \, (1 + \ln p^2/\bar{M}^2)}{p^2}  + (a_{IR} \ln \frac{M^2_{IR}}{M^2} +  b_{IR})\delta(p)
\label{identityUVIR}
\end{equation}
This expression differs from the usual one by scale-dependent local terms proportional to $\ln^2 M^2/M_{IR}^2$ (apart from the explicit local terms with coefficients $a_{IR}$ and $b_{IR}$). It should be used whenever the ``new'' scale is to be treated as independent from the ``old'' one, for consistency of the loop expansion. It has to be noted that when we consider a purely UV expression as in (\ref{High_loop_nested}), we do not have to take care of all of this, because the extra term is cancelled in the RG equation when we derive wrt. the UV scales. With this, the result for $I^0$ (not taking care of the local terms $a_{IR}$ and $b_{IR}$) is
\br
I^0_R (x-y) &=& \frac{1}{32 (4 \pi^2)^3} \int d^4 p \; \Box^p \frac{ - \ln^2 p^2/M^2_{IR} + 2 \ln p^2 /M^2_{IR} \left( 1 + \ln p^2/M^2 \right) }{p^2} e^{-i p (x-y)}   \nonumber \\
&=& - \frac{(x-y)^2}{32 ( 4 \pi^2)^3} \int d^4 p \; \frac{ - \ln^2 p^2/M^2_{IR} + 2 \ln p^2/M^2_{IR} \left( 1 + \ln p^2 /M^2 \right)}{p^2} e^{-i p (x-y)} \nonumber \;, \\
\er
That in position space is
\br
I^0_R (x) &=& \frac{1}{32 (4 \pi^2)^2} \left[ \ln^2 x^2 M^2_{IR} + 2 \ln x^2 M^2_{IR} ( 1 - \ln x^2 M^2) \right] \;. \label{I0_integral_diffR}
\er
Observe that the UV scale $M$ only appears in~(\ref{identityUVIR}), and hence in the above expression for $I^0$, in single logarithms. This is fine, for double logarithms of $M$ are expected to appear only when the bare expression contains both a UV subdivergence and a UV overall divergence. Finally, once we have obtained $I^0$, we can straightforwardly evaluate $G$ to be
\br
G (x) = \frac{1}{32 ( 4 \pi^2)^3} \frac{ \ln^2 x^2 M^2_{IR} + 2 \ln x^2 M^2_{IR} ( 1 - \ln x^2 M^2)}{x^2} \;.
\er

\subsection{Symmetries with DiffR}
\label{DiffR_and_symmetries}
One of the important properties that is required to every sensible renormalization procedure is not to break gauge symmetry when it is applied to a gauge theory. For DiffR it is found that gauge symmetry is preserved as long as Ward identities can be always satisfied with the renormalized amplitudes (except anomalies). However, at the same time we find that we have to make always explicit use of these identities to fix all the ambiguities that have appeared in the calculations; in particular with Ward identities we relate the different scales that we have to use when renormalizing different amplitudes related by a symmetry(i.e. we fix a renormalization scheme).

As an example of this, consider the case of the one-loop renormalization of the photon self-energy in QED \cite{Haagensen:1992vz}. The bare expression is
\br
\Pi_{\m \n} (x) |_{bare} = - 4 e^2 \left[ 2 ( \pa_{\m} \D(x) ) ( \pa_{\n} \D(x) ) - \d_{\m \n} (\pa_{\b} \D(x) ) (\pa_{\b} \D(x) ) \right] \;, \label{photon_1l}
\er 
and renormalizing, this becomes
\br
\Pi_{\m \n} (x) |_R &=& - \frac{e^2}{12 \pi^4} \left[ \pa_{\m} \pa_{\n} \frac{1}{x^4} - 8 \d_{\m \n} \frac{1}{x^6} \right]_R \nonumber \\
&=& - \frac{e^2}{12 \pi^4} \left[ - \frac{1}{4} \pa_{\m} \pa_{\n} \Box \frac{ \ln x^2 M_1^2}{x^2} + \frac{1}{4} \d_{\m \n} \Box \Box \frac{ \ln x^2 M^2_2}{x^2} - 8 \d_{\m \n} \m^2 \d (x) \right] \;.  \nonumber \\
\label{dljlasd}
\er

In this expression we have renormalized with an independent scale the logarithmic divergence ($M_1$) and the quadratic one ($M_2$). At the same time, related to the latter, we have added a possible local term with a parameter of mass dimensions ($\m$). The Ward identity satisfied by this expression imposes that this has to be transverse, so that we have to fix $M_1 = M_2$ and $\m^2 = 0$. 

When going to higher loop computations, the Ward identities play a non-trivial r\^ole, as they influence part of the divergences that are obtained in the next step. The reason is that these identities relate all the relevant mass scales found. So, they allow us to write the one-loop renormalized subdiagrams that made up the two-loop expressions in terms of the same scale, say $M$, and fixed local terms, which then are promoted to logarithms of the scale. Going back to the example at two loops that was solved in (\ref{High_loop_nested}), suppose that after imposing the Ward identities we have the inner divergence written as $- \frac{1}{4} \Box \ln x^2 M^2 /x^2  + a \d(x)$, with $a$ a fixed coefficient. Thus, in the two-loop expression we find  
\br
\left[ \D_{xy} \int d^4 u \D_{xu} [\D^2_{yu}]_{R}\right]_{R}
&=&
\left[\frac{1}{4 (4 \pi^2)^4} \frac{\ln (y-u)^2 M^{2} + a}{(x-y)^4}  \right]_{R} \nonumber\\
&=& 
 - \frac{1}{32(4 \pi^2)^3} \Box \frac{ \ln^2 x^2 M^{2} +2 \ln x^2 M_2^{\prime 2} + 2a \ln x^2 M_2^{\prime \prime 2}}{x^2} \;, \nonumber \\
\er
where $M_2^{\prime}$ and $M_2^{\prime \prime}$ are two-loop scales. Hence, as we have anticipated, the one-loop Ward identities have fixed the coefficients of the logarithms of the scales in the two-loop final expression. Concerning the new two-loop scales $M_2^{\prime}$ and $M_2^{\prime \prime}$, it is clear that both can be set also equal to $M$ modulo a {\em  local ambiguity} that will depend on quotients $M_2'/M$ and $M_2''/M$ (like in \eqref{ambigu}). Again, use of the Ward identities would set these quotients to certain computable values. In other words, after use has been made of the symmetry, in the two-loop expression the only scale that remains can be chosen to be $M$ and sits only inside the terms with logarithms {\em whose coefficients were determined from the one-loop Ward identities}. This observation is at the heart of the present work an permeates implicitly all the calculations contained in it. So we repeat it here for full clarity: if one is interested in computing a physical amplitude at two loops, a concrete value of the local terms is essential and use Ward identities at two loops is unavoidable. If however, as is the case of the present work, one is looking for the RG equations, then all the relevant information on the scale $M$ resides in the terms with logarithms, whose coefficients only need one-loop Ward identities to be fixed.

\section{Constrained Differential Renormalization}
\label{CDR_rules}
 
Constrained Differential Renormalization (CDR) was developed in \cite{delAguila:1997kw,delAguila:1997su,delAguila:1998nd,Perez-Victoria:PhD} to avoid the necessity of imposing Ward identities in each calculation to fix the renormalization scheme, as we have seen in the previous section. The idea is to give a procedure that allows us to fix the scheme {\em a priory}. Central to the fulfilment of the Ward identities (and the action principle, from which they can be derived) is that the application of the kinetic differential operator to some propagator line inside a Feynman graph is equivalent to the contraction of the line to a point \cite{delAguila:1997kw}. This statement is guaranteed to hold if we apply the following set of rules

\begin{enumerate}
\item {\em Differential reduction}
\begin{itemize}
\item Functions with singular behaviour worse than logarithmic are reduced to derivatives of (at most) logarithmically divergent
functions without introducing extra dimensionful constants.
\item Logarithmically divergent expressions are written as derivatives of regular functions, 
introducing one single constant $M$, which has dimensions of mass and plays the r\^ole of the renormalization group scale.
\end{itemize}
\item {\em Formal integration by parts}. We do not take care of the divergent surface terms that appear when we integrate by parts. Related to this, differentiation and renormalization must be two commutative operations: let $F$ an arbitrary function, then 
$[ \pa F ]_R = \pa [F]_R$.
\item {\em Renormalization rule of the delta function}:
\begin{equation} [ F (x, x_1, \ldots , x_n ) \d (x-y) ]_R = [ F ( x, x_1, \ldots , x_n)]_R \d (x-y)
\end{equation}
\item {\em Validity of the propagator equation}
\begin{equation} [F(x,x_1,\ldots,x_n) ( \Box - m^2) \D_{m}(x)]_R = - [F(x,x_1,\ldots,x_n) \d(x)]_R
\end{equation} 
where $\D_{m}$ is the propagator of a particle of mass $m$ and $F$ an arbitrary function.
\end{enumerate}

The upshot is a basic set of renormalized expressions (basic functions) with different numbers of propagators and various differential operators acting only on one of them, involving a single scale $M$. Therefore the CDR program amounts to the following two step operation:
\begin{itemize}
\item
Express the Feynman diagram in terms of these basic functions performing all the index contractions (this is an important point, because CDR does not commute with index contraction) and, by means of the Leibniz rule, moving all the derivatives to make them act on one of the propagators.
\item 
Replace the basic functions with their renormalized version.
\end{itemize}

Let us now obtain as an example some of these functions. Consider the one-point basic function $\D (x) \d (x)$ (this corresponds to the one-loop correction to the two-point function in $\l \Phi^4$ theory). Power counting and the locality of the expression implies that the most general renormalized value that we have for this is
\br
[ \D(x) \d(x) ]_R = ( c \Box + \mu^2 ) \d(x) \;,
\er
where $\mu$ is a mass-dimension constant and $c$ an adimensional constant. However, rule $1$ implies that $\mu = 0$. Now, considering $[ \D(x) \d(x) ]_R \d(y) $ and using rule $3$ we find
\br
[ \D(x) \d(x) ]_R \d(y) &=& [ \D(x) \d(x) \d(x+y) ]_R = [ \D(x) \d(x) ]_R \d(x+y) \nonumber \\
\er
and integrating over $x$ we arrive to
\br
\d (y) \int d^4 x \; [ \D(x) \d(x) ]_R = [ \D(y) \d(y) ]_R \;.
\er
Finally, with this result rule $2$ implies that $c=0$. Proceeding in a similar way we find that all the massless one-point functions in CDR vanish.

As two-point function examples, we will consider $\D \pa_{\m} \D$ and $ \D \Box \D$. In the first case we have to apply the Leibniz rule to find
\br
[ \D(x) \pa_{\m} \D(x) ]_R &=& \pa_{\m} [ \D(x) \D(x)]_R - [ (\pa_{\m} \D (x) ) \D(x) ]_R \nonumber \\
&=& \frac{1}{2} \pa_{\m} [ \D^2 (x)]_R = - \frac{1}{8 (4 \pi^2)^2} \pa_{\m} \Box \frac{ \ln x^2 M^2}{x^2} \;,
\er
where we have used the result of (\ref{ren_D2}). If we study now $\D \Box \D$, we have only to use rule 4 to arrive at
\br
[ \D(x) \Box \D(x) ]_R &=& - [ \D(x) \d(x) ]_R = 0 \;.
\er

Here we present as a summary the most relevant CDR identities that are used in this work. We only list the massless examples, although a complete list including massive propagators can be found in \cite{delAguila:1998nd,Perez-Victoria:PhD} 
\br
\left[ \D^2 \right]_R (x) &=& - \frac{1}{4 (4 \pi^2)^2} \Box \frac{\ln x^2 M^2}{x^2} \nonumber \\
\left[ \D \pa_{\m} \D \right]_R (x) &=& - \frac{1}{8 (4 \pi^2)^2}\pa_{\m} \Box \frac{ \ln x^2 M^2 }{x^2}  \nonumber \\
\left[ \D \pa_{\m} \pa_{\n} \D \right]_R (x) &=& - \frac{1}{12 (4 \pi^2)^2} (\pa_{\m} \pa_{\n} - \frac{1}{4} \d_{\m \n} \Box) \Box \frac{ \ln x^2 M^2}{x^2} + \nonumber \\
& & + \frac{1}{288 
\pi^2} (\pa_{\m} \pa_{\n} - \d_{\m \n} \Box ) \d (x) \nonumber \\
\left[ \D \Box \D \right]_R (x) &=& 0 \;. \label{basic_CDR_fun}
\er

CDR can be applied to more than two propagators. In particular, when dealing with three propagators, defining $T[{\cal{O}}] = \D \D {\cal{O}} \D $, we can find the following relation to hold when making a decomposition into trace and traceless parts \cite{{delAguila:1997kw},{delAguila:1998nd}}
\br
T^R[\pa_{\m} \pa_{\n}] &=& T^R[\pa_{\m} \pa_{\n} - \frac{1}{4} \d_{\m \n} \Box] + \frac{1}{4} \d_{\m \n} T^R [\Box] - 
\frac{1}{128 \pi^2} \d_{\m \n} \d (x) \d (y) \;. \label{CDR_T}
\er

When using other gauges different from Feynman gauge, some bare expressions are written in terms of a quantity we define as $\bar{\D} (x) = \frac{1}{4 (4 \pi^2)} \ln x^2 s^2$, where $s$ is an irrelevant constant with mass dimension. For this structure, CDR prescribes \cite{delAguila:1997kw}
\br
\left[ \D \Box \bar{\D} \right]_R (x) &=& - \frac{1}{4 (4 \pi^2)^2} \Box \frac{ \ln x^2 M^2}{x^2} \nonumber \\
\left[ \D \pa_{\m} \pa_{\n} \bar{\D} \right]_R (x) &=& \frac{1}{4} \left( - \d_{\m \n} \frac{1}{4(4 \pi^2)^2} \Box \frac{ \ln x^2 M^2}{x^2}  - \frac{1}{32 \pi^2} \pa_{\m} \pa_{\n} \frac{1}{x^2} \right) \;. \label{CDR_rules_other_gauge}
\er 

CDR has been checked in abelian and non-abelian gauge theories \cite{delAguila:1997kw,Perez-Victoria:1998fj} and in supersymmetric calculations \cite{delAguila:1997ma,delAguila:1997yd}. As an example of its use, we will re-obtain the one-loop renormalization of the photon self-energy of QED that we have renormalized with DiffR in the previous section. From the bare expression (\ref{photon_1l}) we apply rule number 2 to write it in terms of the CDR basic functions as
\br
\Pi_{\m \n }(x) |_R &=& - 4 e^2 \left[ 2 \pa_{\m} ( \D \pa_{\n} \D) - 2 \D \pa_{\m} \pa_{\n} \D - \d_{\m \n} \pa_{\b} ( \D \pa_{\b} \D ) + \d_{\m \n} ( \D \Box \D ) \right]_R \;. \nonumber \\
\er

Now, we have to replace each basic expression with its renormalized value, and straightforwardly we arrive to
\br
\Pi_{\m \n }(x) |_R &=& ( \pa_{\m} \pa_{\n} - \d_{\m \n} \Box ) \left[ \frac{e^2}{3(4 \pi^2)^2} \Box \frac{ \ln x^2 M^2}{x^2} + \frac{e^2}{36 \pi^2} \d(x) \right] \;.
\er

As we have remarked, CDR has fixed all the ambiguities {\em{a priori}}, obtaining a direct final result that is transverse, as it has to be to fulfill the Ward identity. 

Finally, it is worth to mention that this method is equivalent to a momentum-space regularization method defined also in four dimensions: Constrained Implicit Regularization (CIR). Implicit Regularization \cite{Battistel:1998sz,BaetaScarpelli:1998fd} is a regularization method based in the assumption of a regulating function as part of the integrand of divergent amplitudes, and the extension of the properties of regular integrals to regularized ones. As in differential renormalization, this procedure generates arbitrary parameters that with CIR are fixed {\em{a priori}} \cite{Pontes:2007fg}. 

\section{Two-loop uses of one-loop CDR results}
\label{2loop_CDR}

As we have seen, one of the drawbacks of DiffR is the plethora of scales that pop up at each step of the calculation. In symmetric theories, at fixed order in the perturbative expansion, these scales should reduce to a single one upon use of the Ward identities. In the previous section we have explained how CDR paves the way to this reduction of scales at the one-loop level. So far, CDR has not been fully developed at loop-order higher than one, and therefore it is not useful for computing, say, scattering amplitudes. However, as is mentioned at the end of section \ref{DiffR_and_symmetries}, as long as we are interested in the RG equations, all that we need are the terms with logarithms, and to obtain them the knowledge of the local terms at one loop-level is enough. Hence, we will discuss both the way the logarithms are generated from one loop to the next, and the implementation of the CDR rules in such diagrams \cite{Seijas:2006vt}.
 
\subsection{Nested divergences}
\label{Nested_div}
This case is particularly simple because CDR can be applied in a systematic way. Starting from the ``inner'' divergence, its regularization according to CDR gives an expression with logarithms of a single scale ($ \ln x^2 M^2 $) and
fixed local terms. The one-loop Ward identities are fulfilled. In the next step, when tackling the outer part of the diagram, a simple logarithm like the one shown above is promoted to an expression of the form $ \ln^2 x^2 M^2 + C \ln x^2 M^{\prime 2}$, with $C$ a calculable coefficient and $M^{\prime}$ a two-loop scale; at the same time, the local terms that multiply outer divergences will produce additional logarithms of new scales. CDR does not yet prescribe what the different two-loop scales should be; hence, we may take all of them the same, and equal to $M$, at the price of leaving undetermined local terms which are irrelevant when obtaining the RG equation.

This simple scheme has some subtleties when considering diagrams with indices because, even at one-loop, index contraction does not commute with CDR. Therefore, the correct order is to first insert into the outer diagram the non-renormalized expression for the ``inner'' one-loop diagram, perform all the index contractions, and then renormalize. This crucial observations looks as the first one in a list of rules that eventually would setup the implementation of CDR at higher loops.

Let us consider now the two-loop example discussed in sections \ref{Higher_Loops} and \ref{DiffR_and_symmetries}. If we had imposed CDR in the first step, $M_1=M$ is the only scale generated  upon renormalizing the most internal divergence, and the local one-loop ambiguity will be fixed to zero, as can be seen from (\ref{basic_CDR_fun}). We express this by stating that the renormalization of $ I^1 (x-y) = \int d^4 u \D_{xu} \D^2_{yu} $ according to CDR rules is given by
\begin{equation}
I^{1}_R (x) = \frac{1}{4 (4 \pi^2)^2} \frac{\ln x^2 M^2}{x^2} \;.
\end{equation}
Once we have this, to renormalize the complete two-loop expression $\D I^1$, we have to apply usual differential renormalization and set, modulo local terms, the two-loop mass-scale $M_2^{\prime} = M$. So, we arrive to an expression of the form of
\br
\left[ \D I^1 \right]_R (x) = - \frac{1}{32(4 \pi^2)^3} \Box \frac{ \ln^2 x^2 M^{2} + 2 \ln x^2 M^2}{x^2} \label{2loop_CDR_ex2} + \ldots \;, \label{2loop_DR_id1}
\er 
where $\ldots$ stand for the two-loop local terms that we are not taking into account. Notice that in the rest of the work (unless explicitly stated otherwise) $\ldots$ in a two-loop renormalized expression like the one shown above will have the same meaning: local terms not considered. With this procedure we have renormalized all the different structures made of $I^1$ that we have encountered in our calculations. Apart from the previous one, we have found the following relevant expressions
\br
\left[ \D \pa_{\m} I^{1} \right]_R (x) &=& - \frac{1}{64 (4 \pi^2)^3} \pa_{\m} \Box \frac{ \ln^2 x^2 M^2 + \ln x^2 M^2}{x^2} +\ldots    \label{2loop_DR_id2}\\
\left[ \D \pa_{\m} \pa_{\n} I^1 \right]_R (x) &=& - \frac{1}{96 (4 \pi^2)^3} \left[\pa_{\m} \pa_{\n} \Box \frac{ \ln^2 x^2 M^2 + \frac{2}{3} \ln x^2 M^2}{x^2} \right. \nonumber \\
& & - \left. \frac{1}{4}\d_{\m \n} \Box \Box \frac{\ln^2 x^2 M^2 + \frac{11}{3} \ln x^2 M^2}{x^2} \right] + \ldots  \label{2loop_DR_id3}  \\
\left[ \D \Box I^1 \right]_R (x) &=& \frac{1}{32 ( 4 \pi^2)^2} \Box \Box \frac{ \ln x^2 M^2}{x^2} +\ldots
 \label{2loop_DR_id4}
\er
To obtain each of these results, we only have to consider the CDR renormalization of $I^1$, and apply afterwards usual DiffR, setting all the two-loop mass scales equal to $M$. Let us illustrate the simplicity of the procedure with $\D\pa_{\m} I^1$. Given the renormalized form of $I^1$ we find
\br
\left[ \D \pa_{\m} I^1 \right]_R (x) &\stackrel{CDR}{=}& \frac{1}{4(4 \pi^2)^3} \left[ \frac{1}{x^2} \pa_{\m} \frac{\ln x^2 M^2}{x^2} \right]_R \nonumber \\
&\stackrel{DiffR}{=}& - \frac{1}{16(4 \pi^2)^3} \left[ \pa_{\m} \frac{1 - 2 \ln x^2 M^2}{x^4} \right]_R \nonumber \\
&=& - \frac{1}{64 (4 \pi^2)^3} \pa_{\m} \Box \frac{ \ln^2 x^2 M^2 + \ln x^2 M^2}{x^2} + \ldots
\er

\subsection{Overlapping divergences}
\label{overlap_integrals}
Diagrams with overlapping divergences are more complex as it is sometimes difficult to recognize the one-loop subdivergences that need to be treated with CDR to start with. Our approach will be to obtain through different methods (that we will explain later in  detail), a list of renormalized two-loop integrals with overlapping divergences, where in each calculation one-loop CDR rules have been maintained in every step. Although this list is restricted to integrals with at most four derivatives acting on the propagators and two free indices, it is found to be very useful, as serves as a basis that we can use to express the renormalized overlapping contributions to two-point functions in theories with derivative couplings at two loops. As we detail in appendix \ref{ap_BFM}, these two-point functions are what we need to obtain the beta function if we use the background field method. This list will be applied in our work to renormalize and obtain the two-loop beta function of  (Super)QED and (Super)Yang-Mills. 
We use the conventions of $z = x-y$ and  $\pa_{\m} \equiv \pa_{\m}^x$. We also define $H(x-y) \equiv H(z)$ as
\br
H[{\cal{O}}_1,{\cal{O}}_2 \; ; \; {\cal{O}}_3,{\cal{O}}_4] = \int d^4 u d^4 v \; ( {\cal{O}}_1^{x} \D_{xu})( {\cal{O}}_2^{x} \D_{xv})( {\cal{O}}_3^{y} \D_{yu} ) ({\cal{O}}_4^{y} \D_{yv}) \D_{uv} \;, \label{H_definition}
\er
with ${\cal{O}}_i$ a differential operator. 
\br
H^R[1,1 \; ; \; 1,1] &=& \frac{6 \pi^4 \xi(3) }{ ( 4 \pi^2)^4} \D  \equiv a \D \label{int1} \\
H^R[\pa_{\m},1 \; ; \; 1,1] &=&  \frac{ 3 \xi(3)}{16 (4 \pi^2)^2} ( \pa_{\m} \D) \equiv \frac{a}{2} \pa_{\m} \D \label{int2} \\
H^R[1,\pa_{\l} \; ; \; 1,\pa_{\l}] &=&  - \frac{1}{16(4 \pi^2)^3} \Box \frac{\ln z^2 M^2}{z^2} + \ldots \label{int3} \\
\pa_{\l} H^R[1,\pa_{\m} \; ; \; 1,\pa_{\l}] &=&  - \frac{1}{32 (4 \pi^2)^3} \pa_{\m} \Box \frac{ \frac{1}{2} \ln z^2 M^2}{z^2} + \dots \label{int4}\\
\pa_{\l} H^R[1,1 \; ; \; \pa_{\l} \pa_{\n},1] &=& \frac {1}{32(4 \pi^2)^3} \pa_{\n} \Box \frac{ \frac{1}{4} \ln^2 z^2 M^2 + \frac{3}{4} \ln z^2 M^2 }{z^2} + \ldots \label{int5}\\ 
H^R[1,\pa_{\l} \; ; \; \pa_{\l} \pa_{\m},1] &=&  \frac{1}{32 (4 \pi^2)^3} \pa_{\m} \Box \frac{\frac{1}{8} \ln^2 z^2 M^2 - \frac{7}{8} \ln z^2 M^2 }{z^2} + \ldots \label{int6} \\
H^R[\pa_{\m} \pa_{\l},\pa_{\l} \; ; \; 1,1] &=&  \frac{1}{32(4 \pi^2)^3} \pa_{\m} \Box \frac{ - \frac{1}{2} \ln^2 z^2 M^2 - \ln z^2 M^2}{z^2}+ \ldots  \label{int7} \\
\pa_{\l} H^R[1,\pa_{\m} \; ; \; \pa_{\n} \pa_{\l},1] &=& \frac{1}{32(4 \pi^2)^3} \left[ \pa_{\m} \pa_{\n} \Box \frac{ \frac{1}{8} \ln^2 z^2 M^2 + \frac{1}{8} \ln z^2 M^2}{z^2}  \right. \nonumber \\
& & \left. + \d_{\m \n} \Box \Box \frac{-\frac{1}{4} \ln z^2 M^2}{z^2} \right] + \ldots  \label{int8}\\
H^R[1,\pa_{\m} \; ; \; 1,\pa_{\n}] &=& \frac{1}{32 (4 \pi^2)^3} \d_{\m \n} \Box \frac{- \frac{1}{2} \ln z^2 M^2}{z^2} + \ldots \label{int9} \\ 
\pa_{\l} H^R[1,\pa_{\l} \; ; \; \pa_{\m} \pa_{\n},1] &=& \frac{1}{32 (4 \pi^2)^3} \left[ \pa_{\m} \pa_{\n} \Box \frac{ - \frac{1}{2} \ln z^2 M^2}{z^2}  \right. \nonumber \\
& & \left. + \d_{\m \n} \Box \Box \frac{\frac{1}{8} \ln^2 z^2 M^2 + \frac{3}{8} \ln z^2 M^2}{z^2} \right] + \ldots  \label{int10} \\
\pa_{\l} H^R[1,\pa_{\l} \; ; \; 1, \pa_{\m} \pa_{\n}] &=& \frac{1}{32 (4 \pi^2)^3} \left[ \pa_{\m} \pa_{\n} \Box \frac{ \frac{1}{2} \ln z^2 M^2}{z^2}  \right. \nonumber \\
& & \left. + \d_{\m \n} \Box \Box \frac{\frac{1}{8} \ln^2 z^2 M^2 + \frac{3}{8} \ln z^2 M^2}{z^2} \right] + \ldots \label{int10a}
\er
\br
H^R[1,1 \; ; \; \pa_{\m} \pa_{\n},1] &=&  \frac{1}{32(4 \pi^2)^3} \d_{\m \n} \Box \frac{ \frac{1}{4} \ln^2 z^2 M^2 + \frac{3}{4} \ln z^2 M^2}{z^2} + \ldots \label{int11} \\
\pa_{\l} H^R[1,1 \; ; \; \pa_{\l} \pa_{\n},\pa_{\m}] &=& \frac{1}{32 (4 \pi^2)^3} \d_{\m \n} \Box \Box \frac{ \frac{1}{8} \ln^2 z^2 M^2 + \frac{3}{8} \ln z^2 M^2}{z^2} + \ldots \label{int12} \\
\pa_{\l} H^R[1,1 \; ; \; \pa_{\m} \pa_{\n},\pa_{\l}] &=& \frac{1}{32 (4 \pi^2)^3} \pa_{\m} \pa_{\n} \Box \frac{\frac{1}{8} \ln^2 z^2 M^2 + \frac{3}{8} \ln z^2 M^2}{z^2} + \ldots \label{int13} \\
H^R[1,\pa_{\m} \pa_{\l} \; ; \; \pa_{\n} \pa_{\l},1] &=& \frac{1}{32 (4 \pi^2)^3} \left[ \pa_{\m} \pa_{\n} \Box \frac{ \frac{1}{6} \ln^2 z^2 M^2 - \frac{5}{36} \ln z^2 M^2}{z^2}  \right. \nonumber \\
& & \left. + \d_{\m \n} \Box \Box \frac{ - \frac{1}{24} \ln^2 z^2 M^2 - \frac{29}{72} \ln z^2 M^2}{z^2} \right] + \ldots \label{int14}\\
H^R[1,\pa_{\m} \pa_{\l} \; ; \; 1,\pa_{\n} \pa_{\l}] &=&  \frac{1}{32 (4 \pi^2)^3} \left[ \pa_{\m} \pa_{\n} \Box \frac{\frac{1}{6} \ln^2 z^2 M^2 + \frac{49}{36} \ln z^2 M^2}{z^2}  \right. \nonumber \\
& & \left. + \d_{\m \n} \Box \Box \frac{- \frac{1}{24} \ln^2 z^2 M^2 - \frac{11}{72} \ln z^2 M^2}{z^2} \right] + \ldots \nonumber \\ \label{int15}
\er

These integrals are obtained basically applying two properties:

\begin{itemize}
\item Integral relations presented in appendix \ref{ap_calc}. These exact relations allow us to put some of the integrals in terms of others that have an explicit d'alembertian acting on one of the propagators. Once we have done that, using $\Box \D = - \d$ we can put these integrals in terms of the previously defined $I^1$. Then, we can straightforwardly apply the procedure for nested divergences that we have just presented in the previous section.\footnote{Also this is the reason why we have not listed here the cases where the differential operator is a d'alembertian. For example, it is obvious that $ H[ \Box,1 \; ; \; 1,1] = - \D I^1$. }

\item The decomposition into trace part, traceless part and fixed local term imposed by CDR to $T[\pa_{\m} \pa_{\n}]$ as (\ref{CDR_T}).

\end{itemize} 

As in the previous section, let us illustrate the procedure with an explicit example. Considering integral (\ref{int5}), this can be evaluated with both methods. First, we will make use of integral relation (\ref{rel_int2}) and put this integral as sum of different integrals that have the divergences nested 
\br
\pa_{\l}^x \int &d^4 u d^4 v & \D_{xu} \D_{xv} ( \pa_{\l}^y \pa_{\n}^y \D_{yu} ) \D_{yv} \D_{uv} = \nonumber \\
&=& - \frac{1}{2} \pa_{\n}^y \int d^4 u d^4 v \; \D_{xu} \D_{xv} ( \Box \D_{yu} ) \D_{yv} \D_{uv}  \nonumber \\
& & + \int d^4 u d^4 v \; \D_{xu} \D_{xv} ( \Box \D_{yu} ) ( \pa_{\n}^y \D_{yv}) \D_{uv}  \nonumber \\
& & - \frac{1}{2} \pa_{\n}^y \pa_{\l}^y \int d^4 u d^4 v \; \D_{xu} \D_{xv} ( \pa_{\l}^y \D_{yu} ) \D_{yv} \D_{uv} \;. \nonumber \\
\er

Now, we have to apply as usual $\Box \D = - \d $ and rewrite these integrals in terms of $I^1$. Note that the third integral can be easily shown to be finite, and its value is obtained in appendix \ref{ap_integrales} to be $a/4 \pa_{\n} ( \Box \D)$ with $a=\frac{6 \pi^4 \xi(3) }{ ( 4 \pi^2)^4}$) .
\br
\pa_{\l}^x \int &d^4 u d^4 v & \D_{xu} \D_{xv} ( \pa_{\l}^y \pa_{\n}^y \D_{yu} ) \D_{yv} \D_{uv} = \nonumber \\
&=& - \frac{1}{2} \pa_{\n} ( \D I^1 ) + \frac{1}{2} ( \D \pa_{\n} I^1 ) + \frac{a}{4} \pa_{\n} ( \Box \D) \;. \nonumber \\
\er

Applying the results found in section \ref{Nested_div} for the $I^1$ expression, the renormalized value of this is
\br
\pa_{\l}^x \int &d^4 u d^4 v & \D_{xu} \D_{xv} ( \pa_{\l}^y \pa_{\n}^y \D_{yu} ) \D_{yv} \D_{uv} = \nonumber \\
&=& \frac{1}{32 (4 \pi^2)^3} \pa_{\n} \Box \frac{ \frac{1}{4} \ln^2 z^2 M^2 + \frac{3}{4} \ln z^2 M^2}{z^2} + \ldots
\er 
where $\ldots$ stands for the finite terms that we are not taking into account and $z=x-y$. 

We can also obtain this integral making use of the CDR relation (\ref{CDR_T}) and perform a trace-traceless decomposition of $( \pa_{\l}^y \pa_{\n}^y \D_{yu} ) \D_{yv} \D_{uv}$ as
\br
\pa_{\l}^x \int &d^4 u d^4 v & \D_{xu} \D_{xv} ( \pa_{\l}^y \pa_{\n}^y \D_{yu} ) \D_{yv} \D_{uv} = \nonumber \\
&=& \frac{1}{4} \pa_{\n}^x \int d^4 u d^4 v \; \D_{xu} \D_{xv} ( \Box \D_{yu}) \D_{yv} \D_{uv}  \nonumber \\
& & + \pa_{\l}^x \int d^4 u d^4 v \; \D_{xu} \D_{xv} \left[ ( \pa_{\l}^y \pa_{\n}^y - \frac{1}{4} \d_{\l \n} \Box ) \D_{yu} \right] \D_{yv} \D_{uv}  \nonumber \\
& & - \frac{1}{128 \pi^2} \pa_{\n}^x \int d^4 u d^4 v \; \D_{xu} \D_{xv} \d (y-u) \d (y-v) \nonumber \\
&=& - \frac{1}{4} \pa_{\n} ( \D I^1 )_R - \frac{1}{128 \pi^2} \pa_{\n} \D^2_R + \pa_{\l} I_{\l \n \; R} \\
&=& \frac{1}{32 (4 \pi^2)^3} \pa_{\n} \Box \frac{ \frac{1}{4} \ln^2 z^2 M^2 + \frac{3}{4} \ln z^2 M^2}{z^2} + \pa_{\l} I_{\l \n \; R} \;,
\er
where $\pa_{\l} I_{\l \n \; R}$ is the traceless part, that is finite. As we can see, both results agree. Although in this example we can perform the calculation with both methods with the same effort, with other integrals the situation is different, and we shall have to study each case in order to choose the best one. The explicit evaluation of all the integrals is presented in section \ref{ap_integrales} of appendix \ref{ap_calc} .

\chapter{Abelian QFT applications}
\label{chap_abelian_examples}
In this chapter we apply the ideas and methods we have just presented to two of the most relevant examples of abelian gauge theories: QED and its supersymmetric extension, SuperQED. Although both theories have already been treated in \cite{Haagensen:1992vz,Song} using DiffR, we will show that our procedure simplifies the calculations, avoiding the use of Ward identities. 
\section{QED}

QED is one of the simplest examples of a gauge theory, as the gauge symmetry group is an abelian one, $U(1)$. Hence, is a good theory to start with, as we can clearly see all the key points of our renormalization procedure. 

\subsection{The model}

We use the same conventions as \cite{Haagensen:1992vz}. The $d=4$ massless QED lagrangian is 
\br
{\cal{L}} &=& \frac{1}{4} F^{\m \n} F_{\m \n} + \bar{\psi} \g^{\m} ( \pa_{\m} + i e A_{\m} ) \psi \;,
\er
where $\psi$ is the fermion field, $A_{\m}$ is the $U(1)$ gauge field and $F_{\m \n}$ is the field strength made up with $A_{\m}$ as $F_{\m \n} (x) = \pa_{\m} A_{\n}(x) - \pa_{\n} A_{\m} (x)$. The $\gamma$ matrices satisfy the Clifford algebra $\anticomm{\gamma_{\m}}{\gamma_{\n}} = 2 \d_{\m \n}$.

With $w$ an infinitesimal parameter, the QED action is invariant under the following $U(1)$ transformations 
\br
A_{\m} \rightarrow A_{\m} - \frac{1}{e} \pa_{\m} w \nonumber \\
\psi \rightarrow \psi + i \psi w \nonumber \\
\bar{\psi} \rightarrow \bar{\psi} - i \bar{\psi} w \;.
\er

Hence, when quantizing QED we have to take care of this invariance. We have an infinite number of different gauge field configurations (those obtained through gauge transformations from a given one) that correspond to the same physical state. In a path integral approach, we want to integrate only over the relevant gauge field configurations; hence, we have to pick up only one field from each gauge orbit. To accomplish this there is a well-known procedure \cite{Faddeev:1967fc,Peskin:1995ev} which implies that we have to add to the action a gauge fixing term that depends on a new parameter $\a$ and possible auxiliary fields (Faddeev-Popov ghosts fields, that in the concrete case of QED and SuperQED are not relevant). Different values for $\a$ correspond to different gauge choices. In particular, in our calculation we will use $\a = 1$ (Feynman gauge). The complete lagrangian is then
\br
{\cal{L}} &=& \frac{1}{4} F^{\m \n} F_{\m \n} + \frac{1}{2 \a} ( \pa_{\m} A_{\m} )^2 + \bar{\psi} \g^{\m} ( \pa_{\m} + i e A_{\m} ) \psi \;.
\er

With this action the gauge field and fermion propagators (in a generic gauge) are 
\br
\D_{\m \n} (x-y) &=& \frac{1}{16 \pi^2} \left( \d_{\m \n} \Box + (\a - 1) \pa_{\m} \pa_{\n} \right) \ln (x-y)^2 s^2 \nonumber \\
S(x-y) &=& - \g^{\l} \pa_{\l} \D (x-y) \;, \label{QED_propagators}
\er 
where $s$ stands for an irrelevant constant with mass dimension. Also, considering the expansion of the effective action, we write the terms corresponding to the vacuum polarization $\Pi_{\m \n}$ and fermion self-energy $\Sigma$ as

\br
\Gamma_{eff} &=& \frac{1}{2} \int d^4 x d^4 y \; A_{\m}(x) \left[ \left( \left(1- \frac{1}{\a} \right) \pa_{\m} \pa_{\n} - \d_{\m \n} \Box \right) \d^{(4)}(x-y) - \Pi_{\m \n}(x-y) \right] A_{\n}(y)  \nonumber \\
& & + \int d^4 x d^4 y \; \bar{\psi}(x) ( \g^{\m} \pa_{\m} \d^{(4)}(x-y) - \S (x-y))\psi(y) + \ldots
\er

\subsubsection{Background field method}

Fixing the gauge is necessary in order to quantize the theory, but has the drawback to make us lose explicit gauge invariance in the intermediate results. In order to avoid this, the background field method was developed \cite{DeWitt:1967ub}. As the method is detailed in appendix \ref{ap_BFM}, we only briefly outline it here. The key point is the splitting of the gauge field in two parts: the quantum and the background fields ($A_{\m}$ and $B_{\m}$ respectively)
\br
A_{\m} \rightarrow A_{\m} + B_{\m} \;.
\er

\begin{figure}[ht]
\centerline{\epsfbox{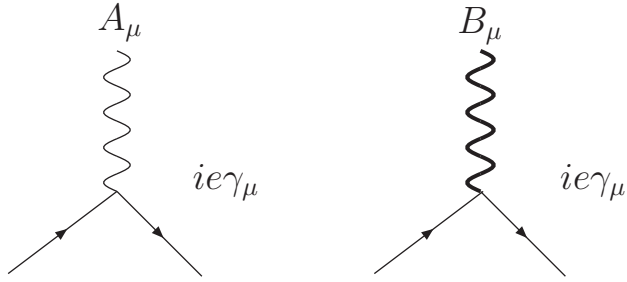}}
\caption{QED interaction vertices. Thick wavy lines represent background external fields and thin wavy lines correspond to quantum gauge field propagators. Solid lines represent fermion fields.}
\label{QED_Feynman_rules}
\end{figure}

We can use $A_{\m}$ as the integration variable of the partition function, which implies that the gauge has to be fixed only for this field. As a result, we retain explicit gauge invariance in $B_{\m}$. Along with this, as is shown in appendix \ref{ap_BFM}, this procedure has other relevant consequences: the coupling constant and the background field renormalizations are related, which implies that the beta function can be obtained only from the two-point function contribution. Hence, we have a background effective action of the form
\br
\Gamma_{eff}[B] &=& \frac{1}{2} \int d^4 x d^4 y \; B_{\m}(x) \left[ \left( \pa_{\m} \pa_{\n} - \d_{\m \n} \Box \right) \d^{(4)}(x-y) - \Pi^{BB}_{\m \n}(x-y) \right] B_{\n}(y)  + \ldots  \;, \nonumber \\
\er
and our aim is to calculate the two-loop expansion of the two-point 1PI function $\Pi^{BB}_{\m \n}$. Notice also that, as the infinitesimal gauge transformation of the unsplit theory does not depend on the quantum gauge field, we can choose in the split theory the usual gauge fixing condition $G = \pa^{\m} A_{\m}$. So, we have a split lagrangian of the form of
\br
{\cal{L}} &=& \frac{1}{4} F^{\m \n} F_{\m \n} + \frac{1}{4} B^{\m \n} B_{\m \n} +\frac{1}{2 \a} ( \pa_{\m} A_{\m} )^2 + \bar{\psi} \g^{\m} ( \pa_{\m} + i e A_{\m} ) \psi + ie \bar{\psi} \g^{\m} B_{\m} \psi \;,
\er
with $B_{\m \n} = \pa_{\m} B_{\n} - \pa_{\n} B_{\m}$. From this lagrangian, we have two relevant interaction vertices, which are shown in figure \ref{QED_Feynman_rules}.

\subsection{One-loop level}
\begin{figure}[ht]
\centerline{\epsfbox{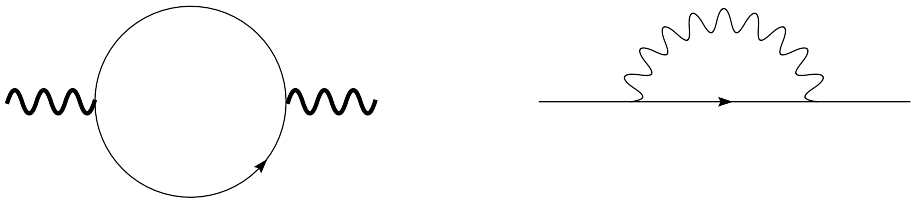}}
\caption{One-loop QED diagrams.}
\label{1loopQED}
\end{figure}

We first consider the one-loop renormalization of the background photon and fermion self-energies. The diagrams that correspond to these contributions are those of figure \ref{1loopQED}, where thick lines represent $B_{\m}$ fields, and thin ones correspond to $A_{\m}$. The correction to the background self-energy is what we need to obtain the one-loop coefficient of the beta function, although it will not be used in two-loop calculations. On the other hand, the fermion self-energy will be not relevant for the one-loop beta function study, but it will be used afterwards as a one-loop insertion in a two-loop diagram.

\subsubsection{Photon self-energy}
We consider now the one-loop contribution to the photon self-energy $\Pi_{\m \n}$. The expression for this diagram is
\br
\Pi_{\m \n \;(1)}^{BB} (x-y) &=& - (i e)^2 Tr \left[ \g_{\m} \g^{\l} \pa_{\l}^x \D \g_{\n} \g^{\sigma} \pa_{\sigma}^y \D \right] \nonumber \\
&=& - e^2 Tr\left[ \g_{\m} \g^{\l} \g_{\n} \g^{\sigma} \right] \left( \pa_{\l} ( \D \pa_{\sigma} \D ) - \D \pa_{\l} \pa_{\sigma} \D \right) \;. \label{QED1loop_bare}
\er

To proceed with the CDR program, we first have to perform all the index contractions, writing this diagram in terms of the basic CDR functions (\ref{basic_CDR_fun}). To do that, we have to apply in \ref{QED1loop_bare} the following Clifford algebra result  
\br
Tr\left[ \g_{\m} \g_{\l} \g_{\n} \g_{\s} \right] = 4 ( \d_{\m \l} \d_{\n \s} - \d_{\m \n} \d_{\l \s} + \d_{\m \s} \d_{\n \l} ) \;.
\er

which allows us to find the fully expanded expression as
\br
\Pi_{\m \n \;(1) }^{BB} (x) &=& - 4 e^2 \left[ 2 \pa_{\m} ( \D \pa_{\n} \D ) - 2 \D \pa_{\m} \pa_{\n} \D - \d_{\m \n} \pa_{\l} ( \D \pa_{\l} \D ) - \d_{\m \n} \D \Box \D \right] \;.
\er

Now, CDR renormalization of this expression entails only to replace these basic functions with their renormalized values. The full one-loop renormalized background self-energy is given by \cite{Haagensen:1992vz,delAguila:1997kw}
\br
\left. \Pi_{\m \n \;(1)}^{BB} (x) \right|_R &=& - ( \pa_{\m} \pa_{\n} - \d_{\m \n} \Box ) \left[ - \frac{e^2}{12 \pi^2 ( 4 \pi^2 )} \Box \frac{ \ln x^2 M^2}{x^2} - \frac{e^2}{36 \pi^2} \d (x) \right] \;. \label{QED_1loop_r}
\er

 As was guaranteed by the use of CDR, this result is transverse (as is required by the Ward identities), and has the ambiguity (local term) fixed.

\subsubsection{Fermion self-energy}
We will also consider the renormalization of the fermion self-energy $\S$. In a general gauge, the bare expression for this diagram is 
\br
\S_{(1)} (x) &=& e^2 \g_{\m} \D_{\m \n} (x) \g^{\l} \pa_{\l} \D (x) \g_{\n} \;.
\er

By making use of the expression for the photon propagator in a generic gauge (\ref{QED_propagators}), we can replace the basic functions by their CDR values and obtain the renormalized fermion self-energy as \cite{Haagensen:1992vz,delAguila:1997kw}
\br
\left. \S_{(1)} (x) \right|_R &=& e^2 \g^{\l} \left[ \frac{1}{4(4 \pi^2)^2} \pa_{\l} \Box \frac{\ln x^2 M^2}{x^2} + (\a -1) \left( \frac{1}{4(4 \pi^2)^2} \pa_{\l} \Box \frac{\ln x^2 M^2}{x^2} + \frac{1}{16 \pi^2} \pa_{\l} \d (x) \right)\right]. \nonumber \\
\er

\subsection{Two-loop level}
\label{QED_two_loop}
\begin{figure}[ht]
\centerline{\epsfbox{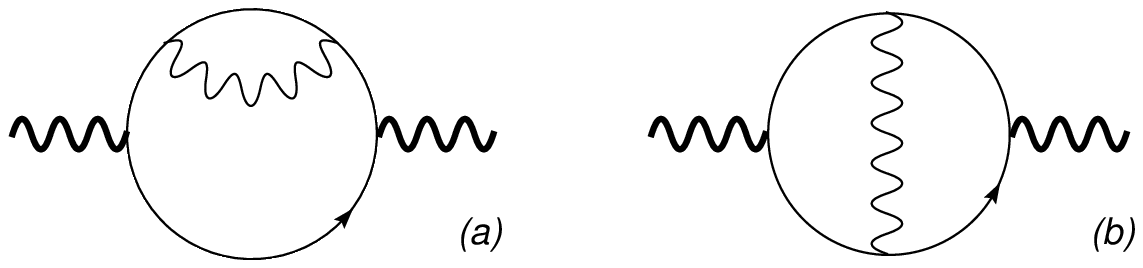}}
\caption{Two-loop QED diagrams.}
\end{figure}

Now we proceed with the two-loop case. There are two relevant graphs with external background fields. Diagram (a) has the divergences nested, whereas diagram (b) has overlapping divergences. 

\subsubsection{Diagram (a)}
The expression for this diagram is 
\br
\Pi^{BB}_{\m \n \;(2\;a)} (x-y) &=& - (ie)^2 \int d^4 u d^4 v \; Tr \left[ \g_{\m} \g^{\l} (- \pa_{\l}^x \D_{xu}) \S^{(1)} (u-v) \g^{\eps} (- \pa_{\eps}^v \D_{vy}) \g_{\n} \g^{\s}  \right. \nonumber \\
& & \left. \times ( - \pa_{\s}^y \D_{yx} ) \right].  
\er
where $\S^{(1)}$ is the one-loop fermion self-energy. In the following we will restrict ourselves to Feynman gauge, as the term that takes care of the running of the gauge parameter in the RG equation will be shown not to be relevant for the two-loop beta function \cite{Haagensen:1992vz}. We will discuss this later in detail, when we apply the RG equation. So, in this gauge, the bare fermion self-energy is
\br
\S_{(1)}(x) &=& - 2 e^2 \g^{\l} \left[ \D \pa_{\l} \D \right] (x) \;.
\er

As we stated in section \ref{CDR_rules}, CDR imposes a strict order to the operations of index contraction and renormalization: first all the indices should be contracted, and only after that we can renormalize. Inserting here the bare fermion self-energy we are keeping this order.  

Expanding the expression of $\Pi^{BB}_{\m \n \; (2\;a)}$ we find 
\br
\Pi^{BB}_{\m \n \;(2 \; a)} (x-y) &=&  - 2 e^4 Tr[\g_{\m} \g_{\l} \g_{\r} \g_{\eps} \g_{\n} \g_{\s}] ( \pa_{\s}^x \D_{xy} ) \pa_{\l}^x \pa_{\eps}^x \int d^4 u d^4 v \; \D_{xu} \D_{yv} ( \D_{uv} \pa^u_{\r} \D_{uv} ) \;. \nonumber \\ \label{QED_2loop_aext}
\er

In order to simplify the notation, we define an integral expression of the form \begin{equation}
I^0_{\m} (x-y) = \int d^4 u d^4 v \; \D_{xu} \D_{yv} ( \D_{uv} \pa_{\m} \D_{uv}) \;. 
\end{equation}

Then, with standard Clifford algebra, we can write (\ref{QED_2loop_aext}) as
\br
\Pi^{BB}_{\m \n \;(2\;a)} (x) &=& e^4 \left[ -32 (\pa_{\m} \D) \pa_{\l} \pa_{\n} I^0_{\l} + 16 \d_{\m \n} ( \pa_{\s} \D ) \pa_{\l} \pa_{\s} I^0_{\l} + 16 ( \pa_{\m} \D ) \Box I^0_{\n} - 8 \d_{\m \n} ( \pa_{\r} \D ) \Box I^0_{\r} \right] \;. \nonumber \\
\er

The renormalization of $I^0_{\m}$ is studied in section \ref{ap_UV_IR_I0m} of appendix \ref{ap_calc}. It is found there that this integral expression verifies $\pa^{\m} I^0_{\m \; R} = - \frac{1}{2} I^1_R$ and $\Box I^0_{\m \;R} = - \frac{1}{2} \pa_{\m} I^1_R$. Thus, we need only the renormalized value that we have previously obtained for $I^1$. We have finally 
\br
\left. \Pi^{BB}_{\m \n\;(2\;a)} (x) \right|_R &=& \frac{e^4}{24(4 \pi^2)^3} \left[ \pa_{\m} \pa_{\n} \frac{- \ln^2 x^2 M^2 - \frac{5}{3} \ln x^2 M^2 }{x^2} + \d_{\m \n} \Box \Box \frac{\ln^2 x^2 M^2 + \frac{8}{3} \ln x^2 M^2}{x^2} \right]  \nonumber  \\ & & + \ldots \nonumber \\
\er

\subsubsection{Diagram (b)}

This diagram, opposite to the previous one, has overlapping divergences. Following the procedure presented in section \ref{overlap_integrals}, we will express this in terms of the integrals listed in that section. We begin by considering the bare contribution 

\br
\Pi_{\m \n \;(2\;b)}^{BB} (x-y) &=& - ( i e )^4 \int d^4 u d^4 v \; Tr \left[ \g_{\m} ( \g^{\a} \pa^x_{\a} \D_{xu} ) \g^{\rho} ( \g^{\b} \pa_{\b}^u \D_{uy} ) \g_{\n}  \nonumber \right. \\
& & \times \left. ( \g^{\l} \pa_{\l}^y \D_{yv} ) \g_{\rho} ( \g^{\s} \pa_{\s} \D_{vx}) \D_{uv} \right] \;,\nonumber \\
\er
or, written in terms of the expressions we defined as $H$ in (\ref{H_definition})
\br
\Pi_{\m \n \;(2\;b)}^{BB} (x-y) &=& 2 e^4 Tr[ \g_{\m} \g_{\a} \g^{\rho} \g_{\b} \g_{\n} \g_{\l} \g_{\rho} \g_{\sigma}] H[ \pa_{\a}, \pa_{\s} \; ; \; \pa_{\b}, \pa_{\l}] \;. \nonumber \\
\er

If we use the identity for $\g$ matrices $\g^{\m} \g_{\n} \g_{\rho} \g_{\sigma} \g_{\m} = -2 \g_{\sigma} \g_{\rho} \g_{\n}$ \cite{Peskin:1995ev} and integrate by parts the derivatives acting over $\D_{xu}$ and $\D_{yv}$, we find that this diagram can be put as
\br
\Pi_{\m \n}^{BB \; (2 \; b)} (x-y) &=& 2 e^4 Tr[ \g_{\m} \g_{\a} \g_{\l} \g_{\n} \g_{\b} \g_{\s}] \left( - \pa_{\a}^x \pa_{\l}^x H [1, \pa_{\s} \; ; \; \pa^x_{\b}, 1] - \pa^x_{\a} H[1, \pa_{\s} \; ; \; \pa_{\b} \pa_{\l},1]  \right. \nonumber \\
& & \left. + \pa^x_{\l} H[1, \pa_{\a} \pa_{\s} \; ; \; \pa_{\b},1] + H[1, \pa_{\a} \pa_{\s} \; ; \; \pa_{\b} \pa_{\l},1] \right) \;. \label{QED_diag_b_bare_H}
\er

Using the properties of the trace, the clifford algebra definition $\anticomm{\g_{\m}}{\g_{\n}} = 2 \d_{\m \n}$ and usual $\g$ matrices results as $Tr[\g_{\m} \g_{\l} \g_{\n} \g_{\b} ] = 4 ( \d_{\m \l} \d_{\n \b} - \d_{\m  \n} \d_{\l \b} + \d_{\m \b} \d_{\n \l})$, we can write the right hand-side of the previous equation as
\br
e^4 &\left[\right.& - 8 \d_{\m \n} \Box H[ 1 , \pa_{\l} \; ; \; \pa_{\l}, 1] + 16 \pa_{\m}^x H[ 1, \pa_{\n} \; ; \; \Box, 1] - 8 \d_{\m \n} \pa_{\l}^x H[ 1, \pa_{\l} \; ; \; \Box , 1]  \nonumber \\
& & - 16 \pa_{\m}^x H[ 1, \pa_{\l} \; ; \; \pa_{\l} \pa_{\n}, 1] + 16 \pa_{\l}^x H [ 1, \pa_{\l} \; ; \; \pa_{\m} \pa_{\n} ,1] - 16 \pa_{\l}^x H [ 1, \pa_{\m} \; ; \; \pa_{\l} \pa_{\n},1]  \nonumber \\
& & - 16 \pa_{\m} H [ 1, \Box \; ; \; \pa_{\n}, 1] + 8 \d_{\m \n} \pa_{\l}^x H[ 1, \Box \; ; \; \pa_{\l}, 1] + 16 \pa_{\l}^x H[ 1, \pa_{\l} \pa_{\m} \; ; \; \pa_{\n},1]  \nonumber \\
& & - 16 \pa_{\l}^x H[ 1, \pa_{\m} \pa_{\n} \; ; \; \pa_{\l}, 1] + 16 \pa_{\n}^x H[ 1, \pa_{\l} \pa_{\m} \; ; \; \pa_{\l}, 1] - 16 H[ 1, \Box \; ; \; \pa_{\m} \pa_{\n},1]  \nonumber \\
& & \left. + 8 \d_{\m \n} H[ 1, \Box \; ; \; \Box ,1 ] + 32 H[ 1 , \pa_{\m} \pa_{\l} \; ; \; \pa_{\n} \pa_{\l}, 1] - 16 H[ 1 , \pa_{\m} \pa_{\n} \; ; \; \Box , 1] \; \right]. \nonumber \\
\er

As we have discussed in section \ref{overlap_integrals}, contributions containing a $\Box$ can be reduced by means of the propagator equation $\Box \D = - \d$ and expressed in terms of $I^1$ as given in section \ref{Nested_div}. The remaining contributions can be found in the list of renormalized expressions with overlapping divergences given in section \ref{overlap_integrals} (or can be easily expressed in terms of integrals of the list). Hence, we obtain the renormalized result as
\br
\left. \Pi^{BB}_{\m \n \;(2\;b) } (x) \right|_R &=& \frac{e^4}{12 (4 \pi^2)^3} \left[ \pa_{\m} \pa_{\n} \Box \frac{ \ln^2 x^2 M^2 + \frac{14}{3} \ln x^2 M^2}{x^2} - \d_{\m \n} \Box \Box \frac{ \ln^2 x^2 M^2 + \frac{17}{3} \ln x^2 M^2}{x^2} \right]  \nonumber \\
& & + \ldots 
\er

\subsubsection{Final expression}

Adding the two previous results, the final two-loop renormalized expression is
\br
\left. \Pi_{ \; \m \n \;(2)}^{BB} (x) \right|_R &=& \left. 2 \left. \Pi_{\m \n \;(2\;a)}^{BB} (x) \right|_R + \Pi_{\m \n\;(2\;b)}^{BB} (x) \right|_R \nonumber \\
&=& \frac{e^4}{4(4 \pi^2)^3} ( \pa_{\m} \pa_{\n} - \d_{\m \n} \Box ) \Box \frac{ \ln x^2 M^2}{x^2} + \ldots
\er
up to local terms.

\subsection{RG equation}

When renormalizing the two-loop diagrams we have restricted ourselves to the Feynman gauge. We are allowed to do that as the term in the RG equation that takes into account the running of the gauge parameter ($\g_{\a} \pa / \pa \a$) will be shown not to be relevant when verifying the two-loop RG equations. To prove this, in appendix \ref{ap_Gauge} we evaluate the one-loop RG equation for quantum gauge fields. We find there the expansion of $\g_{\a}$ to be
\br
\g_{\a} (e) &=& - \frac{2 \a}{3 (4 \pi^2)} e^2 + \ldots
\er 

Along with this, notice that the tree level background effective action and the one-loop correction do not depend on the gauge parameter (in this theory we have no quantum-background coupling). Hence, as the first gauge corrections arise at the two-loop level, we do not have to take them into account in order to verify the two-loop RG equations ($\g_{\a} \pa / \pa \a$ acting on them is two orders higher in $e$). So, we are allowed to perform our calculation in the Feynman gauge.

We define the background field two-point function to be 
\br
\G_{\m \n}^{BB}(x-y) &=& \left(\pa_{\m} \pa_{\n} - \d_{\m \n} \Box \right) \d^{(4)}(x-y) - \Pi^{BB}_{\m \n} (x-y) \;.
\er
As is detailed in appendix \ref{ap_BFM}, in the background field method the charge and background field renormalizations are related: $Z_{e} \sqrt{Z_B} = 1$. Hence, we will redefine the background field to be $B_{\m} = \frac{1}{e} B_{\m}^{\prime}$ as for this new field the anomalous dimension $\g_{B^{\prime}}$ cancels\footnote{With our definition $B^{\prime}_{\m 0} = e_{0} B_{\m 0} = Z_{e} \sqrt{Z_{B}} e B_{\m} = B^{\prime}_{\m}$ and then $\g_{B^{\prime}} = \frac{1}{2} M \frac{\pa}{\pa M} \ln Z_{B^{\prime}} = 0$}. Hence, the background two-point function up to two loops is found to be
\br
\G_{\m \n}^{BB} (x) &=& ( \pa_{\m} \pa_{\n} - \d_{\m \n} \Box ) \left[ \frac{1}{e^2} \d^{(4)}(x) - \frac{1}{9( 4 \pi^2)} \d^{(4)}(x) - \frac{1}{3( 4 \pi^2)^2} \Box \frac{\ln x^2 M^2}{x^2}  \right. \nonumber \\
& & \left. - \frac{e^2}{4 (4 \pi^2)^3} \Box \frac{\ln x^2 M^2}{x^2} \right] + \ldots \;,
\er
that verifies the following RG equation
\br
\left( M \frac{\pa}{\pa M} + \b(e) \frac{\pa}{\pa e} \right) \G_{\m \n}^{BB} = 0 \;,
\er
where, again, $\b(e)$ is the QED $\b$-function and we have dropped out the term corresponding to the running of the gauge parameter, as it is of higher order in $e$. Then, we obtain the following two-loop expansion for $\b(e)$
\br
\b (e) &=& \frac{1}{3(4 \pi^2)} e^3 + \frac{1}{4 (4 \pi^2)^2} e^5 + {\cal{O}}(e^7) \;.
\er

These results agree with previous ones found in the literature \cite{Haagensen:1992vz,Schwinger:1973rv,Berestetsky:1982aq,Itzykson:1980rh,Arnone:2005vd}. 

\subsection{Comparison with DiffR}
To stress the key points of our calculation, let us compare this procedure with usual DiffR \cite{Haagensen:1992vz}. With $M_{\S}$ and $M_{V}$ the one-loop renormalization scales of the fermion self-energy and the three point vertex $V_{\m}$ respectively, the Ward identity 
\br
\frac{\pa}{\pa z^\m} V_{\m} (x-z,y-z) = - i e \left[ \d^{(4)} (z-x) - \d^{(4)}(z-y) \right] \S (x-y) 
\er 
imposes that these scales are related as 
\br
\ln \frac{M^2_{\S}}{M^2_{V}} = \frac{1}{2} \;. \label{QED_mass_rel} 
\er

When dealing with two-loop contributions, in each case one has to make use of the corresponding one-loop scale ($M_{\S}$ or $M_{V}$). After a lengthy calculation we find the final values for $\Pi_{\m \n \;(2a)}^{BB}$ and $\Pi_{\m \n \;(2b)}^{BB}$ to be
\br
\left. \Pi_{\m \n \;(2a)}^{BB} (x) \right|_R &=& - \frac{e^4}{24 (4 \pi^2)^3} \left[ ( \pa_{\m} \pa_{\n} - \d_{\m \n} \Box ) \Box \left( \frac{ \ln^2 x^2 M^2_{\S} + \frac{5}{3} \ln x^2 M^2}{x^2}\right) - \d_{\m \n} \Box \Box \frac{\ln x^2 M^2}{x^2}\right] \nonumber \\
\left. \Pi_{\m \n \;(2b)}^{BB} (x) \right|_R &=& - \frac{e^4}{12 (4 \pi^2)^3} \left[ - ( \pa_{\m} \pa_{\n} - \d_{\m \n} \Box) \Box \left( \frac{\ln^2 x^2 M^2_V + \frac{17}{3} \ln x^2 M^2}{x^2}\right) \right. \nonumber \\
& & \left. + \d_{\m \n} \Box \Box \frac{\ln x^2 M^2}{x^2} \right] \;, 
\er
and to obtain the entire two-loop vacuum polarization, we have to use the mass relation (\ref{QED_mass_rel}) to put one of the scales in terms of the other 
\br
\ln^2 x^2 M^2_{\Sigma} = \ln^2 x^2 M^2_{V} + \ln x^2 M^2_{V} + \frac{1}{4} \;.
\er

\section{SuperQED}
In this section we will deal with the supersymmetric extension of QED, SuperQED. As the gauge group is abelian, this is one of the simplest examples of supersymmetric gauge theory we can consider. This theory was yet renormalized using standard DiffR in \cite{Song}, where as usual, explicit evaluation of Ward identities played a central r\^ole. We will re-obtain those results applying our procedure.
\subsection{Supersymmetry}
Coleman and Mandula \cite{Coleman:1967ad} showed that the commutators of the generators of any internal bosonic symmetry group and the generators of the Poincare group vanish. Hence, space-time and internal symmetry groups can not be mixed in a non-trivial way. However, this {\em{no-go}} theorem can be avoided by allowing fermionic symmetry generators \cite{Haag:1974qh}, and the algebra that we obtain is the so-called {\em{supersymmetry algebra}}. Therefore, supersymmetric transformations are generated by traslationally invariant quantum operators which change fermionic states into bosonic states and vice versa. Hence, for each particle we have another one with the same mass and opposite statistic (of course, as this is not observed in nature we conclude that if supersymmetry is a fundamental symmetry of nature it is necessarily broken). Other relevant consequences of supersymmetry are the positivity of the energy \cite{Gates:1983nr,Sohnius:1985qm,West:1990tg,Wess:1992cp} and, due to the relations imposed to the coupling constants and the equality of the bosonic and fermionic degrees of freedom, some cancellations that occur between different Feynman diagrams that make supersymmetric theories to be more convergent \cite{Gates:1983nr,West:1990tg,Wess:1992cp}.

In order to work efficiently with supersymmetric theories, an extension of the usual space-time with additional anticommuting coordinates was developed \cite{Salam:1974yz}. With this space (called superspace) and the extended fields defined in it (superfields), we can perform all the calculations with supersymmetry being manifest. In particular, perturbation theory can be extended to superspace, which allow us to simplify the calculations as component graphs related by supersymmetry are automatically cancelled out. 

In appendix \ref{ap_SUSY} we give a brief introduction to supersymmetry, superspace and the conventions that we use. Also, we give there a list of references where the reader can find a complete treatment of the subject.

\subsection{The SuperQED model}
A supersymmetric abelian gauge theory can be defined in terms of a field strength  $W_{\a}$ \cite{Gates:1983nr} which is a chiral superfield ($\bar{D}_{\dot{\a}} W_{\a} = 0$) that verifies
\br
D^{\a} W_{\a} = - \bar{D}^{\dot{\a}} \bar{W}_{\dot{\a}} \;.
\er
Hence, this field can be expressed in terms of an unconstrained real scalar superfield $V = \bar{V}$ as
\br
W_{\a} &=& i \bar{D}^2 D_{\a} V \nonumber \\
\bar{W}_{\dot{\a}} &=& -i D^2 \bar{D}_{\dot{\a}} V \;.
\er
From the algebra of covariant derivatives (\ref{SUSY_D_algebra}) we can easily conclude that $W_{\a}$ is invariant under the transformations
\br
V^{\prime} &=& V + i ( \bar{\L} - \L) \;,
\er
where $\L$ is a chiral parameter. 

The relevant action that we find is 
\br
S = \int d^4 x d^4 \th \; W^2 = \frac{1}{2} \int d^4 x d^4 \th \; V D^{\a} \bar{D}^2 D_{\a} V \;,
\er 
which is the supersymmetric gauge invariant generalization of the action for a free vector field, as can be seen if we write this expression in component fields \cite{Gates:1983nr}. As the matter part of the action can be expressed in terms of a chiral field $\Phi$ as $\int d^8 z \; \bar{\Phi} e^{V} \Phi$ \cite{Gates:1983nr,Wess:1992cp,Sohnius:1985qm}, the supersymmetric extension of massless QED is an action of the form \cite{Wess:1992cp}
\br
S &=& \int d^4 x d^2 \th \; W^2 + \int d^4 x d^4 \th \; \bar{\Phi}_{+} e^{gV} \Phi_{+} + \int d^4 x d^4 \th \; \bar{\Phi}_{-} e^{-gV} \Phi_{-} \;.
\er

\subsubsection{Background field method}
In section \ref{BFM_SYM} of appendix \ref{ap_BFM} we discuss in detail the application of the background field method to supersymmetric gauge theories. It is found there that when dealing with an abelian theory, we have a linear quantum-background splitting of the form
\br
V \rightarrow V + B \;,
\er
where $V$ and $B$ are the quantum and background gauge fields respectively. The background effective action that we have is then of the form
\br
\G[B] &=& \int d^4 x d^4 y d^4 \th \; B(x, \th) \left[ D^{\a} \bar{D}^2 D_{\a} B(y, \th) \right] \G(x-y) + \ldots
\er

Notice that the background field method allows us to obtain the beta function $\b(g)$ only from the renormalization of the background field two-point function. Hence, we have to obtain the the one- and two-loop coefficients of the expansion of $\G(x)$, as with them we can obtain $\b(g)$ up to two-loop order.

\subsection{One-loop level}
\begin{figure}[ht]
\centerline{\epsfbox{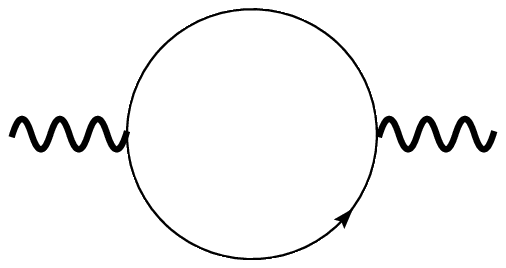}}
\caption{One-loop SuperQED diagrams. Thick lines correspond to external background fields and solid lines represent $\Phi_{+}$ or $\Phi_{-}$ propagators.}
\label{SQED1loop_diag}
\end{figure}

Although we can also consider a diagram which corresponds to a tadpole contribution of a loop of $\Phi_{\pm}$ fields, this diagram gives no contribution as CDR imposes $\D \Box \D |_R = 0$. Hence, the only relevant one-loop contribution from $\Phi_{\pm}$ fields to the background vacuum self-energy is the diagram shown in \ref{SQED1loop_diag}. We will obtain the contribution corresponding to a $\Phi_{+}$ loop, which is denoted as $\G_{+}^{(1)}$, as the other one $\G_{-}^{(1)}$ is exactly the same. With the superspace Feynman rules defined in appendix \ref{ap_SUSY}, we find the expression of the diagram to be (notice that the superspace propagator is $P_{ij} = \D (x_i - x_j) \d^4 (\th_i - \th_j) \equiv \D_{x_i x_j} \d_{ij}$)
\br
\G^{(1)}_{+} &=& \frac{g^2}{2} \int d^8 z_1 d^8 z_2 \; B(z_1) B(z_2) \left[ D^2_1 P_{12} \stackrel{\leftarrow}{D^2}_2 \right] \left[ D^2_2 P_{12} \stackrel{\leftarrow}{D^2}_1 \right] \nonumber \\
&=& \frac{g^2}{2} \int d^8 z_1 d^8 z_2 \; B(z_1) B(z_2) \left[ \bar{D}^2_2 D^2_2 P_{12} \right] \left[ D^2_2 \bar{D}^2_2 P_{21} \right] \;.
\er 
Applying D-algebra we remove all the derivatives from the first propagator and make them act over the external fields and the other propagator. So
\br
\G^{(1)}_{+} &=& \frac{g^2}{2} \int d^8 z_1 d^8 z_2 \; B(z_1) \left[ \bar{D}^2 D^2 B (z_2) \right] P_{12} \left[ \bar{D}^2_2 D^2_2 P_{12} \right]  \nonumber \\ 
& & + \frac{g^2}{2} \int d^8 z_1 d^8 z_2 \; B(z_1) B(z_2) P_{12} \left[ \Box \bar{D}^2_2 \bar{D}^2_2 P_{12} \right]  \nonumber \\
& & - \frac{ig^2}{2} \int d^8 z_1 d^8 z_2 \; B(z_1) \left[ \bar{D}^{\dot{\a}} D^{\a} B (z_2) \right] P_{12} \left[ \pa_{\a \dot{\a}}^2 \bar{D}^2_2 D^2_2 P_{12}\right] \;,
\er 
where we have used the results of the D-algebra $D^2 \bar{D}^2 D^2 = \Box D^2$ and $\comm{D_{\a}}{\bar{D}^2} = - i \pa_{\a \dot{\a}} \bar{D}^{\dot{\a}}$. At this point, we can apply the identity (\ref{SUSY_delta_propagators}) for supercovariant derivatives $\d_{12} D^2 \bar{D}^2 \d_{12} = \d_{12}$. This gives us one free $\th$-space $\d$-function that we can use to evaluate one of the $\th$ integrals. With this, we have (using the identifications $x_1 = x$ and $x_2 = y$)
\br
\G^{(1)}_{+} &=& \frac{g^2}{2} \int d^4 x d^4 y d^4 \th \; B(x, \th) \left[ \bar{D}^2 D^2 B(y, \th) \right] \D^2_{xy}  \nonumber \\
& & + \frac{g^2}{2} \int d^4 x d^4 y d^4 \th \; B(x, \th) B(y, \th) \D_{xy} \left( \Box \D_{xy} \right) \nonumber \\
& & - \frac{i g^2}{2} \int d^4 x d^4 y d^4 \th \; B(x, \th) \left[ \bar{D}^{\dot{\a}} D^{\a} B(y, \th) \right] \D_{xy} \pa_{\a \dot{\a}}^y \D_{xy} \;.
\er

This is the bare expression that we have to renormalize applying CDR rules. It is clear that the second term does not contribute as CDR imposes $\D \Box \D |_R = 0$. The CDR renormalization of the third term $\D \pa_{\a \dot{\a}} \D |_R= \frac{1}{2} \pa_{\a \dot{\a}} \D^2_R$ allows us to integrate by parts the space-time derivative and make it act over the external background fields. Hence, using the identity
\br
\bar{D}^2 D^2 + \frac{i}{2} \pa_{\a \dot{\a}} \bar{D}^{\dot{\a}} D^{\a} = \frac{1}{2} D^{\a} \bar{D}^2 D_{\a} \;,
\er
we find the final renormalized expression to be
\br
\G^{(1)}_{+ \;R} &=& \frac{g^2}{4} \int d^4 x d^4 y d^4 \th \; B(x, \th) \left[ D^{\a} \bar{D}^2 D_{\a} B(y, \th) \right] \left[\D^2 \right]_{R} (x-y) \nonumber \\
&=& - \frac{g^2}{16(4 \pi^2)^2} \int d^4 x d^4 y d^4 \th \; B(x, \th) \left[ D^{\a} \bar{D}^2 D_{\a} B(y, \th) \right] \Box \frac{ \ln (x-y)^2 M^2}{(x-y)^2} \;. \nonumber \\
\er
As we can see, this term is manifestly gauge invariant, as is guaranteed by the use of CDR. The total one-loop effective action is the sum of both contributions corresponding to $\Phi_{+}$ and $\Phi_{-}$
\br
\G^{(1)}_R &=& \G^{(1)}_{+ \; R} + \G^{(1)}_{- \; R} \nonumber \\
&=& - \frac{g^2}{8(4 \pi^2)^2} \int d^4 x d^4 y d^4 \th \; B(x, \th) \left[ D^{\a} \bar{D}^2 D_{\a} B(y, \th) \right] \Box \frac{ \ln (x-y)^2 M^2}{(x-y)^2} \;. \label{SQED_1loop_ren}
\er
\subsection{Two-loop level}
\begin{figure}[ht]
\centerline{\epsfbox{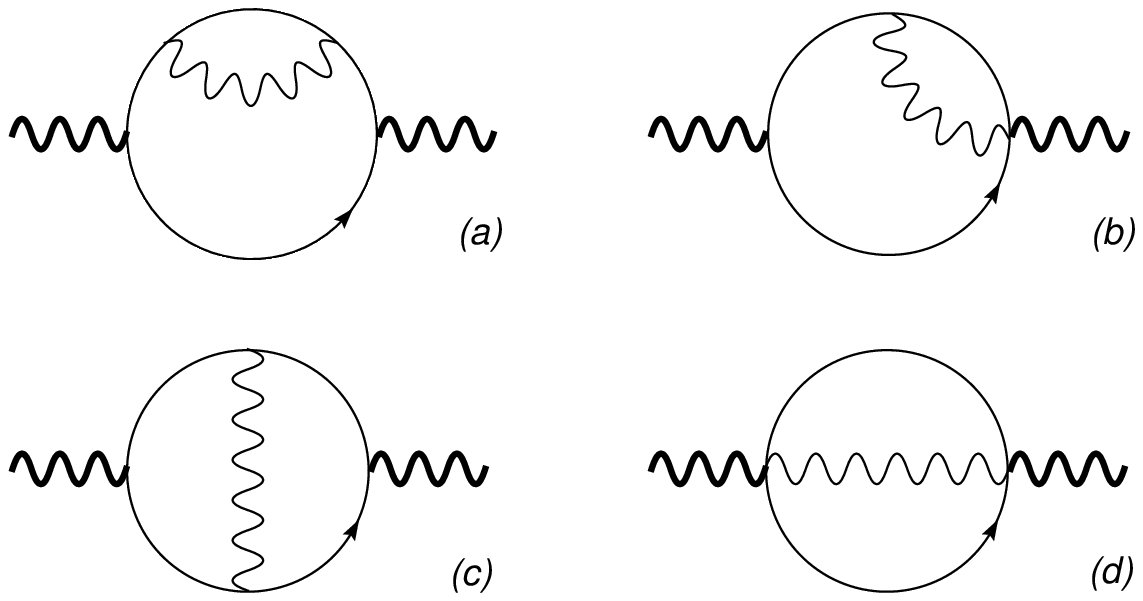}}
\caption{Two-loop SuperQED diagrams. Thick wavy lines correspond to external background gauge fields, thin wavy lines correspond to quantum gauge field propagators and solid lines correspond to $\Phi_{+}$ or $\Phi_{-}$ propagators.}
\label{SQED_two_loop_diag}
\end{figure}
As we can see from figure \ref{SQED_two_loop_diag}, two-loop calculations involve the quantum gauge field propagator. Although this propagator depends on the gauge parameter, we can use it evaluated in Feynman gauge, as the term that takes care of the running of the gauge parameter in the RG equation ($\g_{\a}(g) \pa / \pa \a$), starts its expansion in terms of the coupling constant at order $g^2$. Hence, $\g_{\xi}$ acting over any two-loop diagram will be two orders higher in $g$, and consequently, not relevant when verifying the RG equation. We will detail this later. Also notice that, as in the one-loop case, we will obtain only contributions corresponding to $\Phi_{+}$ fields, as the diagrams with $\Phi_{-}$ fields have the same expression.
  
Before obtaining each diagram, let us discuss an useful identity we will use when we have to apply $D$-algebra to a supergraph \cite{Song}. Let $F_{i}$ be a function of superspace coordinates $z_i = (x_i,\th_i)$, and suppose that we have an expression where the only dependence in $z_i$ is of the form
\br
\int d^8 z_i d^8 z_j d^8 z_k \ldots F_{i} \left[ D^2_i \bar{D}^2_i P_{ij} \right] \left[ \bar{D}^2_i D^2_i P_{ik} \right] \ldots  \label{SQED_algebra_id_paso0}
\er
Applying integration by parts rules in superspace (which can be found in section \ref{SUSY_integration} of appendix \ref{ap_SUSY}) and $D$-algebra, we obtain
\br
(\ref{SQED_algebra_id_paso0})&= & \int d^8 z_i d^8 z_j d^8 z_k \ldots \left[ D^2_i F_i \right] \left[ \bar{D}^2_i P_{ij} \right] \left[ \bar{D}^2_i D^2_i P_{ik} \right] \ldots  \nonumber \\
&+& \int d^8 z_i d^8 z_j d^8 z_k \ldots F_i \left[ \bar{D}^2_i P_{ij} \right] \left[ \Box D^2_i P_{ik} \right] \ldots  \nonumber \\
&+& \int d^8 z_i d^8 z_j d^8 z_k \ldots \left[ D^{\a}_i F_i \right] \left[ \bar{D}^2_i P_{ij} \right] \left[ \pa_{\a \dot{\a}}^i \bar{D}^{\dot{\a}}_i D^2_i P_{ik} \right] \ldots \label{SQED_algebra_id_paso1}
\er
If we integrate by parts again to remove all the superspace derivatives from $P_{ij}$, it is clear that we will obtain some contributions that will vanish by the identities (\ref{SUSY_delta_propagators}), when we use the $\th$-space $\d$-functions to set $j=k$. As an example, consider $\left[ \bar{D}^2_i F_i \right] P_{ij} \left[ \Box D^2_i P_{ik} \right]$, which is obtained from the second term of (\ref{SQED_algebra_id_paso1}). As can be seen, this contribution will cancel when we set $j=k$ and apply $\d_{ij} D^2 \d_{ij} = 0$. So, the final relevant terms of the expansion of (\ref{SQED_algebra_id_paso0}) are
\br
& & \int d^8 z_i d^8 z_j d^8 z_k \ldots \left[ \bar{D}^2_i D^2_i F_i \right] P_{ij} \left[ \bar{D}^2_i D^2_i P_{ik} \right] \ldots  \nonumber \\
&+& \int d^8 z_i d^8 z_j d^8 z_k \ldots F_i P_{ij} \left[ \Box \bar{D}^2_i D^2_i P_{ik} \right] \ldots  \nonumber \\
&-& i \int d^8 z_i d^8 z_j d^8 z_k \ldots \left[ \bar{D}^{\dot{\a}}_i D^{\a}_i F_i \right] P_{ij} \left[ \pa_{\a \dot{\a}}^i \bar{D}^2_i D^2_i P_{ik} \right]  \ldots \nonumber \\
&+& \textrm{(terms~that~vanish~when~j=k)} \;. \label{D_algebra_id}
\er 
\subsubsection{Diagram $(a)$}
The bare expression of this diagram is
\br
\G^{(2a)}_{+} &=& - \frac{g^4}{2} \int d^8 z_1 d^8 z_2 d^8 z_3 d^8 z_4 \; B(z_1) B(z_2) P_{43} \left[ D^2_3 \bar{D}^2_3 P_{43} \right] \left[ \bar{D}^2_2 D^2_2 P_{42} \right]  \nonumber \\
& & ~ \times \left[ D^2_1 \bar{D}^2_1 P_{13} \right] \left[ D^2_2 \bar{D}^2_2 P_{21} \right]  \nonumber \\
& & - \frac{g^4}{2} \int d^8 z_1 d^8 z_2 d^8 z_3 d^8 z_4 \; B(z_1) B (z_2) P_{43} \left[ D^2_4 \bar{D}^2_4 P_{43} \right] \left[ D^2_2 \bar{D}^2_2 P_{24} \right]  \nonumber \\
& & ~ \times \left[ \bar{D}^2_1 D^2_1 P_{31} \right] \left[ D^2_1 \bar{D}^2_1 P_{12} \right] \;. \label{SQED_diag_a_basic}
\er

We will study the first contribution, as the second one, except for having $D$ and $\bar{D}$ interchanged, is the same. Using the identity $P_{43} D^2_3 \bar{D}^2_3 P_{43} = \D^2_{43} \d_{43}$, integrating by parts and applying $D$-algebra we find
\br
\G^{(2aI)}_{+} &=& - \frac{g^4}{2} \int d^8 z_1 d^8 z_2 d^8 z_3 d^8 z_4 \; B(z_1) B(z_2) \left( \D^2_{43} \d_{43} \right) \left[ \bar{D}^2_2 D^2_2 P_{42} \right] \left[ D^2_1 \bar{D}^2_1 P_{13} \right] \left[ D^2_2 \bar{D}^2_2 P_{21} \right] \nonumber \\
&=& - \frac{g^4}{2} \int d^8 z_1 d^8 z_2 d^8 z_3 d^8 z_4 \; B(z_1) B(z_2) \left( \D^2_{43} \d_{43} \right) \left[ \bar{D}^2_2 D^2_2 P_{42} \right] \left( \Box P_{13} \right) \left[ D^2_2 \bar{D}^2_2 P_{21} \right] \;. \nonumber \\
\er
Making use of (\ref{D_algebra_id}) we can write this a
\br
\G^{(2aI)}_{+} &=& - \frac{g^4}{2} \int d^8 z_1 d^8 z_2 d^8 z_3 d^8 z_4 \; B(z_1) \left[ \bar{D}^2 D^2 B (z_2) \right] \left( \D^2_{43} \d_{43} \right) \left[ \bar{D}^2_2 D^2_2 P_{42} \right] \left( \Box P_{13} \right) P_{21}  \nonumber \\
& & - \frac{g^4}{2} \int d^8 z_1 d^8 z_2 d^8 z_3 d^8 z_4 \; B(z_1) B(z_2) \; \left( \D^2_{43} \d_{43} \right) \left[ \Box \bar{D}^2_2 D^2_2 P_{42} \right] \left( \Box P_{13} \right) P_{21}  \nonumber \\
& & + \frac{i g^4}{2} \int d^8 z_1 d^8 z_2 d^8 z_3 d^8 z_4 \; B(z_1) \left[ \bar{D}^{\dot{\a}} D^{\a} B (z_2) \right] \left( \D^2_{43} \d_{43} \right) \left[ \pa_{\a \dot{\a}}^2 \bar{D}^2_2 D^2_2 P_{42} \right] \left( \Box P_{13} \right) P_{21} \;. \nonumber \\
\er
After using identities (\ref{SUSY_delta_propagators}) we can evaluate three of the $\th$ integrals with the free $\d$-functions. Then, with the identifications $x_1 = x$, $x_2 = y$, $x_3 = u$ and $x_4 = v$, the contribution becomes
\br
\G^{(2aI)}_{+} &=& - \frac{g^4}{2} \int d^4 x d^4 y d^4 \th \; B(x, \th) \left[ \bar{D}^2 D^2 B (y, \th) \right] \D_{xy} \int d^4 u d^4 v \; \D_{yv} ( \Box \D_{xu} ) \D^2_{uv}  \nonumber \\
& & - \frac{g^4}{2} \int d^4 x d^4 y d^4 \th \; B(x, \th) B(y, \th) \D_{xy} \int d^4 u d^4 v \; ( \Box \D_{yv} ) ( \Box \D_{xu} ) \D^2_{uv}  \nonumber \\
& & + \frac{i g^4}{2} \int d^4 x d^4 y d^4 \th \; B(x, \th) \left[ \bar{D}^{\dot{\a}} D^{\a} B(y, \th) \right] \D_{xy} \int d^4 u d^4 v \; ( \pa_{\a \dot{\a}}^y \D_{yv} ) ( \Box \D_{xu} ) \D^2_{uv} \;. \nonumber \\
\er
Remembering $\Box \D = - \d$ and the definition of integral expression $I^1$, this can be put as
\br
\G^{(2aI)}_{+} &=& \frac{g^4}{2} \int d^4 x d^4 y d^4 \th \; B(x, \th) \left[ D^2 \bar{D}^2 B(y, \th) \right] \left[ \D I^1 \right](x-y)  \nonumber \\
& & + \frac{g^4}{2} \int d^4 x d^4 y d^4 \th \; B(x, \th) B(y, \th) \D^3_{xy}  \nonumber \\
& & + \frac{i g^4}{2} \int d^4 x d^4 y d^4 \th \; B(x, \th) \left[ \bar{D}^{\dot{\a}} D^{\a} B(y, \th) \right] \left[ \D \pa_{\a \dot{\a}}^x I^{1} \right](x-y) \;.
\er
The second contribution of (\ref{SQED_diag_a_basic}) only differs from the first one by the interchange of $D$ and $\bar{D}$. Hence, using $D$-algebra identities (\ref{SUSY_D_algebra}), the total bare expression of diagram $(a)$ is found to be
\br
\G^{(2a)}_{+} &=& \G^{(2aI)}_{+} + \G^{(2aII)}_{+} \nonumber \\
&=& \frac{g^4}{2} \int d^4 x d^4 y d^4 \th \; B(x, \th) B(y, \th) \left[ \Box ( \D I^{1} ) - 2 \D^3 - \pa^{\a \dot{\a}} ( \D \pa_{\a \dot{\a}} I^{1} ) \right] (x-y)  \nonumber \\
& & + \frac{g^4}{2} \int d^4 x d^4 y d^4 \th \; B(x, \th) \left[ D^{\a} \bar{D}^2 D_{\a} B(y, \th) \right] \left[ \D I^1 \right] (x-y) \;.
\er

\subsubsection{Diagram $(b)$}
The bare expression of this contribution is 
\br
\G^{(2b)}_{+} &=& - g^4 \int d^8 z_1 d^8 z_2 d^8 z_3 \; B(z_1) B(z_2) P_{31} \left[ D^2_3 \bar{D}^2_3 P_{31} \right] \left[ D^2_2 \bar{D}^2_2 P_{23} \right] \left[ D^2_1 \bar{D}^2_1 P_{12} \right]  \nonumber \\
& & - g^4 \int d^8 z_1 d^8 z_2 d^8 z_3 \; B(z_1) B(z_2) P_{31} \left[ D^2_3 \bar{D}^2_3 P_{23} \right] \left[ D^2_1 \bar{D}^2_1 P_{31} \right] \left[ D^2_2 \bar{D}^2_2 P_{21} \right] \;. \nonumber \\ \label{SQED_2loop_diag_b_bare}
\er
As in the previous diagram, the two terms that form (\ref{SQED_2loop_diag_b_bare}) differ only by the interchange of $D$ and $\bar{D}$, so we will obtain the first one, that we name $\G^{(2bI)}_{+}$. With identities (\ref{D_algebra_id}) and (\ref{SUSY_delta_propagators}) we find the relevant expansion of $\G^{(2bI)}_{+}$ to be
\br
\G^{(2bI)}_{+} &=& - g^4 \int d^8 z_1 d^8 z_2 d^8 z_3 \; B(z_1) B(z_2) \left( \D^2_{31} \d_{31} \right) \left[ D^2_2 \bar{D}^2_2 P_{32} \right] \left[ \bar{D}^2_2 D^2_2 P_{12} \right] \nonumber \\
&=& - g^4 \int d^8 z_1 d^8 z_2 d^8 z_3 \; B(z_1) \left[ D^2_2 \bar{D}^2_2 B(z_2) \right] \left( \D^2_{31} \d_{31} \right) \left[ D^2_2 \bar{D}^2_2 P_{32} \right] P_{12}  \nonumber \\
& & - g^4 \int d^8 z_1 d^8 z_2 d^8 z_3 \; B(z_1) B(z_2) \left( \D^2_{31} \d_{31} \right) \left[ \Box D^2_2 \bar{D}^2_2 P_{32} \right] P_{12}  \nonumber \\
& & + i g^4 \int d^8 z_1 d^8 z_2 d^8 z_3 \; B(z_1) \left[ D^{\a} \bar{D}^{\dot{\a}} B(z_2) \right] \left( \D^2_{31} \d_{31} \right) \left[ \pa_{\a \dot{\a}}^2 D^2_2 \bar{D}^2_2 P_{32} \right] P_{12} \;. \nonumber \\   
\er
Using identities (\ref{SUSY_delta_propagators}) we can get rid of the superspace derivatives and obtain free grassmanian $\d$-functions. After evaluating the grassmanian integrals, the identifications $x_1 = x$, $x_2 = y$ and $x_3 = u$ allow us to obtain
\br
\G^{(2bI)}_{+} &=& - g^4 \int d^4 x d^4 y d^4 \th \; B(x, \th) \left[ D^2 \bar{D}^2 B(y, \th) \right] \D_{xy} \int d^4 u \; \D_{yu} \D^2_{xu}  \nonumber \\
& & - g^4 \int d^4 x d^4 y d^4 \th \; B(x, \th) B(y, \th) \D_{xy} \int d^4 u \; ( \Box \D_{yu} ) \D^2_{xu}  \nonumber \\
& & + i g^4 \int d^4 x d^4 y d^4 \th \; B(x, \th) \left[ D^{\a} \bar{D}^{\dot{\a}} B(y, \th) \right] \D_{xy} \int d^4 u \; ( \pa_{\a \dot{\a}}^y \D_{yu} ) \D^2_{xu}\;, \nonumber \\
\er 
or, in terms of the $I^1$ integral expression
\br
\G^{(2bI)}_{+} &=& - g^4 \int d^4 x d^4 y d^4 \th \; B(x, \th) \left[ D^2 \bar{D}^2 B(y, \th) \right] \left[ \D I^1 \right] (x-y)  \nonumber \\
& & + g^4 \int d^4 x d^4 y d^4 \th \; B(x, \th) B(y, \th) \D^3_{xy}  \nonumber \\
& & - i g^4 \int d^4 x d^4 y d^4 \th \; B(x, \th) \left[ D^{\a} \bar{D}^{\dot{\a}} B(y, \th) \right] \left[ \D \pa_{\a \dot{\a}}^x I^1 \right] (x-y) \;.
\er
Finally, adding up the other contribution and using $D$-algebra relations (\ref{SUSY_D_algebra}) we find the total bare contribution to $\G^{(2b)}_{+}$ to be
\br
\G^{(2b)}_{+} &=& g^4 \int d^4 x d^4 y d^4 \th \; B(x, \th) B(y,\th) \left[ - \Box ( \D I^1) + 2 \D^3 + \pa^{\a \dot{\a}} \left( \D \pa_{\a \dot{\a}} I^1 \right) \right] (x-y)  \nonumber \\
& & - g^4 \int d^4 x d^4 y d^4 \th \; B(x, \th) \left[ D^{\a} \bar{D}^2 D_{\a} B(y, \th) \right] \left[ \D I^1 \right] (x-y) \;.
\er

\subsubsection{Diagram $(c)$}
This diagram is given by
\br
\G^{(2c)}_{+} &=& - \frac{g^4}{2} \int d^8 z_1 d^8 z_2 d^8 z_3 d^8 z_4 \; B(z_1) B(z_2) P_{34} \left[ D^2_1 \bar{D}^2_1 P_{14} \right] \left[ \bar{D}^2_1 D^2_1 P_{13} \right]  \nonumber \\
& & \times \left[ D^2_2 \bar{D}^2_2 P_{23} \right] \left[ \bar{D}^2_2 D^2_2 P_{24} \right] \;.
\er
Applying identity (\ref{D_algebra_id}) we can split this expression in three contributions as
\br
\G^{(2c)}_{+} &=& \G^{(2cI)}_{+} + \G^{(2cII)}_{+} + \G^{(2cIII)}_{+} \;,
\er
with
\br
\G^{(2cI)}_{+} &=& - \frac{g^2}{2} \int d^8 z_1 d^8 z_2 d^8 z_3 d^8 z_4 \; B(z_1) \left[ \bar{D}^2 D^2 B(z_2) \right] P_{23} P_{34} \left[ \bar{D}^2_2 D^2_2 P_{24}\right] \left[ D^2_1 \bar{D}^2_1 P_{14}\right]  \nonumber \\
& & ~ \times \left[ \bar{D}^2_1 D^2_1 P_{13}\right] \nonumber \\
\G^{(2cII)}_{+} &=& - \frac{g^4}{2} \int d^8 z_1 d^8 z_2 d^8 z_3 d^8 z_4 \; B(z_1) B(z_2) P_{23} P_{34} \left[ \Box \bar{D}^2_2 D^2_2 P_{24} \right] \left[ D^2_1 \bar{D}^2_1 P_{14} \right] \left[ \bar{D}^2_1 D^2_1 P_{13} \right]  \nonumber \\
\G^{(2cIII)}_{+} &=& - \frac{g^4}{2} \int d^8 z_1 d^8 z_2 d^8 z_3 d^8 z_4 \; B(z_1) \left[ \bar{D}^{\dot{\a}} D^{\a} B (z_2) \right] P_{23} P_{34} \left[ \pa_{\a \dot{\a}}^2 \bar{D}^2_2 D^2_2 P_{24} \right]   \nonumber \\
& & ~ \times \left[ D^2_1 \bar{D}^2_1 P_{14} \right] \left[ \bar{D}^2_1 D^2_1 P_{13} \right] \;.
\er
We will evaluate each contribution separately. Starting by $\G^{(2cI)}_{+}$ we can apply again (\ref{D_algebra_id}) and write this as
\br
\G^{(2cI)}_{+} &=& - \frac{g^4}{2} \int d^8 z_1 d^8 z_2 d^8 z_3 d^8 z_4 \; \left[ \bar{D}^2 D^2 B (z_1) \right] \left[ \bar{D}^2 D^2 B (z_2) \right] P_{34} P_{23} P_{24} P_{14} \left[ \bar{D}^2_1 D^2_1 P_{13} \right]  \nonumber \\
& & - \frac{g^4}{2} \int d^8 z_1 d^8 z_2 d^8 z_3 d^4 z_4 \; B (z_1) \left[ \bar{D}^2 D^2 B(z_2) \right] P_{34} P_{23} P_{24} P_{14} \left[ \Box \bar{D}^2_1 D^2_1 P_{13} \right]  \nonumber \\
& & + \frac{i g^4}{2} \int d^8 z_1 d^8 z_2 d^8 z_3 d^8 z_4 \; \left[ \bar{D}^{\dot{\a}} D^{\a} B(z_1) \right] \left[ \bar{D}^2 D^2 B(z_2) \right] P_{34} P_{23} P_{24} P_{14}  \nonumber \\
& & ~ \times \left[ \pa_{\a \dot{\a}}^1 \bar{D}^2_1 D^2_1 P_{13} \right] \;. \nonumber \\
\er 
Integrating by parts and using the anticommutative nature of the superspace derivatives, we find that an expression of the form $\int dx dy d \th [\bar{D}^{\dot{\a}} A (x, \th)][\bar{D}^2 B(y, \th) ]f(x-y)$ vanishes. Hence, the first and third expressions automatically cancel. Using the identifications $x_1 = x$, $x_2 = y$, $x_3 = u$ and $x_4 = v$ we have for this contribution
\br
\G^{(2cI)}_{+} &=& \frac{g^4}{2} \int d^4 x d^4 y d^4 \th \; B(x, \th) \left[ \bar{D}^2 D^2 B(y, \th) \right] \D_{xy} \int d^4 v \; \D_{yv} \D^2_{xv} \nonumber \\
&=& \frac{g^4}{2} \int d^4 x d^4 y d^4 \th \;  B(x, \th) \left[ \bar{D}^2 D^2 B(y, \th) \right] \left[ \D I^1 \right] (x-y) \;.
\er
We continue now evaluating $\G^{(2cII)}_{+}$. As in this case we have the product of the superpropagators $P_{23} P_{34}$, we can use one of the free grassmanian $\d$-functions and evaluate the integral over $\th_3$. With this, we can write this expression as
\br
\G^{(2cII)}_{+} &=& - \frac{g^4}{2} \int d^8 z_1 d^8 z_2 d^4 x_3 d^8 z_4 B(z_1) B(z_2) \D_{23} \D_{34} \left( \Box \D_{24} \right) \d_{24} \left[ D^2_1 \bar{D}^2_1 P_{14} \right] \left[ \bar{D}^2_1 D^2_1 \D_{13} \d_{12} \right] \nonumber \\
&=& \frac{g^4}{2} \int d^8 z_1 d^6 z_2 d^4 x_3 \; B(z_1) B(z_2) \D_{23}^2 \left[ D^2_1 \bar{D}^2_1 P_{12} \right] \left[ \bar{D}^2_1 D^2_1 \D_{13} \d_{12} \right] \;.
\er
We have performed the integral over $z_4$ applying $\Box \D(x_2-x_4) = - \d (x_2-x_4)$ and the grassmanian $\d$-function $\d_{24}$. Introducing again the grassmanian coordinate $\th_3$ with a $\d$-function and integrating by parts the superspace derivatives that act over $\D_{13} \d_{13} (\equiv P_{13})$ we find
\br
\G^{(2cII)}_{+} &=& \frac{g^4}{2} \int d^8 z_1 d^8 z_2 d^8 z_3 \; B(z_1) B(z_2) \left[ D^2_2 \bar{D}^2_2 \D^2_{23} \d_{23} \right] \left[ \bar{D}^2_2 D^2_2 P_{12} \right] P_{13} \;.
\er
At this point, applying (\ref{D_algebra_id}), we can integrate by parts the superspace derivatives acting over $P_{12}$ and arrive to 
\br
\G^{(2cII)}_{+} &=& \frac{g^4}{2} \int d^8 z_1 d^8 z_2 d^8 z_3 \; B(z_1) \left[ D^2 \bar{D}^2 B(z_2) \right] \left[ D^2_2 \bar{D}^2_2 \D^2_{23} \d_{23} \right] P_{12} P_{13}  \nonumber \\
& & + \frac{g^4}{2} \int d^8 z_1 d^8 z_2 d^8 z_3 \; B(z_1) B(z_2) \left[ \Box D^2_2 \bar{D}^2_2 \D^2_{23} \d_{23} \right] P_{12} P_{13}  \nonumber \\
& & - i \frac{g^4}{2} \int d^8 z_1 d^8 z_2 d^8 z_3 \; B(z_1) \left[ D^{\a} \bar{D}^{\dot{\a}} B(z_2) \right] \left[ \pa_{\a \dot{\a}}^2 D^2_2 \bar{D}^2_2 \D^2_{23} \d_{23} \right] P_{12} P_{13} \;. \nonumber \\
\er
These expressions can be evaluated straightforwardly with the identities (\ref{SUSY_delta_propagators}). Using also the coordinate identifications $x_1 = x$, $x_2 = y$, $x_3 = u$ and $x_4 = v$, we have this contribution written in terms of the integral expression $I^1$ as
\br
\G^{(2cII)}_{+} &=& \frac{g^4}{2} \int d^4 x d^4 y d^4 \th \; B(x, \th) \left[ D^2 \bar{D}^2 B(y, \th) \right] \left[ \D I^1 \right] (x-y)  \nonumber \\
& & - \frac{g^4}{2} \int d^4 x d^4 y d^ \th \; B(x, \th) B(y, \th) \D^3_{xy}  \nonumber \\
& & + i \frac{g^4}{2} \int d^4 x d^4 y d^4 \th \; B(x, \th) \left[ D^{\a} \bar{D}^{\dot{\a}} B(y, \th) \right] \left[ \D \pa_{\a \dot{\a}}^x I^{1} \right] (x-y) \;.
\er 
Finally, we take care of $\G^{(2cIII)}_{+}$. With (\ref{D_algebra_id}), taking into account that the superspace derivatives anticommute and making the usual identifications $x_1 = x$, $x_2 = y$, $x_3 = u$ and $x_4 = v$, we find
\br
\G^{(2cIII)}_{+} &=& i \frac{g^4}{2} \int d^4 x d^4 y d^4 \th \; B(x, \th) \left[ \bar{D}^{\dot{\a}} D^{\a} B(y, \th) \right]  \nonumber \\
& & ~\times \int d^4 u d^4 v \; \D_{yu} \D_{uv} ( \pa_{\a \dot{\a}}^y \D_{yv}) \D_{xv} ( \Box \D_{xu})  \nonumber \\
& & + \frac{g^4}{2} \int d^4 x d^4 y d^4 \th \; \left[ \bar{D}^{\dot{\b}} D^{\b} B(x, \th)\right] \left[ \bar{D}^{\dot{\a}} D^{\a} B(y, \th) \right]  \nonumber \\
& & ~\times \int d^4 u d^4 v \; \D_{yu} \D_{uv} ( \pa_{\a \dot{\a}}^y \D_{yv} ) \D_{xv} ( \pa_{\b \dot{\b}}^x \D_{xu} ) \;,
\er
or which is the same, integrating  by parts the superspace derivatives of the last integral\footnote{$\int [ \bar{D}^{\dot{\b}} D^\b B(z_1) ] [ \bar{D}^{\dot{\a}} D^{\a} B(z_2) ] f_{\a \dot{\a}, \b \dot{\b}} (z_1 - z_2 ) = - \int B(z_1) [D^{\b} \bar{D}^{\dot{\b}} \bar{D}^{\dot{\a}} D^{\a} B(z_2)] f_{\a \dot{\a},\b \dot{\b}}(z_1 -z_2)$. Also it has to be noted that due to the anticommutative nature of the superspace derivatives $\bar{D}^{\dot{\a}} \bar{D}^{\dot{\b}} = - D^2 C^{\dot{\a} \dot{\b}}$} and using the integral expression $H$ defined in (\ref{H_definition})
\br
\G^{(2cIII)}_{+} &=& + \frac{i g^4}{2} \int d^4 x d^4 y d^4 \th \; B(x, \th) \left[ \bar{D}^{\dot{\a}} D^{\a} B(y, \th) \right] \left[ \D \pa_{\a \dot{\a}} I^1 \right](x-y)  \nonumber \\
& & + \frac{g^4}{2} \int d^4 x d^4 y d^4 \th \; B(x,\th) \left[ D^{\b} \bar{D}^2 D^{\a} B(y,\th) \right] C^{\dot{\b} \dot{\a}} H[\pa_{\b \dot{\b}},1 \; ; 1, \pa_{\a \dot{\a}}] \;. \nonumber \\
\er

Adding up the three contributions, we find the final bare expression of $\G^{(2c)}_{+}$ to be 
\br
\G^{(2c)}_{+} &=& \frac{g^4}{2} \int d^4 x d^4 y d^4 \th \; B(x, \th) \left[ D^{\a} \bar{D}^2 D_{\a} B(y, \th) \right] \left[ \D I^1 \right](x-y)  \nonumber \\
& & + \frac{g^4}{2} \int d^4 x d^4 y d^4 \th \; B(x, \th) B(y, \th) \left[ \Box ( \D I^1 ) - \D^3 - \pa^{\a \dot{\a}} ( \D \pa_{\a \dot{\a}} I^1 ) \right] (x-y)  \nonumber \\
& & + \frac{g^4}{2} \int d^4 x d^4 y d^4 \th B(x, \th) \left[D^{\b} \bar{D}^2 D^{\a} B(y, \th) \right] C^{\dot{\b} \dot{\a}} H[\pa_{\b \dot{\b}},1 \; ; 1, \pa_{\a \dot{\a}}] \;, \nonumber \\ \label{SQED_diag_c}
\er 
with $C^{\dot{\b} \dot{\a}}$ given in section \ref{SUSY_Notation} of appendix \ref{ap_SUSY}.
\subsubsection{Diagram $(d)$}
The bare contribution of this diagram is 
\br
\G^{(2d)}_{+} &=& - \frac{g^4}{2} \int d^8 z_1 d^8 z_2 \; B(z_1) B(z_2) \left[ D^2_1 \bar{D}^2_1 P_{12} \right] P_{12} \left[ D^2_2 \bar{D}^2_2 P_{12} \right] \;.
\er
Applying the identity $\d_{12} D^2 \bar{D}^2 \d_{12} = \d_{12}$ it is clear that, with the identifications $x_1 = x$ and $x_2 = y$, the non-renormalized expression of this diagram is
\br
\G^{2d}_{+} &=& - \frac{g^4}{2} \int d^4 x d^4 y d^4 \th \; B(x, \th) B(y, \th) \D^3_{xy} \;.
\er

\subsubsection{Renormalization}
As our renormalization procedure verifies CDR at one loop, in order to obtain the final renormalized result we can replace each bare expression for its renormalized value and simply add them. Even more, as with our procedure each expression always has the same renormalized value, we can first add all the bare expressions, and then perform the renormalization. This is forbidden when we use DiffR, as each expression has to be renormalized with its corresponding scale. From the explicit form of the different contributions, it is clear that all the terms cancel exactly, except the last part of diagram $(c)$ (\ref{SQED_diag_c}). Hence, the two-loop renormalized contribution to the vacuum self-energy is (multiplying by two as we consider both contributions from the chiral matter superfields $\Phi_{+}$ and $\Phi_{-}$)
\br
\G^{2}_R  &=& \left. 2 \left( \G_{+}^{(2a)} + \G_{+}^{(2b)} + \G_{+}^{(2c)} + \G_{+}^{(2d)} \right) \right|_R \nonumber \\
&=& g^4 \int d^4 x d^4 y d^4 \th \; B(x, \th) \left[ D^{\b} \bar{D}^2 D^{\a} B(y, \th) \right] C^{\dot{\b} \dot{\a}} H^R [\pa_{\b \dot{\b}},1 \; ; \; 1, \pa_{\a \dot{\a}} ] \nonumber \\
&=& - \frac{g^4}{16 (4 \pi^2)^3} \int d^4 x d^4 y d^4 \th \; B(x, \th) \left[ D^{\a} \bar{D}^2 D_{\a} B(y, \th) \right] \Box \frac{ \ln (x-y)^2 M^2}{(x-y)^2} + \ldots  \;, \nonumber \\ 
\er
where, when obtaining the final result, we have directly applied the corresponding identity from the list of integrals with overlapping divergences of section \ref{overlap_integrals}. Let us remark again that, as was guaranteed by fulfilling CDR rules at one loop, this expression is directly gauge invariant. 

Before dealing with the background RG equation, we have to justify the use of the Feynman gauge in the calculations. As for the QED case, in appendix $\ref{ap_Gauge}$, by means of the evaluation of the one-loop RG equation for the quantum gauge fields, we obtain the expansion of the term that takes into account the running of the gauge parameter in the RG equation ($\g_{\a} \pa / \pa \a$). This is of the form
\br
\g_{\a} &=& - \frac{\a}{(4 \pi^2)} g^2 + \ldots 
\er
At this point, notice that the first gauge corrections to the background effective action arise at the two-loop level. Hence, when verifying the two-loop RG equation, we may not take into account $\g_{\a} \pa / \pa \a$ acting on them, as this is two orders higher in $g$. This is the reason why we are allowed to use the Feynman gauge in our calculations in both QED and SuperQED models.

Let us now proceed to the evaluation of the RG equation for the background gauge field self-energy. As in the QED case, if we make the redefinition $B \rightarrow \frac{1}{g}B$, the anomalous dimension term of the renormalization group equation cancels (remember that the coupling constant and the background field renormalizations are related: $Z_{g} \sqrt{Z_B} = 1$ ). So, with this definition, the background field two-point function to two-loop order is
\br
\G_R(x) &=& \frac{1}{2g^2} \d^4(x) - \frac{1}{8(4 \pi^2)^2} \Box \frac{\ln x^2 M^2}{x^2} - \frac{g^2}{16 ( 4 \pi^2)^3} \Box \frac{ \ln x^2 M^2}{x^2} + \ldots \;, \label{SQED_2_point_function}
\er
and it fulfills the following RG equation
\br
\left[ M \frac{\pa}{\pa M} + \b(g) \frac{\pa}{\pa g} \right] \G_R (x) = 0 \;,
\er
where we do not consider the term that takes care of the running of the gauge parameter. By solving this equation order by order in $g$ it is clear that the beta function is of the form
\br
\b(g) &=& \b_1 g^3 + \b_2 g^5 + {\cal{O}}(g^7) \nonumber \\
\b_1 &=& \frac{1}{16 \pi^2} \nonumber \\
\b_2 &=& \frac{1}{8 (4 \pi^2)^2} \;.
\er
However, as with our supersymmetry conventions the gauge coupling constant differs from the usual one $g_{SQED}$ ($g = \sqrt{2} g_{SQED}$) \cite{Gates:1983nr}, the expansion of the beta function to two-loop order in terms of the usual coupling constant is
\br
\b(g_{SQED}) &=& \frac{1}{2(4\pi^2)} g^3_{SQED} + \frac{1}{2 (4 \pi^2)^2} g^5_{SQED} + {\cal{O}}(g^7_{SQED}) \;.
\er 
This agrees with previous results found in the literature \cite{Vainshtein:1986ja,Shifman:1985fi,Novikov:1985rd}.
\begin{figure}[ht]
\centerline{\epsfbox{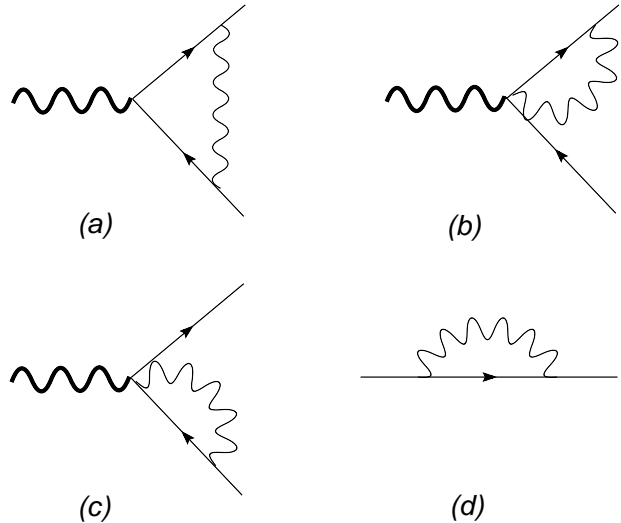}}
\caption{One-loop SQED Ward identities.}
\label{SQED_Ward_id_diag}
\end{figure}

Let us compare this procedure with the steps that we have to take when using usual DiffR. We have to consider first the Ward identities, that can be shown  to relate the 3-point 1PI Green's function $<T \; B \Phi_{+} \bar{\Phi}_{+}>_{1PI}$ and the 2-point 1PI Green's function $<T \; \Phi_{+} \bar{\Phi}_{+}>_{1PI}$ as \cite{Song}
\br
\bar{D}^2(z_1) < T \; B (z_1) \Phi_{-}(z_2) \bar{\Phi}_{-} (z_3) >_{1PI} &=& - g < T \; \Phi_{-} (z_1) \bar{\Phi}_{-}(z_3)>_{1PI} \bar{D}^2(z_1) \d^8(z_1-z_2) \nonumber \\
D^2(z_1) < T \; B(z_1) \Phi_{-}(z_2) \bar{\Phi}_{-}(z_3) >_{1PI} &=& - g < T \; \Phi_{-}(z_2) \bar{\Phi}_{-}(z_1) >_{1PI} D^2(z_1) \d^8(z_1-z_3) \nonumber \\
\bar{D}^2(z_1) < T \; B(z_1) \Phi_{+} (z_2) \bar{\Phi}_{+} (z_3) > &=& g < T \; \Phi_{+}(z_1) \bar{\Phi}_{+}(z_3) >_{1PI} \bar{D}^2(z_1) \d^8(z_1-z_2) \nonumber \\
D^2(z_1) < T \; B(z_1) \Phi_{+}(z_2) \bar{\Phi}_{+}(z_3)>_{1PI} &=& g < T \; \Phi_{+}(z_2) \bar{\Phi}_{+}(z_1)>_{1PI} D^2(z_1) \d^8(z_1 - z_3 ) \;. \nonumber \\ \label{SQED_Ward_id}
\er

With these identities, we can obtain the one-loop relation between the scales that renormalize the $B- \Phi - \bar{\Phi}$ vertex functions (diagrams $(a)$-$(c)$ of figure \ref{SQED_Ward_id_diag} with scales $M_{V_{a}}$, $M_{V_{b}}$ and $M_{V_{c}}$ respectively) and the $\Phi \bar{\Phi}$ self-energy corrections (diagram $(d)$ of figure \ref{SQED_Ward_id_diag} with scale $M_{V_{\Sigma}}$). Performing the explicit renormalization and imposing identities (\ref{SQED_Ward_id}), this relation is found to be \cite{Song}
\br
M^2_{V_{a}} M^2_{V_{\Sigma}} = M^2_{V_{b}} M^2_{V_{c}} \;. \label{SQED_mass_relation}
\er

Hence, when renormalizing each of the two-loop diagrams, we have to use the corresponding one-loop scale, add up all the results and apply (\ref{SQED_mass_relation}). In \cite{Song} is shown that this relation cancels contributions that came from different diagrams and are grouped in an expression multiplied by $\ln [(M^2_{V_{\Sigma}} M^2_{V_{a}})/ (M^2_{V_{b}} M^2_{V_{c}})]$. As can be seen from our procedure, these cancellations take place automatically once we have renormalized the one-loop divergences with the rules of CDR.

\chapter{Non-abelian QFT applications}

\section{Yang-Mills}
\subsection{Conventions and definitions}
\subsubsection{Relevant group theory definitions}
Let $G$ be a continuous symmetry group with generators $T^a$. We can define an associated Lie algebra through the commutation relation
\br
\comm{T^a}{T^b} = i f^{abc} T^c  \;,
\er
where $f^{abc}$ are the structure constants of the Lie algebra, which obey the Jacobi identity $f^{ade} f^{bcd} +  f^{bde} f^{cad} + f^{cde} f^{abd} = 0$. We can have several representations of this Lie algebra in terms of matrices $t^a_r$: one of them is the {\em{adjoint representation}}, denoted by $r=G$, where the representation matrices are given by the structure constants $(t^b_G)_{ac} = i f^{abc}$. These representation matrices are found to satisfy
\br
tr[t^a_r t^b_r] &=& C(r) \d^{ab} \nonumber \\
\sum_{a} t^a_r t^a_r &=& C_2(r) \boldsymbol{1} \;,
\er 
where $C(r)$ and $C_2(r)$ are constants, being the latter called the quadratic Casimir operator. For the concrete case of the adjoint representation, we write the relation for the Casimir operator as 
\br
f^{acd} f^{bcd} = C_A \d^{ab}  \;.
\er
where we define $C_A = C_2(G)$.
\subsubsection{Yang-Mills model}
\label{YM_conventions}
Yang-Mills theory is one of the simplest examples of non-abelian gauge theory \cite{Yang:1954ek}. It is obtained by imposing invariance under a continuous symmetry group. We start by considering $V(x)$ to be an unitary $n \times n$ matrix representing one of the elements of a gauge group. Then, the fields transform according to
\br
\psi (x) &\rightarrow& V(x) \psi(x) \nonumber \\
&=& ( 1 + i w^a(x) t^a + {\cal{O}}(w^2)) \psi(x) \;,
\er 
where we have considered an infinitesimal parameter $w^a$, which has allowed us to expand $V(x)$ in terms of the generators of the symmetry group. Now we have to construct a covariant derivative that, when acting over $\psi(x)$, has the same transformation as the field. This derivative, expressed in terms of a connection (gauge potential) $A_{\m \; ij}= A^a_{\m} t^a_{ij}$, is found to be \cite{Peskin:1995ev}
\br
D_{\m \; ij} &=& \pa_{\m} \d_{ij} - i g A_{\m}^a t^a_{ij} \;.
\er

If we choose the adjoint representation, this becomes $D_{\m}^{ab} = \pa_{\m} \d^{ab} + g f^{abc} A_{\m}^b$. As this derivative has to transform covariantly under the gauge group, the infinitesimal gauge transformation of $A_{\m}^a$ is found to be \cite{Peskin:1995ev}
\br
A_{\m}^a &\rightarrow&  A_{\m}^a + \frac{1}{g} \pa_{\m} w^a - f^{abc} w^b A_{\m}^c + {\cal{O}} (w^2) \nonumber \\
&=& A_{\m}^a + \frac{1}{g} ( D_{\m}w )^a + {\cal{O}} (w^2) \;.
\er

By considering the commutator of covariant derivatives, we can define a field strength as $-i g F_{\m \n}^a t^a = \comm{D_{\m}}{D_{\n}}$ that in terms of the gauge potential has the form
\br
F_{\m \n}^a &=& \pa_{\m} A_{\n}^a - \pa_{\n} A_{\m}^a + g f^{abc} A_{\m}^b A_{\n}^c \;.
\er 

With this field strength is straightforward to define an gauge invariant quantity that is the Yang-Mills action:
\br
S &=& \frac{1}{4} \int d^4 x \; F_{\m \n}^a F_{\m \n}^a \;.
\er

As in the abelian case, when quantizing the action in a path integral approach, we have to fix the gauge in order to suppress all the equivalent field configurations obtained from a given one through gauge transformations. The result is that the gauge-fixed partition function $Z$ is 
\br
Z[J] = \int [dA] \; det \left[ \frac{\d G^a (A^w)}{\d w^b} \right]_{w = 0} exp \left[ - S(A) - \frac{1}{2 \a} \int d^4 x \; G^a G^a + J_{\m}^a A_{\m}^a \right] \;,
\er
where $G^a$ is the gauge-fixing function. Writing the determinant in terms of anticommuting ghost fields\footnote{Being $\th$ an anticommutative variable $ \int \Pi d \th d \bar{\th} e^{\S \bar{\th}_i a_{ij} \th_j} = det[a]$} $\eta$ and choosing for the gauge-fixing function $G^a = \pa^{\m} A_{\m}^a$ we can find the complete Yang-Mills lagrangian to be
\br
\cal{L} &=& \frac{1}{4} F^a_{\m \n} F^a_{\m \n} + \frac{1}{2 \a} ( \pa_{\m} A_{\m} )^a ( \pa_{\n} A_{\n} )^a + ( \pa_{\m} \bar{\eta})^a ( {D}_{\m} \eta )^a \;.
\er

This implies that we have the following gauge field and ghost propagators
\br
<A_{\m}^a(x) A_{\n}^b(y) > &=& \d_{\m \n} \d^{ab} \D (x-y) \nonumber \\
< \eta^a(x) \bar{\eta}^b(y) > &=& \d^{ab} \D(x-y)
\er

\subsubsection{Background field method}

As is detailed in appendix \ref{ap_BFM}, with the standard quantum-background splitting $A_{\m}^a \rightarrow A_{\m}^a + B_{\m}^a$ we can define  two gauge covariant derivatives ${\bf{D}}_{\m}^{ac} = \pa_{\m} \d^{ac} + g f^{abc} B_{\m}^b $ and ${\cal{D}}_{\m}^{ac} = \pa_{\m} \d^{ac} + g f^{abc} ( B_{\m}^b + A_{\m}^b)$. Using them, and with a background covariant gauge-fixing function as $G^{a} = ( {\bf{D}}^{\m} A_{\m})^a$, we find the split lagrangian to be written as
\br
\cal{L} &=& \frac{1}{4} F^a_{\m \n} F^a_{\m \n} + \frac{1}{2 \a} ( {\bf{D}}_{\m} A_{\m} )^a ( {\bf{D}}_{\n} A_{\n} )^a + ( {\bf{D}}_{\m} \bar{\eta})^a ( {\cal{D}}_{\m} \eta )^a \;, \label{YM_background_lagrangian}
\er
with the field strength depending in both quantum and background fields $F_{\m \n}^a = F^a_{\m \n} ( A + B )$. As in the previous abelian examples, we will perform the calculations in Feynman gauge; however, there is one important difference. As can be seen from the lagrangian (\ref{YM_background_lagrangian}), we have an interaction term of the form
\br
g f^{abc} \left[ 2 ( \pa_{\m} B_{\n}^a) A_{\m}^b A_{\n}^c - B_{\m}^a ( \pa_{\m} A_{\n}^b ) A_{\n}^c \right] \;,
\er 
which implies that the one-loop background self-energy depends on the gauge parameter, as we have a loop with quantum gauge fields propagators. Hence, although the term in the RG equation that takes care of the running of the gauge parameter, $\g_{\a} \pa / \pa \a$, will be shown to be  of order $g^2$ (like in QED and SQED), in this case we can not leave it out, as when acting on the one-loop contribution it will affect the verification of the two-loop RG equation. Then, our procedure will be as follows: first of all, the standard gauge fixing parameter $\a$ will be redefined here as $\frac{1}{\a} = 1 + \xi$, so that usual Feynman gauge ($\a = 1$) will correspond to $\xi = 0$. We will obtain the one-loop contribution to the background self-energy in this gauge. Then, by means of functional methods, we will expand the complete effective action at one loop at second order in the background fields and retain only the linear dependence in $\xi$, that we term $\G_{\xi}^{(1)}$. We apply this procedure as in the renormalization group equation we will first take derivatives with respect to this parameter and after that impose Feynman gauge ($\xi$=0). 

Hence, we have a background effective action of the form
\br
\G_{eff}[B] &=& \frac{1}{2} \int d^4 x d^4 y B_{\m}^a(x) \G_{\m \n}^{BB \; ab} (x-y) B_{\n}^b(y) + \ldots \nonumber \\
&=& \frac{1}{2} \int d^4 x d^4 y B_{\m}^a(x) \left[ \d^{ab} ( \pa_{\m} \pa_{\n} - \d_{\m \n} \Box )\d^{(4)}(x-y) - \Pi_{\m \n \; \xi}^{BB \; ab} (x-y) \right] B_{\n}^b(y) + \ldots \nonumber \\
&=& S_0[B] + \G_{\xi}^{(1)} - \frac{1}{2} \int d^4 x d^4 y \; B_{\m}^a(x) \Pi_{\m \n}^{BB \; ab} (x-y)B_{\n}^b(y) + \ldots \;,
\er
where $S_0$ is the tree-level background two-point function.

\begin{figure}[ht]
\centerline{\epsfbox{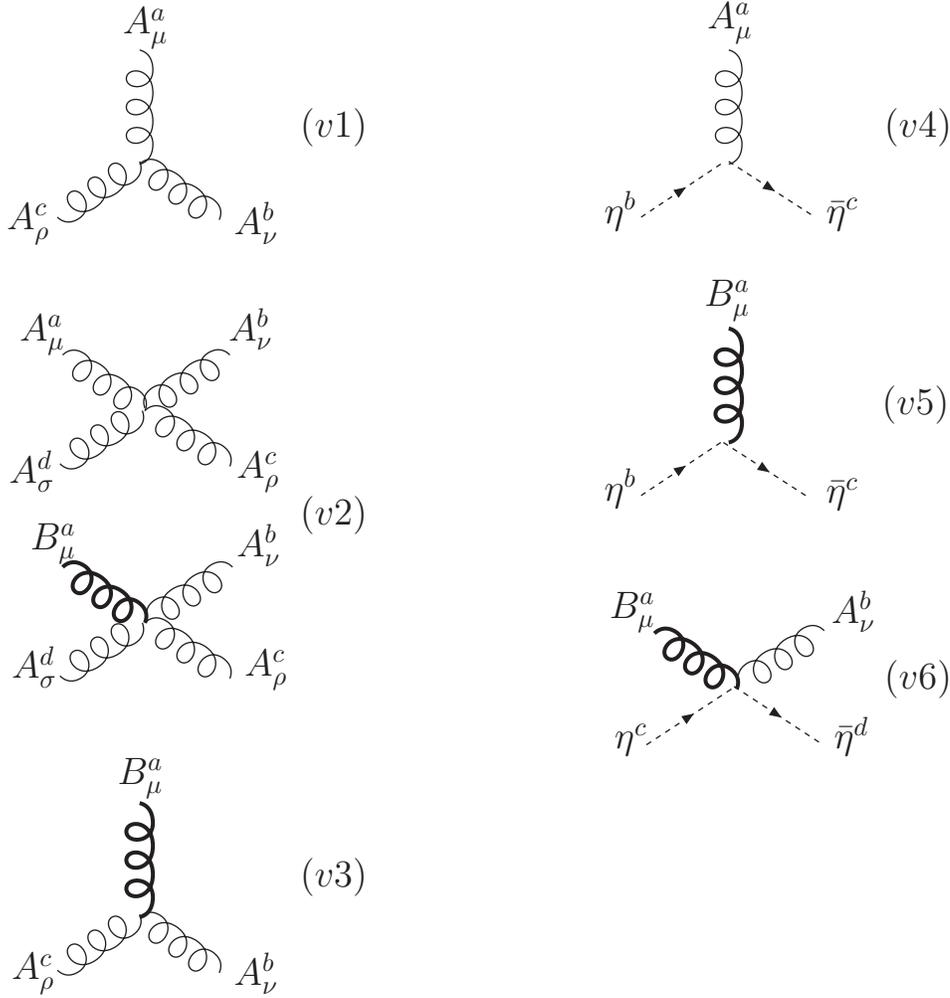}}
\caption{Relevant interaction vertices of the Yang-Mills quantum-background split action. Thick lines represent external background fields, thin lines are quantum gauge propagators and dashed lines correspond to ghost propagators.}
\label{YM_Feynman_rules}
\end{figure}

In figure \ref{YM_Feynman_rules}, the interaction vertices derived from the quantum-background split action which are relevant to this work are shown. We have the following corresponding Feynman rules (evaluated in Feynman gauge)
\br
\textrm{(v1)} &=& g f^{abc} \left[ \d_{\m \n} ( \pa_{\r}^{A_{\m}^a} - \pa_{\r}^{A_{\n}^b}) + \d_{\r \m} ( \pa_{\n}^{A_{\r}^c} - \pa_{\n}^{A_{\m}^a}) + \d_{\n \r} ( \pa_{\m}^{A_{\n}^b} - \pa_{\m}^{A_{\r}^c}) \right] \nonumber \\
\textrm{(v2)} &=& - g^2 \left[ f^{abx} f^{xcd} ( \d_{\m \r} \d_{\n \s} - \d_{\m \s} \d_{\n \r}) + f^{acx} f^{xdb} ( \d_{\m \s} \d_{\r \n} - \d_{\m \n} \d_{\r \s} ) \right. \nonumber \\
& & \left. + f^{adx} f^{xbc} ( \d_{\m \n} \d_{\r \s} - \d_{\m \r} \d_{\n \s} ) \right] \nonumber \\
\textrm{(v3)} &=& g f^{abc} \left[ - 2 \d_{\m \r} \pa_{\n}^{B_{\m}^a} + \d_{\n \r} ( \pa_{\m}^{A_{\n}^b} - \pa_{\m}^{A_{\r}^c} ) + 2 \d_{\m \n} \pa_{\r}^{B_{\m}^a} \right] \nonumber \\
\textrm{(v4)} &=& - g f^{abc} \pa_{\m}^{\bar{\eta}^c} \nonumber \\
\textrm{(v5)} &=& - g f^{abc} ( \pa_{\m}^{\bar{\eta}^c} - \pa_{\m}^{\eta^b} ) \nonumber \\
\textrm{(v6)} &=& - g^2 f^{adx} f^{xbc} \d_{\m \n}
\er 

\subsection{One-loop level}

Although the background self-energy is all that we need to find the one-loop beta function, we will also obtain the linear dependence in the gauge parameter $\xi$ of the effective action calculated in a generic gauge and expanded to second order in background fields. We need this contribution in order to take care of the running of the gauge parameter in the RG equation.

\subsubsection{Correction to the $B_{\m}^a$ propagator}
\begin{figure}
\centerline{\epsfbox{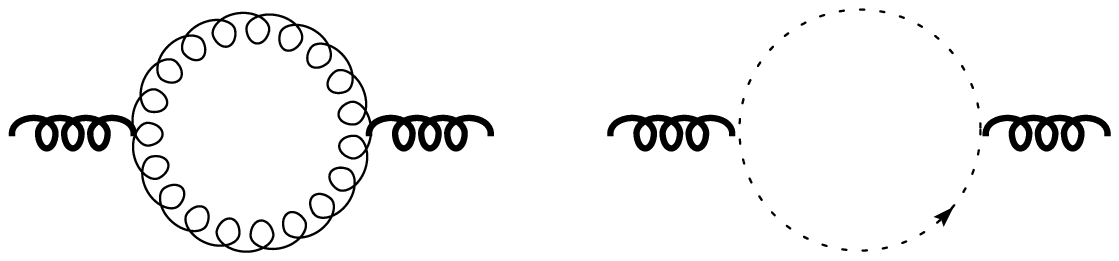}}
\caption{One-loop YM diagrams.}
\label{YM_1loop}
\end{figure}
This is the sum of two different diagrams: one with a loop of quantum gauge fields, and another of ghost fields, as can be seen in figure \ref{YM_1loop}. In these diagrams we have to apply the CDR procedure: First we write the expressions in terms of the basic functions defined in (\ref{basic_CDR_fun}), and after that we replace them with their renormalized values.

Here and in the rest of the diagrams of the Yang-Mills theory, $D_{\m}^{x,y}$ denotes a space-time derivative acting over one external field. Applying the Leibniz rule, $D_{\m}^{x,y}$ becomes a minus derivative acting over the propagators. 
The bare expression for both contributions is 
\begin{itemize}
\item {\bf Gauge loop}
\end{itemize}
\br
& & \frac{ g^2 f^{acd} f^{bdc}}{2} \D_{xy} \left[ - 2 \d_{\m \s} D^x_{\rho} + \d_{\rho \s} (\stackrel{\leftarrow}{\pa_{\m}^{x}} - \pa_{\m}^x) + 2 \d_{\m \rho} D_{\s}^x \right] \nonumber \\
& & \times \left[ - 2 \d_{\n \rho} D_{\s}^y + \d_{\rho \s} (\pa_{\n}^y - \stackrel{\leftarrow}{\pa_{\n}^y}) + 2 \d_{\n \s} D^y_{\rho} \right] \D_{xy} \nonumber \\&=& \frac{g^2 C_A \d^{ab}}{2} \left[ 8 \pa_{\m} \pa_{\n} \D^2 - 8 \d_{\m \n} \Box \D^2 + 8 \pa_{\m}( \D \pa_{\n} \D ) - 16 \D \pa_{\m} \pa_{\n} \D \right]  \;. \nonumber \\
\er
\begin{itemize}
\item {\bf Ghost loop}
\end{itemize}
\br
- g^2 f^{abc} f^{bcd} \D_{xy} ( \stackrel{\leftarrow}{\pa_{\m}^{x}} - \pa_{\m}^x) (\pa_{\n}^y - \stackrel{\leftarrow}{\pa_{\n}^y}) \D_{xy} = - g^2 C_A \d^{ab} \left[ 2 \pa_{\m} ( \D \pa_{\n} \D ) - 4 \D \pa_{\m} \pa_{\n} \D \right] \;. \nonumber \\
\er

Adding the two previous results we find the total non-renormalized contribution to be
\br
\Pi_{\m \n\;(1)}^{BB\;ab} (x) &=& g^2 C_A \d^{ab} \left[ 4 \pa_{\m} \pa_{\n} \D^2 - 4 \d_{\m \n} \Box \D^2 + 2 \pa_{\m} ( \D \pa_{\n} \D ) - 4 \D \pa_{\m} \pa_{\n} \D \right] \;.  \nonumber \\
\er

It is worth to mention here again that we are allowed to do the previous step (adding up the expression even before renormalizing) because we are using CDR, as we pointed out previously. With CDR the basic functions are renormalized always with the same expression, despite their origin. In an usual DiffR procedure, we first have to renormalize each diagram in a separate way, relate the different scales that appeared via the Ward identities, and only after that we can add up the results.

Replacing the values of CDR for $\D^2$ and $\D \pa_{\m} \pa_{\n} \D$, the renormalized one-loop contribution to the $B_{\m}^a$ propagator is obtained as
\br
\left. \Pi_{\m \n \;(1)}^{BB\;ab} (x) \right|_R &=&  g^2 C_A \d^{ab} (\pa_{\m} \pa_{\n} - \d_{\m \n} \Box) \left[ \frac{11}{3} \D^2_R (x) - \frac{1}{72 \pi^2} \d (x) \right] \nonumber \\
&=& g^2 C_A \d^{ab} (\pa_{\m} \pa_{\n} - \d_{\m \n} \Box) \left[ - \frac{11}{48 \pi^2 (4 \pi^2)} \Box \frac{\ln x^2 M^2}{x^2} - \frac{1}{72 \pi^2} \d (x) \right]  \;. \nonumber \\ \label{1_loop}
\er

As a check, the result found here is automatically transverse, fulfilling the corresponding Ward identity.

\subsubsection{Effective action in a generic gauge}
As we have discussed previously, in order to take care of the running of the gauge parameter in the RG equation, we will obtain the linear dependence in $\xi$ of the one-loop background effective action expanded to second order in the background fields. To perform this calculation we consider a functional approach: to obtain the exact one-loop effective action it is well known that we have to consider only the part of the lagrangian quadratic in the quantum $A_{\m}^a$ fields \cite{Jackiw:1974cv,Peskin:1995ev}. This part is
\br
{\cal{L}}^{(2)}_{gauge} &=& g f^{abc} B_{\m \n}^a A_{\m}^b A_{\n}^c + \frac{1}{2} ({\bf{D}}_{\m} A_{\n})^a ({\bf{D}}_{\m} A_{\n})^a + \frac{\xi}{2} ({\bf{D}}_{\m} A_{\m})^a ({\bf{D}}_{\n} A_{\n})^a \nonumber \\
&=& - \frac{1}{2} A_{\m}^a \left[ \d_{\m \n} \Box^{ab} - 2 g f^{cab} B_{\m \n}^c + \xi ({\bf{D}}_{\m} {\bf{D}}_{\n} )^{ab} \right] A_{\n}^b \;,
\er
where ${\Box}^{ab} = ( {\bf{D}}^{\m} {\bf{D}}_{\m})^{ab}$ and $B_{\m \n}^a = \pa_{\m} B_{\n}^a - \pa_{\n} B_{\m}^a + g f^{abc} B_{\m}^b B_{\n}^c$. Then, the generating functional for connected Green functions can be put as
\br
W &=& - \frac{1}{2} tr \ln \left[ \d_{\m \n} \Box^{ab} - 2 g f^{cab} B_{\m \n}^c + \xi ({\bf{D}}_{\m} {\bf{D}}_{\n})^{ab} \right] \;.
\er

At first order in $\xi$ and second order in $B$ fields, this is expressed as
\br
W &=& W_0 + \xi C_A g^2 tr \left[ \frac{1}{2} \D B_{\m \n}^a \D B_{\m \n}^a - 2 \D B_{\m \n}^a \D B_{\n \l}^a \D \pa_{\l} \pa_{\m} \right]  + {\cal{O}}(\xi^2, B^3)\;, \label{effective_1loop}
\er
where as usual $\Box = \pa^{\m} \pa_{\m}$ and $\D = - \Box^{-1}$. We can write the renormalized expression of the first term of (\ref{effective_1loop}) as
\br
(A) &=& \frac{1}{2} \int d^4 x d^4 y \; B_{\m \n}^a (x) B_{\m \n}^a (y)  \D^2 |_R \;, \nonumber \\
\er
whereas the second one is of the following form
\br
(B) &=& - 2 \int d^4 x d^4 y d^4 u \; ( \pa_{\l}^u \pa_{\m}^u \D_{ux}) B_{\m \n}^a (x) B_{\n \l}^a (y) \D_{xy} \D_{yu} \nonumber \\
&=& - 2 \int d^4 x d^4 y \; B_{\m \n}^a (x) B_{\n \l}^a(y) \D_{xy} \pa_{\l}^x \pa_{\m}^x \int d^4 u \; \D_{xu} \D_{yu} \;.
\er

In order to evaluate the latter expression, we must apply CDR in momentum space
\br
\int d^4 u \; \D_{xu} \D_{uv} &=& \frac{1}{(4 \pi^2)^2} \int d^4 u \; \frac{1}{(x-u)^2} \frac{1}{(u-y)^2} \nonumber \\
&=& \frac{1}{(4 \pi^2)^4} \int d^4 u d^4 p d^4 q \; \frac{1}{p^2 q^2} e^{-i p(x-u)} e^{-iq(u-y)} \nonumber \\
&=& \frac{1}{(4 \pi^2)^2} \int d^4 p \; \frac{1}{p^4} e^{-i p(x-y)} \nonumber \\
&\stackrel{R}{\rightarrow}& - \frac{1}{4 (4 \pi^2)^2} \int d^4 p \; \Box^p \frac{\ln p^2/m^2}{p^2} e^{-ip(x-y)} \nonumber \\
&=& - \frac{1}{4(4 \pi^2)} \ln (x-y)^2 m^2 \nonumber \\
&\equiv& - \bar{\D} (x-y) \;. \label{CDR_momentum_space}
\er
\br
(B) &=& 2 \int d^4 x d^4 y \; B_{\m \n}^a (x) B_{\n \l}^a (y) \left( \D_{xy} \pa^x_{\l} \pa^x_{\m} \bar{\D}_{xy} \right)|_R  \;.
\er

Hence, remembering the CDR renormalization of $\D \pa_{\l} \pa_{\m} \bar{\D}$ (\ref{CDR_rules_other_gauge}) we can obtain the renormalized expression for $(B)$ as
\br
(B) &=& - \frac{1}{2} \int d^4 x d^4 y \; B_{\m \n}^a (x) B_{\m \n}^a (y) \D^2 |_R - \frac{1}{16 \pi^2} \int d^4 x d^4 y B_{\m \n}^a (x) B_{\n \l}^a (y) \pa^x_{\m} \pa^x_{\l} \D_{xy} \;. \nonumber \\
\er

Adding up the two contributions we have
\br
(A)+(B) &=& - \frac{\xi C_A g^2}{4(4 \pi^2)} \int d^4 x d^4 y \; B_{\m \n}^a(x) B_{\n \l}^a (y) \pa^x_{\m} \pa^x_{\l} \D_{xy} \;,
\er
which can be written in a more familiar form at explicit second order in the $B$ fields as
\br
\int d^4 x d^4 y B_{\m \n}^a (x) B_{\n \l}^a (y) \pa^x_{\m} \pa^x_{\l} \D_{xy} &=&  \int d^4 x d^4 y (\pa_{\m} B_{\n}^a(x) - \pa_{\n} B_{\m}^a(x) )(\pa_{\n} B_{\l}^a (y) - \pa_{\l} B_{\n}^a (y) )  \nonumber \\
& & \times  \pa^x_{\m} \pa^x_{\l} \D_{xy} +{\cal{O}}(B^3) \nonumber \\
&=& - \int d^4 x d^4 y \; B_{\m}^a (x) B_{\n}^a (y) ( \pa^x_{\m} \pa^x_{\n} - \d_{\m \n} \Box) \Box \D_{xy} + {\cal{O}}(B^3) \;. \nonumber \\
\er

With this result we obtain the previously defined $\G_{\xi}^{(1)}$ to be
\br
\G_{\xi}^{(1)} = - \frac{\xi C_A g^2}{4(4 \pi^2)} \int d^4 x d^4 y \; B_{\m}^a (x) B_{\n}^a (y) (\pa^x_{\m} \pa^x_{\n} - \d_{\m \n} \Box) ( \Box \D(x-y) ) \;. \label{gauge_fix_ren}
\er

\subsection{Two-loop level}

\begin{figure}[t]
\centerline{\epsfbox{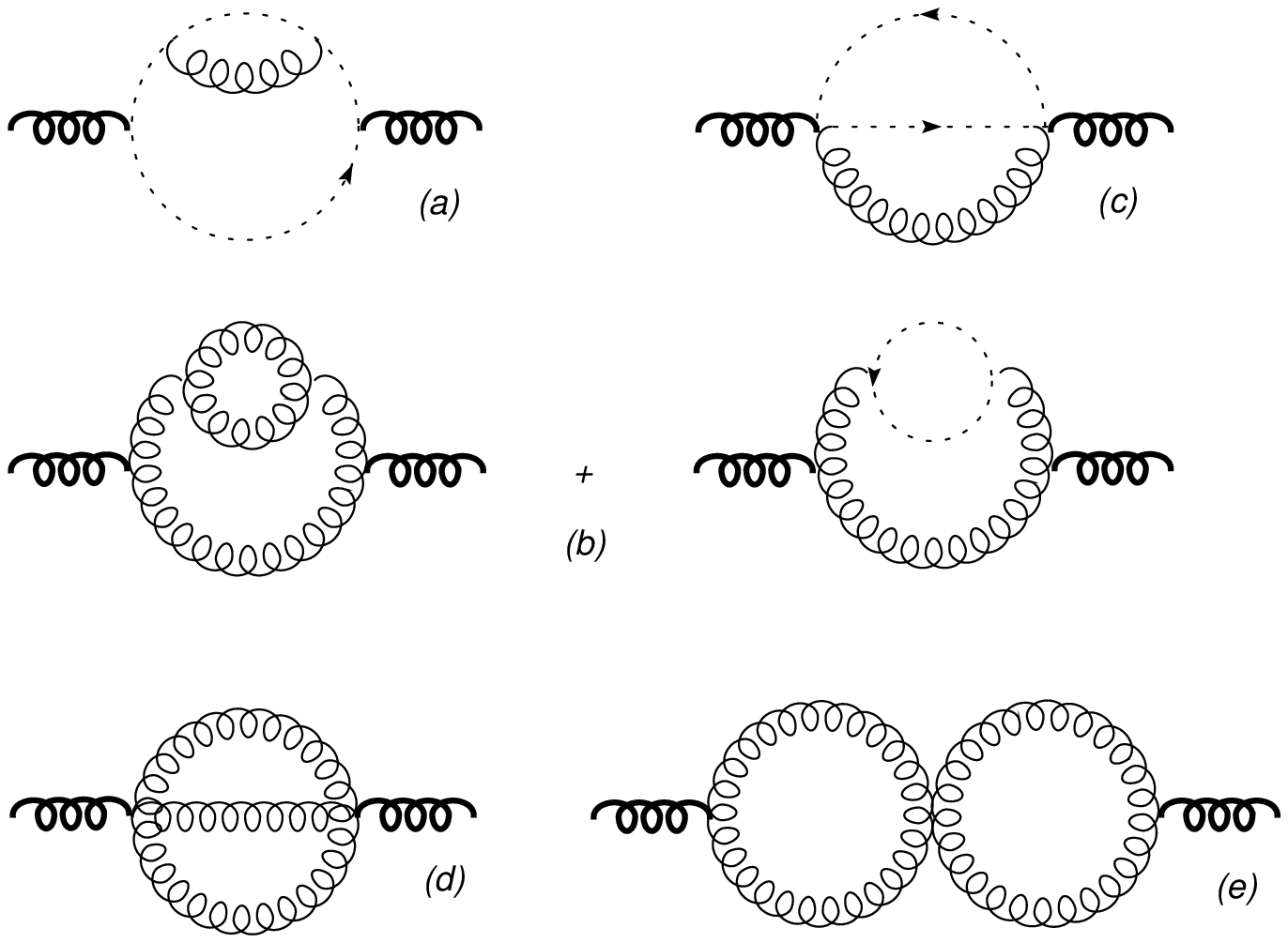}}
\caption{Two-loop YM diagrams (a)-(e).}
\label{2loop_1}
\end{figure}

\begin{figure}[t]
\centerline{\epsfbox{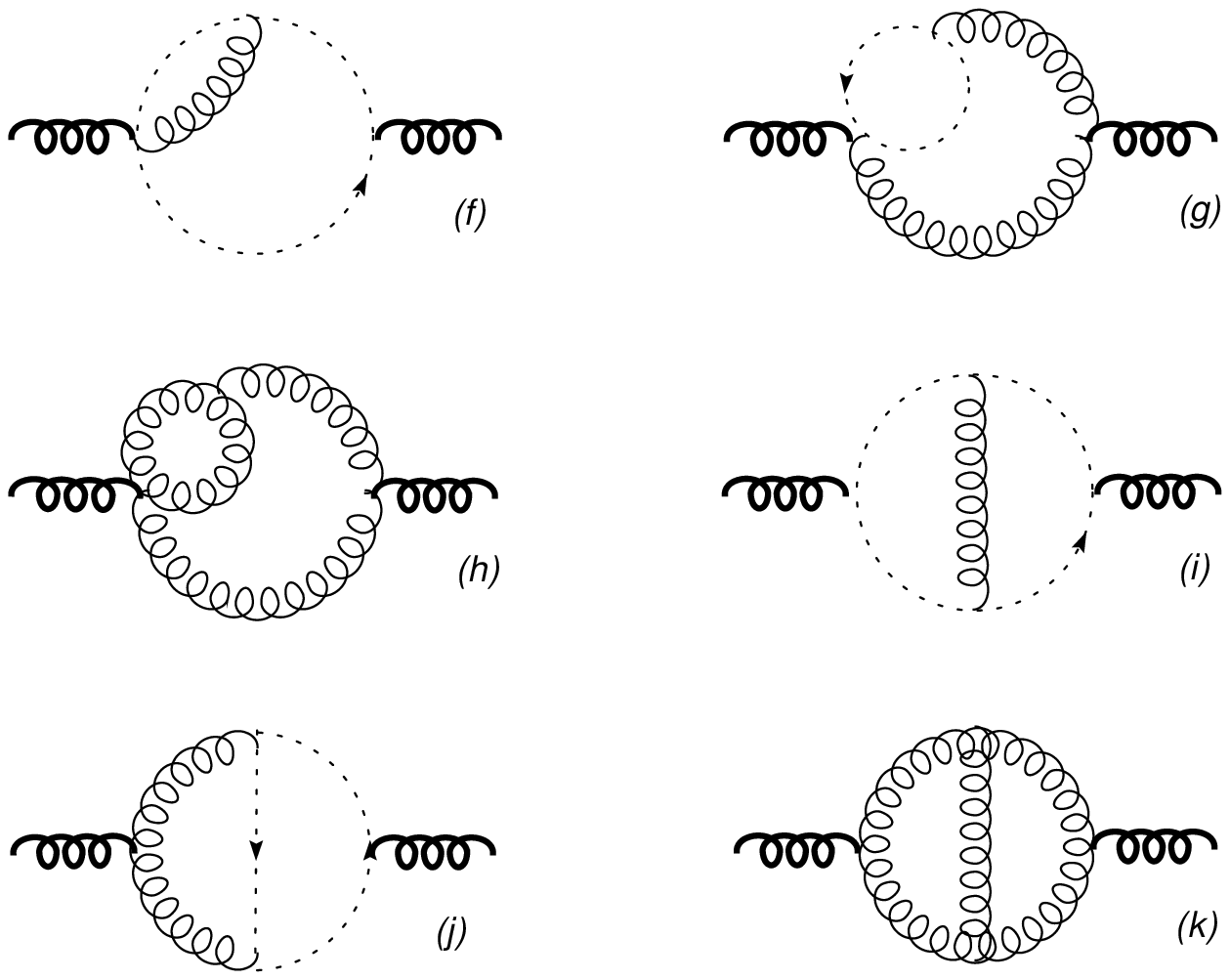}}
\caption{Two-loop YM diagrams (f)-(k).}
\label{2loop_2}
\end{figure}

Now we follow with the two-loop contribution to the background field self-energy. The relevant diagrams are those of figures \ref{2loop_1} and \ref{2loop_2} ((a) to (k)). Diagrams (a) to (h) have nested divergences, whereas diagrams (i), (j) and (k) have overlapping divergences.

\subsubsection{Diagram (a)}
This diagram has the following bare expression
\br
\Pi_{\m \n \; (2a)}^{BB\;ab} (x-y) &=& - 2 g^4 f^{aec} f^{bcd} f^{gdf} f^{gfe} \int  d^4 u d^4 v \; \D_{xy} ( \stackrel{\leftarrow}{\pa_{\m}^x} - \pa_{\m}^x) ( \pa_{\n}^y - \stackrel{\leftarrow}{\pa_{\n}^y}) \D_{yv}  \nonumber \\
& & \times ( \pa_{\l}^v \D_{uv}) \D_{uv} ( \pa_{\l}^u \D_{xu}) \;, \nonumber \\ 
\er
which can be rearranged in terms of the integral $I^1$ 
\br
\Pi_{\m \n \;(2a)}^{BB\;ab} (x) &=& - g^4 C_A^2 \d^{ab} \left[ 4 \pa_{\n} ( \D \pa_{\m} I^1 ) - \pa_{\m} \pa_{\n} ( \D I^1 ) - 4 \D \pa_{\m} \pa_{\n} I^1 \right] \;.
\er

In order to renormalize, we have to replace these expressions with their renormalized values, arriving to
\br
\left. \Pi_{\m \n\;(2a)}^{BB\;ab}(x)\right|_R &=& \frac{g^4 C_A^2 \d^{ab}}{32(4 \pi^2)^3} \left[ \pa_{\m} \pa_{\n} \Box \frac{ - \frac{1}{3} \ln^2 x^2 M^2 - \frac{8}{9} \ln x^2 M^2}{x^2} \right. \nonumber \\
& & \left. + \d_{\m \n} \Box \Box \frac{\frac{1}{3}  \ln^2 x^2 M^2 + \frac{11}{9} \ln x^2 M^2}{x^2} \right] + \ldots \nonumber \\
\er

\subsubsection{Diagram (b)}
This diagram is of the form
\br
\Pi_{\m \n \;(2b)}^{BB\;ab} (x-y) &=& g^2 f^{ace} f^{bed} \int d^4 u d^4 v \; \D_{xu} \left[ - 2 \d_{\m \l} D_{\rho}^x + \d_{\l \rho} ( \stackrel{\leftarrow}{\pa_{\m}^x} - \pa_{\m}^x) + 2 \d_{\m \rho} D_{\l}^x \right]  \nonumber \\
& & \times \Pi^{AA\;cd}_{\rho \s\;(1)} (u-v) \D_{vy} \left[ - 2 \d_{\n \s} D_{\l}^y + \d_{\s \l} ( \pa_{\n}^y - \stackrel{\leftarrow}{\pa_{\n}^y }) + 2 \d_{\n \l} D_{\s}^y \right] \D_{xy} \;, \nonumber \label{YM_2loop_diagb_bare}
\er
where $\Pi^{AA\;ab}_{\m \n\;(1)} (x-y) = \d^{ab} \Pi_{\m \n\;(1)}^{AA}(x-y)$ is the one-loop correction to the quantum gauge field propagator. Its bare and renormalized expressions are found in section \ref{ap_Gauge_YM} of appendix \ref{ap_Gauge}, where it is used to obtain the leading term of the expansion of the function that takes care of the running of the gauge parameter in the RG equation. It has to be noted again that, in contrast with dimensional regularization, the renormalized one-loop expression for the quantum gauge field propagator (\ref{YM1loop_quantum}) can not be used in the two-loop diagram. The reason is that the indices of the one-loop insertion the will be contracted in a second step, and one of the rules of CDR is to make first all the index contractions before performing the renormalization. Hence, only the bare one-loop contribution (\ref{YM1loop_quantum_bare_prop}) can be inserted. Therefore, expanding (\ref{YM_2loop_diagb_bare}) we find
\br
\Pi_{\m \n \; (2b)}^{BB\;ab}(x-y) =  - g^2 C_A \d^{ab} \int & d^4 u d^4 v&  - 4 \pa_{\m}^x \pa_{\rho}^x \left[ \D_{xu} \Pi^{AA}_{\rho \n\;(1)} (u-v) \D_{vy} \D_{xy} \right] \nonumber \\
& & + 4 \d_{\m \n} \pa_{\rho}^x \pa_{\s}^x \left[ \D_{xu} \Pi^{AA}_{\rho \s\;(1)} (u-v) \D_{vy} \D_{xy} \right] \nonumber \\
& & + \D_{xu} ( \stackrel{\leftarrow}{\pa_{\m}^x} - \pa_{\m}^x )\Pi^{AA}_{\rho \rho\;(1)} (u-v) \D_{vy} ( \pa_{\n}^y - \stackrel{\leftarrow}{\pa_{\n}^y} ) \D_{xy} \nonumber \\
& & + 4 \Box \left[ \D_{xu} \Pi^{AA}_{\m \n \;(1)} (u-v) \D_{vy} \D_{xy} \right] \nonumber \\
& & - 4 \pa_{\n}^x \pa_{\s}^x \left[ \D_{xu} \Pi^{AA}_{\m \s\;(1)} (u-v) \D_{vy} \D_{xy}  \right]. \nonumber 
\er

If we use the bare result (\ref{YM1loop_quantum_bare_prop}) for $\Pi^{AA}_{\m \n\;(1)}$, straightforward operations lead us to write this in terms of the previously defined $I^0$ and a new integral expression of the form 
\br
I_{\m \n}^0 (x-y)= \int d^4 u d^4 v \; \D_{xu} \D_{yv} ( \D_{uv} \pa_{\m}^u  \pa_{\n}^u  \D_{uv} ) \;. 
\er
So, we have
\br
\left.\Pi_{\m \n \;(2b)}^{BB\;ab} (x) \right|_R &=& g^4 C_A^2 \d^{b a} \left[ - 24 \pa_{\m} \pa_{\s} ( \D \pa_{\n} \pa_{\s} I^0) + 11 \pa_{\m} \pa_{\n} ( \D \Box I^0) + 32 \pa_{\m} \pa_{\s} ( \D I^0_{\n \s})  \right. \nonumber \\
& & + 12 \d_{\m \n} \pa_{\s} \pa_{\rho} ( \D \pa_{\rho} \pa_{\s} I^0) - 16 \d_{\m \n} \Box ( \D \Box I^0) - 16 \d_{\m \n} \pa_{\rho} \pa_{\s} ( \D I^0_{\rho \s} )   \nonumber \\
& & + \left. 20 \pa_{\m} ( \D \pa_{\n} \Box I^0) - 20 \D \pa_{\m} \pa_{\n} \Box I^0 + 12 \Box ( \D \pa_{\m} \pa_{\n} I^0) - 16 \Box (\D I_{\m \n}^0) \right]_R \;. \nonumber \\
\er

It is clear that, as $\Box I^0 = -I^1$, with the renormalized values found for $I^1$ we can obtain all the expressions made up with $\Box I^0$. In appendix \ref{ap_UV_IR} we study the renormalization of the rest of the expressions made up with $I^0$ and $I^0_{\m \n}$ that appear in this diagram. It is found there for $\D \pa_{\m} \pa_{\n} I^0$ and $\D I_{\m \n}^0$ the following renormalized values
\br
\left[ \D \pa_{\m} \pa_{\n} I^0 \right]_R (x) &=& \frac{1}{32(4 \pi^2)^3} \left[ \pa_{\m} \pa_{\n} \frac{ \ln x^2 M^2}{x^2} + \d_{\m \n} \Box \frac{ \frac{1}{4} \ln^2 x^2 M^2 + \frac{1}{4} \ln x^2 M^2}{x^2} \right] +~\ldots \nonumber \\
\left[ \D I_{\m \n}^0 \right]_R (x)&=& \frac{1}{32(4 \pi^2)^3} \left[ \pa_{\m} \pa_{\n} \frac{ \frac{1}{3} \ln x^2 M^2}{x^2} + \d_{\m \n} \Box \frac{ - \frac{1}{6} \ln x^2 M^2}{x^2} \right] + \ldots \nonumber \\
\er

With this, it is easy to arrive at the following renormalized expression
\br
\left.\Pi_{\m \n\;(2b)}^{BB\;ab}(x) \right|_R &=& \frac{g^4 C_A^2 \d^{a b}}{32 (4 \pi^2)^3} \left[ \pa_{\m} \pa_{\n} \Box \frac{- \frac{25}{3} \ln^2 x^2 M^2 - \frac{86}{9} \ln x^2 M^2}{x^2} \right. \nonumber \\
&+& \left.  \d_{\m \n} \Box \Box \frac{ \frac{25}{3} \ln^2 x^2 M^2 + \frac{71}{9} \ln x^2 M^2}{x^2} \right] + \ldots \nonumber \\
\er

\subsubsection{Diagram (c)}

This diagram is easily renormalized as its expression is
\br
\Pi_{\m \n\;(2c)}^{BB\;ab} (x) &=& -  g^4 f^{acx} f^{xed} f^{bdy} f^{yec} \d_{\m \n} \D^3  \nonumber \\
&=& - \frac{1}{2} g^4 C_A^2 \d^{ab} \d_{\m \n} \D^3  \nonumber \\
&\stackrel{R}{\rightarrow}& \frac{g^4 C_A^2 \d^{a b}}{32 (4 \pi^2)^3} \d_{\m \n} \Box \Box \frac{ \frac{1}{2} \ln x^2 M^2}{x^2} \;.
\er

\subsubsection{Diagram (d)}

This diagram is similar to the previous one, and we find
\br
\left. \Pi_{\m \n\;(2d)}^{BB\;ab} (x) \right|_R &=& \frac{9}{2} g^4 C_A^2 \d^{ab} \d_{\m \n} \D^3_R \nonumber \\
&=& \frac{g^4 C_A^2 \d^{a b}}{32 (4 \pi^2)^3} \d_{\m \n} \Box \Box \frac{ -\frac{9}{2} \ln x^2 M^2}{x^2} \;.
\er

\subsubsection{Diagram (e)}

The bare expression of this diagram is
\br
\Pi_{\m \n\;(2e)}^{BB\;ab}(x) &=& - \frac{1}{4} g^4 f^{acd} f^{bge} \int d^4 u \;  \D_{xu} [ - 2 \d_{\m \s} D_{\rho}^x + \d_{\rho \s} ( \stackrel{\leftarrow}{\pa_{\m}^x} - \pa_{\m}^x ) + 2 \d_{\m \rho} D^x_{\s} ]  \nonumber \\
& &\times  \D_{xu} [ f^{cex} f^{xgd} ( \d_{\rho \l} \d_{\eps  \s } - \d_{\rho \s} \d_{\eps \l} ) + f^{cgx} f^{xde} ( \d_{\rho \s} \d_{\eps \l} - \d_{\rho \eps} \d_{\s \l} )  \nonumber \\
& & +   f^{cdx} f^{xeg} (\d_{\rho \eps} \d_{\l \s} - \d_{\rho \l} \d_{\eps \s} ) ] \D_{yu} [ - 2 \d_{\n \eps} D_{\l}^y + \d_{\eps \l} ( \pa_{\n}^y - \stackrel{\leftarrow}{\pa_{\n}^y})   \nonumber \\
& & +  2 \d_{\n \l} D_{\eps}^y ] \D_{yu} \;,
\er
which, making all the index contractions can be written as
\br
\Pi_{\m \n \;(2e)}^{BB\;ab}(x-y)  &=& - 6 g^4 C_A^2 \d^{ab} ( \pa_{\m}^x \pa_{\n}^x - \d_{\m \n} \Box ) \int d^4 u \; \D^2_{xu} \D^2_{yu} \;. \nonumber  
\er

The renormalized expression of the integral is easily obtained as
\br
\int d^4 u \; \frac{1}{(x-u)^4} \frac{1}{u^4} \rightarrow - \frac{\pi^2}{4} \Box \frac{\ln^2 x^2 M^2}{x^2} \;, \nonumber 
\er
so that
\br
\left. \Pi_{\m \n\;(2e)}^{BB\;ab}(x) \right|_R &=& \frac{3}{8 (4 \pi^2)^3} g^4 C_A^2 \d^{ab} (\pa_{\m} \pa_{\n} - \d_{\m \n} \Box ) \Box \frac{ \ln^2 x^2 M^2}{x^2} + \ldots
\er

\subsubsection{Diagram (f)}

This diagram is of the following form
\br
\Pi_{\m \n \;(2f)}^{BB\;ab}(x-y) &=& 2 g^4 f^{acx} f^{xdf} f^{dce} f^{bef} \int d^4 u \; \D^2_{xu} ( \pa_{\m}^u \D_{uy} )( \pa_{\n}^y - \stackrel{\leftarrow}{\pa_{\n}^y} ) \D_{xy}  \nonumber \\
& &+ 2 g^4 f^{afc} f^{ecd} f^{bfx} f^{xed} \int d^4 u \; \D_{xu} ( \stackrel{\leftarrow}{\pa_{\m}^x} - \pa_{\m}^x ) \D_{xy} ( \pa_{\n}^u \D_{yu} ) \D_{uy} \;. \nonumber
\er

Operating, this can be written in terms of $I^1$, which allows us to write
\br
\left. \Pi_{\m \n \;(2f)}^{BB\;ab} (x) \right|_R &=& - g^4 C_A^2 \d^{ab} \left[ - 2 \pa_{\m} ( \D \pa_{\n} I^1 ) + 4 \D \pa_{\m} \pa_{\n} I^1 \right]_R \nonumber \\
&=& \frac{g^4 C_A^2 \d^{a b}}{32 (4 \pi^2)^3} \left[ \pa_{\m} \pa_{\n} \Box \frac{ \frac{1}{3} \ln^2 x^2 M^2 - \frac{1}{9} \ln x^2 M^2}{x^2}  \right. \nonumber \\
& &+ \left. \d_{\m \n} \Box \Box \frac{ - \frac{1}{3} \ln^2 x^2 M^2 - \frac{11}{9} \ln x^2 M^2}{x^2} \right] + \ldots \nonumber \\
\er

\subsubsection{Diagram (g)}
Contracting the indices of the bare expression
\br
\Pi_{\m \n\;(2g)}^{BB\;ab} (x-y) &=& - 2 g^4 f^{acx} f^{xfd} f^{ecd} f^{bfe} \int d^4 u \; \d_{\m \s} \D_{xu} ( \pa_{\l}^u \D_{xu} ) \D_{uy} \left[ - 2 \d_{\n \l} D_{\s}^y   \right. \nonumber \\
& &+ \left. \d_{\l \s} ( \pa_{\n}^y - \stackrel{\leftarrow}{\pa_{\n}^y} ) + 2 \d_{\n \s} D_{\l}^y \right] \D_{xy} 
\er
it is easy to write this diagram in terms of $I^1$, which implies that the renormalized form is
\br
\left. \Pi_{\m \n\;(2g)}^{BB\;ab}(x) \right|_R &=& - g^4 C_A^2 \d^{ab} \left[ \frac{3}{2} \pa_{\m} ( \D \pa_{\n} I^1) - \D \pa_{\m} \pa_{\n} I^1 - \d_{\m \n} \pa_{\l} ( \D \pa_{\l} I^1)  \right]_R \nonumber \\
&=& \frac{g^4 C_A^2 \d^{ab}}{32 (4 \pi^2)^3} \left[ \pa_{\m} \pa_{\n} \Box \frac{ \frac{5}{12} \ln^2 x^2 M^2 + \frac{19}{36} \ln x^2 M^2}{x^2}  \right. \nonumber \\
& &+ \left. \d_{\m \n} \Box \Box \frac{ - \frac{5}{12} \ln^2 x^2 M^2 - \frac{7}{36} \ln x^2 M^2}{x^2} \right] + \ldots \nonumber \\
\er

\subsubsection{Diagram (h)}
From the bare expression
\br
\Pi_{\m \n\;(2h)}^{BB\;ab}(x-y) = &- g^4 & \int d^4 u  \; \D^{(c)}_{xu} \left[ f^{acx} f^{xdf} ( \d_{\m \s} \d_{\rho \eps} - \d_{\m \eps} \d_{\rho \s} ) + f^{adx} f^{xfc}   \right. \nonumber \\ 
&\times& \left. ( \d_{\m \eps} \d_{\s \rho} - \d_{\m \rho} \d_{\eps \s} ) + f^{afx} f^{xcd} ( \d_{\m \rho} \d_{\s \eps} - \d_{\m \s} \d_{\rho \eps} ) \right]  \nonumber \\ 
&\times& \D^{(d)}_{xu} f^{edc} \left[ \d_{\l \s} ( \stackrel{e}{\pa_{\rho}^{u}}- \stackrel{d}{\pa_{\rho}^u} )+ \d_{\l \rho} ( \stackrel{c}{\pa_{\s}^u} - \stackrel{e}{\pa_{\s}^u} ) + \d_{\s \rho} ( \stackrel{d}{\pa_{\l}^u} - \stackrel{c}{\pa_{\l}^u} ) \right]  \nonumber \\ 
&\times& \D_{uy}^{(e)} f^{bfe} \left[ - 2 \d_{\n \l} D_{\eps}^y + \d_{\eps \l} ( \pa_{\n}^y - \stackrel{\leftarrow}{\pa_{\n}^y}) + 2 \d_{\n \eps} D_{\l}^y \right] \D_{xy} \;, \nonumber 
\er 
where $\D^{(i)} \stackrel{i}{\pa_{\m}} \D^{(j)} = ( \pa_{\m} \D^{(i)} ) \D^{(j)}$, evaluating all the index contractions we can also express this contribution in terms of $I^1$. Hence, the renormalized result is 
\br
\left. \Pi_{\m \n\;(2h)}^{BB\;ab} (x) \right|_R &=& - g^4 C_A^2 \d^{ab} \left[ \frac{45}{2} \pa_{\m} ( \D \pa_{\n} I^1 ) - 27 \D \pa_{\m} \pa_{\n} I^1 - 9 \d_{\m \n} \pa_{\l} ( \D \pa_{\l} I^1 ) \right]_R \nonumber \\
&=& \frac{g^4 C_A^2 \d^{a b}}{32 (4 \pi^2)^3} \left[ \pa_{\m} \pa_{\n} \Box \frac{ \frac{9}{4} \ln^2 x^2 M^2 + \frac{21}{4} \ln x^2 M^2}{x^2}  \right. \nonumber \\
& &+ \left. \d_{\m \n} \Box \Box \frac{ - \frac{9}{4} \ln^2 x^2 M^2 + \frac{15}{4} \ln x^2 M^2}{x^2} \right] + \ldots
\er

\subsubsection{Diagram (i)}

This diagram and the two following ones have overlapping divergences. In order to renormalize them, we will make use of the list of integrals obtained in section \ref{overlap_integrals}. 

The bare expression for this diagram which is
\br
\Pi_{\m \n\; (2i)}^{BB\;ab}(x-y) &=& - g^4 f^{afc} f^{gcd} f^{bde} f^{gef} \int d^4 u d^4 v \; \D_{xu} ( \stackrel{\leftarrow}{\pa_{\m}^x} - \pa_{\m}^x ) ( \pa_{\l}^u \D_{uy})  \nonumber \\
&\times& ( \pa_{\n}^y - \stackrel{\leftarrow}{\pa_{\n}^y}) \D_{yv} ( \pa_{\l}^v \D_{vx} ) \D_{uv} \;.
\er

Expanding this expression, we can write it in terms of the $H$ integrals defined in (\ref{H_definition}). The bare contribution is then found to be
\br
\Pi_{\m \n\; (2i)}^{BB\;ab}(x-y)  = - \frac{1}{2} g^4 C_A^2 \d^{ab} &\left[ \right.& \pa_{\m}^x \pa_{\n}^y H[1, \pa_{\l} \; ; \; \pa_{\l},1] - 2 \pa_{\m}^x H[1, \pa_{\l} \; ; \; \pa_{\l} \pa_{\n} , 1]  \nonumber \\
& & \left. - 2 \pa_{\n}^y H[1, \pa_{\l} \pa_{\m} \; ; \; \pa_{\l},1] + 4 H[ 1, \pa_{\m} \pa_{\l} \; ; \; \pa_{\n} \pa_{\l} , 1] \; \right] \;. \nonumber \\
\er

At this point, we have to straightforwardly use the list of overlapping divergences of section \ref{overlap_integrals}. In concrete, with (\ref{int3}), (\ref{int6}), and (\ref{int15}) we find  the renormalized result to be
\br
\left. \Pi_{\m \n\; (2i)}^{BB\;ab}(x) \right|_R &=& \frac{g^4 C_A^2 \d^{a b}}{32 (4 \pi^2)^3} \left[ \pa_{\m} \pa_{\n} \Box \frac{ - \frac{1}{12} \ln^2 x^2 M^2 - \frac{17}{36} \ln x^2 M^2 }{x^2} \right.  \nonumber \\
& &+ \left. \d_{\m \n} \Box \Box \frac{ \frac{1}{12} \ln^2 x^2 M^2 + \frac{29}{36} \ln x^2 M^2}{x^2} \right] + \ldots
\er

\subsubsection{Diagram (j)}
The basic form of this diagram is 
\br
\Pi_{\m \n\; (2j)}^{BB\;ab}(x-y) &=& 2 g^4 f^{acf} f^{cgd} f^{bde} f^{feg} \int d^4 u d^4 v \; \D_{xu} \left[ - 2 \d_{\m \s} D_{\rho}^x + \d_{\rho \s} ( \stackrel{\leftarrow}{\pa_{\m}^x} - \pa_{\m}^x )  \right. \nonumber \\
& & + \left. 2 \d_{\m \rho} D_{\s}^x \right] ( \pa_{\rho}^u \D_{uy} ) ( \pa_{\n}^y - \stackrel{\leftarrow}{\pa_{\n}^y} ) \D_{yv} ( \pa_{\s}^v \D_{uv}) \D_{vx} \;.
\er

Evaluating the index contractions this becomes
\br
\Pi_{\m \n\; (2j)}^{BB\;ab}(x-y) = - g^4 C_A^2 \d^{ab} &\left[ \right.& - 4 \pa_{\n}^x \pa_{\l}^x H[1, \pa_{\m} \pa_{\l} \; ; \; 1 , 1] - 4 \pa_{\n}^x \pa_{\l}^x H[ 1, \pa_{\l} \; ; \; 1, \pa_{\m} ]  \nonumber \\
& & - 4 \pa_{\l}^x H[ 1, \pa_{\m} \pa_{\l} \; ; \; 1, \pa_{\n} ] - 4 \pa_{\l}^x H[ 1, \pa_{\l} \; ; \; 1, \pa_{\m} \pa_{\n} ]  \nonumber \\
& & - 4 \pa_{\n}^x H[ 1, \pa_{\m} \pa_{\l} \; ; \; 1, \pa_{\l} ] + \pa_{\m}^x \pa_{\n}^x H[ 1, \pa_{\l} \; ; \; 1, \pa_{\l}]  \nonumber \\
& & - 4 H[ 1, \pa_{\m} \pa_{\l} \; ; \; 1, \pa_{\n} \pa_{\l} ] + 4 \Box H[ 1, \pa_{\m} \; ; \; 1, \pa_{\n} ]  \nonumber \\
& & + 4 \Box H[ 1, 1 \; ; \; 1, \pa_{\m} \pa_{\n} ] - 2 \pa_{\n}^x H[ 1, \pa_{\m} \; ; \; 1, \Box]  \nonumber \\
& & + \pa_{\m}^x \pa_{\n}^x H[1, 1 \; ; \; 1, \Box] - 4 H[1, \pa_{\n} \Box \; ; \; 1, \pa_{\m} ]  \nonumber \\
& & \left. - 2 \pa_{\m}^x H[ 1, \pa_{\n} \Box \; ; \; 1, 1] \; \right] \;.
\er

From this list, the expressions that have a $\Box$ (remember $\Box \D (x) = -\d(x)$) can be written in terms of the integral $I^1$, and their renormalization is straightforward. The rest of the integrals can be found in the list of integrals with overlapping divergences. So, we arrive at the following renormalized result 
\br
\left. \Pi_{\m \n\; (2j)}^{BB\;ab}(x) \right|_R &=& \frac{g^4 C_A^2 \d^{a b}}{32(4 \pi^2)^3} \left[ \pa_{\m} \pa_{\n} \Box \frac{ \frac{1}{2} \ln^2 x^2 M^2 + \frac{1}{2} \ln x^2 M^2}{x^2}  \right. \nonumber \\
& &+ \left. \d_{\m \n} \Box \Box \frac{ - \frac{1}{2} \ln^2 x^2 M^2 - \frac{1}{2} \ln x^2 M^2}{x^2} \right] + \ldots
\er

\subsubsection{Diagram (k)}

In order to obtain all the contributions that form this diagram, the Mathematica package 'FeynCalc' was used, so that all the index contractions were performed by the computer. The output of this process are the final relevant expressions that need to be renormalized. The contributions shown here are those that have a divergent part, omitting those terms that are finite.
\br
\Pi_{\m \n\; (2k)}^{BB\;ab}(x-y) = \frac{1}{4} C_A^2 \d^{ab} &\left[\right.& + 16 \d_{\m \n} \pa_{\l}^x \pa_{\s}^x H[1,\pa_{\l} \pa_{\s} \; ; \; 1,1] - 20 \pa_{\n}^x \pa_{\l}^x H[1, \pa_{\m} \pa_{\l} \; ; \; 1,1]  \nonumber \\
& & - 124 \pa_{\n}^x \pa_{\l}^x H [ 1, \pa_{\l} \; ; \; 1, \pa_{\m}] + 72 \pa_{\l}^x H[1, \pa_{\m} \pa_{\l} \; ; \; 1, \pa_{\n}]  \nonumber \\
& & + 56 \d_{\m \n} \pa_{\l}^x \pa_{\s}^x H[1, \pa_{\l} \; ; \; 1, \pa_{\s}] - 72 \pa_{\n}^x H[ 1, \pa_{\m} \Box \; ; \; 1,1]  \nonumber \\
& & + 20 \pa_{\m}^x \pa_{\n}^x H [ 1 , \Box \; ; \; 1, 1] - 144 H[ 1, \pa_{\m} \Box \; ; \; 1, \pa_{\n} ]  \nonumber \\
& & + 72 \pa_{\m}^x H[1, \Box \; ; \; 1, \pa_{\n}] - 72 \pa_{\n}^x H[1, \pa_{\m} \pa_{\l} \; ; \; 1, \pa_{\l}]  \nonumber \\
& & + 34 \pa_{\m}^x \pa_{\n}^x H[ 1, \pa_{\l} \; ; \; 1, \pa_{\l} ] - 72 H[ 1, \pa_{\m} \pa_{\l} \; ; \; 1, \pa_{\n} \pa_{\l} ]  \nonumber \\
& & + 40 \Box H [ 1, \pa_{\m} \pa_{\n} \; ; \; 1, 1] + 32 \Box H [ 1, \pa_{\m} \; ; \; 1, \pa_{\n}]  \nonumber \\
& & \left. + 16 \d_{\m \n} \Box H[ 1, \Box \; ; \; 1, 1] - 16 \d_{\m \n} \Box H[ 1, \pa_{\l} \; ; \; 1, \pa_{\l}] \; \right]  \nonumber \\
& & +~\textrm{(finite~terms)} \;.
\er

Proceeding in the same way as in the two previous diagrams, we can easily found the renormalized form of this contribution to be 
\br
\left. \Pi_{\m \n\; (2k)}^{BB\;ab}(x) \right|_R  &=& \frac{g^4 C_A^2 \d^{ab}}{32 (4 \pi^2)^3} \left[ \pa_{\m} \pa_{\n} \Box \frac{- \frac{27}{4} \ln^2 x^2 M^2 - \frac{45}{4} \ln x^2 M^2}{x^2}  \right. \nonumber \\
& & \left. + \d_{\m \n} \Box \Box \frac{ \frac{27}{4} \ln^2 x^2 M^2 + \frac{33}{4} \ln x^2 M^2}{x^2} \right] + \ldots \nonumber \\
\er
\subsubsection{Two-loop final results}
In order to obtain the total two-loop renormalized contribution to the background gauge field self-energy, we only have to add all the renormalized expressions for the diagrams that we have obtained. So, we have
\br
\left. \Pi_{\m \n\;(2)}^{BB\;ab} (x) \right|_R &=& - \frac{g^4 C_A^2 \d^{ab}}{2(4 \pi^2)^3} ( \pa_{\m} \pa_{\n} - \d_{\m \n} \Box ) \Box \frac{ \ln x^2 M^2}{x^2} + \ldots \label{2_loop}
\er

By fulfilling one-loop CDR rules, we have fixed the renormalization scheme {\em{a priori}}, which implies that one-loop local terms have defined values. Hence, as is imposed by gauge invariance, we have obtained directly a transverse result. 

\subsection{RG equation}
\label{sec_RG}

With the previously obtained expressions for the one- and two-loop corrections of the background gauge field propagator, we can obtain the first two coefficients of the expansion of the beta function of this theory. If we define 
\begin{equation}
\G^{BB \; ab}_{\m \n} (x) = (\pa_{\m} \pa_{\n} - \d_{\m \n} \Box)\d^{ab} \G^{(2)} (x) \;,
\end{equation}
then, with the one-loop contribution (\ref{1_loop}), the gauge fixing renormalization (\ref{gauge_fix_ren}) and the two-loop contribution (\ref{2_loop}), the effective action for the background gauge fields is
\br
\G^{(2)}(x) &=& \frac{1}{g^2} \d (x) + \frac{11 C_A}{48 \pi^2 (4 \pi^2)}\Box \frac{\ln x^2 M^2}{x^2} + \frac{ C_A}{72 \pi^2} \d (x) +  \frac{\xi C_A}{8 \pi^2} \d (x)  \nonumber \\
& &+ \frac{g^2 C_A^2}{2 (4 \pi^2)^3} \Box \frac{ \ln x^2 M^2}{x^2} + \ldots \label{YM_2loop_eff_action}
\er

With this definition, the equation we need to consider is
\br
\left[ M \frac{\pa}{\pa M} + \b (g) \frac{\pa}{\pa g} + \gamma_{\xi} \frac{\pa}{\pa \xi} - 2 \g_{B} \right] \G^{(2)}|_{\xi=0} =0 \;. \label{YM_RG_eq}
\er

Notice that $\g_{\xi}$ is the coefficient that takes care of the running of the gauge parameter. The expansion of this function to order $g^2$ is obtained in appendix \ref{ap_Gauge}. To do so, we consider there the one-loop RG equation for quantum gauge fields. With this, we find for $\g_{\xi}$ 
\br
\g_{\xi} &=& - \frac{5 C_A}{24 \pi^2} g^2 + \cdots \label{g_xi}
\er

Also notice that if the background gauge field is redefined as $B^{\prime} = g B$, this implies $\gamma_{B} =0$ (the charge and background field renormalizations are related: $Z_{g} = Z_{B}^{-1/2}$). So, with (\ref{YM_2loop_eff_action}), (\ref{YM_RG_eq}) and (\ref{g_xi}), we evaluate the first two coefficients of the expansion of the beta function to be
\br
\b (g) &=& \b_1 g^3 + \b_2 g^5 + {\cal{O}}(g^7) \nonumber \\
\b_1 &=& - \frac{11 C_A}{48 \pi^2} \nonumber \\
\b_2 &=& - \frac{17 C^2_A}{24 (4 \pi^2)^2} \;.
\er

These results agree with those previously obtained in the literature \cite{Caswell:1974gg,Jones:1974mm,Abbott:1980hw,Morris:2005tv}.

\section{$N=1$ Super Yang-Mills}
In this section we consider the two-loop differential renormalization of the supersymmetric extension of the previous model, $N=1$ Super Yang-Mills \cite{Ferrara:1974pu,Salam:1974ig}. With this calculation we revisit an old controversy: the origin of higher-order perturbative contributions to the beta function in supersymmetric gauge theories \cite{Novikov:1983uc,Grisaru:1985tc,Shifman:1986zi,Arkani-Hamed:1997mj,Shifman:1999kf}. Differential renormalization has one important advantage over usual renormalization methods (as dimensional reduction), as in this case we have UV and IR divergences. With dimensional methods both renormalizations mix (we need to subtract the IR part in the final result), but differential renormalization clearly distinguishes between UV and IR divergences as both are renormalized with independent scales.

\subsection{$N=1$ Super Yang-Mills model}
As is detailed in appendix \ref{ap_SUSY}, in order to formulate a supersymmetric gauge theory we can follow two different approaches. In the first one, named chiral representation, we begin by considering a multiplet of unconstrained gauge superfields ($V = V^A T_A$, with $T_A$ the group generators), which are a generalization of the results found after studying the off-shell representations of the linear free theory. We use these gauge superfields to construct covariant derivatives $\nabla_A^c =( e^{-gV} D_{\a} e^gV, \bar{D}_{\dot{\a}}, - i \anticomm{\nabla^c_{\a}}{\nabla^c_{\dot{\a}}})$ that allow us to obtain gauge invariant expressions. On the other hand, with the second approach, called vector representation, we begin by considering covariant derivatives (termed $\nabla_A^v =( \nabla^v_{\a}, \bar{\nabla}^v_{\dot{\a}},\nabla^v_{\a \dot{\a}})$) that, after imposing covariant constraints on them, can be expressed in terms of prepotentials. With both approaches we find field strengths defined in terms of an spinorial field. In particular, for chiral representation we have 
\br
W_{\a} &=& i \bar{D}^2( e^{-gV} D_{\a} e^gV) \nonumber \\
W_{\dot{\a}} &=& e^{-gV} \bar{W}_{\dot{\a}} e^gV = e^{-gV} ( - W_{\a})^{+} e^{gV} \;,
\er
which allow us to define a gauge invariant action as
\br
S_0 &=& \frac{1}{g^2} tr \int d^4 x d^2 \th \; W^2 = - \frac{1}{2 g^2} tr \int d^4 x d^4 \th \; ( e^{-gV} D^{\a} e^gV) \bar{D}^2 ( e^{-gV} D_{\a} e^{gV}) \;.
\er

For vector representation, the field strength is defined as 
\br
-i C_{\dot{\b} \dot{\a}} W_{\b} &=& \comm{\bar{\nabla}^v_{\dot{\a}}}{i \nabla^v_{\b \dot{\b}}} ~,~\textrm{with}~\anticomm{\nabla^v_{\a}}{\bar{\nabla}^v_{\dot{\b}}} = i \nabla^v_{\a \dot{\b}} \;,
\er
and with this we write the gauge action as
\br
S_0 = \frac{1}{g^2} tr \int d^4 x d^2 \th \; W^2 = \frac{1}{2 g^2} tr \int d^4 x d^2 \th \; \left( \frac{1}{2} \comm{\bar{\nabla}^{v \; \dot{\a}}}{\anticomm{\bar{\nabla}^v_{\dot{\a}}}{\nabla^v_{\a}}} \right)^2 \;.
\er

Quantization of a supersymmetric gauge theory is also discussed in appendix \ref{ap_SUSY}. We have to add to the action a gauge-fixing term that depends on a gauge parameter $\a$ ($S_{GF}$) and anticommuting chiral ghost fields $c$, $c^{\prime}$ ($S_{FP}$). Their explicit expressions are 
\br
S_{GF} &=& - \frac{1}{\a} tr \int d^8 z \; ( D^2 V ) ( \bar{D}^2 V ) \nonumber \\
S_{FP} &=& tr \int d^4 x d^4 \th \; ( c^{\prime} + \bar{c}^{\prime} ) L_{\frac{1}{2} gV} \left[ ( c + \bar{c})+coth L_{\frac{1}{2} g V} (c - \bar{c}) \right] ~~,~~ L_X Y = \comm{X}{Y} \;. \nonumber \\
\er

One of the relevant features of the supergraph techniques applied to this theory is the appearance, along with the usual on-shell infrared divergences of Yang-Mills theory, of additional infrared divergences due to the form of the gauge propagator in a general covariant gauge \cite{Juer:1982mp,Clark:1977pq,Piguet:1981hh,Howe:1984xq,Abbott:1984pz}. This can be clearly seen if we consider the expression for this propagator in momentum space \cite{Abbott:1984pz}
\br
\D (k) &=& \frac{1 + (1 - \a)( D^2 \bar{D}^2 + \bar{D}^2 D^2) k^{-2}}{k^2} \d^4 (\th - \th^{\prime}) \;.
\er

As the leading term in $\bar{D}^2D^2 + D^2 \bar{D}^2$ is constant when $k^2 \rightarrow 0$, this propagator goes as $1/k^4$ at small $k$, which is the origin of the infrared divergences. Although Feynman gauge ($\a = 1$) seems to be the solution, this is not the case: the one-loop correction of the gauge propagator takes us out of the Feynman gauge; hence, when we consider two-loop diagrams that have as an insertion the one-loop corrected propagator, the infrared divergences reappear \cite{Abbott:1984pz}. With dimensional methods this situation represents a severe problem, as both UV and IR divergences are renormalized with the same dimensional parameter $\eps$ ($d = 4 - 2 \eps$). So, we have to subtract the IR contribution\footnote{In \cite{Abbott:1984pz} this was achieved at two-loops by choosing a non-local gauge-fixing term that cancels exactly the contributions that takes us out of Feynman gauge. In \cite{Grisaru:1985tc} the procedure was to define a $\tilde{R}-$operation \cite{Chetyrkin:1982nn}.}. However, life gets simpler if we use differential renormalization. As we have seen in section \ref{IR_divergences}, with differential renormalization UV and IR divergences are renormalized with different and independent scales.

\subsubsection{Background field method}
We discuss the application of the background field method to supersymmetric gauge theories in appendix \ref{ap_BFM}. As is explained there, due to the non-linear gauge transformations of SYM, a linear quantum-background splitting is unsuitable. The accurate spitting is achieved if we replace $e^{gV}$ with
\br
e^{g V_{(split)}} = e^{\boldsymbol{\Omega}} e^{g V} e^{\bar{\boldsymbol{\Omega}}} \;,
\er   
where $V$ is the quantum gauge superfield and ${\boldsymbol{\Omega}}$ is the background prepotential. It is worthwhile to mention that we have redefined the usual background field $B$ ($\bOmega = \bar{\bOmega} = \frac{1}{2} B$) as $g B \rightarrow B$. This splitting implies that we write the covariant derivatives in a quantum-chiral but background-vector representation as
\br
\nabla_{\a} = e^{- g V} \bnabla_{\a} e^{g V} ~~,~~ \bar{\nabla}_{\dot{\a}} = \bar{\bnabla}_{\dot{\a}} ~~,~~ \nabla_{\a \dot{\a}} = -i \anticomm{ \nabla_{\a}}{\bar{\nabla}_{\dot{\a}}} \;,
\er
where $\bnabla_{\a}$ and $\bar{\bnabla}_{\dot{\a}}$ are the background covariant derivatives. Relative to the quantization of the theory, we remark one of the particular features of SuperYM: the appearance of the Nielsen-Kallosh ghost \cite{Nielsen:1978mp}. In the usual quantization procedure we gauge-average with a simple exponential factor; however, if we use a more complicated function e.g., $ exp \int f M f$ with $M$ an operator, we have to normalize the procedure dividing by $detM$. As we have this situation when we use background chiral superfields, we introduce a new ghost field $b$ (Nielsen-Kallosh ghost), which have opposite statistics to $f$ and allow us to properly normalize the gauge-averaging procedure. As this new field does not interact with the quantum fields and enters quadratically in the action, we find that it only contributes at the one-loop level. Hence, all the relevant terms that form the split Super Yang-Mills action are
\br
S_{YM} &=& - \frac{1}{2 g^2} tr \int d^4 x d^4 \th \; ( e^{-g V} \bnabla^{\a} e^{g V}) \bar{\bnabla}^2 ( e^{- g V} \bnabla_{\a} e^{g V} ) \nonumber \\
S_{GF} &=& - ( 1 + \xi) tr \int d^4 x d^4 \th ( \bnabla^2 V ) ( \bar{\bnabla}^2 V ) \nonumber \\
S_{FP} &=& tr \int d^4 x d^4 \th \; \left[ \bar{c}^{\prime} c - c^{\prime} \bar{c} + \frac{1}{2} ( c^{\prime} + \bar{c}^{\prime} ) \comm{gV}{c+\bar{c}} + \ldots\right] \nonumber \\
S_{NK} &=& (1 + \xi) tr \int d^4 x d^4 \th \; \bar{b} b  \;. \label{SYM_BFM_actions}
\er
Notice that we have redefined the usual gauge parameter $\a$ as $\frac{1}{\a} = 1 + \xi$. Hence, as is discussed in appendix \ref{ap_BFM}, we have a background effective action of the form
\br
\G[B] &=& tr \int d^4 x d^4 y d^2 \th \; \left[ \bW^{\a} (x, \th) \bW_{\a} (y, \th) \right] \G^{(2)}_B (x-y) + \ldots \nonumber \\
&=& S_0[B] +  \G_{\xi} + \nonumber \\
& & + exp S_{int} \left[\frac{\d}{\d J}, \frac{\d}{\d j}, \frac{\d}{\d \bar{j}} \right] exp \left[ \frac{1}{2} J \hat{\dal}^{-1} J - \bar{j} \Box_{+}^{-1} j \right]_{J=j= \bar{j} =0} \;,
\er
where $S_0[B]$ is the ``free'' part of the background action, $\G_{\xi}$ stands for the one-loop contribution in the gauge $\xi$ and $J$, $j$ and $\bar{j}$ are the sources. $\Box_{+}$ and $\hat{\dal}$ are operators defined in appendix \ref{ap_BFM} as
\br
\hat{\dal} &=& \bfBox - i \bW^{\a} \bnabla_{\a} - i \bar{\bW}^{\dot{\a}} \bar{\bnabla}_{\dot{\a}} \nonumber \\
\Box_{+} &=& \Box - i \bW^{\a} \bnabla_{\a} - \frac{i}{2} ( \bnabla^{\a} \bW_{\a} )
\er

To apply supergraph techniques to this theory, we can follow two procedures: In the first one we expand all the background covariant derivatives in terms of the explicit background connections $\bnabla_{\a} = D_{\a} - i \boldsymbol{\G}_{\a}$, so that we can employ usual D-algebra. This procedure is applied in \cite{Abbott:1984pz}, and we will use it in the one-loop level. However, another approach was considered in \cite{Grisaru:1984ja,Grisaru:1984jc}. In this case the spinorial background connection is not explicitly extracted from the background covariant derivative, and covariant D-algebra is used in the diagrams. At the end, we have all the diagrams expressed in terms of the background space-time connections $\boldsymbol{\G}_{\a \dot{\a}}$ or field strengths $\boldsymbol{W}_{\a}$. Thus, as we do not have explicit spinor connections $\boldsymbol{\G}_{\a}$ (which are of lower dimension), the diagrams are more convergent and fewer in number. For the two-loop calculations we will follow this procedure. 

\subsection{One-loop level}

We will proceed with the one-loop case. As the previous examples considered, we will obtain the one- and two-loop corrections evaluated in Feynman gauge (with our conventions, $\xi = 0$). Therefore, at this level we have not only to consider the background gauge field self-energy correction, but as we will have to take care of the running of the gauge parameter in the RG equation ($\g_{\xi} \pa / \pa \xi$), we will obtain the linear contribution in $\xi$ of the one-loop background gauge field two-point function.
 
\subsubsection{Background gauge field self-energy}

\begin{figure}[h]
\centerline{\epsfbox{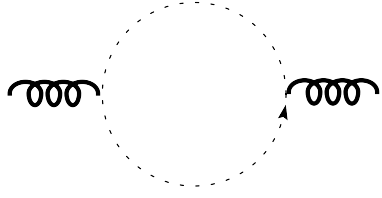}}
\caption{One-loop background gauge field two-point function contribution. Thick lines correspond to external background fields and thin lines represent ghost propagators.}
\label{SYM_1loop}
\end{figure}

To obtain the one-loop contribution, we begin expressing the covariantly chiral ghost fields in terms of ordinary chiral fields by
\br
c \rightarrow e^{ B /2}\; c \;e^{- B /2} \;,
\er
where $B$ is the background gauge field. Hence, the one-loop relevant interaction terms are (see appendix \ref{ap_BFM})
\br
tr \int d^4 x d^4 \th \; \left[ - \frac{1}{2} V \left( \bfBox - \bW^{\a} \bnabla_{\a} - \bar{\bW}^{\dot{\a}} \bar{\bnabla}_{\dot{\a}} \right) V  + \bar{c}^{\prime} B c + c^{\prime} B \bar{c} + \bar{b} B b \right] \;.
\er

Note that in the previous expression, from the part which corresponds to the interaction between the quantum and background gauge superfields, it is clear that we do not have enough covariant derivatives to obtain a non-vanishing contribution to the background gauge field two-point function (remember that at least we need two $\bar{D}^2$ and two $D^2$). Hence, we conclude that the only non-vanishing contribution comes from ghost superfields (as they are chiral superfields, we have at the vertices additional superspace covariant derivatives \cite{Gates:1983nr}). With the definition of the superspace propagator $P_{ij} = \d_{ij} \D_{ij}$, the total contribution of the ghost superfields is straightforwardly evaluated as \cite{Song,Gates:1983nr}
\br
\G^{1\;loop} &=& - \frac{3 C_A \d^{ab}}{2} \int d^8 z_1 d^8 z_2 \; B^a (z_1) B^{b} (z_2) \left[ D^2_1 P_{12} \stackrel{\leftarrow}{\bar{D}^2_2} \right] \left[ D^2_2 P_{12} \stackrel{\leftarrow}{\bar{D}^2_1} \right] \nonumber \\
&=& - \frac{3 C_A}{2} \int d^8 z_1 d^8 z_2 B^a(z_1) B^a(z_2) \left[ \bar{D}^2_2 D^2_2 P_{12} \right] \left[ D^2_2 \bar{D}^2_2 P_{12} \right] \;.
\er
Applying the identity (\ref{D_algebra_id}), we can write this expression as
\br
\G^{1\;loop} &=& - \frac{3 C_A}{2} \int d^8 z_1 d^8 z_2 \; B^a(z_1) \left[ \bar{D}^2 D^2 B^a(z_2) \right] P_{12} \left[ \bar{D}^2_2 D^2_2 P_{12} \right]  \nonumber \\
& & - \frac{3 C_A}{2} \int d^8 z_1 d^8 z_2 \; B^a(z_1) B^a(z_2) P_{12} \left[ \Box \bar{D}^2_2 D^2_2 P_{12} \right]  \nonumber \\
& & + \frac{i 3 C_A}{2} \int d^8 z_1 d^8 z_2 \; B(z_1) \left[ \bar{D}^{\dot{\a}} D^{\a} B^a(z_2) \right] P_{12} \left[ \pa_{\a \dot{\a}}^2 \bar{D}^2_2 D^2_2 P_{12} \right] \;.
\er

And finally, with the usual superspace $\d$-function property $\d_{12} \bar{D}^2_2 D^2_2 \d_{12} = \d_{12}$ and the identifications $x_1 = x$, $x_2 = y$ we arrive at
\br
\G^{1 \;loop} &=& - \frac{3 C_A}{2} \int d^4 x d^4 y d^4 \th \; B^a(x, \th) \left[ \bar{D}^2 D^2 B^a (y, \th) \right] \D^2_{xy}  \nonumber \\
& & - \frac{3 C_A}{2} \int d^4 x d^4 y d^4 \th \; B^a(x, \th) B^a (y, \th) \D_{xy} \Box \D_{xy}   \nonumber \\
& & + \frac{i  3 C_A}{2} \int d^4 x d^4 y d^4 \th \; B^a(x, \th) \left[ \bar{D}^{\dot{\a}} D^{\a} B^a (y, \th) \right] \D_{xy} \pa_{\a \dot{\a}}^y \D_{xy} \;.
\er

Applying CDR rules, we find the renormalized form of this expression to be
\br
\G^{1\;loop} &=& \frac{ 3 C_A}{16 (4 \pi^2)^2} \int d^4 x d^4 y d^4 \th \; B^a(x,\th) \left[ D^{\a} \bar{D}^2 D_{\a} B^a (y, \th) \right] \Box \frac{ \ln (x-y)^2 M^2}{(x-y)^2} \;, \nonumber \\ \label{SYM_1loop_eff_action_background}
\er
where we have used the superspace derivatives identity $\bar{D}^2 D^2 + (i/2) \pa_{\a \dot{\a}} \bar{D}^{\dot{\a}} D^{\a} = 1/2 D^{\a} \bar{D}^2 D_{\a}$.

\subsubsection{Effective action in a generic gauge}
Now we proceed with the additional result that we have to obtain in order to deal with the running of the gauge parameter: the contribution to the background effective action of quantum gauge fields evaluated in a generic gauge. We have to follow a procedure similar to that used in the Yang-Mills case. First, we have to consider the one-loop effective action contribution at second order in background gauge fields evaluated in a generic gauge. Then, we have to expand this in terms of the gauge parameter $\xi$, retaining the linear part. The reason for this is the same as in the non-supersymmetric case: in the background gauge field RG equation, after considering the term that takes care of the running of the gauge parameter $\g_{\xi} (g) \pa/ \pa \xi$, we will impose Feynman gauge ($\xi = 0$); hence, the only relevant term for us in the $\xi$-expansion is the linear one. As in the Yang-Mills case, to perform this calculation we consider a functional approach. The quadratic action that we have in this case implies that we find  the following contributions from $V$ fields and Nielsen-Kallosh ghosts $b$ to the one-loop effective action \cite{Abbott:1984pz}
\br
\G_{eff} &=& - \frac{1}{2} tr \ln \left[ \hat{\dal} + \xi \left( \bnabla^2 \bar{\bnabla}^2 + \bar{\bnabla}^2 \bnabla^2 \right) \right] + tr \ln \left[ \Box_{-} + \xi \bnabla^2 \bar{\bnabla}^2 \right] \;, \label{SYM_1loop_eff_gen_gauge}
\er
which is written in terms of the operators defined in appendix \ref{ap_BFM} as
\br
\hat{\dal} &=& \bfBox - i \bW^{\a} \bnabla_{\a} - i \bar{\bW}^{\dot{\a}} \bar{\bnabla}_{\dot{\a}} \nonumber \\
\Box_{+} &=& \Box - i \bW^{\a} \bnabla_{\a} - \frac{i}{2} ( \bnabla^{\a} \bW_{\a} ) \nonumber \\
\Box_{-} &=& \Box - i \bar{\bW}^{\dot{\a}} \bar{\bnabla}_{\dot{\a}} - \frac{i}{2} ( \bar{\bnabla}^{\dot{\a}} \bar{\bW}_{\dot{\a}}) \;.
\er

Expanding (\ref{SYM_1loop_eff_gen_gauge}) in $\xi$ we find
\br
\G_{eff} &=&  - \frac{1}{2} tr \ln \hat{\dal} + tr \ln \Box_{-} + \G_{\xi} \nonumber \\
&=& - \frac{1}{2} tr \ln \hat{\dal} + tr \ln \Box_{-} + \frac{\xi}{2} \G^{(1)}_{\xi} + {\cal{O}}(\xi^2) \;,
\er
and the linear part is 
\br
\G_{\xi}^{(1)} &=& - tr \left[ \frac{1}{\hat{\dal}} ( \bnabla^2 \bar{\bnabla}^2 +  \bar{\bnabla}^2 \bnabla^2 ) \right] +  tr \left[ \frac{1}{\Box_{-}} \bnabla^2 \bar{\bnabla}^2  \right] \nonumber \\
&=& tr \left[ \frac{1}{\hat{\dal}} ( \hat{\dal} - \Box_{-} ) \frac{1}{\Box_{-}} \bnabla^2 \bar{\bnabla}^2 \right] + tr \left[ \frac{1}{\hat{\dal}} ( \hat{\dal} - \Box_{+}) \frac{1}{\Box_{+}} \bar{\bnabla}^2 \bnabla^2 \right] \nonumber \\
&=& tr \left[ \frac{1}{\hat{\dal}} \left( - i \bW^{\a} \bnabla_{\a} + \frac{i}{2} ( \bar{\bnabla}^{\dot{\a}} \bar{\bW}_{\dot{\a}}) \right) \frac{1}{\Box_{-}} \bnabla^2 \bar{\bnabla}^2 \right]  \nonumber \\
& & + tr \left[ \frac{1}{\hat{\dal}} \left( - i \bar{\bW}^{\dot{\a}} \bar{\bnabla}_{\dot{\a}} + \frac{i}{2} ( \bnabla^{\a} \bW_{\a}) \right) \frac{1}{\Box_{+}} \bar{\bnabla}^2 \bnabla^2 \right] \;,
\er
where in the second step we have used the property $tr \Box_{-}^{-1} \bnabla^2 \bar{\bnabla}^2 = tr \Box_{+}^{-1} \bar{\bnabla}^2 \bnabla^2$ \cite{Grisaru:1984ja}. At this point, applying $\Box_{-}^{-1} \bnabla^2 \bar{\bnabla}^2 = \bnabla^2 \Box_{+}^{-1} \bar{\bnabla}^2$ \cite{Grisaru:1984ja}, the anticonmutative nature of the covariant derivatives ($\bnabla_{\a} \bnabla^2 = 0$) and the Bianchi identity $\bnabla^{\a} \bW_{\a} = - \bar{\bnabla}^{\dot{\a}} \bar{\bW}_{\dot{\a}}$ we arrive at
\br
\G_{\xi}^{(1)} &=& \frac{i}{2} tr \left[ \frac{1}{\hat{\dal}} ( \bar{\bnabla}^{\dot{\a}} \bar{\bW}_{\dot{\a}} ) \frac{1}{\Box_{-}} \bnabla^2 \bar{\bnabla}^2 \right] - \frac{i}{2} tr \left[ \frac{1}{\hat{\dal}} ( \bar{\bnabla}^{\dot{\a}} \bar{\bW}_{\dot{\a}}) \frac{1}{\Box_{+}} \bar{\bnabla}^2 \bnabla^2  \right] \;.
\er

So, considering the inverse of the operators,
\br
\frac{1}{\Box_{+}} &=& \frac{1}{\Box} + \frac{i}{\Box} \left( \bW^{\a} \bnabla_{\a} + \frac{1}{2} ( \bnabla^{\a} \bW_{\a} ) \right) \frac{1}{\Box} + \ldots \nonumber \\
\frac{1}{\Box_{-}} &=& \frac{1}{\Box} + \frac{i}{\Box} \left( \bar{\bW}^{\dot{\a}} \bar{\bnabla}_{\dot{\a}} + \frac{1}{2} ( \bar{\bnabla}^{\dot{\a}} \bar{\bW}_{\dot{\a}} ) \right) \frac{1}{\Box} + \ldots
\er
where $\Box = 1/2 \bnabla^{\a \dot{\a}} \bnabla_{\a \dot{\a}}$, the contribution at second order in the background gauge fields is  
\br
\G_{\xi}^{(1)} &=& - \frac{1}{2} tr \left[ \frac{1}{\Box_0} (\bar{\bnabla}^{\dot{\a}} \bar{\bW}_{\dot{\a}}) \frac{1}{\Box_0} \left( \bar{\bW}^{\dot{\a}} \bar{\bnabla}_{\dot{\a}} + \frac{1}{2} ( \bar{\bnabla}^{\dot{\a}} \bar{\bW}_{\dot{\a}}) \right) \frac{1}{\Box_0} \bnabla^2 \bar{\bnabla}^2  \right] \nonumber \\
& & + \frac{1}{2} tr \left[ \frac{1}{\Box_0} ( \bar{\bnabla}^{\dot{\a}} \bar{\bW}_{\dot{\a}}) \frac{1}{\Box_0} \left( \bW^{\a} \bnabla_{\a} + \frac{1}{2} ( \bnabla^{\a} \bW_{\a}) \right) \frac{1}{\Box_0} \bar{\bnabla}^2 \bnabla^2  \right] + {\cal{O}}(B^3) \;, \nonumber \\
\er
with $\Box_0 = 1/2 \pa^{\a \dot{\a}} \pa_{\a \dot{\a}}$ being the usual d'alembertian. Hence, as the terms that corresponds to $\bar{\bW}^{\dot{\a}} \Box_0^{-1} \bar{\bnabla}_{\dot{\a}} \bnabla^2 \bar{\bnabla}^2$ and $\bW^{\a} \Box_0^{-1} \bnabla_{\a} \bar{\bnabla}^2 \bnabla^2$ have not enough covariant derivatives to give a non-vanishing result at second order in the background gauge fields (remember $\comm{\bnabla_{\a}}{\bar{\bnabla}^2} = - i \bnabla_{\a \dot{\a}} \bar{\bnabla}^{\dot{\a}} + i \bW_{\a}$), the contribution is found to be (once we have imposed the Bianchi identities)
\br
\G_{\xi}^{(1)} &=& - \frac{1}{2} tr \left[ \frac{1}{\Box_0} ( D^{\a} \bW_{\a} ) \frac{1}{\Box_0} ( D^{\b} \bW_{\b} ) \frac{1}{\Box_0} D^2 \bar{D}^2 \right] + {\cal{O}}(B^3) \nonumber \\
&=& \frac{1}{2} tr \int d^8 z_1 d^8 z_2 d^8 z_3 \; [ D^{\a} \bW_{\a} (z_2) ] [ D^{\b} \bW_{\b}(z_3) ] \left[ \bar{D}^2_1 D^2_1 P_{12} \right] P_{23} P_{13} + {\cal{O}}(B^3) \;. \nonumber \\
\er

After simplifying this result with the usual superspace $\d$-function identity (\ref{SUSY_delta_propagators}) and using the identifications $x_2 = x$, $x_3 = y$ and $x_1 = u$ we have
\br
\G_{\xi}^{(1)} &=& \frac{1}{2} tr \int d^4 x d^4 y d^4 \th \; [ D^{\a} \bW_{\a}(x, \th) ] [ D^{\b} \bW_{\b} (y, \th) ] \D_{xy} \int d^4 u \; \D_{xu} \D_{yu} \;. 
\er 

This expression is IR divergent, and it has been evaluated in (\ref{CDR_momentum_space}). The renormalized result found there is
\br
\G_{\xi \; R}^{(1)} &=& - \frac{1}{8(4 \pi^2)^2} tr \int d^4 x d^4 y d^4 \th \; [ D^{\a} \bW_{\a} (x, \th) ] [ D^{\b} \bW_{\b} (y, \th) ] \frac{\ln (x-y)^2 M^2_{IR}}{(x-y)^2} + {\cal{O}}(B^3) \nonumber \\
&=& - \frac{1}{8(4 \pi^2)^2} tr \int d^4 x d^4 y d^2 \th \; \bW^{\a} (x, \th)  \bW_{\a} (y, \th) \Box \frac{\ln (x-y)^2 M^2_{IR}}{(x-y)^2} + {\cal{O}}(B^3) \;, \nonumber \\
\er
where in the last step we have applied the identity\footnote{$\displaystyle\int d^4 x d^4 y d^4 \th (D^{\a} \bW_{x \; a}) (D^{\b} \bW_{y \; \b}) f(x-y) = - \int d^4 x d^4 y d^4 \th \bW_{x \; \a} ( \d_{\b}^{\;\a} D^2 \bW^{\b}_y ) f(x-y) \\ = - \int d^4 x d^4 y d^2 \th \bW_{x \; \a} (\comm{\bar{D}^2}{D^2} \bW^{\a}_y) f (x-y) + {\cal{O}}(B^3) = \int d^4 x d^4 y d^2 \th \bW^{\a}_x ( \Box \bW_{y \; \a}) f (x-y) + {\cal{O}}(B^3) \\ = \int d^4 x d^4 y d^2 \th \bW_{x}^{\a} \bW_{y \; \a} \Box f(x-y) +{\cal{O}}(B^3)$} 
\br
& &\int d^4 x d^4 y d^4 \th \; [ D^{\a} \bW_{\a} (x,\th) ] [ D^{\b} \bW_{\b} (y, \th) ] f (x-y) \nonumber \\
& & = \int d^4 x d^4 y d^2 \th \; \bW^{\a}(x,\th) \bW_{\a}(y, \th) \Box f(x-y) + {\cal{O}}(B^3) \;.
\er

Hence, the linear term of the expansion in the gauge parameter of the one-loop effective action evaluated at second order in the background gauge fields is
\br
\G_{\xi} &=& - \frac{\xi}{16 (4 \pi^2)^2} tr \int d^4 x d^4 y d^2 \th \; \bW^{\a}(x, \th) \bW_{\a} (y, \th) \Box \frac{ \ln (x-y)^2 M^2_{IR}}{(x-y)^2} + {\cal{O}}(\xi^2 ; B^3) \;. \nonumber \\ \label{SYM_1loop_eff_action_gen_gauge}
\er 

\subsection{Two-loop level}

We proceed now with the calculation of the two-loop contribution to the background gauge field self-energy. As we have stated previously, we will use here covariant D-algebra, which simplifies and reduces the number of the diagrams we have to consider. We begin with the pure contribution of quantum gauge fields, and leave ghost contributions for a later section.
 
\subsubsection{Quantum gauge field contribution}
\begin{figure}[ht]
\centerline{\epsfbox{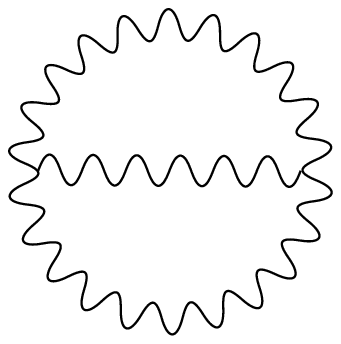}}
\caption{Two-loop contribution to the background effective action. Wavy lines correspond to quantum gauge field propagators.}
\label{SYM_2loop_vacuum}
\end{figure}
In order to obtain these contributions, we have to expand the gauge action $S_0$ of (\ref{SYM_BFM_actions}) and obtain the different interaction terms. With them we are instructed by the background field method and covariant Feynman rules to consider diagrams with external background fields and covariant propagators $\hat{\dal}^{-1}$ inside loops. From this expansion, the requirement of having at least four $\bnabla$ and four $\bar{\bnabla}$ to get a non-vanishing contribution imposes that the only relevant interaction term in our problem is \cite{Grisaru:1984ja}
\br
\frac{g}{2} tr \left[  V \anticomm{\bnabla^{\a} V}{ \bar{\bnabla}^2 \bnabla_{\a} V} \right] \;,
\er
which allow us to construct a vacuum diagram like the one shown in figure \ref{SYM_2loop_vacuum}, where the wavy lines correspond to covariant propagators $\hat{\dal}^{-1}$. Note that we have implicit interactions with the background field in the covariant derivatives and the covariant propagators. 

To evaluate this diagram, we have to rearrange the covariant derivatives at the vertices, where their explicit expression is $(i g / 2) f^{abc} V^a \bnabla^{\a} V^b \bar{\bnabla}^2 \bnabla_{\a} V^c$. At the left vertex we choose a given configuration of derivatives which, after using the commutation relations of the covariant derivatives and integration by parts, can be rewritten as
\br
\frac{ig}{2} f^{abc} V^a \bnabla^{\a} V^b \bar{\bnabla}^2 \bnabla_{\a} V^c &=& \frac{ig }{2} f^{abc} (- 2 V^a \bnabla^2 V^b \bar{\bnabla}^2 V^c + V^a \bnabla^{\a} V^b i \bnabla_{\a \dot{\a}} \bar{\bnabla}^{\dot{\a}} V^c   \nonumber \\
& &  - V^a \bnabla^{\a} V^b \comm{i \bW_{\a}}{V}^c  ) \;. \label{SYM_2loop_Lvertex}
\er  

Once we have this fixed arrangement, at the right vertex we have to choose the six permutations that are possible. Then, we integrate by parts in each of them so that one specific line if free of any operators. This implies that this vertex is written as \cite{Grisaru:1984ja}
\br
& &\frac{ig}{2} f^{abc} V^a ( - 2 \bar{\bnabla}^2 V^b \bnabla^2 V^c + 2 \bnabla^{\a} V^b \bar{\bnabla}^2 \bnabla_{\a} V^c - 2 \bar{\bnabla}^{\dot{\a}} V^b \bnabla^2 \bar{\bnabla}_{\dot{\a}} V^c - \bnabla^{\a} V^b i \bnabla_{\a \dot{\a}} \bar{\bnabla}^{\dot{\a}} V^c  \nonumber \\
& & + \bar{\bnabla}^{\dot{\a}} V^b i \bnabla_{\a \dot{\a}} \bnabla^{\a} V^c - i \bnabla^{\a \dot{\a}} V^b \bar{\bnabla}_{\dot{\a}} \bnabla_{\a} V^c + i \bnabla^{\a} V^b \comm{\bW_{\a}}{V}^c - 2 i \bar{\bnabla}^{\dot{\a}} V^b \comm{\bar{\bW}_{\dot{\a}}}{V}^c ) \;. \nonumber \\ \label{SYM_2loop_Rvertex}
\er

It can be shown that most of the different combinations of terms at each vertex either cancel by not having enough covariant derivatives, produce pairs of divergent contributions that cancel each other after using the Bianchi identity $\bnabla ^{\a} \bW_{\a} = - \bar{\bnabla}^{\dot{\a}} \bW_{\dot{\a}}$ or give finite Feynman integrals \cite{Grisaru:1984ja}. The only non-vanishing divergent contribution that we find is given by the first terms of (\ref{SYM_2loop_Lvertex}) and (\ref{SYM_2loop_Rvertex}). At this stage, we can obtain explicit background gauge fields by expanding each of the covariant propagators to second order in background gauge fields as
\br
\frac{1}{\hat{\dal}} - \frac{1}{\Box} &=&  \frac{i}{\Box} \left(  \bW^{\a} \frac{1}{\Box} \bnabla_{\a} + \bar{\bW}^{\dot{\a}} \frac{1}{\Box} \bar{\bnabla}_{\dot{\a}} \right) + \frac{i}{\Box} \left( \bW^{\a} \comm{\nabla_{\a}}{\frac{1}{\Box}} +\bar{\bW}^{\dot{\a}} \comm{\bar{\bnabla}_{\dot{\a}}}{\frac{1}{\Box}} \right)  \nonumber \\
& & - \frac{1}{\Box} ( \bW^{\a} \bnabla_{\a} + \bar{\bW}^{\dot{\a}} \bar{\bnabla}_{\dot{\a}} ) \frac{1}{\Box} ( \bW^{\b} \bnabla_{\b} + \bar{\bW}^{\dot{\b}} \bar{\bnabla}_{\dot{\b}} ) \frac{1}{\Box} + \ldots \label{SYM_2loop_expansion_prop}
\er
Notice that the first term of the expansion gives no contribution, as when combined with a similar term from another line is finite, and when combined with a space-time background connection $\bG_{\a \dot{\a}}$\footnote{Recall that there is another implicit dependence in background fields, as we have $\Box = 1/2 \bnabla^{\a \dot{\a}} \bnabla_{\a \dot{\a}} = 1/2 ( \pa^{\a \dot{\a}} - i \bG^{\a \dot{\a}} )( \pa_{\a \dot{\a}} - i \bG_{\a \dot{\a}})$, where $\bG_{\a \dot{\a}}$ is the background space-time connection.} it gives a contribution of the form of $\bG( \bnabla^{\a} \bW_{\a} + \bar{\bW}^{\dot{\b}} \bar{\bnabla}_{\dot{\b}})$ that clearly vanishes by the Bianchi identities \cite{Grisaru:1984ja}.

\begin{figure}[ht]
\centerline{\epsfbox{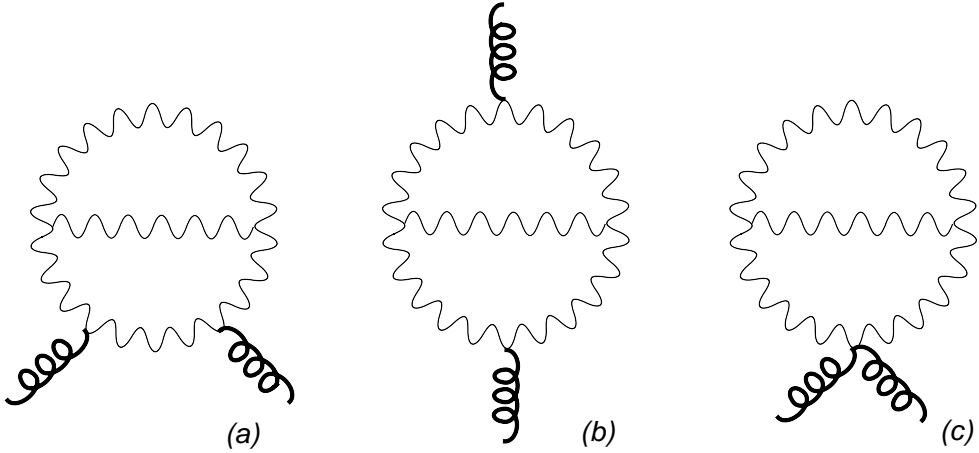}}
\caption{Diagrams corresponding to the expansion of $\hat{\dal}^{-1}$. Thick lines correspond to external background fields.}
\label{SYM_2loop}
\end{figure}
Let us consider first the third term of (\ref{SYM_2loop_expansion_prop}), which generates a diagram of the form of diagram $(a)$ of figure \ref{SYM_2loop}. Notice that all of these contributions have a common symmetry factor of $\frac{1}{2}$ that we will take into account at the end. As we have two explicit background field strengths, at second order in the background gauge fields we have $\Box^{-1} = \Box_0^{-1}$, with $\Box_0 = (1/2) \pa^{\a \dot{\a}} \pa_{\a \dot{\a}}$ the usual d'alembertian. Therefore, we have an explicit expression of the form 
\br
\G^{2 \; loop}_1 &=& - 3 g^2 C_A^2 tr \int d^8 z_1 d^8 z_2 d^8 z_3 d^8 z_4 \; \left[ \bnabla^2_1 P_{12} \right] \left[  \bW^{\a} (z_2) \bnabla_{2 \; \a} + \bar{\bW}^{\dot{\a}} (z_2) \bar{\bnabla}_{2 \; \dot{\a}} \right] P_{23}  \nonumber \\
& & \times \left[ \bW^{\b} (z_3) \bnabla_{3 \; \b} + \bar{\bW}^{\dot{\b}} (z_3) \bar{\bnabla}_{3 \; \dot{\b}} \right] \left[ \bar{\bnabla}^2_3 P_{34} \right] P_{14} \left[ \bnabla^2_4 \bar{\bnabla}^2_4 P_{41} \right] \;. \nonumber \\
\er
Due to the anticommutative nature of the covariant derivatives $\bar{\bnabla}_{3 \; \dot{\b}} \bar{\bnabla}^2_3 = 0$. Hence, we have
\br
\G^{2\;loop}_1 &=& - 3 g^2 C_A^2 tr \int d^8 z_1 d^8 z_2 d^8 z_3 d^8 z_4 \; \left[ \bnabla^2_1 P_{12} \right] \bar{\bW}^{\dot{\a}}(z_2) \bar{\bnabla}_{2 \; \dot{\a}} P_{23} \bW^{\b}(z_3) \left[ \bnabla_{\b \; 3} \bar{\bnabla}^2_3 P_{34} \right] P_{14}  \nonumber \\
& & \times \left[ \bnabla^2_4 \bar{\bnabla}^2_4 P_{14} \right] \nonumber \\
& & - 3 g^2 C_A^2 tr \int d^8 z_1 d^8 z_2 d^8 z_3 d^8 z_4 \; \left[ \bnabla^2_1 P_{12} \right] \bW^{\a} (z_2) \bnabla_{2 \; \a} P_{23} \bW^{\b} (z_3) \left[ \bnabla_{3 \; \b} \bar{\bnabla}^2_3 P_{34} \right] P_{14}  \nonumber \\
& & \times \left[ \bnabla^2_4 \bar{\bnabla}^2_4 P_{41} \right] \;. \label{SYM_2loop_g1_bare}
\er
As can be seen, we have divided this expression in two contributions. When dealing with the first one, as $\bW^{\b}$ is a covariantly chiral superfield ($\bar{\bnabla}_{\dot{\a}} \bW^{\b} = 0$) and taking into account the basic relation between the covariant derivatives $\anticomm{\bnabla_{\dot{\a}}}{\bnabla_{\a}} = i \bnabla_{\a \dot{\a}}$, we find 
\br
\G^{2\;loop}_{1.1} &=& 3 i g^2 C_A^2 tr \int d^8 z_1 d^8 z_2 d^8 z_3 d^8 z_4  \left[ \bnabla^2_1 P_{12} \right] \bar{\bW}^{\dot{\a}}(z_2) P_{23} \bW^{\a} (z_3) \left[ \bnabla_{\a \dot{\a}}^3 \bar{\bnabla}^2_3 P_{34} \right] P_{14}  \nonumber \\
& & \times \left[ \bnabla^2_4 \bar{\bnabla}^2_4 P_{41} \right] \;.
\er 
As we need four $\bnabla$ and four $\bar{\bnabla}$ to get a non-vanishing result, and realizing that at second order in the background fields in this expression $\bnabla_{\a}$ and $\bnabla_{\a \dot{\a}}$ commute ($\bW_{\a} = - \frac{1}{2} \comm{\bar{\bnabla}^{\dot{\a}}}{\bnabla_{\a \dot{\a}}}$), it is clear that integrating by parts we obtain the non-vanishing contribution to be
\br
\G^{2\;loop}_{1.1} &=& 3 i g^2 C_A^2 tr \int d^8 z_1 d^8 z_2 d^8 z_3 d^8 z_4 \; P_{12} \bar{\bW}^{\dot{\a}} (z_2) P_{23} \bW^{\a}(z_3) \left[ \bnabla_{\a \dot{\a}}^3 \bnabla^2_3 \bar{\bnabla}^2_3 P_{34} \right] P_{14}  \nonumber \\
& & \times \left[ \bnabla^2_4 \bar{\bnabla}^2_4 P_{14} \right] + {\cal{O}}(B^3) \;.
\er
At this point, replacing the covariant derivatives by the usual ones (as we have already two explicit background field strengths), using the usual $\d$-function superspace identity and the identifications $x_1 = u$, $x_2 = x$, $x_3 =y$, $x_4 = v$, we have
\br
\G^{2\;loop}_{1.1} &=& 3 i g^2 C_A^2 tr \int d^4 x d^4 y d^4 \th \; \bar{\bW}^{\dot{\a}}(x, \th) \bW^{\a}(y, \th) \D_{xy} \int d^4 u d^4 v \; \D_{xu} ( \pa_{\a \dot{\a}}^y \D_{yv}) \D_{uv}^2 \nonumber \\
& & + {\cal{O}} (B^3) \;.  
\er

To obtain the second contribution to $\G^{2\;loop}_1$, we consider the covariant derivatives that are acting over $P_{12}$ and, integrating by parts, make them act over $P_{34}$
\br
\G^{2\;loop}_{1.2} &=& - 3 g^2 C_A^2 tr \int d^4 x d^4 y d^4 \th \; P_{12} \bW^{\a} (z_2) \bnabla_{2 \; \a} P_{23} \bW^{\b} (z_3) \left[ \bnabla_{3 \; \b} \bar{\bnabla}^2_3 \bnabla^2_3 P_{34} \right] P_{41}  \nonumber \\
& & \times \left[ \bnabla^2_4 \bar{\bnabla}^2_4 P_{41} \right] \;.
\er
Now, with the relation $\comm{\bnabla_{\a}}{\bar{\bnabla}^2} = - \bnabla_{\a \dot{\a}} \bar{\bnabla}^{\dot{\a}} + i \bW_{\a}$, we conclude that this contribution vanishes, as we do not have enough covariant derivatives.

We now consider the contributions that come from the second term of (\ref{SYM_2loop_expansion_prop}). In this case we study the commutator between $\Box^{-1}$ and the covariant derivatives, finding out
\br
\comm{\bnabla_{\a}}{\frac{1}{\Box}} &=& - \frac{1}{\Box} \comm{\bnabla_{\a}}{\Box} \frac{1}{\Box} \nonumber \\
&=& - \frac{1}{\Box} \left( \frac{1}{2} \bar{\bW}^{\dot{\a}} \bnabla_{\a \dot{\a}} + \frac{1}{2} \bnabla_{\a \dot{\a}} \bar{\bW}^{\dot{\a}} \right) \frac{1}{\Box} \nonumber \\
&=& - \frac{1}{\Box} \left( \bar{\bW}^{\dot{\a}} \bnabla_{\a \dot{\a}} + \frac{1}{2} ( \bnabla_{\a \dot{\a}} \bar{\bW}^{\dot{\a}}) \right) \frac{1}{\Box} \;, 
\er
and
\br
\comm{\bar{\bnabla}_{\dot{\a}}}{\frac{1}{\Box}} &=& - \frac{1}{\Box} \left( \bW^{\a} \bnabla_{\a \dot{\a}} + \frac{1}{2} ( \bnabla_{\a \dot{\a}} \bW^{\a}) \right) \frac{1}{\Box}  \;.
\er
Hence, for this case we also find diagrams of the form of diagram $(a)$ of figure \ref{SYM_2loop}. As we have two contributions that are identical, except for having $\bW_{\a}$ and $\bar{\bW}_{\dot{\a}}$ interchanged, we detail the calculation of only one of them.
\br
\G^{2\;loop}_{2.1} &=& 3 i g^2 C_A^2 tr \int d^8 z_1 d^8 z_2 d^8 z_3 d^8 z_4 \; \left[ \bnabla^2_1 P_{12} \right] \bW^{\a} (z_2) P_{23}  \nonumber \\
& & \times \left[ \bar{\bW}^{\dot{\a}}(z_3) \bnabla_{\a \dot{\a}}^3 + \frac{1}{2} ( \bnabla_{\a \dot{\a}} \bar{\bW}^{\dot{\a}} (z_3)) \right] \left[ \bar{\bnabla}^2_4 P_{34} \right] P_{14} \left[ \bnabla^2_4 \bar{\bnabla}^2_4 P_{14} \right] \;.
\er
As when obtaining $\G^{2\;loop}_{1.2}$, we can integrate by parts the covariant derivatives acting over $P_{12}$ to find 
\br
\G^{2\;loop}_{2.1} &=& 3 i g^2 C_A^2 tr \int d^8 z_1 d^8 z_2 d^8 z_3 d^8 z_4 \; P_{12} \bW^{\a}(z_2) P_{23} \left[ \bar{\bW}^{\dot{\a}}(z_3) \bnabla_{\a \dot{\a}}^3 + \frac{1}{2} ( \bnabla_{\a \dot{\a}} \bar{\bW}^{\dot{\a}} (z_3)) \right] \nonumber \\
& & \times \left[ \bar{\bnabla}^2_4 \bnabla^2_4 P_{34} \right] P_{14} \left[ \bnabla^2_4 \bar{\bnabla}^2_4 P_{14} \right] \;,
\er
that is an expression that can be directly simplify with the application the $\d$-function identity (\ref{cov_SUSY_delta_id}) obtained in appendix \ref{ap_BFM}. With the same identifications as in the $\G^{2\;loop}_1$ contributions, we find the total $\G^{2\;loop}_2$ contribution to be
\br
\G^{2\; loop}_2 &=& - 3 i g^2 C_A^2 tr \int d^4 x d^4 y d^4 \th \; \left[ \bW^{\a}(x,\th) \bar{\bW}^{\dot{\a}}(y,\th) + \bar{\bW}^{\dot{\a}}(x,\th) \bW^{\a}(y,\th) \right]  \nonumber \\
& & \times \D_{xy} \int d^4 u d^4 v \; \D_{xu} \pa_{\a \dot{\a}}^y \D_{yv} \D^2_{uv} + {\cal{O}} (B^3) \nonumber \\
& & - \frac{3i}{2} g^2 C_A^2 tr \int d^4 x d^4 y d^4 \th \; \left[ \bW^{\a}(x,\th) \pa_{\a \dot{\a}}^y \bar{\bW}^{\dot{\a}}(y,\th)  + \bar{\bW}^{\dot{\a}} (x, \th) \pa_{\a \dot{\a}}^y \bW^{\a} (y, \th) \right]  \nonumber \\
& & \times \D_{xy} \int d^4 u d^4 v \; \D_{xu} \D_{yv} \D^2_{uv} + {\cal{O}} (B^3) \;.
\er 

Hence, the sum of the contributions $\G^{2 \; loop}_1$ and $\G^{2 \;loop}_2$ (written in terms of the $I^0$ integral expression defined in section \ref{IR_divergences}) is
\br
\G^{2 \;loop}_1 + \G^{2 \; loop}_2 &=&  3 i g^2 C_A^2 tr \int d^4 x d^4 y d^4 \th \; \bW^{\a}(x, \th) \bar{\bW}^{\dot{\a}}(y, \th) \left[ \D \pa_{\a \dot{\a}} I^0 \right] (x-y)  \nonumber \\
& & - \frac{3 i g^2 C_A^2}{2} tr \int d^4 x d^4 y d^4 \th \; \left[ \bW^{\a}(x,\th) \pa_{\a \dot{\a}}^y \bar{\bW}^{\dot{\a}} + \bar{\bW}^{\dot{\a}}(x,\th) \pa_{\a \dot{\a}}^y \bW^{\a}(y,\th) \right]  \nonumber \\
& & \times \left[ \D I^0 \right] (x-y) +{\cal{O}}(B^3) \;.
\er

Once we have finished the study of the expansion of (\ref{SYM_2loop_expansion_prop}), we proceed now to consider diagrams with background space-time connections $\bG$. We begin by considering the expansion of the inverse of the $\Box$ operator to second order in $\bG$, which is obtained as
\br
\frac{1}{\Box} - \frac{1}{\Box_0} &=& \frac{i}{\Box_0} \left[ \frac{1}{2} ( \pa^{\a \dot{\a}} \bG_{\a \dot{\a}}) + \bG^{\a \dot{\a}} \pa_{\a \dot{\a}}\right] \frac{1}{\Box_0}  + \frac{1}{2} \frac{1}{\Box_0} \left[ \bG^{\a \dot{\a}} \bG_{\a \dot{\a}} \right] \frac{1}{\Box_0}  \nonumber \\
& & - \frac{1}{\Box_0} \left[ \frac{1}{2} ( \pa^{\a \dot{\a}} \bG_{\a \dot{\a}}) + \bG^{\a \dot{\a}} \pa_{\a \dot{\a}} \right] \frac{1}{\Box_0} \left[ \frac{1}{2} ( \pa^{\b \dot{\b}} \bG_{\b \dot{\b}}) + \bG^{\b \dot{\b}} \pa_{\b \dot{\b}} \right] \frac{1}{\Box_0} + \ldots \nonumber \\ \label{SYM_2loop_spacetime_exp}
\er

If we start with the third term of (\ref{SYM_2loop_spacetime_exp}), what we have is a contribution of the form of diagram $(a)$ of figure \ref{SYM_2loop}. So, this is written as
\br
\G^{2 \;loop}_3 &=& 3 g^2 C_A^2 tr \int d^8 z_1 d^8 z_2 d^8 z_3 d^8 z_4 \; \left[ \bnabla^2_1 P_{12} \right] \left[ \bG^{\a \dot{\a}}(z_2) \pa^2_{\a \dot{\a}} + \frac{1}{2} ( \pa^{\a \dot{\a}} \bG_{\a \dot{\a}})(z_2) \right] P_{23}  \nonumber \\
& & \times \left[ \bG^{\b \dot{\b}}(z_3) \pa^3_{\b \dot{\b}} + \frac{1}{2} ( \pa^{\b \dot{\b}} \bG_{\b \dot{\b}})(z_3) \right] \left[ \bar{\bnabla}^2_3 P_{34} \right] P_{14} \left[ \bar{\bnabla}^2_4 \bar{\bnabla}^2_4 P_{14} \right] \nonumber \\
&=& 3 g^2 C_A^2 tr \int d^8 z_1 d^8 z_2 d^8 z_3 d^8 z_4 \; P_{12} \left[ \bG^{\a \dot{\a}}(z_2) \pa^2_{\a \dot{\a}} + \frac{1}{2} ( \pa^{\a \dot{\a}} \bG_{\a \dot{\a}})(z_2) \right] P_{23}  \nonumber \\
& & \times \left[ \bG^{\b \dot{\b}}(z_3) \pa^3_{\b \dot{\b}} + \frac{1}{2} ( \pa^{\b \dot{\b}} \bG_{\b \dot{\b}})(z_3) \right] \left[ \bar{\bnabla}^2_3 \bnabla^2_3 P_{34} \right] P_{14} \left[ \bar{\bnabla}^2_4 \bar{\bnabla}^2_4 P_{14} \right] \;,
\er
where in the last step we have applied the same procedure as with $\G^{(1)}_{1.2}$ and $\G^{2 \; loop}_2$, integrating by parts the covariant derivatives that are acting over $P_{12}$. Hence, after the usual identifications and some simple algebra, we find for this contribution
\br
\G^{2 \; loop}_3 &=& - \frac{3}{4} g^2 C_A^2 tr \int d^4 x d^4 y d^4 \th \; \bG^{\a \dot{\a}} (x,\th) \bG^{\b \dot{\b}}(y, \th) \pa_{\a \dot{\a}}^x \pa_{\b \dot{\b}}^x \left[ \D I^0 \right] (x-y)  \nonumber \\
& & + 3 g^2 C_A^2 tr \int d^4 x d^4 y d^4 \th \; \bG^{\a \dot{\a}}(x, \th) \bG^{\b \dot{\b}}(y, \th) \pa_{\a \dot{\a}}^x \left[ \D \pa_{\b \dot{\b}}^x I^0 \right] (x-y)  \nonumber \\
& & -3 g^2 C_A^2 tr \int d^4 x d^4 y d^4 \th \; \bG^{\a \dot{\a}}(x, \th) \bG^{\b \dot{\b}}(y, \th) \left[ \D \pa_{\a \dot{\a}}^x \pa_{\b \dot{\b}}^x I^0 \right](x-y) + {\cal{O}}(B^3) \;. \nonumber \\
\er

Let us now consider the first term of the $\Box^{-1}$ expansion (\ref{SYM_2loop_spacetime_exp}). At second order in the background fields, we find the relevant contribution to be obtained when we expand two different lines, obtaining a diagram of the form of diagram $(b)$ of \ref{SYM_2loop}. Explicitly, we have 
\br
\G^{2 \; loop}_4 &=& \frac{3}{2} g^2 C_A^2 tr \int d^8 z_1 d^8 z_2 d^8 z_3 d^8 z_4 \; \left[ \bnabla^2_1 P_{12} \right] \left[ \bG^{\a \dot{\a}}(z_2) \pa^2_{\a \dot{\a}} + \frac{1}{2} ( \pa^{\a \dot{\a}} \bG_{\a \dot{\a}})(z_2) \right] \left[ \bar{\bnabla}^2_3 P_{23} \right]  \nonumber \\
& & \times \left[ \bnabla^2_3 P_{34} \right]  \left[ \bG^{\b \dot{\b}}(z_4) \pa_{\b \dot{\b}}^4 + \frac{1}{2} (\pa^{\b \dot{\b}} \bG_{\b \dot{\b}})(z_4) \right] \left[ \bar{\bnabla}^2_4 P_{41} \right] P_{13} \;.
\er
In this expression we have only four $\bnabla$ and four $\bar{\bnabla}$, so it is obvious how we can obtain the non-vanishing contribution: we integrate by parts the covariant derivatives to move four of them (for example those that are acting over $P_{12}$ and $P_{34}$), and made them act on the other four ($ \bar{\bnabla}^2_3 P_{23}$ and $\bar{\bnabla}^2_4 P_{41}$ in our example). After some simple algebra, we find for this contribution
\br
\G^{2 \; loop}_4 &=& \frac{3}{2} g^2 C_A^2 tr \int d^4 x d^4 y d^4 \th \; \bG^{\a \dot{\a}}(x,\th) \bG^{\b \dot{\b}}(y,\th) H[1, \pa_{\a \dot{\a}} \; ; \; 1, \pa_{\b \dot{\b}} ]  \nonumber \\
& & + \frac{3}{8} g^2 C_A^2 tr \int d^4 x d^4 y d^4 \th \; \bG^{\a \dot{\a}} (x, \th) \bG^{\b \dot{\b}}(y, \th) \pa_{\a \dot{\a}}^x \pa_{\b \dot{\b}}^x H[ 1,1 \; ; \; 1,1] \;,
\er 
where we have used the $H$ integrals defined in (\ref{H_definition}).

Finally, we consider the second term of the $\Box^{-1}$ expansion. It is clear that this generates a contribution of the form of diagram $(c)$ of figure \ref{SYM_2loop}, which is written as
\br
\G^{2 \; loop}_5 &=& - \frac{3}{2} g^2 C_A^2 tr \int d^8 z_1 d^8 z_2 d^8 z_3 d^8 z_4 \; \left[ \bnabla^2_1 P_{12} \right] \bG^{\a \dot{\a}}(z_2) \d^8_{23} \bG_{\a \dot{\a}}(z_3) \left[ \bar{\bnabla}^2_4 P_{34} \right]  \nonumber \\
& & \times P_{14} \left[ \bnabla^2_4 \bar{\bnabla}^2_4 P_{14} \right] \;.
\er

After freeing $P_{12}$ of covariant derivatives and make them act over $P_{34}$, the expression can be treated with the usual steps ($\d$-function identity and point identifications). Hence, we find the bare expression to be
\br
\G^{2 \; loop}_{5} &=& \frac{3}{4} g^2 C_A^2 tr \int d^4 x d^4 y d^4 \th \; \bG^{\a \dot{\a}}(x, \th) \bG^{\b \dot{\b}}(y, \th) (2 C_{\a \b} C_{\dot{\a} \dot{\b}}) \left[ \Box( \D I^0) - \pa^{\; \a \dot{\a}}_x ( \D \pa_{\dot{\a}}^x I^0 )  \right. \nonumber \\
& & \left. + \D \Box I^0  \right](x-y) + {\cal{O}}(B^3) \;.
\er

\subsubsection{Ghost contribution}
\begin{figure}[ht]
\centerline{\epsfbox{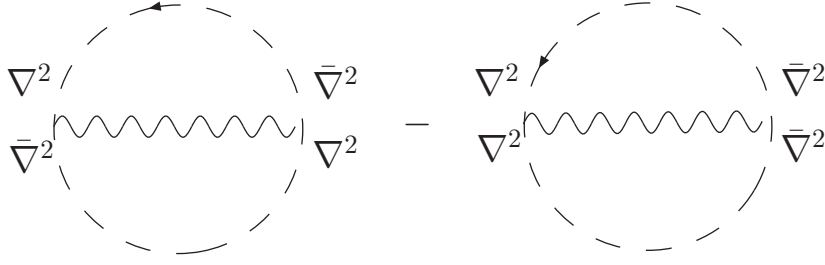}}
\caption{Two-loop ghost contribution to the background effective action}
\label{SYM_2loop_ghosts}
\end{figure}

As the Nielsen-Kallosh ghosts only interact with the background gauge field and they enter quadratically in the action, it is clear that they only contribute at the one-loop level. Hence at two-loops we have contributions from $c$ and $c^{\prime}$ ghosts, which are proportional to the difference of the two graphs shown in figure \ref{SYM_2loop_ghosts}. Expanding the propagators as for the quantum gauge field, we easily find that terms with two $\bG$ cancel \cite{Grisaru:1984ja}. For $\bW$-terms, we obtain that the only divergent contributions come from factors acting on the same line (either $\bW \bnabla \Box^{-1} \bar{\bnabla}$ or $\bW \comm{\bnabla}{\Box^{-1}}$). However, all of these terms can be listed and showed to cancel each other or produce a combination that vanishes once we impose the Bianchi identities \cite{Grisaru:1984ja}. Hence, we conclude that the two-loop ghost contribution vanishes \cite{Abbott:1984pz,Grisaru:1984ja}.

\subsubsection{Total renormalized contribution}
As we have seen, all the divergent contributions have been written in terms of the integral expression we have defined as $I^0$ and two of the $H$ overlapping integrals listed in section \ref{overlap_integrals}. Hence, we only have to directly use the renormalized results previously found and replace the bare expressions with the renormalized ones. Thus, we find for the first two contributions
\br
\sum^{2}_{i=1}\G^{2 \; loop}_i |_R &=& - \frac{3 i g^2 C_A^2}{2} tr \int d^4 x d^4 y d^4 \th \; \left[ \bW^{\a}(x,\th) \pa_{\a \dot{\a}}^y \bar{\bW}^{\dot{\a}} + \bar{\bW}^{\dot{\a}}(x,\th) \pa_{\a \dot{\a}}^y \bW^{\a}(y,\th) \right] \nonumber \\
& & \times \left[ \D I^0 \right]_R (x-y) \nonumber \\ 
& & + \frac{3 i g^2 C_A^2}{16 (4 \pi^2)^3} tr \int d^4 x d^4 y d^2 \th \; \bW^{\a} (x, \th) \bar{\bW}^{\dot{\a}}(y, \th) \pa_{\a \dot{\a}}^x \frac{\ln (x-y)^2 M^2}{(x-y)^2} + {\cal{O}}(B^3) \;. \nonumber \\ \label{SYM_G1_G2}
\er
Appliying the Bianchi identities we find an useful relation to simplify this result and express it in a explicit gauge invariant form. Let us consider an expression of the form
\br
\int d^4 x d^4 y d^4 \th \; \bW^{\a}(x,\th) ( \pa_{\a \dot{\a}}^y \bar{\bW}^{\dot{\a}}(y, \th)) f(x-y) \;. \label{SYM_2loop_bianchi}
\er
Then, at second order in the background gauge fields, we can replace $\bW^{\a}_x \pa_{\a \dot{\a}}^y \bar{\bW}^{\dot{\a}}_y$ with $\bW^{\a}_x \bnabla_{\a \dot{\a}}^y \bar{\bW}^{\dot{\a}}_y $. If we write the space-time covariant derivative as the anti-commutator of the spinorial covariant derivatives, taking into account that $\bW^{\dot{\a}}$ is a covariantly chiral superfield and using the Bianchi identity $\bnabla \bW + \bar{\bnabla} \bW = 0$, we find that the expression we are considering can be written as 
\br
\bW^{\a}_x \pa_{\a \dot{\a}}^y \bar{\bW}^{\dot{\a}}_y &=& i \bW^{\a}_x \bnabla_{\a} \bnabla_{\b} \bW_{y}^{\b} + {\cal{O}}(B^3) \nonumber \\
&=& i  \bW^{\a}_x C_{\a \b} \bnabla^2 \bW_y^{\b} + {\cal{O}}(B^3) \nonumber \\
&=& i \bW^{\a}_x \bnabla^2 \bW_{y \; \a} + {\cal{O}}(B^3) \nonumber \\
&=& i \bW^{\a}_x D^2 \bW_{y \; \a} + {\cal{O}}(B^3) \;.
\er
Hence, the integral expression (\ref{SYM_2loop_bianchi}) can be put as
\br
\int d^4 x d^4 y d^4 \th \; \bW^{\a}_x ( \pa_{\a \dot{\a}}^y \bar{\bW}^{\dot{\a}})_y f(x-y) &=& i \int d^4 x d^4 y d^4 \th \; \bW^{\a}_x ( D^2 \bW_{\a})_y f(x-y) + {\cal{O}}(B^3) \nonumber \\
&=& i \int d^4 x d^4 y d^2 \th \; \bW^{\a}_x ( \bar{D}^2 D^2 \bW_{\a})_y f(x-y) + {\cal{O}}(B^3) \nonumber \\
&=& i \int d^4 x d^4 y d^2 \th \; \bW^{\a}_x \bW_{y \; \a} \Box f(x-y) + {\cal{O}}(B^3) \;.  \nonumber \\ \label{SYM_int_bianchi_id}
\er
Therefore, we have a gauge invariant expression for the $\bW$-contributions of the form
\br
\sum^{2}_{i=1}\G^{2 \; loop}_i |_R &=& 3 g^2 C_A^2 tr \int d^4 x d^4 y d^2 \th \; \bW^{\a} (x, \th) \bW_{\a} (y,\th) \Box \left[ \D I^0 \right]_R (x-y)  \nonumber \\
& & - \frac{3 g^2 C_A^2}{16(4 \pi^2)^3} tr \int d^4 x d^4 y d^2 \th \; \bW^{\a}(x,\th) \bW_{\a}(y, \th) \Box \frac{ \ln (x-y)^2 M^2}{(x-y)^2} + {\cal{O}}(B^3) \;. \nonumber \\
\er

The $\bG$ contributions are also added up and renormalized as
\br
\sum^{5}_{i=3} \G^{2 \; loop}_i |_R &=& - 3 g^2 C_A^2 tr \int d^4 x d^4 y d^4 \th \; \bG^{\a \dot{\a}} (x, \th) \bG^{\b \dot{\b}}(y,\th) \left( \pa_{\a \dot{\a}}^x \pa_{\b \dot{\b}}^x - (2 C_{\a \b} C_{\dot{\a} \dot{\b}}) \Box \right) \nonumber \\
& & \times \left[ \frac{1}{4} [ \D I^0 ]_R (x-y) - \frac{1}{32(4 \pi^2)^3} \frac{\ln (x-y)^2 M^2}{(x-y)^2} \right] + {\cal{O}}(B^3) \;. \label{SYM_G3_G4_G5}
\er

As this expression is transverse we can rewrite it in terms of the background field strength. To do so, we have to use the following property: Let $f$ be a generic function, then, to second order in the background gauge fields we have
\br
& & tr \int d^4 x d^4 y d^4 \th \; \bG^{\a \dot{\a}} (x, \th) \bG^{\b \dot{\b}} (y, \th) \left( \pa_{\a \dot{\a}}^x \pa_{b \dot{\b}}^x - 2 C_{\a \b} C_{\dot{\a} \dot{\b}} \Box \right) f(x-y) \nonumber \\
& & = - 3 tr \int d^4 x d^4 y d^4 \th \; \left[ D^{\a} B(x, \th) \right] \left[ \bar{D}^2 D_{\a} B(y, \th) \right] \Box f(x-y) + {\cal{O}}(B^3) \nonumber \\
& & = 3 tr \int d^4 x d^4 y d^2 \th \; \bW^{\a} (x, \th) \bW_{\a} (y, \th) \Box f(x-y) + {\cal{O}}(B^3) \;. \label{SYM_2loop_st_conn_W}
\er

To prove this relation we have only to write the connection in terms of the background gauge field as $\bG_{\a \dot{\a}} = \left(\bar{D}_{\dot{\a}} D_\a - (i/2) \pa_{\a \dot{\a}} \right) B + {\cal{O}}(B^2)$. As a consequence of having a transverse structure, the integral is then written as
\br
& & tr \int d^4 x d^4 y d^4 \th \; \left[ \bar{D}^{\dot{\a}} D^{\a} B(x,\th) \right] \left[ \bar{D}^{\dot{\b}} D^{\b} B(y,\th) \right] \left( \pa_{\a \dot{\a}}^x \pa_{b \dot{\b}}^x - 2 C_{\a \b} C_{\dot{\a} \dot{\b}} \Box \right) \Box f(x-y) \nonumber \\
&=& tr \int d^4 x d^4 y d^4 \th \; \left[ D^{\a} B(x,\th) \right] \left[ \bar{D}^2 D^{\b} B(y,\th) \right] C^{\dot{\b} \dot{\a}} \left( \pa_{\a \dot{\a}}^x \pa_{b \dot{\b}}^x - 2 C_{\a \b} C_{\dot{\a} \dot{\b}} \Box \right) \Box f(x-y) \;, \nonumber \\
\er
where we have integrated by parts the superspace derivative $D^{\a}$. Then, as $C^{\dot{\b} \dot{\a}} \pa_{\a \dot{\a}} \pa_{\b \dot{\b}} = \d_{\a}^{~\b} \Box$, $C^{\dot{\b} \dot{\a}} C_{\a \b} C_{\dot{\a} \dot{\b}} = - 2 C_{\a \b}$ and $\bW_{\a} = (1/i) \bar{D}^2 D_{\a} B + {\cal{O}}(B^3)$ we straightforwardly obtain the relation (\ref{SYM_2loop_st_conn_W}). With this, we can express the space-time connection contribution (that is transverse as is required) in terms of the background field strength. Adding up all the results and taking into account the symmetry factor, we find the total two-loop renormalized contribution to the background gauge field self-energy to be
\br
\frac{1}{2}\sum^{5}_{i=1} \G^{2 \; loop}_i &=& tr \int d^4 x d^4 y d^2 \th \bW^{\a} (x, \th) \bW_{\a} (y, \th) \G^{(2) \; 2\;loop}_B(x-y) \;,
\er
with
\br
\G^{(2) \; 2 \; loop}_B(x) &=& \frac{3 g^2 C_A^2}{64(4 \pi^2)^3} \Box \frac{\frac{1}{4} \ln^2 x^2 M^2_{IR} + \frac{1}{2} \ln x^2 M^2_{IR} ( 1 - \ln x^2 M^2) + \ln x^2 M^2}{x^2} + \ldots \;,  \nonumber \\ \label{SYM_2loop_eff_action_background}
\er
where $M_{IR}$ corresponds to the IR divergence and $M$ to the UV one. As is expected, we have found a result that has both types of divergences. However, opposite to the case of dimensional regularization, we are able to clearly distinguish them.

\subsection{RG equation}

With our conventions, remembering that in the background field method the background field and the coupling constant renormalizations are related ($Z_g \sqrt{Z_B} = 1$), the anomalous dimension of the background field cancels. Hence, the RG equation for this fields is\footnote{Note also that in the RG equation we have not included a term $\g_{IR} = M_{IR} \pa/ \pa M_{IR}$ because in the renormalization we have required that both IR and UV scales were independent, which implies that $\g_{IR} = M / M_{IR} \pa M_{IR} / \pa M = 0$.}
\br
\left. \left[ M \frac{\pa}{\pa M} + \b(g) \frac{\pa}{\pa g} + \g_{\xi} (g) \frac{\pa}{\pa \xi} \right] \G^{(2)}_B \right|_{\xi = 0} = 0 \;.
\er
with $\G_B^{(2)}$ containing the free action, the one- and two-loop contributions ((\ref{SYM_1loop_eff_action_background}) and (\ref{SYM_2loop_eff_action_background}) respectively) and the linear expansion in $\xi$ of the effective action in a generic gauge (\ref{SYM_1loop_eff_action_gen_gauge}):
\br
\G_B^{(2)} (x) &=& \frac{1}{2g^2} \d(x) + \frac{3 C_A}{16(4 \pi^2)^2} \Box \frac{\ln x^2 M^2}{x^2} - \frac{\xi C_A}{16 (4 \pi^2)^2} \Box \frac{\ln x^2 M_{IR}^2}{x^2}  \nonumber \\
& & + \frac{3 g^2 C_A^2}{256 (4 \pi^2)^3} \Box \frac{\ln^2 x^2 M^2_{IR} + 2 \ln x^2 M^2_{IR} ( 1 - \ln x^2 M^2) + 4 \ln x^2 M^2}{x^2} + \ldots \nonumber \\ \label{SYM_2loop_eff}
\er

As in all the previously considered models, the gauge-running term $\g_{\xi}$ is evaluated in appendix \ref{ap_Gauge} with the one-loop RG equation for the quantum gauge field self-energy. We find there that this function has an expansion to second order in $g^2$ of the form 
\br
\g_\xi &=& - \frac{3 C_A}{4 (4 \pi^2)}g^2 + \ldots 
\er
 
Hence, inserting (\ref{SYM_2loop_eff}) in the RG equation, and using the result obtained for $\g_{\xi}$, we find the following value for the one- and two-loop coefficients of the expansion of the beta function
\br
\b(g) &=& - \frac{3}{4} \left( \frac{C_A}{8 \pi^2} \right)g^3 - \frac{3}{4} \left(\frac{C_A}{8 \pi^2} \right)^2 g^5 \;.
\er

Let us remark again that as our supersymmetric conventions differ in a $\sqrt{2}$ factor in the coupling constant wrt. the usual ones \cite{Gates:1983nr} ($g = \sqrt{2} g_{SYM}$), this results matches the standard beta function expansion $\b(g_{SYM}) = - (3/2) [ C_A/ (8\pi^2)] g^3_{SYM} - (3/2) [ C_A / (8 \pi^2) ]^2 g^5_{SYM} + {\cal{O}}(g^7_{SYM})$.

\subsubsection{Discussion of the result}
There has been some controversy about the origin (UV or IR) of the higher-order perturbative contributions to the beta function in supersymmetric gauge theories. The ``exact beta function'' of $N=1$ SYM was discovered by Novikov, Shifman, Vainshtein and Zakharov (NSVZ) in \cite{Novikov:1983uc} (although the expression was first derived in \cite{Jones:1983ip}), and it is of the form (in this discussion, we will use the usual coupling constant, thus implicitly we have $g \equiv g_{SYM}$)
\br
\b(g) &=& - \frac{3 C_A}{16 \pi^2} \frac{g^3}{1 - \frac{C_A g^2}{8 \pi^2}} \;. \label{beta_NSVZ} 
\er

In the NSVZ derivation of the function, instanton analysis was used, showing that the higher-loop corrections to the one-loop result were due an imbalance in the number of fermionic and bosonic zero modes. Thus, these correction had and IR origin. Afterwards, the two-loop coefficient of the expansion of the beta function was obtained using dimensional reduction \cite{Abbott:1984pz,Grisaru:1985tc,Grisaru:1984ja}, matching (\ref{beta_NSVZ}). However, as we have pointed out previously, in this calculation IR divergences appear and dimensional reduction regularizes both types of divergences with the same parameter \cite{Abbott:1984pz}. In \cite{Grisaru:1985tc} this is solved by subtracting the IR divergences by a $\tilde{R}$ operation \cite{Chetyrkin:nn}, obtaining that the two-loop correction of the beta function has its origin in an specific operator of dimensional reduction that is not available in a renormalization procedure that stays in four dimensions\footnote{This operator is written in terms of the background space-time connection $\bG^{\a \dot{\a}}$ and the Kronecker delta functions in four ($\d_{\a \dot{\a}}^{~\b \dot{\b}}$) and $n$ ($\hat{\d}_{\a \dot{\a}}^{~\b \dot{\b}}$) dimensions as $tr \int d^4 x d^4 \th \; \bG^{\a \dot{\a}} \bG_{\b \dot{\b}} ( \d_{\a \dot{\a}}^{~\b \dot{\b}} - \hat{\d}_{\a \dot{\a}}^{~\b \dot{\b}})$, which, by means of a Bianchi identity can be put in terms of the classical action as $- \eps tr \int d^4 x d^2 \th \; \bW^{\a} \bW_{\a}$}, which seems to imply, as is pointed out in \cite{Abbott:1984pz}, that no divergence should occur beyond one-loop. 

This situation is cleared if we distinguish between the running of the physical coupling constant (the constant that is used in perturbative calculations when obtaining the 1PI effective action), and the running of the coupling constant in the Wilson's effective action approach\footnote{In this method initially we consider a theory defined with a cutoff scale $M$, and study how the different terms of the lagrangian flow when we integrate over momentum slices down to another scale $M^{\prime}$. This flow implies that the coupling constants verify RG equations of the form $M \pa / \pa M \l = \b (\l)$}. In the latter case, using different arguments as the holomorphic dependence on the complexified coupling constant or the fact that the relevant domain of the non-local operators responsible of the higher-loop corrections to be virtual momenta of the order of external momenta (and therefore excluded by definition in the Wilson action) \cite{Novikov:1985rd,Shifman:1986zi}, we can conclude that the flow of the coupling constant is exhausted at one loop. However, in the case of the physical coupling constant, we have higher-order contributions that appear when we take the expectation value (in a external field) of the operators in the Wilson action, being the relevant IR pole related to an anomaly \cite{Shifman:1986zi}.

This IR origin of the higher-order corrections is questioned in \cite{Arkani-Hamed:1997ut,Arkani-Hamed:1997mj}. In these works, in a purely wilsonian framework, an NSVZ flow is obtained. The key idea is the differentiation of the flow of an ``holomorphic'' coupling constant and a ``canonical'' coupling constant. The first ``holomorphic'' coupling constant corresponds to a lagrangian which is normalized at a scale M as
\br
{\cal{L}}_h (M) &=& \frac{1}{g^2_h} tr \int d^4 x d^2 \th W^{\a} (V_h) W_{\a} (V_h) +~h.c. 
\er
with $1/g^2_h = 1/g^2 + i \th / (8 \pi^2)$ being the complexified coupling constant, whereas the ``canonical'' coupling constant corresponds to a lagrangian of the form
\br
{\cal{L}}_c (M) &=& ( \frac{1}{g^2_c} + i \frac{\th}{8 \pi^2}) tr \int d^4 x d^2 \th W^{\a} (g_c V_c) W_{\a} (g_c V_c) +~h.c.
\er
Although the $g_h$ running can be shown to be one-loop, when we consider the running of the coupling constant $g_c$ we found that obeys a NSVZ flow. The reason is that in order to maintain ``canonical'' normalization in the lagrangian, we are forced to perform a rescaling of the fields, being this rescaling anomalous and the origin of the higher-order terms. In these works it is also argued that the flow of the 1PI coupling constant is closely related to this ``canonical'' coupling constant flow, being this confirmed in \cite{Bonini:1996bk}, where it is found that the first two coefficients of the 1PI beta function coincide (in any mass-independent scheme) with the first two coefficients of the ``canonical'' wilsonian beta function \cite{Mas:2002xh}. As the construction we have just described is made {\em{a la}} Wilson, it is claimed in \cite{Arkani-Hamed:1997ut,Arkani-Hamed:1997mj} that it only depends on the UV properties of the theory. However, in \cite{Shifman:1999kf}, this interpretation is criticized, as it is pointed out that the IR degrees of freedom must be included in the derivation of the anomaly, if we want to maintain low-energy physics unchanged under rescaling \cite{Shifman:1999kf,Shifman:1988zk}.

We proceed to the discussion of our result. We have taken the following steps
\begin{enumerate}
\item We first renormalize the one-loop UV subdivergences with a scale M.
\item If we have had an overall UV divergence, we could have renormalized it with a different scale, say $M^{\prime}$. By power counting only $\bG \bG$ diagrams could have overall UV divergences; nevertheless, as the traceless parts multiplying $\bG \bG$ are finite, they can only depend on the one-loop scale M, which implies that we can not have any $M^{\prime}$ dependence, as gauge invariance imposes transversality in the $\bG \bG$ expressions. 
\item Although we could have when taking the derivative wrt. the UV scale in the RG equation a non-local dependence in $M$ (see (\ref{SYM_G1_G2}) and (\ref{SYM_G3_G4_G5})), after integration over half of the supercoordinates these contributions become local.
\item The one-loop scale is cancelled by the two-loop coefficient of the beta function in the RG equation, playing the off-shell IR scale a passive r\^ole, as it is exactly cancelled in this equation.  
\end{enumerate}

Hence, we conclude that the scale associated to the one-loop renormalization of the quantum superfield is the one that gives rise to the two-loop coefficient of the beta function. The fact that no overall UV scale appears, is directly related to the conclusion obtained in \cite{Abbott:1984pz} that in a four-dimensional regularization method there are no superficial divergences. However, as we have seen, this not implies that the two-loop coefficient of the beta function cancels. The mechanism presented here agrees with previous calculations in which the corrections to the one-loop result arise from a one-loop anomaly \cite{Shifman:1986zi,Arkani-Hamed:1997ut,Arkani-Hamed:1997mj,Kraus:2001tg,Kraus:2001id}. In our case, the anomaly is to be associated with the external loop, and is responsible of the promotion of the $M$ dependence into a non-vanishing non-local structure that eventually generates the two-loop coefficient of the beta function. So, we have found that {\em{the two-loop coefficient of the beta function arises from a one-loop UV scale which survives at two loops when IR effects are included}}.

\chapter{Conclusions}
\label{Conclusions}
We conclude that differential renormalization is a useful method we can employ when renormalizing a gauge theory. Although the strengths of the method are well-known (e.g., gauge invariance is not broken, and we stay all the time in four dimensions, which is crutial when studying supersymmetric theories), in its original formulation it has, at least for us, one important drawback: the necessity of imposing explicitly the Ward identities in each calculation with gauge theories. However, this point was solved for the one-loop case by the introduction of Constrained Differential Renormalization.

We have shown that we can made fruitful use of the one-loop CDR results in two-loop calculations, due to the fact that CDR fixes all the ambiguities related to the logarithms of the scales at the two-loop order. We have distinguished two cases: in the first one we have diagrams where the divergences are ``nested''. This implies that we can directly apply CDR to the ``inner'' divergence, which straightforwardly fixes all the coefficients of the logarithms of the scales in the total two-loop expression. The second case corresponds to diagrams with overlapping divergences. To deal with them we have obtained a list of renormalized two-loop integrals where in each calculation one-loop CDR rules have been maintained in every step. Although the problem is not solved for the general case, as the list is restricted to two-point functions with at most four derivatives acting on the propagators and two free indices, these integrals are the expression we have to deal with when we use the background field method.

We have discussed also the application of DiffR to IR divergent expressions. Although the renormalization procedure in momentum space resembles the usual one, we have one subtle point: the co-existence in the same expression of UV and IR divergences. In this case, as both renormalizations should decouple (UV and IR divergences are local in position and momentum space respectively, so that Bogoliubov's UV $R$ and IR $\tilde{R}$ operations commute), the scales related to each type of divergence must be independent. In order to guarantee this, we have found that we have to modify the usual renormalization relations by means of an adjustment of the local terms involving both scales. 

In this work we have re-obtained two results that where previously derived with usual differential renormalization and Ward identities: the two-loop beta function of QED and its supersymmetric extension, SuperQED. In both cases, we have shown that using one-loop CDR simplifies the calculations, as expressions that vanish by symmetry automatically cancel, and we do not have to relate the different scales {\em{via}} Ward identities in the renormalized results.

However, we have not only re-obtained previous results, but we have also performed two of the relevant calculations that were pending with differential renormalization: the two-loop renormalization of Yang-Mills and $N=1$ SuperYang-Mills. With the first one we have found no difficult when performing the calculation, as our use of one-loop CDR results allow us to perform the renormalization with the same ease as with standard dimensional methods. With Super Yang-Mills theory, the use of differential renormalization has one clear advantage over dimensional reduction (which is the usual regularization method employed with supersymmetric theories): in this case we have both UV and IR divergences, and with dimensional reduction they become mixed (both are renormalized with the same infinitesimal dimensional parameter), being necessary to subtract the IR contribution in the final result. Differential renormalization, however, clearly distinguishes between both divergences as they are renormalized with different scales. This feature allows us to give new insight on the origin (UV or IR) of the higher-loop contributions to the beta function, which has been a controversial point. We have found that higher order corrections to the beta function come from the one-loop UV scale, which survives in the higher-loop expression by the presence of the IR divergences.

Finally, among the different open problems that we have, it is clear that the principal one is the extension to higher-loop order of CDR. To achieve this, we have to obtain a complete set of rules that fix the local ambiguity of the higher-order expressions as it has been derived for the one-loop case. We think that the results we have found are a step ahead in this direction, as the complete CDR renormalized expressions must coincide, at least in parts corresponding to the logarithms of the scales, with the renormalized results that we have presented here.

\appendix
\chapter{Conventions for supersymmetric calculations}
\label{ap_SUSY}

In this section we will briefly review the most relevant results and conventions of supersymmetry and superspace that are used in this work. Although this topic is covered with great detail in various references (e.g. \cite{Sohnius:1985qm,West:1990tg,Wess:1992cp}), we will follow closely \cite{Gates:1983nr}, where the reader can found a complete treatment of the subject. 

\section{Notation}
\label{SUSY_Notation}
Setting up the notation, we express the vectors (representations $(\frac{1}{2}, \frac{1}{2})$ of the Lorentz group) with two pairs of two-valued spinor indices, one dotted and another undotted $V^{\a \dot{a}}$. To relate a vector in an arbitrary basis with index $\underline{a}$ we use the Pauli matrices:

\begin{tabular}{lcc}
\\
Fields: &  $V^{\a \dot{\a}} = \frac{1}{\sqrt{2}} \s_{\underline{b}}^{\a \dot{\a}} V^{\underline{b}}$ & $V^{\underline{b}} = \frac{1}{\sqrt{2}} \s^{\underline{b}}_{\a \dot{\a}} V^{\a \dot{\a}} $ \\ \\
Derivatives: & $ \pa_{\a \dot{\a}} = \s^{\underline{b}}_{\a \dot{\a}} \pa_{\underline{b}} $ & $ \pa_{\underline{b}} = \frac{1}{2} \s_{\underline{b}}^{\a \dot{\a}} \pa_{\a \dot{\a}}$ \\ \\
Coordinates: & $ x^{\a \dot{\a}} = \frac{1}{2} \s_{\underline{b}}^{\a \dot{\a}} x^{\underline{b}}$ & $ x^{\underline{b}} = \s^{\underline{b}}_{\a \dot{\a}} x^{\a \dot{\a}}$ \\ \\
\end{tabular}

Pauli matrices satisfy
\be
\s_{\underline{b}}^{\a \dot{\a}} \s^{\underline{c}}_{\a \dot{\a}} = 2 \d_{\underline{b}}^{~\underline{c}} ~~~,~~~ \s^{\underline{b}}_{\a \dot{\a}} \s_{\underline{b}}^{\b \dot{\b}} = 2 \d_{\a}^{~ \b} \d_{\dot{\a}}^{~ \dot{\b}}
\ee
With this conventions, the Super Yang-Mills coupling constant ($g$) that we use is related to the usual one ($g_{SYM}$) by $g = \sqrt{2} g_{SYM}$ \cite{Gates:1983nr}.

A graded commutator $[\Omega_A, \Omega_B \} \equiv \Omega_A \Omega_B - (-)^{AB} \Omega_B \Omega_A $ is defined as the anticommutator $\anticomm{\Omega_A}{\Omega_B}$ when both $\Omega_A$ and $\Omega_B$ are fermionic operators and the commutator $\comm{\Omega_A}{\Omega_B}$ otherwise. Symmetrization and antisymmetrization (sum over all permutation of indices with the corresponding sign in the case of antisymmetrization) are indicated by $(~~ )$ and $[~~ ]$ respectively. Indices between vertical lines $|~~ |$ are not taken into account in the previous operations. We also define a graded antisymmetrization  $[~~ )$ so that $[ \Omega_A, \Omega_B \} \equiv \Omega_{[A} \Omega_{B)}$.
 
For raising and lowering spinor indexes, we use the matrices $C_{\a \b}$ that have the following properties
\br
\bar{C_{\a \b}} = C_{\dot{\b} \dot{\a}} & & C_{\a \b} C^{\g \d} = \d_{\left[ \a \right. }^{~~\g} \d_{\left. \b \right]}^{~~\d} \nonumber \\
C_{\a \b} = C^{\dot{\b} \dot{\a}} & & C_{\a \b} = C_{\dot{\a} \dot{\b}} 
\er
With these matrices, if $\psi_{\a}$ denotes a spinor we define
\br
& &\psi^2 = \frac{1}{2} C_{\a \b} \psi^{\a} \psi^{\b}= \frac{1}{2} \psi^{\a} \psi_{\a} \\
& & \psi_{\a} = \psi^{\b} C_{\b \a} ~~~~,~~~~ \psi^{\a} = C^{\a \b} \psi_{\b}
\er
Similar relations hold for the hermitian conjugate $\bar{\psi}^{\dot{\a}}$. In the case of a vector $V^{\a \dot{\a}}$, we define the square to be 
\br
V^2 &=& \frac{1}{2} V^{\a \dot{\a}} V_{\a \dot{\a}} \;.
\er

\section{Supersymmetric algebra}

Coleman and Mandula \cite{Coleman:1967ad} showed that any group of {\em{bosonic}} symmetries of the S-matrix is the direct product of the Poincar\'e group with an internal symmetry group. This implies that the commutator of the bosonic generators of the internal symmetry group and the generators of the Poincar\'e group ($J_{\a \b}, \bar{J}_{\dot{\a} \dot{b}}, P_{\a \dot{\b}}$) vanish. In \cite{Haag:1974qh}, Haag, Lopuszanski and Sohnius avoided this no-go theorem by allowing fermionic symmetry generators $Q_{a \a}$ (where $a=1 ,\ldots, N$ is an isospin index). They found the most general super-Poincar\'e algebra to be
\br
\anticomm{Q_{a \a}}{\bar{Q}^b_{\dot{\b}}} &=& \d_{a}^{\; b} P_{\a \dot{\b}} \nonumber \\
\anticomm{Q_{a \a}}{Q_{b \b}} &=& C_{\a \b} Z_{a b} \nonumber \\
\comm{Q_{a \a}}{P_{\b \dot{\b}}} &=& \comm{P_{\a \dot{\a}}}{P_{\b \dot{\b}}} = \comm{\bar{J}_{\dot{\a}{\dot{\b}}}}{Q_{c \g}} = 0 \nonumber \\
\comm{J_{\a \b}}{Q_{c \g}} &=& \frac{i}{2} C_{\g(\a}Q_{c \b)} \nonumber \\
\comm{J_{\a \b}}{P_{\g \dot{\g}}} &=& \frac{i}{2} C_{\g ( \a}P_{\b ) \dot{\g}} \nonumber \\
\comm{J_{\a \b}}{J^{\g \d}} &=& - \frac{i}{2} \d_{( \a}^{\; (\g} J_{\b)}^{\; \d)} \nonumber \\
\comm{J_{\a \b}}{\bar{J}_{\dot{\a} \dot{\b}}} &=& \comm{Z_{a b}}{Z_{c d}} = \comm{Z_{a b}}{\bar{Z}^{cd}} = 0 
\er
where $Z_{a b}$ are $\frac{1}{2} N(N-1)$ complex central charges. The $N=1$ case is called simple supersymmetry, whereas $N>1$ is called extended supersymmetry. We consider only the $N=1$ case. For theories satisfying super-Poincar\'e algebra some remarkable properties can be derived \cite{Gates:1983nr}:
\begin{itemize}
\item Equality of bosonic and fermionic freedom degrees.
\item Energy is positive, and in particular, vacuum energy can be shown to vanish. 
\end{itemize}

\section{Superspace and superfields}
A compact technique for working with supersymmetric theories is what is called superspace \cite{Salam:1974yz}. By means of anticommuting parameters we can integrate the super-Poincar\'e algebra and obtain a group, the super-Poincar\'e group\cite{Gates:1983nr,Sohnius:1985qm}. As usual spacetime can be defined as the coset space Poincar\'e group/Lorentz group, superspace is defined to be the coset space super-Poincar\'e group/Lorentz group. Hence, superspace is a space spanned by the usual real commuting space time coordinates and new anticommuting coordinates $z^{A} = ( x^{\a \dot{\a}}, \th^{\a}, \bar{\th}^{\dot{\a}})$. Supersymmetry generators are realized as coordinate transformations in superspace \cite{Gates:1983nr}. As the usual generators of the Poincar\'e algebra can be represented by differential operators, $Q_{\a}$ and $\bar{Q}_{\dot{\a}}$ are found to have the following expression \cite{Gates:1983nr}
\br
Q_{\a} &=& i \left( \pa_{\a} - \frac{i}{2} \bar{\th}^{\dot{\a}} \pa_{\a \dot{\a}} \right) \nonumber \\
\bar{Q}_{\dot{\a}} &=& i \left( \bar{\pa}_{\dot{\a}} - \frac{i}{2} \th^{\a} \pa_{\a \dot{\a}}  \right) 
\er
It has to be noted that neither usual coordinate derivative $\pa_{\a \dot{\a}}$ nor fermionic coordinate derivatives $\pa_{\a}$, $\bar{\pa}_{\a}$ are invariant under supertranslations (those that are generated by $Q_{\a}$, $\bar{Q}_{\dot{\a}}$). However, supersymmetric covariant derivatives that are invariant under these transformations can be defined as \cite{Gates:1983nr}:\br
D_{\a} &=& \pa_{\a} + \frac{i}{2} \bar{\th}^{\dot{\a}} \pa_{\a \dot{\a}} \nonumber \\
\bar{D}_{\dot{\a}} &=& \bar{\pa}_{\dot{\a}} + \frac{i}{2} \th^{\a} \pa_{\a \dot{\a}} \label{Susy_cov_dev}
\er
We list here some relevant algebraic relations of these derivatives 
\br
\anticomm{D_{\a}}{\bar{D}_{\dot{\a}}} &=& i \pa_{\a \dot{\a}} \nonumber \\
\anticomm{D_{\a}}{D_{\b}} &=& \anticomm{\bar{D}_{\dot{\a}}}{\bar{D}_{\dot{\b}}} = 0 \nonumber \\
\comm{D^{\a}}{\bar{D}^2} &=& i \pa^{\a \dot{\a}} \bar{D}_{\dot{\a}} \nonumber \\\anticomm{D^2}{\bar{D}^2} &=& \Box + D^{\a} \bar{D}^2 D_{\a} \nonumber \\
D^2 \bar{D}^2 D^2 &=& \Box D^2 \nonumber \\
D^2 \th^2 &=& - 1 \;. \label{SUSY_D_algebra}
\er

Also, with these derivatives we can define two projection operators as
\br
\Pi_{\frac{1}{2}} &=& - D^{\a} \bar{D}^2 D_{\a} / \Box \nonumber \\
\Pi_0 &=& ( D^2 \bar{D}^2 + \bar{D}^2 D^2 ) / \Box \;,
\er
which verify that $\Pi_{\frac{1}{2}} + \Pi_0 = 1$.

\subsection{Superfields}
Superfields are defined to be multispinor functions over the superspace $\Phi \equiv \Phi(x, \th, \bar{\th})$. It is clear that Taylor-expansion of these functions in terms of $\th^{\a}$ and $\bar{\th}^{\dot{\a}}$ breaks off at order $\th^2 \bar{\th^2}$, due to the anticommuting nature of these parameters. The different terms of the expansion are called the {\em{component fields}}. We can impose constraints into a superfield and reduce the number of independent component fields. The most relevant ones for our work are the following two:

{\bf{Chiral superfields}}
With the aid of the covariant derivative (\ref{Susy_cov_dev}) we can impose the constraint 
\br
\bar{D}_{\dot{\a}} \Phi = 0 \;.
\er

If $\Phi$ verifies the previous relation it is called a chiral superfield. This constraint implies that this superfield has the following component expansion \cite{Gates:1983nr}
\br
\Phi &=& A + \th^{\a} \psi_{\a} - \th^2 F + \frac{i}{2} \th^{\a} \bar{\th}^{\dot{\a}} \pa_{\a \dot{\a}} A + \frac{i}{2} \th^2 \bar{\th}^{\dot{\a}} \pa_{\a \dot{\a}} \psi^{\a} + \frac{1}{4} \th^2 \bar{\th}^2 \Box A \;.
\er

{\bf{Real superfields}}
A superfield $V$ is called a real superfield if it verifies the constraint $V=V^{+}$. With this constraint, the component expansion is found to be  \cite{Gates:1983nr}
\br
V &=& C + \th^{\a} \chi_{\a} + \bar{\th}^{\dot{\a}} \bar{\chi}_{\dot{\a}} - \th^2 M - \bar{\th}^2 \bar{M} + \th^{\a} \bar{\th}^{\dot{\a}} A_{\a \dot{\a}} - \bar{\th}^2 \th^{\a} \l_{\a}  \nonumber \\
& & - \th^2 \bar{\th}^{\dot{\a}} \bar{\l}_{\dot{\a}} + \th^2 \bar{\th}^2 D \;. \label{SUSY_V_expansion}
\er
\subsection{Superspace integration and superfunctional derivation}
\label{SUSY_integration}
In order to obtain supersymmetric invariant actions, we have to define the integration over the anticommuting coordinates. The basic properties of these integrals are  
\br
\int d \th \; \th &=& 1 \nonumber \\
\int d \th \; 1 &=& 0 
\er
With this definition, the delta function of the anticommuting variables is found to be
\be
\d(\th - \th^{\prime}) = (\th - \th^{\prime}) \;.
\ee
These properties imply that the integration over $\th$ is identical to differentiation, or, which is the same, that inside a $d^4 x$ integration we have $D_{\a} = \int d \th_{\a}$. Also, since supersymmetric variations are total derivatives, if we consider $\Psi$ a general superfield and $\Phi$ a chiral superfield, the following quantities are supersymmetric invariants
\br
S_{\Psi} &=& \int d^4 x d^4 \th \; \Psi  \nonumber \\
S_{\Phi} &=& \int d^4 x d^2 \th \; \Phi 
\er

As it will be necessary when studying perturbation theory in superspace, we have to consider the superfield extension of the functional derivative. It is found \cite{Gates:1983nr} for a general superfield $\Psi$ and a chiral superfield $\Phi$ that the superfunctional derivation is  
\br
\frac{\d \Psi(z)}{\d \Psi(z^{\prime})} &=& \d^{8} (z-z^{\prime}) = \d^4(x-x^{\prime}) \d^{4}(\th - \th^{\prime}) \nonumber \\
\frac{\d \Phi(z)}{\d \Phi(z^{\prime})} &=& \bar{D}^2 \d^8 (z-z^{\prime})
\er 

Finally, this integral definition implies that we have the following integration by part rules for the superspace derivatives $D_{\a}$
\br
\int d^8 z \; A (D_{\a} B ) C &=& \int d^8 z \; \left[ - (D_{\a} A) BC - AB (D_{\a} C) \right] \nonumber \\
\int d^8 z \; A (D_{\a} E_{\b} ) C &=& \int d^8 z \; \left[ - (D_{\a} A) E_{\b} C + A E_{\b} (D_{\a} C) \right] \nonumber \\
\int d^8 z \; E_{\b} ( D_{\a} A) C &=& \int d^8 z \; \left[ ( D_{\a} E_{\b}) AC - E_{\b} A (D_{\a} C) \right] \nonumber \\
\int d^8 z \; E_{\b} (D_{\a} F_{\g} ) C &=& \int d^8 z \; \left[ (D_{\a} E_{\b}) F_{\g} C + E_{\b} F_{\g} (D_{\a} C) \right] \;, \label{SUSY_IntegrationByParts}\er
where $A$, $B$ and $C$ are bosonic (conmmuting) superfields and $E_{\b}$, $F_{\g}$ ferminic (anticommuting) superfields.  

\section{Superspace formulation of supersymmetric gauge theories}
Among the different models we can consider in superspace (Wess-Zumino, nonlinear $\sigma-$models, etc...), we will restrict ourselves to gauge theories. To deal with them there are two different approaches that we detail here, as both will be useful when studying the supersymmetric extension of the background field method. 

\subsection{Chiral representation. Prepotentials}
Starting with the linear case, by studying the field content of the $N=1$ vector multiplet \cite{Gates:1983nr}, we can obtain that the corresponding irreducible off-shell field strength is a chiral superfield $W_{\a}$ that satisfies
\br
D^{\a} W_{\a} = - \bar{D}^{\bar{\a}} \bar{W}_{\dot{\a}} \;.
\er 
Hence, $W_{\a}$ can be expressed in terms of an unconstrained real scalar superfield $V$ by
\br
W_{\a} &=& i \bar{D}^2 D_{\a} V \nonumber \\
\bar{W}_{\dot{\a}} &=& - i D^2 \bar{D}_{\a} V
\er
This definition is clearly invariant under gauge transformations with a chiral parameter $\L$ of the form of
\br
V^{\prime} = V + i( \bar{\L} - \L ) \;.
\er
This linear study has to be generalized to the non-abelian case. To do so, we consider a multiplet of chiral scalars fields transforming according to some representation of a group with generators $T_{A}$. These fields are postulated to transform with chiral parameters $\L = \L^{A} T_{A}$ as
\br
\Phi^{\prime} &=& e^{i g \L} \Phi \;,
\er
with $g$ a coupling constant. Considering an antichiral field $\bar{\Phi}$, we found that it transforms with the complex conjugate representation with antichiral parameter $\bar{\L}$. To obtain an invariant expression we introduce a multiplet of real superfields $V^A$ that transforms as
\br
\left( e^{g V} \right)^{\prime} &=& e^{ig \bar{ \L}} e^{g V} e^{-i g \L} \;,
\er
which implies that $\bar{\Phi} e^{g V} \Phi$ is invariant under gauge transformations. 
With the prepotential $V$ we can construct derivatives gauge covariant with respect to $\L$ transformations as \cite{Gates:1983nr}
\br
\nabla_{A} &=& D_{A} - i \G_{A} = ( \nabla_{\a}, \nabla_{\dot{\a}}, \nabla_{\a \dot{\a}} ) \nonumber \\
&=& ( e^{-g V} D_{\a} e^{g V}, \bar{D}_{\dot{\a}}, - i \anticomm{\nabla_{\a}}{\nabla_{\dot{\a}}} ) \;, \label{SUSY_CR_CovDev}
\er
which satisfy the requirement
\br
( \nabla_A \Phi )^{\prime} = e^{i g \L} ( \nabla_A \Phi ) \hspace{2cm} \nabla^{\prime}_A =  e^{i g \L} \nabla_A e^{-i g  \L}
\er
These derivatives are called gauge chiral representation covariant derivatives. Their conjugates $\bar{\nabla}$, that we call gauge antichiral representation covariant derivatives, are covariant with respect to $\bar{\L}$ transformations
\br
\bar{\nabla}_A &=& ( D_{\a}, e^{gV} \bar{D}_{\dot{\a}} e^{-g V}, -i \anticomm{\bar{\nabla}_{\a}}{\bar{\nabla}_{\dot{\a}}}) \;.
\er
Both representations are related by a nonunitary similarity transformation
\br
\bar{\nabla}_A = e^{gV} \nabla_A e^{-g V} \;.
\er
Field strengths defined by commutation of the covariant derivatives can be expressed in terms of the following fields denoted as $W_{\a}$ and $W_{\dot{\a}}$ \cite{Gates:1983nr}
\br
W_{\a} &\equiv& i \bar{D}^2( e^{-g V} D_{\a} e^{gV} ) \nonumber \\
W_{\dot{\a}} &\equiv& e^{-g V} \bar{W}_{\dot{\a}} e^{gV} \equiv e^{-gV} ( - W_{\a})^{+} e^{gV}
\er
that satisfy Bianchi identities of the form
\br
\nabla^{\a} W_{\a} = - \nabla^{\dot{\a}} W_{\dot{\a}} \;.
\er
With these fields we can construct a gauge invariant action as
\br
S &=& \frac{1}{g^2} tr \int d^4 x d^2 \th \; W^2 \nonumber \\
&=& - \frac{1}{2g^2} tr \int d^4 x d^4 \th \; ( e^{-gV} D^{\a} e^{gV} ) \bar{D}^2 ( e^{-gV} D_{\a} e^{gV} ) \;.
\er 
\subsection{Vector representation. Covariant approach}
\label{SUSY_vector_rep}
In this approach we start defining covariant derivatives and, by means of covariant constraints, express all quantities in terms of a single irreducible representation of supersymmetry.

For a Lie algebra with generators $T_A$ we covariantize the derivatives with the introduction of connection fields $\G_A = \G_A^{B} T_B$ as
\br
\nabla_A = D_A - i \G_A \;.
\er
Under gauge transformations these derivatives are postulated to transform with a real superfield $K= K^A T_A$ as
\br
\nabla_A^{\prime} &=& e^{i gK} \nabla_A e^{-i gK} \;.
\er
By commutation, field strengths $F_{AB}$ are defined in terms of the connections and the flat superspace torsion $T_{AB}^{~~C}$ (where $T_{\a \dot{\b}}^{~~\g \dot{\g}} = i \d_{\a}^{~\g} \d_{\dot{\b}}^{~\dot{\g}}$ is the only nonzero component) as
\br
\left[ \nabla_A, \nabla_B \right\} &=& T_{AB}^{~~C} \nabla_C - i F_{AB} \nonumber \\
F_{AB} &=& D_{[A} \G_{B \}} - i \left[ \G_A , \G_B \right\} - T_{AB}^{~~C} \G_C \;.
\er
Over these field strengths we can impose different constraints.
\subsubsection{Conventional constraints}
Due to the fact that one can always add a covariant term to the connection without changing the transformation of the covariant derivative, we can impose the constraint 
\br
F_{\a \dot{\a}} = 0 \;,
\er
as without imposing it we can define new connections $\G_A^{\prime} = (\G_{\a}, \bar{\G}_{\dot{\a}}, \G_{\a \dot{\a}} - i F_{\a \dot{\a}})$ that identically satisfy the constraints. Hence, the covariant derivatives take the following form
\br
\nabla_A = ( \nabla_{\a}, \bar{\nabla}_{\dot{\a}}, -i\anticomm{\nabla_\a}{\bar{\nabla}_{\dot{\a}}}) \;.
\er

\subsubsection{Representation-preserving constraints}
Let us define a {\em{covariantly chiral}} superfield $\Phi$, ie. a superfield that verifies
\br
\bar{\nabla}_{\dot{\a}} \Phi &=& 0 ~~,~~ \Phi^{\prime} = e^{ig K} \Phi \nonumber \\
\nabla_{\a} \bar{\Phi} &=& 0 ~~,~~ \bar{\Phi}^{\prime} = \bar{\Phi} e^{-ig K}
\er
Hence, consistency implies that we have to impose the constraint
\br
F_{\a \b} = F_{\dot{\a} \dot{\b}} = 0 \;,
\er
as with the previously defined covariantly chiral superfield we have that
\br
0 = \anticomm{\bar{\nabla}_{\dot{\a}}}{\bar{\nabla}_{\dot{\b}}} \Phi = - i F_{\dot{\a} \dot{\b}} \Phi \;.
\er
This constraint is solved with a complex superfield $\Omega = \Omega^A T_A$ which allow us to write the covariant derivatives as
\br
\nabla_{\a} &=& e^{-g \Omega} D_{\a} e^{g \Omega} \nonumber \\
\bar{\nabla}_{\dot{\a}} &=& e^{g \bar{\Omega}} \bar{D}_{\dot{\a}} e^{- g \bar{ \Omega}} \label{SUSY_VR_CovDEv}
\er
With this superfield we have two types of gauge transformations
\begin{itemize}
\item K gauge transformations
\br
(e^{g \Omega})^{\prime} = e^{g \Omega} e^{-ig K} \;.
\er
\item If we consider $\L$ to be an usual chiral superfield $\bar{D}_{\dot{\a}} \L = 0$, the derivatives defined in (\ref{SUSY_VR_CovDEv}) are invariant under
\br
(e^{g \Omega})^{\prime} = e^{i g \bar{\L}} e^{g \Omega} \;.  
\er
\end{itemize}
From the K-invariant hermitian part of $\Omega$ we can define a real superfield $V$ as
\br
e^{V} = e^{g \Omega} e^{g \bar{\Omega}} \;.
\er
We can also use the $\Omega$ superfield to write all the quantities in a gauge chiral representation, where everything transforms only under $\L$ transformations\cite{Gates:1983nr}:
\br
\nabla_{0 A} &=& e^{-g \bar{\Omega}} \nabla_A e^{g \bar{\Omega}}  \nonumber \\
\Phi_0 &=& e^{-g \bar{\Omega}} \Phi \;,
\er
where $\nabla_{0 A}$ are the chiral representation derivatives (\ref{SUSY_CR_CovDev}), $\Phi$ is a covariantly chiral superfield and $\Phi_0$ a chiral superfield.
\subsubsection{Field content. Bianchi identities}
The field content of the theory can be obtained through the Bianchi identities that are derived from the Jacobi identities satisfied by the covariant derivatives. With the aid of these identities and the different constraints imposed on the derivatives, all of the field strengths can be expressed in terms of a spinor superfield $W_{\a}$ \cite{Gates:1983nr} that is defined as 
\br
\comm{\bar{\nabla}_{\dot{\a}}}{i \nabla_{\b \dot{\b}}} &=& - i C_{\dot{\b} \dot{\a}} W_{\b} \;.
\er
This superfield can be shown to be covariantly chiral ($\bar{\nabla}_{\dot{\b}} W_{\a} = 0$) and to satisfy the following identity
\br
\nabla^{\a} W_{\a} + \bar{\nabla}^{\dot{\a}} \bar{W}_{\dot{\a}} = 0 \;.
\er
Finally, with these deriatives we can construct the gauge Lagrangian as
\br
tr W^2 = - \frac{1}{2} tr \left( \comm{\bar{\nabla}^{\dot{\a}}}{\anticomm{\bar{\nabla}_{\dot{\a}}}{\nabla_\a}} \right)^2  \;.
\er

\section{Supergraphs}

Although supersymmetric theories can be quantized at the component level with conventional methods, the use of superfields simplifies all the calculations. Along with the compact notation and automatic cancellation of graphs related by supersymmetry, in a superfield formalism supersymmetry is manifest. We will detail the extension of the usual functional methods to superspace \cite{Gates:1983nr,Grisaru:1979wc} .

Let $\Psi$ be a generic superfield, $S(\Psi)$ the action and $J$ a source of the same type (chiral,etc.) as $\Psi$. The generating functional for Green functions is 
\br
Z[J] &=& \int \; [d \Psi] e^{S(\Psi) + \int J \Psi} \;.
\er 
For connected Green functions, the generating functional is
\br
W[J] &=& \ln Z[J] \;.
\er
Finally, the generating functional of 1PI graphs (the effective action) is 
\br
\G[ \hat{\Psi}] &=& W[J(\hat{\Psi})] - \int \; J(\hat{\Psi}) \hat{\Psi} \;,
\er
where $\hat{\Psi}$ is the expectation value of $\Psi$ in the presence of the source:
\br
\hat{\Psi} &=& \frac{\d W}{\d J} \;.
\er
Let us consider two examples: a real scalar superfield (as in gauge theories) and a chiral superfield. With the first one, the superspace partition function can be written as
\br
Z[J] &=& \int [d V] \; exp \left\{\int [- \frac{1}{2} V \Box V + {\cal{L}}_{int}(V) + J V] \right\} \nonumber \\
&=& exp \left\{ \int {\cal{L}}_{int}(\frac{\d}{\d J}) \right\} exp \left\{ \frac{1}{2} \int J \frac{1}{\Box} J \right\} \;,
\er
whereas in the case of a massless chiral superfield, we have
\br
Z[j,\bar{j}] &=& \int [ d \Phi d \bar{\Phi} ] \; exp \left\{ \int d^8 z \; [ \bar{\Phi} \Phi + {\cal{L}}_{int}(\Phi,\bar{\Phi}) ] \int d^6 z j \Phi + \int d^6 \bar{z} \bar{j} \bar{\Phi} \right\} \nonumber \\
&=& exp \left\{\int {\cal{L}}_{int}(\frac{\d}{\d j},\frac{\d}{\d \bar{j}})\right\} exp \left\{- \int \bar{j} \frac{1}{\Box} j \right\} \;.
\er 
From the expansion of these expressions we can derive the propagators, vertices and symmetry factors.
\subsubsection{Superspace Feynman rules}
\begin{itemize}
\item Propagators: We present here the propagators for massless chiral(antichiral) and real superfields. These propagators are those that we need in our work. A detailed list with more propagators can be found in \cite{Gates:1983nr} or \cite{Grisaru:1979wc}.
\br
 VV: &\hspace{2cm}& -P(z_1 - z_2) \equiv - \D (x_1-x_2) \d^4 (\th_1 - \th_2) \nonumber  \\
\bar{\Phi} \Phi:  &\hspace{2cm}& P(z_1 - z_2) \equiv \D (x_1 - x_2) \d^4 ( \th_1 - \th_2 )
\er
\item Vertices:
For each chiral(antichiral) line there is a $\bar{D}^2$($D^2$) factor acting on the propagator. If all the lines are purely chiral or antichiral, one of the factors is omitted.
\item Apart of the usual space time integrals (or momentum space integrals), there is a $d^4 \th$ integration at each vertex.
\item When computing 1PI graphs (in order to obtain the effective action), for each external line we multiply by the corresponding  superfield. In the case of a chiral (or antichiral) line, no $\bar{D}^2$($D^2$) factor appears.
\item Some diagrams have symmetry factors.
\end{itemize}
A superspace diagram which corresponds to a contribution to the effective action is an expression formed by  some external fields, supercoordinate integrals $\int d^4 x_i d^4 \th_i$, propagators ($P_{ij}$) and superspace covariant derivatives acting on them \cite{Gates:1983nr,Grisaru:1979wc}. The propagators are of the form $P_{ij} = \D_{ij} \d (\th_i - \th_j) \equiv \D_{ij} \d_{ij}$ with $\D_{ij}$ the usual spacetime propagator and $\d_{ij}$ the $\d$-function on the anticommuting coordinates. The covariant derivatives can be integrated by parts, obey the Leibnitz rule and a ``transfer'' rule of the form of $\d_{ij} \stackrel{\leftarrow}{D_j} = - D_i \d_{ij}$. So, we can choose a propagator that links two vertices and remove all the $D$'s  from its $\d$-function. Then, if we have other propagators that link these two vertices, we can apply the following properties
\br
& & \d_{ij}\d_{ij} = \d_{ji}\d_{ij} = \d_{ij} D^{\a}_i \d_{ij} = \d_{ij} D^2_i \d_{ij} = \d_{ij} D^{\a}_i \bar{D}^{\dot{\a}}_i \d_{ij} = \d_{ij} D^{\a}_i \bar{D}^2_i \d_{ij} = 0 \nonumber \\
& & \d_{ij} D^2_i \bar{D}^2_i \d_{ij} = \d_{ij} \bar{D}^2_i D^2_i \d_{ij} =  \d_{ij} D^{\a}_i \bar{D}^2_i D_{i \a} \d_{ij} = \d_{ij} \label{SUSY_delta_propagators}
\er
Now, with the free superspace $\d$-function, we can contract the propagator between the two vertices to a point in $\th$-space. As this procedure can be repeated for other pair of vertices, we conclude that we can write the effective action as
\br
\G &=& \sum_{n} \int d^4 x_i \ldots d^4 x_n d^4 \th \; \G(x_1, \ldots, x_n) \P (x1, \th) \ldots V(x_i, \th) \ldots
\er
This expression has one important consequence. We have found that in a perturbative calculation a contribution to the effective action with a purely chiral(antichiral) integral $d^2 \th$($d^2 \bar{\th}$) never gets generated. Hence, if the original action had purely chiral terms (as mass terms $\P^2$ or cubic interactions $\P^3$) they can not be  modified with radiative corrections \cite{Abbott:1980jk,Haagensen:1991vd}. This is called the no-renormalization theorem for chiral superfields. 
\subsection{Quantization of supersymmetric gauge theories}
As in usual gauge theories, when we quantize a supersymmetric gauge theory with functional methods we have to fix the gauge, as in the path integral we do not have to integrate over the physically equivalent gauge field configurations related by gauge transformations \cite{Peskin:1995ev}. Hence, we will present here the generalization of the usual gauge fixing procedure to superspace. We start with the previously defined SUSY Yang-Mills \cite{Sohnius:1985qm} action
\br
S_{0} = - \frac{1}{2 g^2} tr \int d^4 x d^4 \th \; (e^{-g V} D^{\a} e^{g V}) \bar{D}^2 ( e^{-g V} D_{\a} e^{g V}) \;.
\er   
Then, with $\L$ (the chiral parameter of the gauge transformation), an arbitrary function $f$ and a gauge-variant function $F$ such that $F=f$ for some value of $\L$, we define a functional determinant as 
\br
\D (V) &=& \int [ d \L d \bar{\L}] \; \d[ F(V,\L,\bar{\L})-f] \; \d[ \bar{F}(V,\L,\bar{\L})-\bar{f}] \nonumber \\
&=& \int [ d \L d \bar{\L} d \L^{\prime} d \bar{\L}^{\prime}] e^{ \int d^6 z \; \L^{\prime} \left( \frac{\d F}{\d \L} \L + \frac{\d F}{\d \bar{\L}} \bar{\L} \right) + \int d^6 \bar{z} \; \bar{\L}^{\prime} \left( \frac{\d \bar{F}}{\d \L} \L + \frac{\d \bar{F}}{\d \bar{\L}} \bar{\L} \right)} \;, \label{SUSY_det_gauge_fix}
\er
where the variational derivatives are evaluated at $\L = \bar{\L} = 0$.

Then, we introduce inside the partition function the unity in the form of the functional determinant and its inverse. With a change of variables that is a gauge transformation we set the $\L$ integral as a constant infinite factor reabsorbed into the normalization \cite{Gates:1983nr}. Also, in order to get a result independent of $f$ and $\bar{f}$, we average with a gaussian weighting factor of the form $\int [d f d \bar{f}] \; exp(-\frac{1}{\a} tr \int d^8 z \bar{f}f)$. Hence, the partition function is written as
\br
Z &=& \int [dV] \; (\D (V))^{-1} \; exp\left\{ S_0 - \frac{1}{\a} tr \int d^8 z \; \bar{F} F\right\} \;.
\er
Using as gauge fixing function $F = \bar{D}^2 V$ and replacing the parameters $\L$, $\L^{\prime}$ of (\ref{SUSY_det_gauge_fix}) by anticommuting chiral ghost fields $c$, $c^{\prime}$ (superfield extension of Faddeev-Popov ghosts  \cite{Faddeev:1967fc}) we find \cite{Gates:1983nr}
\br
Z &=& \int [d V d c d c^{\prime} d \bar{c} d \bar{c}^{\prime}] e^{S_0 + S_{GF} + S_{FP} } \;,
\er
where
\br
S_{GF} &=& - \frac{1}{\a} tr \int d^8 z \; ( D^2 V ) ( \bar{D}^2 V ) \nonumber \\
S_{FP} &=& tr \int d^4 x d^4 \th \; ( c^{\prime} + \bar{c}^{\prime} ) L_{\frac{1}{2}g V} \left[ ( c + \bar{c})+coth L_{\frac{1}{2} g V} (c - \bar{c}) \right] ~~,~~ L_X Y = \comm{X}{Y} \nonumber \\
\er

\chapter{Background (super)field method}
\label{ap_BFM}
In order to quantize a gauge theory with functional methods, a gauge fixing procedure has to be used, so as no gauge field configurations related to a given one by gauge transformations are taken into account in the path integral (all of them correspond to the same physical state). However, as a result of this procedure, explicit gauge invariance is lost. The background field method was developed (see references from \cite{DeWitt:1967ub} to \cite{Grisaru:1975ei}) to allow us to fix a gauge without losing explicit gauge invariance. Although originally the method was developed for the one-loop case, it was soon extended to include multi-loop effects \cite{Abbott:1980hw,DeWitt:1980jv,'tHooft:1975vy,Capper:1982tf,Abbott:1983zw}. The basic idea is the splitting of the gauge field in two parts: the quantum field, which is the variable of integration in the functional integral, and the background field. Thus, we are allowed to fix the gauge for the quantum field whereas we can maintain explicit gauge invariance in the background one. We will first discuss the non-supersymmetric case and then obtain the generalization of the method to superspace \cite{Gates:1983nr,Grisaru:1979wc}.

\section{Yang-Mills theory}   
We will use the conventions of \cite{Zinn-Justin:1993wc} which are those detailed in section \ref{YM_conventions}. We begin defining the splitting of the gauge field in two parts as
\br
A_{\m}^a \rightarrow A_{\m}^a + B_{\m}^a \;,
\er
where $A_{\m}^a$ is the quantum field and $B_{\m}^a$ is the background field. With this splitting, let us consider the functional ${\bf{Z}}[B] = \int [dA] e^{-S_0(A+B)}$
where $S_0 (A) = 1/4 \int d^4 x F_{\m \n}^a (A) F_{\m \n}^a (A) $ is the usual Yang-Mills action. After the usual gauge fixing procedure \cite{Abbott:1980hw} this functional becomes ($c$, $\bar{c}$ are Faddeev-Popov ghost fields)
\br
{\bf{Z}}[B] &=& \int [dA dc d \bar{c}] exp \left\{ - S_0 (A + B) - \frac{1}{2 \a} tr \int d^4 x \;  F(A,B)^2  + tr \int d^4 x \; \bar{c} \left. \frac{\d F(B)}{\d w} \right|_{w = 0} c \right\} \;. \nonumber \\
\er

Notice now that $S_0 (A +B)$ is invariant under two types of transformations
\begin{enumerate}
\item Quantum
\br
\d B_{\m}^a &=& 0 \nonumber \\
\d A_{\m}^a &=& \frac{1}{g} \left[ \pa_{\m} w^a + g f^{abc} B_{\m}^b w^c \right]+ f^{abc} A_{\m}^b w^c \nonumber \\
&=& \frac{1}{g} ({\bf{D}}_{\m}w)^a + f^{abc} A_{\m}^b w^c
\er
where ${\bf{D}}_{\m}$ is the background covariant derivative.
\item Background
\br
\d B_{\m}^a &=& \frac{1}{g} \pa_{\m} w^a + f^{abc} B_{\m}^b w^c \nonumber \\
\d A_{\m}^a &=& f^{abc} A_{\m}^b w^c
\er
\end{enumerate}
Our aim is to fix the quantum gauge invariance and at the same time maintain the background gauge invariance. Thus, the gauge fixing function has to transform covariantly with respect to background gauge transformations. Hence, we choose as gauge fixing function $F^a = ({\bf{D}}_{\m} A_{\m})^a$, which implies that ${\bf{Z}}[B]$ becomes
\br
{\bf{Z}}[B] &=& \int [dA dc d \bar{c}] e^{-S (A,B)} \nonumber \\
&=& \int [dA dc d \bar{c}] exp \left\{ - S_0(A+B) - \frac{1}{2 \a} tr \int d^4 x \; ( {\bf{D}}_{\m} A_{\m} )^2 + tr \int d^4 x \; \bar{c} [{\bf{D}}_{\m} {\cal{D}}_{\m}] c \right\} \;, \nonumber \\
\er
with $({\cal{D}}_{\m} w)^a = \pa_{\m} w^a + g f^{abc} (A_{\m}^b + B_{\m}^b) w^c$. 

As can be seen, ${\bf{Z}}[B]$ is manifestly invariant under background gauge transformations. In order to make the connection with the usual functionals ($Z$,$W= \ln Z$ and $\G[\bar{A}] = \int J \bar{A} - W$ with $\bar{A} = \d W / \d J$), we define another functional like ${\bf{Z}}[B]$ but with the quantum field coupled to a source 
\br
\tilde{Z}[J,B] &=& \int [d A dc d \bar{c}] e^{- S (A,B) + \int d^4 x \; J_{\m}^a  A_{\m}^a} \;.
\er 
We remark again that by construction, this is explicitly invariant with respect to background gauge transformations. Starting with $\tilde{Z}[J,B]$ we can define analogous functionals as the usual ones like $\tilde{W} = \ln \tilde{Z}$ and $\tilde{\G}[\tilde{A},B] = \int J \tilde{A} - \tilde{W}[J,B]$, with $\tilde{A} = \d \tilde{W}/\d J$. If we perform the change of variables in the partition function $A_{\m}^a \rightarrow A_{\m}^a - B_{\m}^a$ is straightforward to arrive to \cite{Abbott:1980hw}
\br
\tilde{Z}[J,B] &=& e^{ - \int JB} Z[J,B] \;,
\er
where $Z[J,B]$ is the usual partition function with the gauge fixing and ghost terms evaluated in an unusual but nevertheless valid gauge that depends on the background gauge field. So, we have for the other functionals
\br
\tilde{W}[J,B] &=& - \int J B + W[J,B]
\er
and, with $\bar{A} = \d W / \d J$ being the usual classical field
\br
\tilde{A} &=& - B + \bar{A} \nonumber \\
\tilde{\G}[\tilde{A},B] &=& \int J ( \tilde{A} + B ) - W[J,B] \nonumber \\
&=& \int J \bar{A} - W[J,B] \nonumber \\
&=& \G [ \bar{A},B]
\er
We have found is that $\tilde{\G}$ and the usual effective action $\G$ are related by
\br
\tilde{\G} [\tilde{A},B] &=& \G[ \tilde{A}+B,B] \;.
\er
If we restrict ourselves to diagrams with no external $\tilde{A}$ we have a relevant identity: $\tilde{\G} [0,B] = \G [B] $. This implies that the usual effective action can be obtained through the evaluation of $\tilde{\G}[0,B]$. This quantity is computed by summing all 1PI diagrams with $B$ fields on external legs ($\tilde{A} = 0$ implies that no $A$ field propagators appears on external lines) and $A$ fields inside loops (as the functional integral is only evaluated on $A$ fields).

One of the consequences of the background field method is that the renormalization of the gauge parameter ($g_0 = Z_g g$) and the background field ($B_0 = Z_B^{1/2} B$) are related. As we have explicit background gauge invariance, the infinites appearing in $\tilde{\G}[0,B]$ must take the form of a divergent constant times $(F_{\m \n}^a )^2$. At the same time, $F_{\m \n}^a$ is renormalized as
\br
(F_{\m \n}^a)_0 &=& Z_B^{1/2} \left[ \pa_{\m} B_{\n}^a - \pa_{\n} B_{\m}^a + g Z_g Z_B^{1/2} f^{abc} B_{\m}^b B_{\n}^c \right] \;.
\er
Hence, in order to get explicit background gauge invariance, the following relation must hold
\br
Z_g &=& Z_B^{-1/2} \;.
\er 
\section{Super Yang-Mills theory}
\label{BFM_SYM}
In this section we apply the conventions for superspace discussed in appendix \ref{ap_SUSY} (those of reference \cite{Gates:1983nr}, which we also follow in this section). As the gauge transformation is non-linear in the supersymmetric Yang-Mills theory, a linear splitting in the gauge field is unsuitable \cite{Gates:1983nr,Grisaru:1979wc}. In order to define a splitting, we will re-examine the Yang-Mills case from a different point of view. The Yang-Mills action is invariant under {\em{local}} transformations of the form $\d A_{\m}^a = 1/g \; \pa_{\m} w^a + f^{abc} A_{\m}^b w^c$. If the transformation is {\em{global}} we still have invariance, with the gauge field transforming as a matter field $\d A_{\m}^a = f^{abc} A_{\m}^b w^c$. Then, considering again a local $w$, we can gauge the global transformation using the background field to covariantize the derivatives. This covariantization is of the form
\br
(D_{\m}w)^a = ( \pa_{\m} w)^a + g f^{abc} A_{\m}^b w^c &\rightarrow& ( {\bf{D}}_{\m} w)^a + g f^{abc} A_{\m}^b w^c \nonumber \\
&=& \pa_{\m} w^a + g f^{abc} ( A_{\m}^b + B_{\m}^b ) w^c \;.
\er
Hence, we have a linear splitting because the gauge field is linear in the covariant derivative. The procedure for the supersymmetric case is completely analogous \cite{Gates:1983nr}. At the end, we have to covariantize the derivatives with the background field, which implies that we have to replace $D_A \rightarrow {\bnabla}_A$ with ${\bnabla}_A$ a background covariant derivative. However, we have to take also into account that the covariant derivatives in a supersymmetric gauge theory can be formulated in two representations:
\begin{itemize}
\item Chiral representation: This is more suitable for quantization. Hence, in the background field method, we work with a quantum $V$ in a chiral representation.
\item Vector representation: We do not have to quantize the background fields. Hence, vector representation is useful for these fields as background covariance will be manifest. In fact, we will show that we can work with the background covariant derivatives without introducing explicitly the background prepotentials.
\end{itemize}
So, we define the supersymmetric splitting writing the covariant derivatives in a quantum chiral but background vector representation:
\br
\nabla_{\a} &=& e^{-g V} \bnabla_{\a} e^{g V} \nonumber \\
\nabla_{\dot{\a}} &=& \bar{\bnabla}_{\dot{\a}} \nonumber \\
\nabla_{\a \dot{\a}} &=& -i \anticomm{\nabla_\a}{\nabla_{\dot{\a}}} 
\er
with $g$ the coupling constant.
  
If we go to a background chiral representation (as can be seen in \ref{SUSY_vector_rep} this is achieved by multiplying with $e^{-g \bar{\boldsymbol{\Omega}}} (\ldots) e^{g \bar{\boldsymbol{\Omega}}}$, where  $\boldsymbol{\Omega}$ is the background prepotential), we can straightforwardly see that this splitting is equivalent to \cite{Gates:1983nr}
\br
e^{g V} \rightarrow e^{g \boldsymbol{\Omega}} e^{g V} e^{g \bar{\boldsymbol{\Omega}}} \;. \label{BFM_SYM_splitting} 
\er
The split derivatives $\nabla_A$ transform covariantly under two sets of transformations:
\begin{enumerate}
\item Quantum:
\br
e^{g V} &\rightarrow& e^{i g \bar{\L}} e^{g V} e^{-ig  \L} ~~,~~ \bnabla_{\a} \bar{\L} = \bar{\bnabla}_{\dot{\a}} \L = 0 \nonumber \\
\bnabla_A &\rightarrow& \bnabla_A
\er
Which implies that $\nabla_A$ transforms as
\br
\nabla_A \rightarrow e^{ig  \L} \nabla_A e^{-ig \L} \;.
\er
\item Background:
\br
e^{g V} &\rightarrow& e^{ig K} e^{g V} e^{- ig K} ~~,~~ K = \bar{K} \nonumber \\
\bnabla_A &\rightarrow& e^{ig K} \bnabla_A e^{-ig K}
\er
Which implies
\br
\nabla_A \rightarrow e^{ig K} \nabla_A e^{-ig K} \;.
\er
\end{enumerate}
Although $\nabla_A$ has different transformations under background and quantum transformations, is not difficult to show that the transformation of the unsplit gauge field is the same in both cases \cite{Gates:1983nr}.

If we study the background splitting of an abelian theory (SuperQED), the situation is simpler. In concrete, we have a linear quantum-background splitting of the form
\br
V \rightarrow V + B \;,
\er
where $V$ and $B$ are the quantum and background gauge fields respectively. This can be seen from the general supersymmetric quantum-background splitting expressed in terms of the background prepotential $\boldsymbol{\O}$ (\ref{BFM_SYM_splitting})
\br
e^{V} &\rightarrow& e^{\boldsymbol{\O}} e^{V} e^{\bar{\boldsymbol{\O}}} \nonumber \\
B &=& \boldsymbol{\O} + \bar{\boldsymbol{\O}} \;.
\er 
With this splitting the two sets of transformations are 
\begin{enumerate}
\item Quantum:
\br
V &\rightarrow& V + i ( \bar{\L} - \L ) \nonumber \\
B &\rightarrow& B \;.
\er
\item Background:
\br
V &\rightarrow & V  \nonumber \\
B &\rightarrow & B + i( \bar{\L} - \L ) \;.
\er
\end{enumerate}

Let us now consider the background field quantization. We start defining, as in the Yang-Mills case, a partition function ${\bf{Z}}$ with the gauge field split. After the gauge fixing procedure with a background covariantly chiral gauge fixing function of the form of $F = \bar{\bnabla}^2 V$ (which implies that the Faddeev-Popov ghosts are also background covariantly chiral) this functional becomes \cite{Gates:1983nr}
\br
{\bf{Z}} &=& \int [dV dc dc^{\prime} d \bar{c} d \bar{c}^{\prime}] \; \d( \bar{\bnabla}^2 V - f) \d( \bnabla^2 V - \bar{f}) e^{S_0 + S_{FP}} \;.
\er
In this case, due to the fact that we are dealing with constrained background chiral superfields rather than usual chiral superfields, when we gauge-average we have to consider a more sophisticated function. If we average with a factor like $exp \int f M f$, with $M$ an operator, in order to normalize we have to divide by $det M$. Hence, if $M$ is a function of the background field we must average with an expression of the form
\br
\int [d f d b] e^{ f M f} e^{b M b} \;,
\er
with $b$ a field of opposite statistics to $f$ that is called Nielsen-Kallosh ghost \cite{Nielsen:1978mp}. As this field only interacts with the background field and enters quadratically in the action, it only contributes at the one-loop level. In our case we gauge-average with a factor as
\br
\int [d f d \bar{f} d b d \bar{b}] e^{- \int d^8 z \; [ \bar{f}f + \bar{b}b]} \;,
\er
where it is clear that $b$,$\bar{b}$ are background covariantly chiral ghost fields.
All of this implies that the partition function can be written as
\br
{\bf{Z}} &=& \int [ d V d c d c^{\prime} d \bar{c} d \bar{c}^{\prime} d b d \bar{b} ] e^{S_{eff}} \nonumber \\
S_{eff} &=& S_0 + S_{GF} + S_{FP} + \int \bar{b}b \;. \label{BFM_SYM_Seff}
\er
As in the usual Yang-Mills case, we can relate the background field functional with the usual effective action (in a special background gauge) once we set the sources to zero and consider diagrams with external background lines and internal quantum propagators \cite{Gates:1983nr}. 

\subsection{Covariant Feynman rules}
After the supersymmetric quantum-background splitting, two approaches can be followed to perform the calculations. The first one is to use explicitly the background connections of the covariant derivatives in the calculations ($\bnabla_{\a} = D_{\a} + {\bf{\G}}_{\a}$ and $\bnabla_{\a \dot{\a}} = \pa_{\a \dot{\a}} + {\bf{\G}}_{\a \dot{\a}}$). So, in this situation we can apply usual D-algebra \cite{Abbott:1984pz}. On the other approach, we do not extract the spinor connection of the covariant derivatives. Therefore, instead of using usual D-algebra, we apply the covariant D-algebra defined for these derivatives. Supergraphs obtained in this way give contributions that are only functions of the space-time connection ${\bf{\G}}_{\a \dot{\a}}$ or the field strength ${\bf{W}}_{\a}$. Not only these diagrams are simpler and fewer in number than those of the first procedure, but they are more convergent, as we do not have contributions with $\bG_{\a}$, which is of lower dimension that $\bW_{\a}$ and $\bG_{\a \dot{\a}}$. In this section we will detail this second approach.

Let us consider a quantum-background split action of the form of
\br
S &=& - \frac{1}{2 g^2} tr \int d^4 x d^4 \th \; ( e^{-g V} \bnabla^{\a} e^{g V}) \bar{\bnabla}^2 ( e^{- g V} \bnabla_{\a} e^{g V} )  \nonumber \\
& & + \int d^4 x d^4 \th \; \bar{\phi} e^{g V} \phi + \int d^4 x \left[ d^2 \th \; P( \phi )+ h.c. \right] \;,
\er
where $\phi$ is background covariantly chiral superfield.

After the gauge fixing procedure, we add to the action a gauge-fixing term ($S_{GF}$), Faddeev-Popov ($S_{FP}$) and Nielsen-Kallosh ghosts terms ($S_{NK}$). All of them have the following expressions
\br
S_{GF} &=& - \frac{1}{\a} tr \int d^4 x d^4 \th ( \bnabla^2 V ) ( \bar{\bnabla}^2 V ) \nonumber \\
S_{FP} &=& tr \int d^4 x d^4 \th ; \left[ \bar{c}^{\prime} c - c^{\prime} \bar{c} + \frac{1}{2} ( c^{\prime} + \bar{c}^{\prime} ) \comm{gV}{c+\bar{c}} + \ldots\right] \nonumber \\
S_{NK} &=& \frac{1}{\a} tr \int d^4 x d^4 \th \; \bar{b} b 
\er
The quantum gauge quadratic action can be written as \cite{Gates:1983nr,Grisaru:1979wc,Grisaru:1984ja,Grisaru:1984jc}
\br
& & - \frac{1}{2} tr \int d^4 x d^4 \th V \left[ {\boldsymbol{\Box}} - i \bW^{\a} \bnabla_{\a} - i \bar{\bW}^{\dot{\a}} \bar{\bnabla}_{\dot{\a}} \right] V \nonumber ~~~,~~~ \bfBox = \frac{1}{2} \bnabla^{\a \dot{\a}} \bnabla_{\a \dot{\a}} 
\er
We denote this kinetic operator as $\hat{\dal} = \bfBox - i \bW^{\a} \bnabla_{\a} - i \bar{\bW}^{\dot{\a}} \bar{\bnabla}_{\dot{\a}}$. Also, for the background covariantly chiral superfields, we define operators $\Box_{\pm}$ by
\br
\bar{\bnabla}^2 \bnabla^2 \phi &=& \Box_{+} \phi ~~~,~~~ \Box_{+} = \Box - i \bW^{\a} \bnabla_{\a} - \frac{i}{2} ( \bnabla^{\a} \bW_{\a} ) \nonumber \\
\bnabla^2 \bar{\bnabla}^2 \bar{\phi} &=& \Box_{-} \bar{\phi} ~~~,~~~ \Box_{-} = \Box - i \bar{\bW}^{\dot{\a}} \bar{\bnabla}_{\dot{\a}} - \frac{i}{2} ( \bar{\bnabla}^{\dot{\a}} \bar{\bW}_{\dot{\a}})
\er
The covariant Feynman rules can be obtained from the partition function once we have introduced real and chiral sources $J, j$. This partition function can be written as \cite{Gates:1983nr}
\br
Z = \D_{+} \hat{\D} exp \left[ S_{int}\left( \frac{\d}{\d J},\frac{\d}{\d j},\frac{\d}{\d \bar{j}}\right) \right] exp \left[ \int d^4 x d^4 \th ( \frac{1}{2} J \hat{\dal}^{-1} J - \bar{j} \bfBox^{-1}_{+} j ) \right] \;,
\er
where $\hat{\D}$, $\D_{+}$ are one-loop contributions from real and chiral superfields (including ghosts). From this it is clear that covariant Feynman rules are similar to the usual ones except that for V-lines we have $\hat{\dal}^{-1}$ as the propagator, $\bar{\phi}$ and $\phi$ fields in vertices are joined by $-\Box_{+}^{-1}$, and usual $D^2$ factors at the vertices are replaced by $\bnabla^2$ factors. Making use of these covariant Feynman rules, in the background field approach we consider vacuum graphs with quantum vertices derived from $S_{int}$, $\hat{\dal}^{-1}$ and $-\Box_{+}^{-1}$ propagators and $\bnabla^2$, $\bar{\bnabla}^2$ factors. The idea is to use the algebra of covariant derivatives and Bianchi identities to push, in each propagator, the covariant spinor derivatives to a given vertex. At this point, using the anticommutation relations, they can be integrated by parts or eliminated in favour of space-times derivatives. Finally, we have to apply the relation \cite{Grisaru:1984ja}
\br
\d_{12} \bnabla^2 \bar{\bnabla}^2 \d_{12} = \d_{12} \;, \label{cov_SUSY_delta_id}
\er   
in order to obtain free grassmanian $\d$-functions that allow us to evaluate $\th$ integrals.

\chapter{Gauge parameters and the RG equations}
\label{ap_Gauge}

In order to fix the gauge when quantizing the different theories that we have treated, a gauge fixing term that depends in an additional parameter $\a$ has been added to the action. Therefore, we have to take care of the running of this gauge parameter in the RG equations. In general, these equations applied to two-point functions are of the form
\br
\left[ M \frac{\pa}{\pa M} + \b \frac{\pa}{\pa g} + \g_{\a} \frac{\pa}{\pa \a} - 2 \g \right] \G_2 = 0 \;, \nonumber 
\er 
where $\g$ is the anomalous dimension, $\b$ the beta function and $\g_{\a}$ the function that takes care of the running of the gauge parameter. So, we need $\g_{\a}$ to verify the two-loop background RG equations. However, as in the models that we have considered the first dependence on $\a$ of the background two-point function can only arise at the one-loop level, we can obtain the relevant terms of the $\g_{\a}(g)$ expansion by means of the evaluation of the one-loop RG equations for the quantum gauge field self-energies.
\section{Abelian examples}

\subsection{QED}

\begin{figure}[ht]
\centerline{\epsfbox{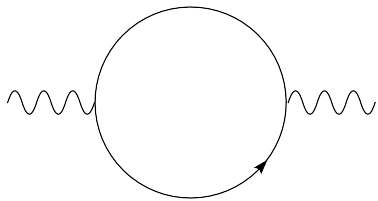}}
\caption{One-loop QED diagram. Wavy lines correspond to gauge fields and solid lines to fermion fields.}
\label{QED1loop_quantum}
\end{figure}
The one-loop contribution to the quantum photon self-energy is shown in figure \ref{QED1loop_quantum}. This has an explicit expression of the form
\br
\Pi_{\m \n}^{AA \;(1 \; loop)} &=& - (i e)^2 Tr \left[ \g_{\m} \g^{\l} \pa_{\l}^x \D \g_{\n} \g^{\sigma} \pa_{\sigma}^y \D \right] \;. \nonumber \\
\er
which is the same that we have found for the background gauge fields in (\ref{QED1loop_bare}). Hence, with the same procedure as for the background fields, we find the following renormalized value 
\br
\Pi_{\m \n R}^{AA \; (1)} (x) &=& - ( \pa_{\m} \pa_{\n} - \d_{\m \n} \Box ) \left[ - \frac{e^2}{12 \pi^2 ( 4 \pi^2 )} \Box \frac{ \ln x^2 M^2}{x^2} - \frac{e^2}{36 \pi^2} \d (x) \right] \;. 
\er

With this, the quantum gauge field two-point function expanded to one-loop order in a general gauge is 
\br
\G_{\m \n \;R}^{A A}(x) &=& \left( \pa_{\m} \pa_{\n} - \d_{\m \n} \Box \right) \left[\d (x) - \frac{e^2}{9 (4 \pi^2)} \d (x) - \frac{e^2}{3 (4 \pi^2)^2} \Box \frac{ \ln x^2 M^2}{x^2} \right] - \frac{1}{\a} \pa_{\m} \pa_{\n} \d (x) + {\cal{O}}(e^4) \;. \nonumber \\ \label{QED1loop_Q_r}
\er

As this function satisfies the usual RG equation
\br
\left[ M \frac{\pa}{\pa M} + \b(e) \frac{\pa}{\pa e} + \g_{\a}(e) \frac{\pa}{\pa \a} - 2 \g_A (e) \right] \G_{\m \n \;R}^{A A} = 0 \;,
\er
using the one-loop expression (\ref{QED1loop_Q_r}) we find the following one-loop values for $\g_A$ and $\g_{\a}$
\br
\g_A(e) &=& \frac{1}{3(4 \pi^2)} e^2 + \ldots \\
\g_{\a} (e) &=& - \frac{2 \a}{3 (4 \pi^2)} e^2 + \ldots \;. \label{QED1loop_q_alfa} 
\er 

\subsection{Super QED}
\begin{figure}[ht]
\centerline{\epsfbox{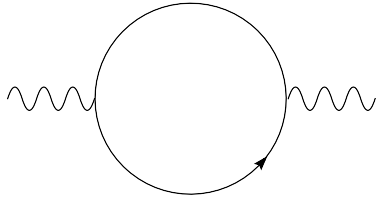}}
\caption{One-loop Super QED diagram. Wavy lines correspond to gauge fields and solid lines to $\Phi_{+}$ or $\Phi_{-}$ propagators.}
\end{figure}
For the supersymmetric extension of the previous model, the situation is very similar. We start by considering the kinetic part of the action for the quantum gauge field in a generic gauge. This is of the form 
\br
S_0^{(2)} &=& \frac{1}{2} \int d^4 x d^4 \th \; V D^{\a} \bar{D}^2 D_{\a} V - \frac{1}{\a} \int d^4 x d^4 \th \; ( D^2 V) ( \bar{D}^2 V) \nonumber \\
&=& - \frac{1}{2} \int d^4 x d^4 \th \; V \Box \Pi_{\frac{1}{2}} V - \frac{1}{2 \a} \int d^4 x d^4 \th \; V \Box \Pi_0 V \;,
\er
where we have used the projection operators $\Pi_{\frac{1}{2}} = - D^{\a} \bar{D}^2 D_{\a} / \Box$ and $\Pi_0 = ( D^2 \bar{D}^2 + \bar{D}^2 D^2 ) / \Box$. As in QED, the one-loop renormalized contribution to the quantum gauge field self-energy is the same that we have evaluated for the background case (\ref{SQED_1loop_ren}). So, the complete expansion to one-loop order of this self-energy is
\br
\G(x) &=& - \frac{1}{2} \Box \Pi_{\frac{1}{2}} \d (x) - \frac{1}{2 \a} \Box \Pi_0 \d(x) + \frac{g^2}{4 ( 4 \pi^2)^2} \Box \Pi_{\frac{1}{2}} \Box \frac{ \ln x^2 M^2}{x^2} + {\cal{O}}(g^4)\;.
\er

Thus, considering that this amplitude satisfies a RG equation of the form
\br
\left[ M \frac{\pa}{\pa M} + \b (g) \frac{\pa}{\pa g} + \g_{\a}(g) \frac{\pa}{\pa \a} - 2 \g_V(g) \right] \G = 0 \;,
\er
the values that we find for $\g_V$ and $\g_{\a}$ are
\br
\g_V &=& \frac{1}{2(4 \pi^2)} g^2 + \ldots \nonumber \\
\g_{\a} &=& - \frac{\a}{(4 \pi^2)} g^2 + \ldots 
\er

\section{Non-abelian examples} 

\subsection{Yang-Mills}
\label{ap_Gauge_YM}
\begin{figure}[ht]
\centerline{\epsfbox{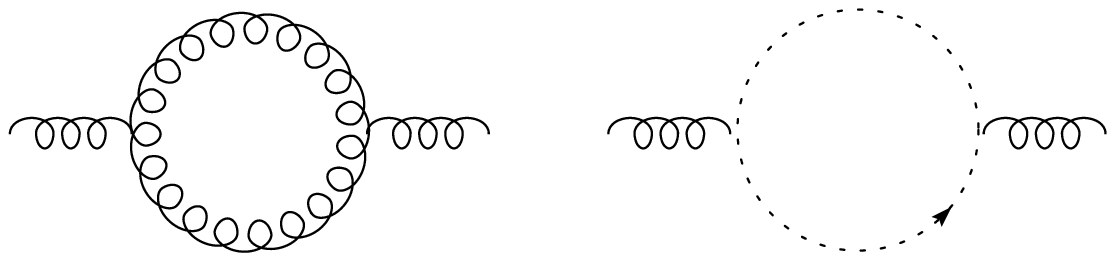}}
\caption{One-loop Yang-Mills diagrams. Curvy lines correspond to gauge fields and dashed lines to ghosts.}
\label{YM1loop_quantum_diag}
\end{figure}
We begin writing the effective action as 
\br
\G &=& \frac{1}{2} \int d^4 x d^4 y \; A_{\m}^a (x) A_{\n}^b (y) \G^{AA \; ab}_{\m \n}(x-y) + {\cal{O}} (A^3) \;.
\er

If we consider the part of the Yang-Mills lagrangian which depends only in the quantum fields $A_{\m}^a$, in a generic gauge this is of the form
\br
\frac{1}{4} F_{\m \n}^a F_{\m \n}^a + \frac{1}{2} (1 + \xi) ( \pa_{\m} A_{\m}^a)(\pa_{\n} A_{\n}^a ) \;.
\er
Notice that we have redefined the usual gauge parameter $\a$ as $\frac{1}{\a} = ( 1 + \xi)$. With this, the effective action can be written as
\br
\G &=& \frac{1}{2} \int d^4 x d^4 y \; A_{\m}^a \left[ \d^{ab}\left( - \d_{\m \n} \Box \d (x-y) - \xi \pa_{\m} \pa_{\n} \d (x-y) \right)- \Pi_{\m \n}^{AA\;ab}(x-y) \right] A_{\n}^a(y)  \nonumber \\
& & + {\cal{O}} (A^3) \;.
\er

At the one-loop level, as is shown in figure \ref{YM1loop_quantum_diag}, we have contributions with gauge and ghost loops. We first obtain the fully expanded bare expressions (in Feynman gauge) and then renormalize them according to CDR rules.
\begin{itemize}
\item {\bf Gauge loop}
\end{itemize}
\br
&& \frac{g^2 f^{acd} f^{bdc}}{2} \D_{xy} \left[ \d_{\m \rho} ( D_{\s}^x - \stackrel{\leftarrow}{\pa_{\s}^{ x}}) + \d_{\s \m} ( \pa_{\rho}^x - D_{\rho}^x ) + \d_{\rho \s} ( \stackrel{\leftarrow}{\pa_{\m}^{x}} -\pa_{\m}^x ) \right] \nonumber \\
& & \times \left[ \d_{\n \s} (D_{\rho}^y - \pa_{\rho}^y) + \d_{\rho \n} ( \stackrel{\leftarrow}{\pa_{\s}^{y}} - D_{\s}^y) + \d_{\rho \s} ( \pa_{\n}^y - \stackrel{\leftarrow}{\pa_{\n}^{y}} ) \right] \D_{xy} \nonumber \\
&=& \frac{g^2 C_A \d^{ab}}{2}  \left[ 2 ( \pa_{\m} \pa_{\n} - \d_{\m \n} \Box ) \D^2 + 10 \pa_{\m} ( \D \pa_{\n} \D ) - 10 \D \pa_{\m} \pa_{\n} \D   \right. \nonumber \\
& & - \left. 4 \d_{\m \n} \pa^{\l} ( \D \pa_{\l} \D ) - 2 \d_{\m \n} \D ( \Box \D ) \right] \;.
\er
\begin{itemize}
\item {\bf Ghost loop}
\end{itemize}
\br
&& - g^2 f^{adc} f^{bcd} \D_{xy} \stackrel{\leftarrow}{\pa_{\m}^{ x}} \pa_{\n}^y \D_{xy} = - g^2 C_A \d^{ab} \left[ \pa_{\m}( \D \pa_{\n} \D) - \D \pa_{\m} \pa_{\n} \D \right] \;.
\er

Adding the two previous results we find the total bare contribution to be
\br
\Pi_{\m \n\;(1)}^{AA\;ab}(x) &=& g^2 C_A \d^{ab} \left[ \pa_{\m} \pa_{\n} \D^2 - \d_{\m \n} \Box \D^2 + 4 \pa_{\m} ( \D \pa_{\n} \D ) - 2 \d_{\m \n} \pa^{\l} ( \D \pa_{\l} \D )  \right. \nonumber \\
& & - \left. 4 \D \pa_{\m} \pa_{\n} \D - \d_{\m \n} \D ( \Box \D ) \right] \;, \label{YM1loop_quantum_bare_prop}
\er
and with CDR identities it is straightforward to obtain the renormalized result as
\br
\left. \Pi_{\m \n\;(1)}^{AA\;ab}(x) \right|_R &=& g^2 C_A \d^{ab} \left[ \frac{5}{3}(  \pa_{\m} \pa_{\n} - \d_{\m \n } \Box) \D^2_R (x) - \frac{1}{72 \pi^2} (\pa_{\m} \pa_{\n} - \d_{\m \n} \Box) \d (x) \right] \nonumber \\
&=& - \frac{g^2 C_A \d^{ab}}{144 \pi^2} (\pa_{\m} \pa_{\n} - \d_{\m \n} \Box) \left[ \frac{15}{4 \pi^2} \Box \frac{ \ln x^2 M^2}{x^2} + 2 \d (x) \right] \;. \label{YM1loop_quantum}
\er

So, $\G^{ ab}_{\m \n}$ can be written as
\br
\G^{AA \; ab}_{\m \n \;R } (x)&=& - \d_{\m \n} \Box \d (x) - \xi \pa_{\m} \pa_{\n} \d (x) + \d^{ab} (\pa_{\m} \pa_{\n} - \d_{\m \n} \Box) \left[ \frac{5 g^2 C_A}{48 \pi^2 (4 \pi^2)} \Box \frac{\ln x^2 M^2}{x^2}  \right. \nonumber \\
& &+ \left. \frac{g^2 C_A}{72 \pi^2 (4 \pi^2)} \d (x) \right] + {\cal{O}} (g^4) \;.
\er

Inserting this in the RG equation
\br
\left[ M \frac{\pa}{\pa M} + \b (g) \frac{\pa}{\pa g} + \g_{\xi} \frac{\pa}{\pa \xi} - 2 \g_A \right] \G^{AA \; ab}_{\m \n \;R} |_{\xi=0} =0 \;,
\er
we can easily obtain the values for $\g_{\xi}$ and $\g_A$ as
\br
\g_{\xi} &=& - \frac{5 C_A}{24 \pi^2} g^2 + \cdots \nonumber \\
\g_A &=& - \frac{5 C_A}{48 \pi^2} g^2 + \cdots \ \label{YM1loop_g_xi} 
\er

\subsection{Super Yang-Mills}
\begin{figure}[ht]
\centerline{\epsfbox{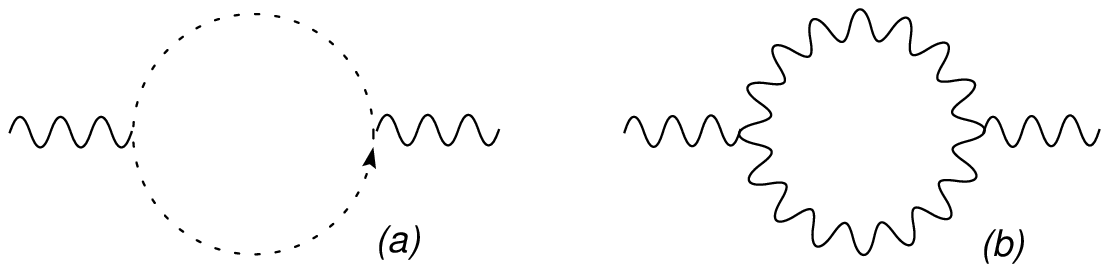}}
\caption{One-loop Super Yang-Mills diagrams. Wavy lines correspond to gauge fields and dashed lines represent ghosts.}
\label{SYM1loop_quantum_diag}
\end{figure}
The kinetic term of the Super Yang-Mills action for quantum gauge fields in a generic gauge is
\br
S^{(2)}_V &=& - \frac{1}{2} tr \int d^4 x d^4 y d^4 \th \; V^{a}(x,\th) \left[ \Box \Pi_{1/2} + (1 + \xi ) \Box \Pi_0 \right] V^{a} (y, \th) \d (x-y) \;, \nonumber \\
\er
where again the usual gauge parameter $\a$ is redefined as $\frac{1}{\a} = 1 +\xi$. We have for these fields an effective action of the form of\br
\G_{V} = \int d^4 x d^4 y d^4 \th \; V^{a}(x, \th) \G^{ab\;(2)}_V (x-y) V^{b}(y,\th) + \ldots
\er

The diagrams that contribute to the one-loop quantum gauge field self-energy are those of figure \ref{SYM1loop_quantum_diag}. As an example, we will detail the calculation of the ghost contribution (diagram $(a)$). This is of the following form
\br
\G^{1 \; loop}_{V \;(a)} &=& \frac{g^2 C_A}{2^3} \int d^8 z_1 d^8 z_2 \; V^{a}(z_1) V^{a}(z_2) \left\{ \left[ D^2_2 \bar{D}^2_2 P_{12} \right] \left[ D^2_2 \bar{D}^2_2 P_{12} \right] + \left[ \bar{D}^2_2 D^2_2 P_{12} \right] \right. \nonumber \\
& & \left. \times \left[ \bar{D}^2_2 D^2_2 P_{12} \right] - 2 \left[ \bar{D}^2_2 D^2_2 P_{12} \right] \left[ D^2_2 \bar{D}^2_2 P_{12} \right] \right\} \nonumber \\
&=& \frac{g^2 C_A}{2^3} \int d^8 z_1 d^8 z_2 \;  V^a(z_1) \left[  \bar{D}^2 D^2 V^{a}(z_2) \right] P_{12} \left[ D^2_2 \bar{D}^2_2 P_{12} \right] \nonumber \\
& & + \frac{g^2 C_A}{2^3} \int d^8 z_1 d^8 z_2 \; V^{a}(z_1) \left[ D^2 \bar{D}^2 V^a (z_2) \right] P_{12}  \left[ D^2_2 \bar{D}^2_2 P_{12} \right]  \nonumber \\
& & - \frac{g^2 C_A}{2^2} \int d^8 z_1 d^8 z_2 \; V^{a}(z_1) \left[\bar{D}^2 D^2 V^{a}(z_2)  \right] P_{12} \left[ \bar{D}^2_2 D^2_2 P_{12} \right]  \nonumber \\
& & - \frac{g^2 C_A}{2^2} \int d^8 z_1 d^8 z_2 \; V^{a}(z_1) V^{a}(z_2) P_{12} \left[ \Box \bar{D}^2_2 D^2_2 P_{12} \right] \nonumber \\
& & + \frac{i g^2 C_A}{2^2} \int d^8 z_1 d^8 z_2 \; V^{a}(z_1) \left[ \bar{D}^{\dot{\a}} D^{\a} V^a(z_2) \right] P_{12} \left[ \pa_{\a \dot{\a}}^2 \bar{D}^2_2 D^2_2 P_{12} \right] \;,
\er
where in the second step we have used identity (\ref{D_algebra_id}). With this expression it is clear that we can apply  the $\d$-function property (\ref{SUSY_delta_propagators}) that leaves a free grassmanian $\d$-function, allowing us to perform one of the $\th$ integrals. At this point, renormalizing and identifying $x_1 = x$, $x_2 = y$  we find for this contribution 
\br
\G^{1\;loop}_{V\;(a)} &=& - \frac{g^2 C_A}{2} \int d^4 x d^4 y d^4 \th \; V^a (x, \th) \left[ - \frac{1}{4} \Box \Pi_0 - \frac{1}{4} \Box \Pi_{1/2} \right] V^a (y, \th) \; \D^2_{xy \; R} \;. \nonumber \\
\er

Diagram $(b)$ can be evaluated in a similar way \cite{Grisaru:1979wc}. The final result is found to be
\br
\G^{1\;loop}_{V\;(b)} &=& - \frac{g^2 C_A}{2} \int d^4 x d^4 y d^4 \th \; V^a(x, \th) \left[ - \frac{5}{4} \Box \Pi_{1/2} + \frac{1}{4} \Box \Pi_0 \right] V^a (y, \th) \; \D^2_{xy \; R} \;. \nonumber \\
\er

The total contribution is then obtained as
\br
\G^{1\;loop}_V &=& \G^{1\;loop}_{V\;(a)} + \G^{1\;loop}_{V\;(b)} \nonumber \\
&=& - \frac{3 g^2 C_A}{16 (4 \pi^2)^2} \int d^4 x d^4 y d^4 \th \; V^a(x, \th) \Box \Pi_{1/2} V^{a} (y, \th) \Box \frac{ \ln (x-y)^2 M^2}{(x-y)^2} \;. \label{SYM1loop_eff_action_quantum}
\er

From the effective action, defining $\G_V^{ab\;(2)}(x) = \d^{ab} \G^{(2)}_V(x)$, we have a RG equation for quantum gauge fields of the form of
\br
\left. \left[ M \frac{\pa}{\pa M} + \b(g) \frac{\pa}{\pa g} + \g_{\xi}(g) \frac{\pa}{\pa \xi} - 2 \g_V \right] \G^{(2)}_V (x) \right|_{\xi = 0} = 0 \;,
\er
and we find $\G^{(2)}_V (x)$ to be evaluated as 
\br
\G^{(2)}_V (x) = - \frac{1}{2} \d(x) \Box \Pi_{1/2} - \frac{1}{2} \d(x) (1 + \xi) \Box \Pi_0 - \frac{3 g^2 C_A}{16 ( 4 \pi^2)^2} \Box \frac{ \ln x^2 M^2}{x^2} \Box \Pi_{1/2} + {\cal{O}}(g^4) \;. \nonumber \\
\er

Straightforward operations lead us to obtain the following values for $\g_V$ and $\g_\xi$
\br
\g_V &=& - \frac{3 C_A}{8 (4 \pi^2)} g^2 + \ldots \nonumber \\
\g_\xi &=& - \frac{3 C_A}{4 (4 \pi^2)}g^2 + \ldots \label{SYM1loop_RG_quantum}
\er
\chapter{Explicit calculations}
\label{ap_calc}
\section{Integrals with overlapping divergences}
\label{ap_integrales}
Here we will show how we can obtain the different expressions for the integrals with overlapping divergences presented in section \ref{overlap_integrals}.
\subsection{Notation}
Be begin by discussing some notation that we use in these calculations. As we did when we listed integrals of section \ref{overlap_integrals}, we will write the final results in terms of a variable $z=x-y$. Some intermediate local results will be found to be multiplied by a constant termed $a$ which value is $a=6 \pi^4 \xi(3) / ( 4 \pi^2)^4$.

\subsubsection{Integral relations}
To obtain some of the results, we use the following exact integral relations
\br
\int & d^4 u d^4 v &  \D_{xu} ( \pa_{\m}^x \D_{xv} ) ( \pa_{\l}^y \pa_{\n}^y \D_{yu} ) \D_{yv} \D_{uv} \nonumber \\
&=& \pa_{\l}^x \int d^4 u d^4 v \; \D_{xu} ( \pa_{\m}^x \D_{xv} ) \D_{yu} ( \pa_{\n}^y \D_{yv} ) \D_{uv}  \nonumber \\
& & + \int d^4 u d^4 v \; \D_{xu} ( \pa_{\m}^x \D_{xv} ) \D_{yu} ( \pa_{\n}^y \pa_{\l}^y \D_{yv} ) \D_{uv}  \nonumber \\
& & + \pa_{\n}^x \int  d^4 u d^4 v \; \D_{xu} ( \pa_{\m}^x \D_{xv}) \D_{yu} ( \pa_{\l}^y \D_{yv} ) \D_{uv}  \nonumber \\
& & + \pa_{\n}^x \pa_{\l}^x \int d^4 u d^4 v \; \D_{xu} ( \pa_{\m}^x \D_{xv}) \D_{yu} \D_{yv} \D_{uv}  \label{rel_int1}
\er
\br
\pa_{\l}^y \int & d^4 u d^4 v & \D_{xu} \D_{xv} ( \pa_{\l}^y \pa_{\n}^y \D_{yu} ) \D_{yv} \D_{uv} \nonumber \\
& = & \frac{1}{2} \pa_{\n}^y \int d^4 u d^4 v \; \D_{xu} \D_{xv} ( \Box \D_{yu} ) \D_{yv} \D_{uv}  \nonumber \\
& & - \int d^4 u d^4 v \; \D_{xu} \D_{xv} ( \Box \D_{yu} ) \D_{yu} \D_{uv}  \nonumber \\
& & + \frac{1}{2} \pa_{\n}^y \pa_{\l}^y \int d^4 u d^4 v \; \D_{xu} \D_{xv} ( \pa_{\l}^y \D_{yu} ) \D_{yv} \D_{uv} \label{rel_int2}
\er
\br
\pa_{\l}^y \int & d^4 u d^4 v& \D_{xu} \D_{xv} ( \pa_{\l}^y \pa_{\n}^y \D_{uy} ) ( \pa_{\m}^y \D_{yv} )\D_{uv}  \nonumber \\
&=& \frac{1}{4} \pa_{\n}^y \pa_{\m}^y \pa_{\l}^y \int d^4 u d^4 v \; \D_{xu} \D_{xv} ( \pa_{\l}^y \D_{uy} ) \D_{yv} \D_{uv}  \nonumber \\
& & + \frac{1}{4} \pa_{\n}^y \Box \int d^4 u d^4 v \; \D_{xu} \D_{xv} ( \pa_{\m}^y \D_{uy} ) \D_{yv} \D_{uv}  \nonumber \\
& & - \frac{1}{2} \Box \int d^4 u d^4 v \; \D_{xu} \D_{xv} ( \pa_{\m}^y \pa_{\n}^y \D_{yu} ) \D_{yv} \D_{uv} \label{rel_int3}
\er

To prove these relations, one has only to perform the derivatives and expand the different terms.

\subsection{Calculations}

\subsubsection{Evaluation of $H[1,1 \; ; \; 1,1]$, $H[\pa_{\m},1 \; ; \; 1,1]$ and $H[ 1, \pa_{\l} \; ; \; 1, \pa_{\l}]$}

These integrals are obtained by means of Gegenbauer Polynomials \cite{Freedman:1991tk,Song}.

\subsubsection{Evaluation of $\pa_{\l}^x H[ 1, \pa_{\m} \; ; \; 1, \pa_{\l}]$}

Contracting (\ref{rel_int1}) with $\d_{\n \l}$ we obtain
\br
\pa_{\l}^x H[ 1, \pa_{\m} \; ; \; 1, \pa_{\l}] &=& \frac{1}{2} \pa_{\m}^x H [ 1, 1 \; ; \; \Box, 1] - H[ 1, \pa_{\m} \; ; \; 1, \Box ] - \frac{1}{2} \Box H [ 1, \pa_{\m} \; ; \; 1,1] \;. \nonumber \\
\er
In order to renormalize, we have only to write the previous expressions in terms of the integral form $I^1(x)$ (remember $\Box \D = - \d$), and use the results found in section \ref{Nested_div}. Therefore we find
\br
\pa_{\l}^x H[ 1, \pa_{\m} \; ; \; 1, \pa_{\l}] &=& \frac{1}{2} \pa_{\m} ( \D I^1 ) - ( \D \pa_{\m} I^1 ) + \frac{a}{4} ( \pa_{\m} \d ) \nonumber \\
&\stackrel{R}{\rightarrow}& - \frac{1}{32 (4 \pi^2)^3} \pa_{\m} \Box \frac{ \frac{1}{2} \ln z^2 M^2}{z^2} + \ldots
\er

\subsubsection{Evaluation of $H[ \pa_{\m} \pa_{\l}, \pa_{\l} \; ; \; 1,1]$}

In this case, no integral relation is used to write $H[ \pa_{\m} \pa_{\l}, \pa_{\l} \; ; \; 1,1]$ in terms of $I^1$
\br
H[ \pa_{\m} \pa_{\l}, \pa_{\l} \; ; \; 1,1] &=& \frac{1}{2} \pa_{\m}^x H[ \pa_{\l}, \pa_{\l} \; ; \; 1,1] \nonumber \\
&=& \frac{1}{2} \pa_{\m}^x \pa_{\l}^x H[1, \pa_{\l} \; ; \; 1, 1] - \frac{1}{2} \pa_{\m}^x H[1, \Box \; ; \; 1,1] \nonumber \\
&=& \frac{1}{2} \pa_{\m} ( \D I^1 ) -  \frac{a}{4} ( \pa_{\m} \d) \nonumber \\
&\stackrel{R}{\rightarrow}&  \frac{1}{32(4 \pi^2)^3} \pa_{\m} \Box \frac{ - \frac{1}{2} \ln^2 z^2 M^2 - \ln z^2 M^2}{z^2} + \ldots
\er

\subsubsection{Evaluation of $\pa_{\l}^x H[ 1, \pa_{\m} \; ; \; \pa_{\n} \pa_{\l}, 1]$, $\pa_{\l}^x H[ 1, \pa_{\l} \; ; \; \pa_{\m} \pa_{\n}, 1]$ and $H[ 1, \pa_{\l} \; ; \; \pa_{\l} \pa_{\m},1]$}

First of all, the third integral (\ref{int6}) will be evaluated with relation (\ref{rel_int1})
\br
H[ 1, \pa_{\l} \; ; \; \pa_{\l} \pa_{\n},1] &=& \frac{1}{2} \pa_{\l}^x H [1, \pa_{\l} \; ; \; 1, \pa_{\n} ] + \frac{1}{2} \pa_{\l}^x H[ 1, 1 \; ; \; 1, \pa_{\l} \pa_{\n} ] \nonumber \\
& & + \frac{1}{2} \pa_{\n}^x H[ 1, \pa_{\l} \; ; \; 1, \pa_{\l}] + \frac{1}{2} \pa_{\n}^x \pa_{\l}^x H[ 1, \pa_{\l} \; ; \; 1,1] \;.
\er

Using the previous results
\br
H^R[ 1, \pa_{\l} \; ; \; \pa_{\l} \pa_{\n},1] &=& \frac{1}{32 (4 \pi^2)^3} \pa_{\m} \Box \frac{\frac{1}{8} \ln^2 z^2 M^2 - \frac{7}{8} \ln z^2 M^2 }{z^2} + \ldots
\er

However, this integral along with the other two, can be obtained with other method. The idea is to apply the CDR decomposition (\ref{CDR_T}) into trace part, traceless part and additional local terms to the divergent subdiagram  $ (\pa_{\m}^x \pa_{\n}^x \D_{yu}) \D_{yv} \D_{uv}$. Considering  this in the general integral 
\br
 \int &d^4 u d^4 v& \D_{xu} ( \pa_{\rho}^x \D_{xv} ) ( \pa_{\eps}^y \pa_{\s}^y \D_{yu} ) \D_{yv} \D_{uv}  \nonumber \\
&\stackrel{R}{\rightarrow}& - \frac{1}{4} \d_{\eps \s} [ \D \pa_{\rho} I^1 ]_R - \frac{\d_{\eps \rho}}{256 \pi^2} \pa_{\rho} \D^2_R - \frac{16}{(4 \pi^2)^5} I_{\rho \eps \s \; R} \;, \label{I_rho_eps_sig} \\
\er
where $I_{\rho \eps \s}$ stands for the traceless part. The one-loop ambiguity fixed by CDR is reflected in the second term of (\ref{I_rho_eps_sig}) that at two loops has become a logarithm of the scale. In the renormalization of the traceless part normal differential renormalization will be used, leaving ambiguities (local terms) not fixed.

The expression for $I_{\rho \eps \s}$ is 
\br
I_{ \rho \eps \s \; R } &=& B \frac{ x_{\eps} x_{\s} x_{\rho}}{x^8} - \frac{1}{2} A \frac{x_{\rho}}{x^6} \d_{\eps \s} + ( A - \frac{1}{2} B) \left[ \frac{x_{\eps}}{x^6} \d_{\rho \s} + \frac{x_{\s}}{x^6} \d_{\rho \eps} \right] |_R \;,
\er
or in the form of the integrals being discussed
\br
I_{\l \l \m \; R} &=& - \frac{3}{8} ( 4 \pi^2)^2 ( 3A - B ) \pa_{\m} \D^2_R \label{I_lam_lam_mu}\\
\pa_{\l} I_{\m \l \n \; R} &=& - (4 \pi^2)^2 ( 3A - B ) \left[ \frac{1}{24} \pa_{\m} \pa_{\n} \D^2_R + \frac{2}{3} ( 4 \pi^2) \d_{\m \n} \D^3_R \right] \\
\pa_{\l} I_{\l \m \n \; R} &=& (4 \pi^2)^2 (3A - B) \left[ - \frac{1}{6} \pa_{\m} \pa_{\n} \D^2_R + \frac{1}{3} (4 \pi^2) \d_{\m \n} \D^3_R \right] \;.
\er

The value of $(3A-B)$ is easily obtained using (\ref{I_lam_lam_mu}), because this corresponds to integral (\ref{int6}) that was obtained previously. I.e.
\br
\int &d^4 u d^4 v& \left. \D_{xu} ( \pa_{\l}^x \D_{xv}) ( \pa_{\l}^y \pa_{\n}^y \D_{yu}) \D_{yv} \D_{uv} \; \right|_R = \nonumber \\
&=& \frac{1}{32 (4 \pi^2)^3} \pa_{\m} \Box \frac{\frac{1}{8} \ln^2 z^2 M^2 - \frac{7}{8} \ln z^2 M^2 }{z^2} + \ldots \nonumber \\
&=& \frac{1}{32 (4 \pi^2)^3} \pa_{\n} \Box \frac{ \frac{1}{8} \ln^2 z^2 M^2 + \frac{1}{4} \ln z^2 M^2}{z^2}- \frac{16}{(4 \pi^2)^5} I_{\rho \eps \s \; R} \;,
\er 
which implies that
\br
(3A - B) &=& \frac{3 \pi^4}{8} \;.
\er

With this result the evaluation of (\ref{int8}) and (\ref{int10}) are straightforward
\br
\pa_{\l}^x H^R[ 1, \pa_{\m} \; ; \; \pa_{\n} \pa_{\l}, 1] &=& - \frac{1}{4} \pa_{\n} ( \D \pa_{\m} I^1 )_R - \frac{1}{256 \pi^2} \pa_{\m} \pa_{\n} \D^2_R - \frac{16}{(4 \pi^2)^5} \pa_{\l} I_{\m \l \n \; R} \nonumber \\
&=& \frac{1}{32(4 \pi^2)^3} \left[ \pa_{\m} \pa_{\n} \Box \frac{ \frac{1}{8} \ln^2 z^2 M^2 + \frac{1}{8} \ln z^2 M^2}{z^2} + \d_{\m \n} \Box \Box \frac{-\frac{1}{4} \ln z^2 M^2}{z^2} \right] \nonumber \\
& & + \ldots
\er

\br
\pa_{\l}^x H^R[ 1, \pa_{\l} \; ; \; \pa_{\m} \pa_{\n}, 1] &=& - \frac{1}{4} \d_{\m \n} \pa_{\l} ( \D \pa_{\l} I^1 )_R - \frac{\d_{\m \n}}{256 \pi^2} \Box \D^2_R - \frac{16}{(4 \pi^2)^5} \pa_{\l} I_{\l \m \n \; R} \nonumber \\
&=& \frac{1}{32 (4 \pi^2)^3} \left[ \pa_{\m} \pa_{\n} \Box \frac{ - \frac{1}{2} \ln z^2 M^2}{z^2}  + \d_{\m \n} \Box \Box \frac{\frac{1}{8} \ln^2 z^2 M^2 + \frac{3}{8} \ln z^2 M^2}{z^2} \right] \nonumber \\
& &+ \ldots 
\er

\subsubsection{Evaluation of $\pa_{\l}^x H[1, \pa_{\l} \; ; \; 1, \pa_{\m} \pa_{\n}]$}

Using integral relation \ref{rel_int1} we can write this contribution in terms of others previously obtained. Explicitly, we find
\br
\pa_{\l}^x H[1,\pa_{\l} \; ; \; 1, \pa_{\m} \pa_{\n}] &=& \pa_{\l}^x H[ 1, \pa_{\l} \; ; \; \pa_{\m} \pa_{\n}, 1] - \pa_{\l}^x \pa_{\m}^x H [1, \pa_{\l} \; ; \; 1, \pa_{\n}]  \nonumber \\
& & - \pa_{\n}^x \pa_{\l}^x H[ 1, \pa_{\l} \; ; \; 1,\pa_{\m}] - \pa_{\n} \Box H[1, \pa_{\m} \; ; \; 1,1] \nonumber \\
&\stackrel{R}{\rightarrow}&  \frac{1}{32 (4 \pi^2)^3} \left[ \pa_{\m} \pa_{\n} \Box \frac{ \frac{1}{2} \ln z^2 M^2}{z^2} + \d_{\m \n} \Box \Box \frac{\frac{1}{8} \ln^2 z^2 M^2 + \frac{3}{8} \ln z^2 M^2}{z^2} \right] \nonumber \\
& & + \ldots \nonumber \\
\er

\subsubsection{Evaluation of $H[1, \pa_{\m} \; ; \; 1, \pa_{\n}]$}

Considering (\ref{int8}) and applying (\ref{rel_int1}) we can put this as
\br
\pa_{\l}^x H[ 1, \pa_{\m} \; ; \; \pa_{\n} \pa_{\l} , 1] &=& \frac{1}{2} \Box H [1, \pa_{\m} \; ; \; 1, \pa_{\n}] + \frac{1}{2} \pa_{\m}^x \pa_{\l}^x H[ 1, 1 \; ; \; 1, \pa_{\n} \pa_{\l}]  \nonumber \\
& &+ \frac{1}{2} \pa_{\n}^x \pa_{\l}^x H[ 1, \pa_{\m} \; ; \; 1, \pa_{\l}] + \frac{1}{2} \pa_{\n}^x \Box H[ 1, \pa_{\m} \; ; \; 1,1] \;.
\er

Remembering previous results
\br
\pa_{\l}^x H^R [1, \pa_{\m} \; ; \; \pa_{\n} \pa_{\l},1] &=& \frac{1}{32(4 \pi^2)^3} \pa_{\m} \pa_{\n} \Box \frac{ \frac{1}{8} \ln^2 z^2 M^2 + \frac{1}{8} \ln z^2 M^2}{z^2}  \nonumber \\
& & + \frac{1}{2} \Box H^R [1, \pa_{\m} \; ; \; 1, \pa_{\n}] + \textrm{(local terms)}
\er

So that
\br
\Box H^R[1, \pa_{\m}\; ; \; 1, \pa_{\n}] &=& \frac{1}{32 (4 \pi^2)^3} \d_{\m \n} \Box \Box \frac{- \frac{1}{2} \ln z^2 M^2}{z^2} + \ldots
\er

\subsubsection{Evaluation of $H[1,1 \; ; \; \pa_{\m} \pa_{\n}, 1]$} 

Using  (\ref{int10}), (\ref{int10a}) and the identity
\br
\Box H[ 1,1 \; ; \; \pa_{\m} \pa_{\n},1] &=& \pa_{\l}^x H[ 1, \pa_{\l} \; ; \; \pa_{\m} \pa_{\n},1] + \pa_{\l}^x H[ \pa_{\l},1 \; ; \; \pa_{\m} \pa_{\n},1] \;,
\er
we can easily arrive to
\br
\Box H^R[1,1 \; ; \; \pa_{\m} \pa_{\n}, 1] &=& \frac{1}{32(4 \pi^2)^3} \d_{\m \n} \Box \Box \frac{ \frac{1}{4} \ln^2 z^2 M^2 + \frac{3}{4} \ln z^2 M^2}{z^2} + \ldots
\er

\subsubsection{Evaluation of $\pa_{\l}^x H[1,1 \; ; \; \pa_{\l} \pa_{\n} , \pa_{\m}]$}

Using (\ref{rel_int3}) and (\ref{int11}) we can write this as
\br
\pa_{\l}^x H[1,1 \; ; \; \pa_{\l} \pa_{\n} , \pa_{\m}] &=& \frac{1}{2} \Box H[ 1,1 \; ; \; \pa_{\m} \pa_{\n} ,1] - \frac{1}{4} \pa_{\m}^y \pa_{\n}^y \pa_{\l}^y H[ 1,1 \; ; \; \pa_{\l},1]  \nonumber \\
& & - \frac{1}{4} \pa_{\n}^y \Box H[1,1 \; ; \; \pa_{\m} , 1] \;. \nonumber \\
\er

Hence, we have only to use previous results to find
\br
\pa_{\l}^x H^R [1,1 \; ; \; \pa_{\l} \pa_{\n} , \pa_{\m}] &=& \frac{1}{32 (4 \pi^2)^3} \d_{\m \n} \Box \Box \frac{ \frac{1}{8} \ln^2 z^2 M^2 + \frac{3}{8} \ln z^2 M^2}{z^2} + \ldots
\er

\subsubsection{Evaluation of $\pa_{\l}^x H[ 1,1 \; ; \; \pa_{\m} \pa_{\n} , \pa_{\l}]$}

With (\ref{rel_int3}) and (\ref{int5}),this contribution can be evaluated with the same procedure of the previous one. So, we have
\br
\pa_{\l}^x H[ 1,1 \; ; \; \pa_{\m} \pa_{\n} , \pa_{\l}] &=& \frac{1}{4} \pa_{\n}^x \Box H[ 1,1 \; ; \; \pa_{\m},1] + \frac{1}{4} \pa_{\l}^x \pa_{\m}^x \pa_{\n}^x H[ 1,1 \; ; \; \pa_{\l},1]  \nonumber \\
& & + \frac{1}{2} \pa_{\l}^x \pa_{\m}^x H[ 1,1 \; ; \; \pa_{\l} \pa_{\n},1] \nonumber \\
&\stackrel{R}{\rightarrow}& \frac{1}{32 (4 \pi^2)^3} \pa_{\m} \pa_{\n} \Box \frac{\frac{1}{8} \ln^2 z^2 M^2 + \frac{3}{8} \ln z^2 M^2}{z^2} + \ldots
\er

\subsubsection{Evaluation of $H[1, \pa_{\m} \pa_{\l} \; ; \; \pa_{\n} \pa_{\l}, 1]$}

In this case the CDR decomposition into trace+traceless+local terms (\ref{CDR_T}) will be used again, as in (\ref{int8}) and (\ref{int10})
\br
H^R[1, \pa_{\m} \pa_{\l} \; ; \; \pa_{\n} \pa_{\l}, 1] &=& - \frac{1}{4} ( \D \pa_{\m} \pa_{\n} I^1 )_R - \frac{1}{4} [ \D ( \pa_{\m} \pa_{\n} - \frac{1}{4} \d_{\m \n} \Box ) I^1 ]_R  \nonumber \\
& &- \frac{1}{128 \pi^2} (\D \pa_{\m} \pa_{\n} \D)_R - \frac{1}{128 \pi^2} [ \D ( \pa_{\m} \pa_{\n} - \frac{1}{4} \d_{\m \n} \Box ) \D ]_R  \nonumber \\
& &+ \frac{64}{(4 \pi^2)^5} I_{\m \l \n \l \; R} \;,
\er
where $I_{\m \l \n \l}$ stands for the integral with the traceless parts. This was calculated in \cite{Haagensen:1992vz}, and the result was found to be
\br
\frac{64}{(4 \pi^2)^5} I_{\m \l \n \l \; R} &=& \frac{5}{96(4 \pi^2)} \pa_{\m} \pa_{\n} \D^2_R + \frac{13}{48} \d_{\m \n} \D^3_R \;.
\er

Adding up all the terms, it is easy to arrive to
\br
H^R[1, \pa_{\m} \pa_{\l} \; ; \; \pa_{\n} \pa_{\l}, 1] &=& \frac{1}{32 (4 \pi^2)^3} \left[ \pa_{\m} \pa_{\n} \Box \frac{ \frac{1}{6} \ln^2 z^2 M^2 - \frac{5}{36} \ln z^2 M^2}{z^2}  \right. \nonumber \\
& &+ \left. \d_{\m \n} \Box \Box \frac{ - \frac{1}{24} \ln^2 z^2 M^2 - \frac{29}{72} \ln z^2 M^2}{z^2} \right] + \ldots \nonumber \\
\er

\subsubsection{Evaluation of $H[1, \pa_{\m} \pa_{\l} \; ; \; 1 , \pa_{\l} \pa_{\n} ]$}

In this case, applying (\ref{rel_int1}), (\ref{int5}), (\ref{int6}), (\ref{int8}) and (\ref{int14}) we get
\br
H[1, \pa_{\m} \pa_{\l} \; ; \; 1 , \pa_{\l} \pa_{\n} ] &=& H[ 1, \pa_{\m} \pa_{\l} \; ; \; \pa_{\n} \pa_{\l},1] - \pa_{\l}^x H[ 1, \pa_{\m} \pa_{\l} \; ; \; 1, \pa_{\n}]  \nonumber \\ 
& & - \pa_{\n}^x H[1, \pa_{\m} \pa_{\l} \; ; \; 1, \pa_{\l}]- \pa_{\n}^x \pa_{\l}^x H[ 1, \pa_{\m} \pa_{\l} \; ; \; 1,1] \nonumber \\
&\stackrel{R}{\rightarrow}& \frac{1}{32 (4 \pi^2)^3} \left[ \pa_{\m} \pa_{\n} \Box \frac{\frac{1}{6} \ln^2 z^2 M^2 + \frac{49}{36} \ln z^2 M^2}{z^2} \right. \nonumber \\
& & + \left.\d_{\m \n} \Box \Box \frac{- \frac{1}{24} \ln^2 z^2 M^2 - \frac{11}{72} \ln z^2 M^2}{z^2} \right] + \ldots \nonumber \\
\er

\section{UV and IR divergent integrals}
\label{ap_UV_IR}
In this section we will detail the renormalization of three relevant integral expressions that appear when considering two-loop diagrams made up by the insertion of a one-loop propagator. Examples of this are diagram $(a)$ of QED or diagram $(b)$ of the Yang-Mills case. These expressions are 
\br
I^0 (x-y) &=& \int d^4 u d^4 v \; \D_{xu} \D_{yv} \D^2_{uv} \nonumber \\
I^{0}_{\m} (x-y) &=& \int d^4 u d^4 v \; \D_{xu} \D_{yv} ( \D_{uv} \pa_{\m}^u \D_{uv} ) \nonumber \\
I^{0}_{\m \n} (x-y) &=& \int d^4 u d^4 v \; \D_{xu} \D_{yv} ( \D_{uv} \pa_{\m}^u \pa_{\n}^v \D_{uv} ) \;.
\er
\subsection{Renormalization of $I^0$}
The renormalization of $I^0$ is detailed in section \ref{IR_divergences}, and here we only recall the final renormalized result found there.
\br
I^0_R(x-y) &=& \frac{1}{32 (4 \pi^2)^2} \left[ \ln^2 x^2 M^2_{IR} + 2 \ln x^2 M^2_{IR} ( 1 - \ln x^2 M^2) \right]+ \ldots  \nonumber \\ \label{I0_integral}
\er
As in diagram $(b)$ of the Yang-Mills theory we have contributions of the form  $\D \Box I^0$ and $\D \pa_{\m} \pa_{\n} I^0$, we have to evaluate them. For the first one, it is clear that
\br
\Box I^0(x-y) &=& \Box \int d^4 u d^4v \; \D_{xu} \D_{yv} \D^2_{uv} \nonumber \\&=& - \int d^4 v \; \D_{vy} \D^2_{xv} \nonumber \\
&=& - I^1 (x-y) \nonumber \\
&\stackrel{R}{\rightarrow}& - \frac{1}{4 (4 \pi^2)^2} \frac{\ln (x-y)^2 M^2}{(x-y)^2} \;, 
\er
where we have used the renormalized value found for $I^1$. Finally, in order to obtain $\D \pa_{\m} \pa_{\n}I^0$ we have to consider (\ref{I0_integral}) and apply usual DiffR. With this, we find  
\br
[\D \pa_{\m} \pa_{\n} I^0 ]_R (x) &=& \frac{1}{32(4 \pi^2)^3} \left[ \pa_{\m} \pa_{\n} \frac{ \ln x^2 M^2}{x^2} + \d_{\m \n} \Box \frac{ \frac{1}{4} \ln^2 x^2 M^2 + \frac{1}{4} \ln x^2 M^2}{x^2} \right] \nonumber \\
& & + \ldots
\er

\subsection{Renormalization of $I^0_{\m}$}
\label{ap_UV_IR_I0m}
The renormalization of $I^0_{\m}$ is straightforward, once we recall that CDR imposes $I^0_{\m \; R} = \frac{1}{2} \pa_{\m}^x I^0_R$.

\subsection{Renormalization of $I^0_{\m \n}$}
Applying CDR to the subdivergence we find
\br
I_{\m \n}^0 &=& \frac{1}{3} \int d^4 u d^4 v \; \D_{xu} \D_{yv} ( \pa_{\m} \pa_{\n} - \frac{1}{4} \d_{\m \n} \Box ) (\D^2_{uv})_{R} + \nonumber \\
& &+ \frac{1}{288 \pi^2} \int d^4 u d^4 v \; \D_{xu} \D_{yv} ( \pa_{\m} \pa_{\n} - \d_{\m \n} \Box) \d (u-v) \nonumber \\
&=& \frac{1}{3} \pa_{\m} \pa_{\n} I^0_R - \frac{1}{12} \d_{\m \n} \Box I^0_R + \frac{1}{72 (4\pi^2)} \pa_{\m} \pa_{\n} \int d^4 u \; \D_{xu} \D_{yu} + \frac{\d_{\m \n}}{72 (4\pi^2)} \D \;. \nonumber 
\er
With this we can evaluate the expression that appears in diagram $(b)$ of the Yang-Mills case ($\D I^0_{\m \n} $). We find the following result
\br
[\D I_{\m \n}^0]_R (x) &=& \frac{1}{32(4 \pi^2)^3} \left[ \pa_{\m} \pa_{\n} \frac{ \frac{1}{3} \ln x^2 M^2}{x^2} + \d_{\m \n} \Box \frac{ - \frac{1}{6} \ln x^2 M^2}{x^2} \right] + \ldots
\er

\backmatter
\renewcommand{\theequation}{\arabic{equation}}
\chapter{Resumen}

\section*{Renormalizaci\'on diferencial}

Renormalizaci\'on diferencial \cite{Freedman:1991tk} es un m\'etodo de renormalizaci\'on en el espacio de posiciones que consiste en sustituir expresiones que son demasiado divergentes para tener una transformada de Fourier bien definida, por derivadas de otras expresiones menos singulares. As\'i, por ejemplo $1/x^4$ no tiene una transformada de Fourier bien definida, y renormalizaci\'on diferencial proponer reemplazarla por la soluci\'on de la ecuaci\'on diferencial
\br
\frac{1}{x^4} &=& \Box G (x^2) ~~~ x^2 \neq 0 \;,
\er    
que es
\br
\frac{1}{x^4} \rightarrow \left[ \frac{1}{x^4} \right]_R = - \frac{1}{4}\Box \frac{ \ln x^2 M^2}{x^2} 
\er
Notar que se ha introducido una constante con dimensiones de masa $M$ que parametriza la ambiguedad local. Debido a que un cambio en $M$ puede ser reabsorbido en un reescalamiento de la constante de acoplamiento, esto sugiere que las amplitudes renormalizadas satisfacen ecuaciones del grupo de renormalizaci\'on, con $M$ jugando el papel de escala del grupo de renormalizaci\'on.

Aunque en este trabajo se traten s\'olo teor\'ias sin masa, renormalizaci\'on diferencial puede ser aplicada sin ning\'un problema a teor\'ias masivas, ya que las masas s\'olo alteran el comportamiento a larga distancia de los correladores \cite{Freedman:1991tk,Haagensen:1992am}.

Renormalizaci\'on diferencial se puede aplicar para renormalizar diagramas de orden arbitrario en teor\'ia de perturbaciones. En concreto, en \cite{Latorre:1993xh} se expone una implementaci\'on sistem\'atica de renormalizaci\'on diferencial a cualquier orden en teor\'ia de perturbaciones. En general, cuando se aplica renormalizaci\'on diferencial a un c\'alculo a orden superior, aparecen nuevas escalas correspondientes a la renormalizaci\'on de los distintos subdiagramas que forman el diagrama completo.

Tambi\'en es relevante se\~nalar que, aplicando renormalizaci\'on diferencial en espacio de momentos, se pueden renormalizar divergencias IR. As\'i, por ejemplo
\be
\left[ \frac{1}{p^4} \right]_{\tilde{R}} = - \frac{1}{4}{\dal}_p
\frac{\ln p^2/\bar{M}_{IR}^2}{p^2} + a_{IR} \delta(p) \; .
\end{equation}
Sin embargo, a la hora de renormalizar una teor\'ia que tenga divergencias IR y UV, se ha de tener en cuenta que ambas renormalizaciones han de estar desacopladas, implicando por ello que ambas escalas (IR y UV) han de ser independientes. En concreto, en este trabajo se discute una expresi\'on divergente IR de la forma $ \ln p^2 / \bar{M}^2 / p^4$, donde $M$ es una escala UV producto de una renormalizaci\'on previa en espacio de posiciones. En este caso, la independencia de las escalas se consigue en el momento que imponemos la relaci\'on
\be
M \frac{\d}{\d M} \left[\frac{\ln
p^2/\bar{M}^2}{p^4}\right]_{\tilde{R}} =
\left[M \frac{\d}{\d M} \frac{\ln
p^2/\bar{M}^2}{p^4}\right]_{\tilde{R}} \;.
\end{equation}

Esto se satisface ajustando los t\'erminos locales en los que aparecen ambas escalas, obteniendo entonces la siguiente forma para la expresi\'on renormalizada m\'as general \cite{Mas:2002xh}
\be
\left[\frac{\ln p^2/\bar{M}^2}{p^4}\right]_{\tilde{R}} =
-\frac{1}{8} {\dal}_p \frac{-\ln^2 p^2/\bar{M}_{IR}^2 + 2 \ln
p^2/\bar{M}_{IR}^2 \, (1 + \ln p^2/\bar{M}^2)}{p^2} + (a_{IR} \ln
\frac{M_{IR}^2}{M^2} + b_{IR}) \delta(p)
\; .
\end{equation}

Una de las caracter\'isticas m\'as importantes de renormalizaci\'on diferencial es que la invariancia gauge se preserva. Sin embargo, debido a las ambiguedades que se generan en el m\'etodo de renormalizaci\'on, se han de imponer en cada c\'alculo (con una teor\'ia gauge) las identidades de Ward de forma expl\'icita, de tal manera que se fije el esquema de renormalizaci\'on. El hecho de que se preserve invariancia gauge se refleja en que siempre es posible satisfacer estas identidades con las expresiones renormalizadas (excepto por supuesto las anomal\'ias).

\subsection*{Renormalizaci\'on diferencial restringida}
Para evitar la necesidad de imponer las identidades de Ward expl\'icitamente en cada c\'alculo, se desarroll\'o Renormalizaci\'on Diferencial Restringida (RDR) \cite{delAguila:1997kw}. Este m\'etodo consiste en proporcionar una serie de reglas que {\em{a priori}} fijan toda la ambiguedad inherente al proceso, de tal modo que las expresiones renormalizadas sean directamente invariantes gauge (no es necesario imponer las identidades de Ward). Las reglas que impone RDR son:
\begin{enumerate}
\item {\em Reducci\'on diferencial}
\begin{itemize}
\item Funciones con comportamiento peor que el logar\'itmico se reducen a derivadas de (como mucho) funciones logar\'itmicamente divergentes sin introducir constantes dimensionales extra.
\item Expresiones logar\'itmicamente divergentes se escriben como derivadas de funciones regulares, introduciendo una \'unica constante $M$, que tiene dimensiones de masa y juega el papel de escala del grupo de renormalizaci\'on.
\end{itemize}
\item{ \em Integraci\'on por partes formal}. No se tienen en cuenta los t\'erminos de contorno divergentes que aparecen cuando integramos por partes. En relaci\'on a esto, la renormalizaci\'on y la diferenciaci\'on deben ser dos operaciones conmutativas: si $F$ es una funci\'on arbitraria, entonces $[ \pa F ]_R = \pa [F]_R$.
\item {\em Regla de renormalizaci\'on de la funci\'on delta}
\begin{equation} [ F (x, x_1, \ldots , x_n ) \d (x-y) ]_R = [ F ( x, x_1, \ldots , x_n)]_R \d (x-y)
\end{equation}
\item {\em Validez de la ecuaci\'on del propagador}
\begin{equation} [F(x,x_1,\ldots,x_n) ( \Box - m^2) \D_{m}(x)]_R = - [F(x,x_1,\ldots,x_n) \d(x)]_R
\end{equation} 
donde $\D_{m}$ es el propagador de una part\'icula de masa $m$ y $F$ una funci\'on arbitraria.
\end{enumerate}

Aplicando estas reglas, obtenemos un conjunto b\'asico de expresiones renormalizadas. Por lo tanto, el proceso de renormalizaci\'on consta de dos partes: en un primer momento, se realizan todas las contracciones de \'indices (RDR no conmuta con esta operaci\'on) y se escribe la expresi\'on desnuda en t\'erminos de estas funciones b\'asicas. En un segundo paso, se sustituyen estas funciones por sus valores renormalizados.

\subsection*{Aplicaci\'on de RDR a c\'alculos a dos bucles}
Aunque RDR se ha desarrollado s\'olo para c\'alculos a un bucle, proporciona informaci\'on \'util cuando tratamos c\'alculos a dos bucles. Veremos que aplicar RDR fija un\'ivocamente los coeficientes de todos los logaritmos de las escalas en la expresi\'on a dos bucles renormalizada, que son los t\'erminos que necesitamos para evaluar la ecuaci\'on del grupo de renormalizaci\'on. Es por ello que, al obtener las expresiones renormalizadas a dos bucles, no tendremos en cuenta los posibles t\'erminos locales que se generen. Distinguiremos dos situaciones diferentes: diagramas con divergencias anidadas y diagramas con solapamiento.

\subsubsection*{Divergencias anidadas}
En este caso, empezamos imponiendo RDR a la subdivergencia. Al hacer esto, fijamos los t\'erminos locales a un bucle que tenemos en el diagrama, junto con los logaritmos de las escalas a un bucle ($\ln x^2 M^2$). Entonces, al considerar la expresi\'on completa del diagrama y aplicar renormalizaci\'on diferencial normal, nos encontramos que todos los coeficientes que corresponden a logaritmos de las escalas est\'an un\'ivocamente determinados, ya que los t\'erminos locales a un bucle (que se promocionan a logaritmos) han sido fijados por RDR. Con este procedimiento, realizamos la renormalizaci\'on de diferentes expresiones que contienen la siguiente integral que se utiliza a lo largo del trabajo: $I^1 = \int d^4 u \D_{xu} \D^2_{yu}$
\br
\left[ \D I^1 \right]_R (x) &=& - \frac{1}{32(4 \pi^2)^3} \Box \frac{ \ln^2 x^2 M^{2} + 2 \ln x^2 M^2}{x^2}  +~\textrm{(termin.~locales)} \nonumber \\
\left[ \D \pa_{\m} I^{1} \right]_R (x) &=& - \frac{1}{64 (4 \pi^2)^3} \pa_{\m} \Box \frac{ \ln^2 x^2 M^2 + \ln x^2 M^2}{x^2} +~\textrm{(termin.~locales)} \nonumber \\
\left[ \D \pa_{\m} \pa_{\n} I^1 \right]_R (x) &=& - \frac{1}{96 (4 \pi^2)^3} \left[\pa_{\m} \pa_{\n} \Box \frac{ \ln^2 x^2 M^2 + \frac{2}{3} \ln x^2 M^2}{x^2}  \right. \nonumber \\
& & - \left. \frac{1}{4}\d_{\m \n} \Box \Box \frac{\ln^2 x^2 M^2 + \frac{11}{3} \ln x^2 M^2}{x^2} \right] +~\textrm{(termin.~locales)}  \nonumber \\
\left[ \D \Box I^1 \right]_R (x) &=& \frac{1}{32 ( 4 \pi^2)^2} \Box \Box \frac{ \ln x^2 M^2}{x^2} +~\textrm{(termin.~locales)} \;.
\er

\subsubsection*{Divergencias con solapamiento}
En el caso de divergencias con solapamiento, la situaci\'on es m\'as complicada, ya que muchas veces es dif\'icil reconocer las expresiones a un bucle a las que hay que empezar aplicado RDR. Por lo tanto, lo que hemos hecho es obtener una conjunto completo de integrales renormalizadas con solapamiento, con como mucho cuatro derivadas actuando sobre los propagadores y dos \'indices libres. Con esta lista podemos obtener la expresi\'on a dos puntos renormalizada de cualquier teor\'ia con acoplos derivativos; lo que implica que, aplicando el m\'etodo del campo de background, nos permite obtener la funci\'on beta. Para evaluar las integrales hemos empleado b\'asicamente dos m\'etodos:
\begin{itemize}
\item Mediante igualdades integrales rescribimos las integrales en t\'erminos de otras que tengan un d'almbertiano actuando sobre uno de los propagadores. Esto permite obtener la integral como suma de contribuciones de integrales con divergencias anidadas, en las cuales se aplica lo que hemos discutido anteriormente.
\item Utilizamos la descomposici\'on en parte con traza y sin traza que impone RDR (en la que se a\~nade un t\'ermino local fijo).
\end{itemize}
Esta lista est\'a escrita en t\'erminos de una expresi\'on $H$ que hemos definido como
\br
H[{\cal{O}}_1,{\cal{O}}_2 \; ; \; {\cal{O}}_3,{\cal{O}}_4] = \int d^4 u d^4 v \; ( {\cal{O}}_1^{x} \D_{xu})( {\cal{O}}_2^{x} \D_{xv})( {\cal{O}}_3^{y} \D_{yu} ) ({\cal{O}}_4^{y} \D_{yv}) \D_{uv} \;, 
\er
siendo ${\cal{O}}_i$ un operador diferencial. 
\br
H^R[1,1 \; ; \; 1,1] &=& \frac{6 \pi^4 \xi(3) }{ ( 4 \pi^2)^4} \D  \equiv a \D  \\
H^R[\pa_{\m},1 \; ; \; 1,1] &=&  \frac{ 3 \xi(3)}{16 (4 \pi^2)^2} ( \pa_{\m} \D) \equiv \frac{a}{2} \pa_{\m} \D  \\
H^R[1,\pa_{\l} \; ; \; 1,\pa_{\l}] &=&  - \frac{1}{16(4 \pi^2)^3} \Box \frac{\ln z^2 M^2}{z^2} + \ldots  \\
\pa_{\l}^x H^R[1,\pa_{\m} \; ; \; 1,\pa_{\l}] &=&  - \frac{1}{32 (4 \pi^2)^3} \pa_{\m} \Box \frac{ \frac{1}{2} \ln z^2 M^2}{z^2} + \dots \\
\pa_{\l}^x H^R[1,1 \; ; \; \pa_{\l} \pa_{\n},1] &=& \frac {1}{32(4 \pi^2)^3} \pa_{\n} \Box \frac{ \frac{1}{4} \ln^2 z^2 M^2 + \frac{3}{4} \ln z^2 M^2 }{z^2} + \ldots \\ 
H^R[1,\pa_{\l} \; ; \; \pa_{\l} \pa_{\m},1] &=&  \frac{1}{32 (4 \pi^2)^3} \pa_{\m} \Box \frac{\frac{1}{8} \ln^2 z^2 M^2 - \frac{7}{8} \ln z^2 M^2 }{z^2} + \ldots \\
H^R[\pa_{\m} \pa_{\l},\pa_{\l} \; ; \; 1,1] &=&  \frac{1}{32(4 \pi^2)^3} \pa_{\m} \Box \frac{ - \frac{1}{2} \ln^2 z^2 M^2 - \ln z^2 M^2}{z^2}+ \ldots   \\
\pa_{\l}^x H^R[1,\pa_{\m} \; ; \; \pa_{\n} \pa_{\l},1] &=& \frac{1}{32(4 \pi^2)^3} \left[ \pa_{\m} \pa_{\n} \Box \frac{ \frac{1}{8} \ln^2 z^2 M^2 + \frac{1}{8} \ln z^2 M^2}{z^2}  \right. \nonumber \\
& & \left. + \d_{\m \n} \Box \Box \frac{-\frac{1}{4} \ln z^2 M^2}{z^2} \right] + \ldots  
\er
\br
H^R[1,\pa_{\m} \; ; \; 1,\pa_{\n}] &=& \frac{1}{32 (4 \pi^2)^3} \d_{\m \n} \Box \frac{- \frac{1}{2} \ln z^2 M^2}{z^2} + \ldots \\ 
\pa_{\l}^x H^R[1,\pa_{\l} \; ; \; \pa_{\m} \pa_{\n},1] &=& \frac{1}{32 (4 \pi^2)^3} \left[ \pa_{\m} \pa_{\n} \Box \frac{ - \frac{1}{2} \ln z^2 M^2}{z^2}  \right. \nonumber \\
& & \left. + \d_{\m \n} \Box \Box \frac{\frac{1}{8} \ln^2 z^2 M^2 + \frac{3}{8} \ln z^2 M^2}{z^2} \right] + \ldots  \\
\pa_{\l}^x H^R[1,\pa_{\l} \; ; \; 1, \pa_{\m} \pa_{\n}] &=& \frac{1}{32 (4 \pi^2)^3} \left[ \pa_{\m} \pa_{\n} \Box \frac{ \frac{1}{2} \ln z^2 M^2}{z^2}  \right. \nonumber \\
& & \left. + \d_{\m \n} \Box \Box \frac{\frac{1}{8} \ln^2 z^2 M^2 + \frac{3}{8} \ln z^2 M^2}{z^2} \right] + \ldots \\
H^R[1,1 \; ; \; \pa_{\m} \pa_{\n},1] &=&  \frac{1}{32(4 \pi^2)^3} \d_{\m \n} \Box \frac{ \frac{1}{4} \ln^2 z^2 M^2 + \frac{3}{4} \ln z^2 M^2}{z^2} + \ldots  \\
\pa_{\l}^x H^R[1,1 \; ; \; \pa_{\l} \pa_{\n},\pa_{\m}] &=& \frac{1}{32 (4 \pi^2)^3} \d_{\m \n} \Box \Box \frac{ \frac{1}{8} \ln^2 z^2 M^2 + \frac{3}{8} \ln z^2 M^2}{z^2} + \ldots  \\
\pa_{\l}^x H^R[1,1 \; ; \; \pa_{\m} \pa_{\n},\pa_{\l}] &=& \frac{1}{32 (4 \pi^2)^3} \pa_{\m} \pa_{\n} \Box \frac{\frac{1}{8} \ln^2 z^2 M^2 + \frac{3}{8} \ln z^2 M^2}{z^2} + \ldots  \nonumber \\
H^R[1,\pa_{\m} \pa_{\l} \; ; \; \pa_{\n} \pa_{\l},1] &=& \frac{1}{32 (4 \pi^2)^3} \left[ \pa_{\m} \pa_{\n} \Box \frac{ \frac{1}{6} \ln^2 z^2 M^2 - \frac{5}{36} \ln z^2 M^2}{z^2}  \right. \nonumber \\
& & \left. + \d_{\m \n} \Box \Box \frac{ - \frac{1}{24} \ln^2 z^2 M^2 - \frac{29}{72} \ln z^2 M^2}{z^2} \right] + \ldots \\
H^R[1,\pa_{\m} \pa_{\l} \; ; \; 1,\pa_{\n} \pa_{\l}] &=&  \frac{1}{32 (4 \pi^2)^3} \left[ \pa_{\m} \pa_{\n} \Box \frac{\frac{1}{6} \ln^2 z^2 M^2 + \frac{49}{36} \ln z^2 M^2}{z^2}  \right. \nonumber \\
& & \left. + \d_{\m \n} \Box \Box \frac{- \frac{1}{24} \ln^2 z^2 M^2 - \frac{11}{72} \ln z^2 M^2}{z^2} \right] + \ldots 
\er

\section*{Ejemplos abelianos}
\subsection*{QED}
La renormalizaci\'on diferencial de QED a dos bucles fue realizada en \cite{Haagensen:1992vz} empleando identidades de Ward para relacionar las escalas. En este trabajo reharemos este c\'alculo empleando RDR a un bucle, que nos permitir\'a obtener la funci\'on beta de QED sin necesidad de emplear dichas identidades. Utilizaremos para los c\'alculos los convenios de \cite{Haagensen:1992vz}, por lo que la acci\'on que consideramos es
\br
{\cal{L}} &=& \frac{1}{4} F^{\m \n} F_{\m \n} + \bar{\psi} \g^{\m} ( \pa_{\m} + i e A_{\m} ) \psi \;,
\er
donde $\psi$ es el campo fermi\'onico y $F_{\m \n}$ se expresa en t\'erminos del campo gauge $A_{\m}$ como $F_{\m \n} (x) = \pa_{\m} A_{\n}(x) - \pa_{\n} A_{\m} (x)$. 

A diferencia que en \cite{Haagensen:1992vz}, realizaremos los c\'alculos con el m\'etodo del campo de background. Con este m\'etodo, se divide el campo gauge en dos contribuciones $A_{\m} \rightarrow A_{\m} + B_{\m}$: una cu\'antica ($A_{\m}$), que es la variable de integraci\'on en el la funci\'on de partici\'on y por lo tanto sobre la que se fija el gauge, y otra background ($B_{\m}$), en la que se mantiene invariancia gauge expl\'icita. Esto tiene m\'ultiples consecuencias, entre las que destacamos el hecho de poder obtener la funci\'on beta a partir s\'olo de la funci\'on a dos puntos.

\subsubsection*{Un bucle}
La autoenerg\'ia del fot\'on a un bucle tiene una expresi\'on desnuda de la forma
\br
\Pi_{\m \n}^{(1 \; bucle)} &=& - (i e)^2 Tr \left[ \g_{\m} \g^{\l} \pa_{\l}^x \D \g_{\n} \g^{\sigma} \pa_{\sigma}^y \D \right] \nonumber \\
&=& - e^2 Tr\left[ \g_{\m} \g^{\l} \g_{\n} \g^{\sigma} \right] \left( \pa_{\l} ( \D \pa_{\eps} \D ) - \D \pa_{\l} \pa_{\sigma} \D \right) \;,
\er
a partir de la que, de acuerdo a las reglas de RDR, obtenemos el siguiente valor renormalizado
\br
\Pi_{\m \n R}^{(1)} (x) &=& - ( \pa_{\m} \pa_{\n} - \d_{\m \n} \Box ) \left[ - \frac{e^2}{12 \pi^2 ( 4 \pi^2 )} \Box \frac{ \ln x^2 M^2}{x^2} - \frac{e^2}{36 \pi^2} \d (x) \right] \;.
\er

Obtenemos ahora la autoenerg\'ia del electr\'on, ya que es una inserci\'on dentro de uno de los diagramas a dos bucles. La expresi\'on para est\'a contribuci\'on es
\br
\S (x)^{(1)} &=& e^2 \g_{\m} \D_{\m \n} (x) \g^{\l} \pa_{\l} \D (x) \g_{\n} \;,
\er
que utilizando el propagador del fot\'on en un gauge general y las funciones b\'asicas de RDR se renormaliza como 
\br
\S (x)_R^{(1)} (x) &=& e^2 \g^{\l} \left[ \frac{1}{4(4 \pi^2)^2} \pa_{\l} \Box \frac{\ln x^2 M^2}{x^2} + (\a -1) \left( \frac{1}{4(4 \pi^2)^2} \pa_{\l} \Box \frac{\ln x^2 M^2}{x^2} + \frac{1}{16 \pi^2} \pa_{\l} \d (x) \right)\right] \;. \nonumber \\
\er

\subsubsection*{Autoenerg\'ia del fot\'on background a dos bucles}
\begin{figure}[ht]
\centerline{\epsfbox{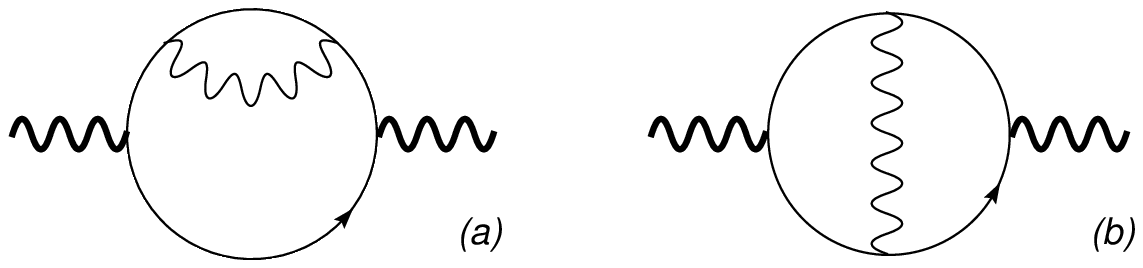}}
\caption{Diagramas a dos bucles de QED.}
\end{figure}
En primer lugar, se\~nalar que los c\'alculos a dos bucles los realizamos en el gauge de Feynman. Esto es debido a que la selecci\'on de este gauge en concreto no afecta a la verificaci\'on de las ecuaciones del grupo de renormalizaci\'on a dos bucles, como se ver\'a al discutir dichas ecuaciones. Por lo tanto, el propagador desnudo del fermi\'on en dicho gauge es $\S^{(1)}(x) = - 2 e^2 \g^{\l} \D \pa_{\l} \D (x)$. 

Comenzando por el diagrama $(a)$, este tiene la siguiente expresi\'on desnuda
\br
\Pi^{(2 \; a)}_{\m \n} (x-y) &=& - (ie)^2 \int d^4 u d^4 v \; Tr \left[ \g_{\m} \g^{\l} (- \pa_{\l}^x \D_{xu}) \S^{(1)} (u-v) \g^{\eps} (- \pa_{\eps}^v \D_{vy}) \g_{\n} \g^{\s} ( - \pa_{\s}^y \D_{yx} ) \right]. \nonumber \\ 
\er
Para simplificar la notaci\'on, definimos $I^0_{\m}$ como $I^0_{\m} = \int d^4 u d^4 v \; \D_{xu} \D_{yv} ( \D_{uv} \pa_{\m} \D_{uv})$, que nos permite escribir esta contribuci\'on como
\br
\Pi^{(2 \; a)}_{\m \n} (x) &=& e^4 \left[ -32 (\pa_{\m} \D) \pa_{\l} \pa_{\n} I^0_{\l} + 16 \d_{\m \n} ( \pa_{\s} \D ) \pa_{\l} \pa_{\s} I^0_{\l} + 16 ( \pa_{\m} \D ) \Box I^0_{\n} - 8 \d_{\m \n} ( \pa_{\r} \D ) \Box I^0_{\r} \right] \;. \nonumber \\
\er
Por lo tanto, la renormalizaci\'on de esta expresi\'on pasa por estudiar la renormalizaci\'on de $I^0_{\m}$. Es f\'acil demostrar que esto se puede escribir en t\'erminos de la renormalizaci\'on de la integral $I^1$ definida previamente. As\'i, $\pa_{\m} I^0_{\m \; R} = - \frac{1}{2} I^1_R$ y  $\Box I^0_{\m \;R} = - \frac{1}{2} \pa_{\m} I^1_R$. Entonces, la contribuci\'on renormalizada de este diagrama es
\br
\Pi^{(2 \; a)}_{\m \n \; R} (x)&=& \frac{e^4}{24(4 \pi^2)^3} \left[ \pa_{\m} \pa_{\n} \frac{- \ln^2 x^2 M^2 - \frac{5}{3} \ln x^2 M^2 }{x^2} + \d_{\m \n} \Box \Box \frac{\ln^2 x^2 M^2 + \frac{8}{3} \ln x^2 M^2}{x^2} \right]  \nonumber  \\ & & +~\textrm{(termin.~locales)} \;. \nonumber \\
\er

Pasamos ahora a renormalizar el diagrama $(b)$. En este caso, lo que tenemos es un diagrama con divergencias de solapamiento. La expresi\'on b\'asica de este diagrama es
\br
\Pi_{\m \n}^{(2 \; b)} (x-y) &=& - ( i e )^4 \int d^4 u d^4 v \; Tr \left[ \g_{\m} ( \g^{\a} \pa^x_{\a} \D_{xu} ) \g^{\rho} ( \g^{\b} \pa_{\b}^u \D_{uy} ) \g_{\n}  \nonumber \right. \\
& & \times \left. ( \g^{\l} \pa_{\l}^y \D_{yv} ) \g_{\rho} ( \g^{\s} \pa_{\s} \D_{vx}) \D_{uv} \right] \;,\nonumber \\
\er
Mediante identidades de las matrices $\g$, e integrando por partes las derivadas que act\'uan sobre $\D_{xu}$ y $\D_{yv}$, podemos rescribir esto en t\'erminos de las expresiones $H$ como
\br
\Pi^{(2 \; b)}_{\m \n} = e^4 &\left[\right.& - 8 \d_{\m \n} \Box H[ 1 , \pa_{\l} \; ; \; \pa_{\l}, 1] + 16 \pa_{\m}^x H[ 1, \pa_{\n} \; ; \; \Box, 1] - 8 \d_{\m \n} \pa_{\l}^x H[ 1, \pa_{\l} \; ; \; \Box , 1]  \nonumber \\
& & - 16 \pa_{\m}^x H[ 1, \pa_{\l} \; ; \; \pa_{\l} \pa_{\n}, 1] + 16 \pa_{\l}^x H [ 1, \pa_{\l} \; ; \; \pa_{\m} \pa_{\n} ,1] - 16 \pa_{\l}^x H [ 1, \pa_{\m} \; ; \; \pa_{\l} \pa_{\n},1]  \nonumber \\
& & - 16 \pa_{\m} H [ 1, \Box \; ; \; \pa_{\n}, 1] + 8 \d_{\m \n} \pa_{\l}^x H[ 1, \Box \; ; \; \pa_{\l}, 1] + 16 \pa_{\l}^x H[ 1, \pa_{\l} \pa_{\m} \; ; \; \pa_{\n},1]  \nonumber \\
& & - 16 \pa_{\l}^x H[ 1, \pa_{\m} \pa_{\n} \; ; \; \pa_{\l}, 1] + 16 \pa_{\n}^x H[ 1, \pa_{\l} \pa_{\m} \; ; \; \pa_{\l}, 1] - 16 H[ 1, \Box \; ; \; \pa_{\m} \pa_{\n},1]  \nonumber \\
& & \left. + 8 \d_{\m \n} H[ 1, \Box \; ; \; \Box ,1 ] + 32 H[ 1 , \pa_{\m} \pa_{\l} \; ; \; \pa_{\n} \pa_{\l}, 1] - 16 H[ 1 , \pa_{\m} \pa_{\n} \; ; \; \Box , 1] \; \right].
\er
Aquellas contribuciones que tienen un d'alamebertiano se pueden rescribir en t\'erminos de la integral $I^1$, por lo que su renormalizaci\'on es inmediata. El resto se encuentran dentro de la lista de integrales con divergencias de solapamiento (o pueden ser f\'acilmente expresadas en t\'erminos de esas integrales), con lo que s\'olo hay que sustituir el valor renormalizado correspondiente. Por lo tanto, la contribuci\'on renormalizada del diagrama $(b)$ es
\br
\Pi^{(2 \; b)}_{\m \n R} (x) &=& \frac{e^4}{12 (4 \pi^2)^3} \left[ \pa_{\m} \pa_{\n} \Box \frac{ \ln^2 x^2 M^2 + \frac{14}{3} \ln x^2 M^2}{x^2} - \d_{\m \n} \Box \Box \frac{ \ln^2 x^2 M^2 + \frac{17}{3} \ln x^2 M^2}{x^2} \right]  \nonumber \\
& & +~\textrm{(termin.~locales)} \;.
\er

Entonces, la contribuci\'on total renormalizada a dos bucles de la autoenerg\'ia del fot\'on es
\br
\Pi_{\m \n \; R}^{(2)} (x) &=& 2 \Pi_{\m \n}^{(2 \; a)} (x) + \Pi_{\m \n}^{(2 \; b)} (x) \nonumber \\
&=& \frac{e^4}{4(4 \pi^2)^3} ( \pa_{\m} \pa_{\n} - \d_{\m \n} \Box ) \Box \frac{ \ln x^2 M^2}{x^2} + \ldots
\er
donde $\ldots$ corresponde a los t\'erminos locales que no estamos teniendo en cuenta.
\subsubsection*{Ecuaci\'on del grupo de renormalizaci\'on}
Estudiando la ecuaci\'on del grupo de renormalizaci\'on a un bucle para la funci\'on a dos puntos de los campos cu\'anticos, obtenemos que la funci\'on que corresponde a la variaci\'on del par\'ametro gauge en dicha ecuaci\'on $\g_{\a}(e) \pa / \pa \a$. Esta funci\'on tiene una expansi\'on de la forma $\g_{\a}(e) = - \frac{2 \a}{3 (4 \pi^2)} e^2 + {\cal{O}}(e^3) $. Esto justifica el poder realizar el c\'alculo a dos bucles en el gauge de Feynman, ya que $\g_{\a}(e) \pa / \pa \a$ actuando sobre cualquier diagrama a dos bucles no afecta a la verificaci\'on de las ecuaciones del grupo de renormalizaci\'on (es de orden superior en $e$, ya que la primera dependencia en $\a$ de la autoenerg\'ia del fot\'on background aparece a dos bucles).

En cuanto a los campos background, si definimos $B_{\m} = \frac{1}{e} B_{\m}^{\prime}$, tenemos que la dimensi\'on an\'omala de este nuevo campo es nula, ya que la renormalizaci\'on de $B_{\m}$ y de la constante de acoplamiento verifican la relaci\'on $Z_{e} \sqrt{Z_B} = 1$. Por lo tanto, la ecuaci\'on del grupo de renormalizaci\'on que verifican estos campos es
\br
\left( M \frac{\pa}{\pa M} + \b(e) \frac{\pa}{\pa e} \right) \G_{\m \n}^{B B \; (2)} = 0 \;,
\er
con 
\br
\G_{\m \n}^{B B}(x-y) &=& \left(\pa_{\m} \pa_{\n} - \d_{\m \n} \Box \right) \d^{(4)}(x-y) - \Pi_{\m \n} (x-y) \;.
\er

Con las contribuciones renormalizadas que tenemos para $\Pi_{\m \n}$, obtenemos que la funci\'on beta a dos bucles de QED es
\br
\b (e) &=& \frac{1}{3(4 \pi^2)} e^3 + \frac{1}{4 (4 \pi^2)^2} e^5 + {\cal{O}}(e^7) \;.
\er

\subsection*{SuperQED}
Pasamos ahora a obtener la renormalizaci\'on de la extensi\'on supersim\'etrica del caso anterior, SuperQED. Aqu\'i, aplicamos los convenios del superespacio definidos en \cite{Gates:1983nr}, con los que la acci\'on de SuperQED es
\br
S &=& \int d^4 x d^2 \th \; W^2 + \int d^4 x d^4 \th \; \bar{\Phi}_{+} e^{gV} \Phi_{+} + \int d^4 x d^4 \th \; \bar{\Phi}_{-} e^{-gV} \Phi_{-} \;,
\er
donde $W_{\a}$ es un supercampo quiral, que se expresa en t\'erminos del supercampo real gauge $V$ y de superderivadas covariales $D_{\a}$ como $W_{\a} = i \bar{D}^2 D_{\a} V$. $\Phi_{\pm}$ son supercampos quirales de materia. Es importante se\~nalar tambi\'en que se puede definir teor\'ia de perturbaciones en el superespacio: en este caso tenemos diagramas definidos en t\'erminos de superpropagadores $P_{ij} = \D_{ij} \d_{ij}$, donde $\D_{ij}$ es el propagador usual y $\d_{ij}$ es la funci\'on delta de variables grassmanianas.

Al igual que en el caso de QED, realizamos los c\'alculos en el gauge de Feynman (se justificar\'a posteriormente su uso) y con el m\'etodo de campo de background. El supercampo gauge $V$ se divide por lo tanto en dos contribuciones $V \rightarrow V + B$: $V$ es el supercampo gauge cu\'antico y $B$ el background. 
\subsubsection*{Funci\'on a dos puntos del campo $B$ a un bucle}
\begin{figure}[ht]
\centerline{\epsfbox{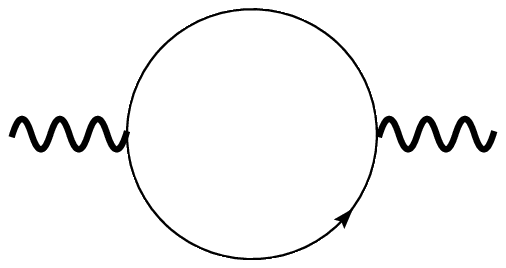}}
\caption{Diagrama a un bucle en SuperQED.}
\end{figure}
En este caso, la expresi\'on desnuda es 
\br
\G^{(1)}_{+} &=& \frac{g^2}{2} \int d^8 z_1 d^8 z_2 \; B(z_1) B(z_2) \left[ D^2_1 P_{12} \stackrel{\leftarrow}{D^2}_2 \right] \left[ D^2_2 P_{12} \stackrel{\leftarrow}{D^2}_1 \right] 
\er
que, con el \'algebra de derivadas covariantes, puede ser rescrita como
\br
\G^{(1)}_{+} &=& \frac{g^2}{2} \int d^4 x d^4 y d^4 \th \; B(x, \th) \left[ \bar{D}^2 D^2 B(y, \th) \right] \D^2_{xy}  \nonumber \\
& & + \frac{g^2}{2} \int d^4 x d^4 y d^4 \th \; B(x, \th) B(y, \th) \D_{xy} \left( \Box \D_{xy} \right) \nonumber \\
& & - \frac{i g^2}{2} \int d^4 x d^4 y d^4 \th \; B(x, \th) \left[ \bar{D}^{\dot{\a}} D^{\a} B(y, \th) \right] \D_{xy} \pa_{\a \dot{\a}}^y \D_{xy} \;.
\er
Aplicando RDR, obtenemos el siguiente valor renormalizado
\br
\G^{(1)}_{+ \;R} &=& - \frac{g^2}{16(4 \pi^2)^2} \int d^4 x d^4 y d^4 \th \; B(x, \th) \left[ D^{\a} \bar{D}^2 D_{\a} B(y, \th) \right] \Box \frac{ \ln (x-y)^2 M^2}{(x-y)^2} \;. \nonumber \\
\er 

\subsubsection*{Funci\'on a dos puntos del campo $B$ a dos bucles}
\begin{figure}[ht]
\centerline{\epsfbox{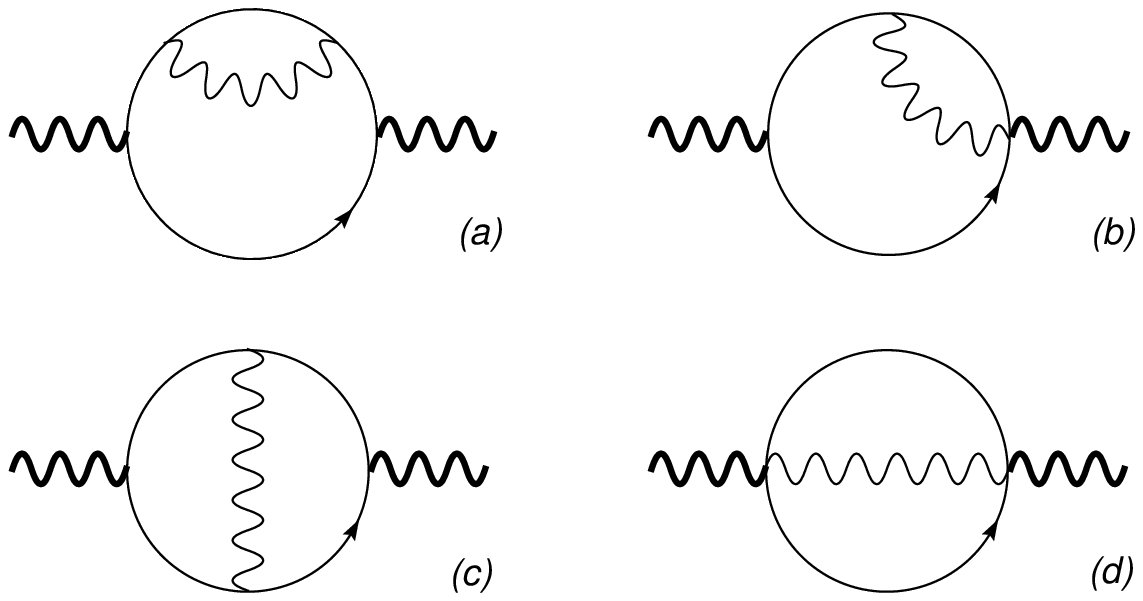}}
\caption{Diagramas de SuperQED a dos bucles.}
\label{Resumen_SQED}
\end{figure}
Omitiendo la expresi\'on desnuda de cada diagrama en t\'erminos del superpropagador $P_{ij}$, tras aplicar el \'algebra de superderivadas covariantes tenemos las siguientes contribuciones no renormalizadas
\br
\G^{(2a)}_{+} &=& \frac{g^4}{2} \int d^4 x d^4 y d^4 \th \; B(x, \th) B(y, \th) \left[ \Box ( \D I^{1} ) - 2 \D^3 - \pa^{\a \dot{\a}} ( \D \pa_{\a \dot{\a}} I^{1} ) \right] (x-y)  \nonumber \\
& & + \frac{g^4}{2} \int d^4 x d^4 y d^4 \th \; B(x, \th) \left[ D^{\a} \bar{D}^2 D_{\a} B(y, \th) \right] \left[ \D I^1 \right] (x-y) \nonumber \\
\G^{(2b)}_{+} &=& g^4 \int d^4 x d^4 y d^4 \th \; B(x, \th) B(y,\th) \left[ - \Box ( \D I^1) + 2 \D^3 + \pa^{\a \dot{\a}} \left( \D \pa_{\a \dot{\a}} I^1 \right) \right] (x-y)  \nonumber \\
& & - g^4 \int d^4 x d^4 y d^4 \th \; B(x, \th) \left[ D^{\a} \bar{D}^2 D_{\a} B(y, \th) \right] \left[ \D I^1 \right] (x-y) \nonumber \\
\G^{(2c)}_{+} &=& \frac{g^4}{2} \int d^4 x d^4 y d^4 \th \; B(x, \th) \left[ D^{\a} \bar{D}^2 D_{\a} B(y, \th) \right] \left[ \D I^1 \right](x-y)  \nonumber \\
& & + \frac{g^4}{2} \int d^4 x d^4 y d^4 \th \; B(x, \th) B(y, \th) \left[ \Box ( \D I^1 ) - \D^3 - \pa^{\a \dot{\a}} ( \D \pa_{\a \dot{\a}} I^1 ) \right] (x-y)  \nonumber \\
& & + \frac{g^4}{2} \int d^4 x d^4 y d^4 \th B(x, \th) \left[D^{\b} \bar{D}^2 D^{\a} B(y, \th) \right] C^{\dot{\b} \dot{\a}} H[\pa_{\b \dot{\b}},1 \; ; 1, \pa_{\a \dot{\a}}]  \nonumber \\ 
\G^{2d}_{+} &=& - \frac{g^4}{2} \int d^4 x d^4 y d^4 \th \; B(x, \th) B(y, \th) \D^3_{xy} \;.
\er
Antes de obtener las expresi\'ones renormalizadas, podemos sumar todas las contribuciones desnudas que forman la expresi\'on a dos bucles, ya que al emplear RDR todas las estructuras se renormalizan siempre con las mismas escalas. En \cite{Song}, donde se obtuvo la renormalizaci\'on diferencial de SuperQED a dos bucles, este c\'alculo simplificado no se pod\'ia realizar, ya que hab\'ia que renormalizar cada estructura con su escala correspondiente, para al final relacionarlas mediante las identidades de Ward. La expresi\'on final renormalizada que encontramos es
\br
\G^{2}_R  &=& \left. 2 \left( \G_{+}^{(2a)} + \G_{+}^{(2b)} + \G_{+}^{(2c)} + \G_{+}^{(2d)} \right) \right|_R \nonumber \\
&=& g^4 \int d^4 x d^4 y d^4 \th \; B(x, \th) \left[ D^{\b} \bar{D}^2 D^{\a} B(y, \th) \right] C^{\dot{\b} \dot{\a}} H^R [\pa_{\b \dot{\b}},1 \; ; \; 1, \pa_{\a \dot{\a}} ] \nonumber \\
&=& - \frac{g^4}{16 (4 \pi^2)^3} \int d^4 x d^4 y d^4 \th \; B(x, \th) \left[ D^{\a} \bar{D}^2 D_{\a} B(y, \th) \right] \Box \frac{ \ln (x-y)^2 M^2}{(x-y)^2} + \ldots \nonumber \\ 
\er
\subsubsection*{Ecuaci\'on del grupo de renormalizaci\'on}
Al igual que en QED, evaluando la ecuaci\'on del grupo de renormalizaci\'on a un bucle de los campos cu\'anticos, obtenemos que el valor de la funci\'on correspondiente a la variaci\'on del par\'ametro gauge es de orden $g^2$. Esto, unido al hecho de que ni el nivel arbol ni la correcci\'on a un bucle de la autoenenerg\'ia del campo B dependen del par\'ametro gauge, justifica que hayamos podido realizar el c\'alculo en el gauge de Feynman. Con la ecuaci\'on del grupo de renormalizaci\'on para campos background, obtenemos el valor de la funci\'on beta a dos bucles como 
\br
\b(g_{SQED}) &=& \frac{1}{8 \pi^2} g^3_{SQED} + \frac{1}{2 (4 \pi^2)^2} g^5_{SQED} + {\cal{O}}(g^7_{SQED}) \;,
\er 
donde empleamos la normalizaci\'on usual de la constante de acoplamiento, $g = \sqrt{2} g_{SQED}$. Estos valores concuerdan con resultados previos encontrados en la literatura \cite{Vainshtein:1986ja,Shifman:1985fi,Novikov:1985rd}.

\section*{Ejemplos no abelianos}
\subsection*{Yang-Mills}
El lagrangiano de esta teor\'ia es
\br
\cal{L} &=& \frac{1}{4} F^a_{\m \n} F^a_{\m \n} + \frac{1}{2 \a} ( \pa_{\m} A_{\m} )^a ( \pa_{\n} A_{\n} )^a + ( \pa_{\m} \bar{\eta})^a ( {D}_{\m} \eta )^a \;,
\er
donde $A_{\m}^a$ es el campo gauge, $\eta$ y $\bar{\eta}$ son los fantasmas de Fadeev-Popov, $F_{\m \n}^a = \pa_{\m} A_{\n}^a - \pa_{\n} A_{\m}^a + g f^{abc} A_{\m}^b A_{\n}^c$ y $f^{abc}$ las constantes de estructura del \'algebra de Lie asociada al grupo de simetr\'ia. Al igual que en los ejemplos previos abelianos, los c\'alculos los realizaremos en el gauge de Feynman y con el m\'etodo del campo de background, obteniendo la funci\'on de dos puntos renormalizada del campo background $B_{\m}^a$.
\subsubsection*{Un bucle}
\begin{figure}
\centerline{\epsfbox{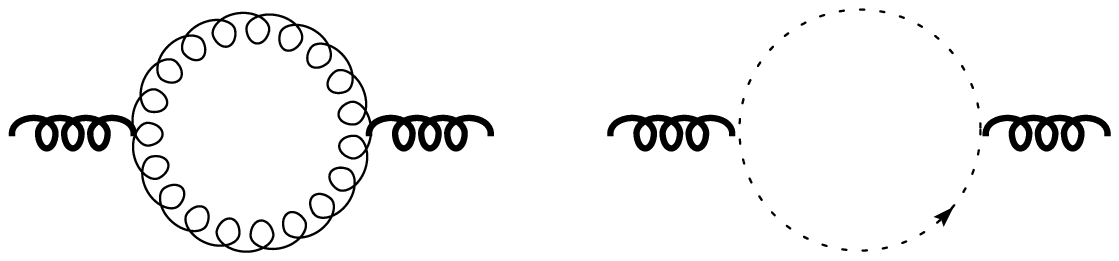}}
\caption{Diagramas de Yang-Mills a un bucle.}
\end{figure}
Si estudiamos la autoenerg\'ia del campo background en el gauge de Feynman, obtenemos la siguiente expresi\'on desnuda (suma de las contribuciones del bucle de gluones y fantasmas)
\br
<B_{\m}^a(x) B_{\n}^b(y)> &=& g^2 C_A \d^{ab} \left[ 4 \pa_{\m} \pa_{\n} \D^2 - 4 \d_{\m \n} \Box \D^2 + 2 \pa_{\m} ( \D \pa_{\n} \D ) - 4 \D \pa_{\m} \pa_{\n} \D \right] \;,  \nonumber \\
\er
que se renormaliza como
\br
<B_{\m}^a(x) B_{\n}^b(0)>_R &=& g^2 C_A \d^{ab} (\pa_{\m} \pa_{\n} - \d_{\m \n} \Box) \left[ - \frac{11}{48 \pi^2 (4 \pi^2)} \Box \frac{\ln x^2 M^2}{x^2} - \frac{1}{72 \pi^2} \d (x) \right]  \;. \nonumber \\ 
\er

Tambi\'en consideramos la funci\'on a dos puntos del campo gauge cu\'antico, ya que ser\'a una inserci\'on en uno de los diagramas a dos bucles. El valor total no renormalizado de esta funci\'on es
\br
<A_{\m}^a (x) A_{\n}^b (y) > &=& g^2 C_A \d^{ab} \left[ \pa_{\m} \pa_{\n} \D^2 - \d_{\m \n} \Box \D^2 + 4 \pa_{\m} ( \D \pa_{\n} \D ) - 2 \d_{\m \n} \pa^{\l} ( \D \pa_{\l} \D )  \right. \nonumber \\
& &- \left. 4 \D \pa_{\m} \pa_{\n} \D - \d_{\m \n} \D ( \Box \D ) \right] \;,
\er
y aplicando RDR
\br
<A_{\m}^a (x) A_{\n}^b (0) >_R &=& g^2 C_A \d^{ab} \left[ \frac{5}{3}(  \pa_{\m} \pa_{\n} - \d_{\m \n } \Box) \D^2_R (x) - \frac{1}{72 \pi^2} (\pa_{\m} \pa_{\n} - \d_{\m \n} \Box) \d (x) \right] \nonumber \\
&=& - \frac{g^2 C_A \d^{ab}}{144 \pi^2} (\pa_{\m} \pa_{\n} - \d_{\m \n} \Box) \left[ \frac{15}{4 \pi^2} \Box \frac{ \ln x^2 M^2}{x^2} + 2 \d (x) \right] \;. 
\er

\subsubsection*{Acci\'on efectiva en un gauge gen\'erico}
A diferencia que en los ejemplos abelianos, el restringirnos al gauge de Feynman no es inocuo en este caso. Para incluir las variaciones del par\'ametro gauge en la ecuaci\'on del grupo de renormalizaci\'on, obtendremos la dependencia lineal en $\xi$ (el par\'ametro gauge definido a partir del usual como $\frac{1}{\a} = 1 + \xi$) de la acci\'on efectiva background expandida a segundo orden en los campos $B_{\m}^a$. Para obtener dicha acci\'on efectiva, consideramos el generador de funciones de Green conectadas $W$
\br
W &=& - \frac{1}{2} tr \ln \left[ \d_{\m \n} \Box^{ab} - 2 g f^{cab} B_{\m \n}^c + \xi ({\bf{D}}_{\m} {\bf{D}}_{\n})^{ab} \right] \;,  
\er
con ${\bf{D}}_{\m}^{ac} = \pa_{\m} \d^{ac} + g f^{abc} B_{\m}^b $ y ${\Box}^{ab} = ( {\bf{D}}^{\m} {\bf{D}}_{\m})^{ab}$. A primer orden en $\xi$ y segundo orden en $B_{\m}^a$, esto se rescribe como
\br
W &=& \xi C_A g^2 tr \left[ \frac{1}{2} \D B_{\m \n}^a \D B_{\m \n}^a - 2 \D B_{\m \n}^a \D B_{\n \l}^a \D \pa_{\l} \pa_{\m} \right] \;.
\er
Renormalizando esta expresi\'on, se obtiene f\'acilmente la acci\'on efectiva como
\br
\G_\xi = - \frac{\xi C_A g^2}{4(4 \pi^2)} \int d^4 x d^4 y \; B_{\m}^a (x) B_{\n}^a (y) (\pa^x_{\m} \pa^x_{\n} - \d_{\m \n} \Box) ( \Box \D(x-y) ) \;.
\er

\subsubsection*{Renormalizaci\'on del propagador background a dos bucles}
\begin{figure}[ht]
\centerline{\epsfbox{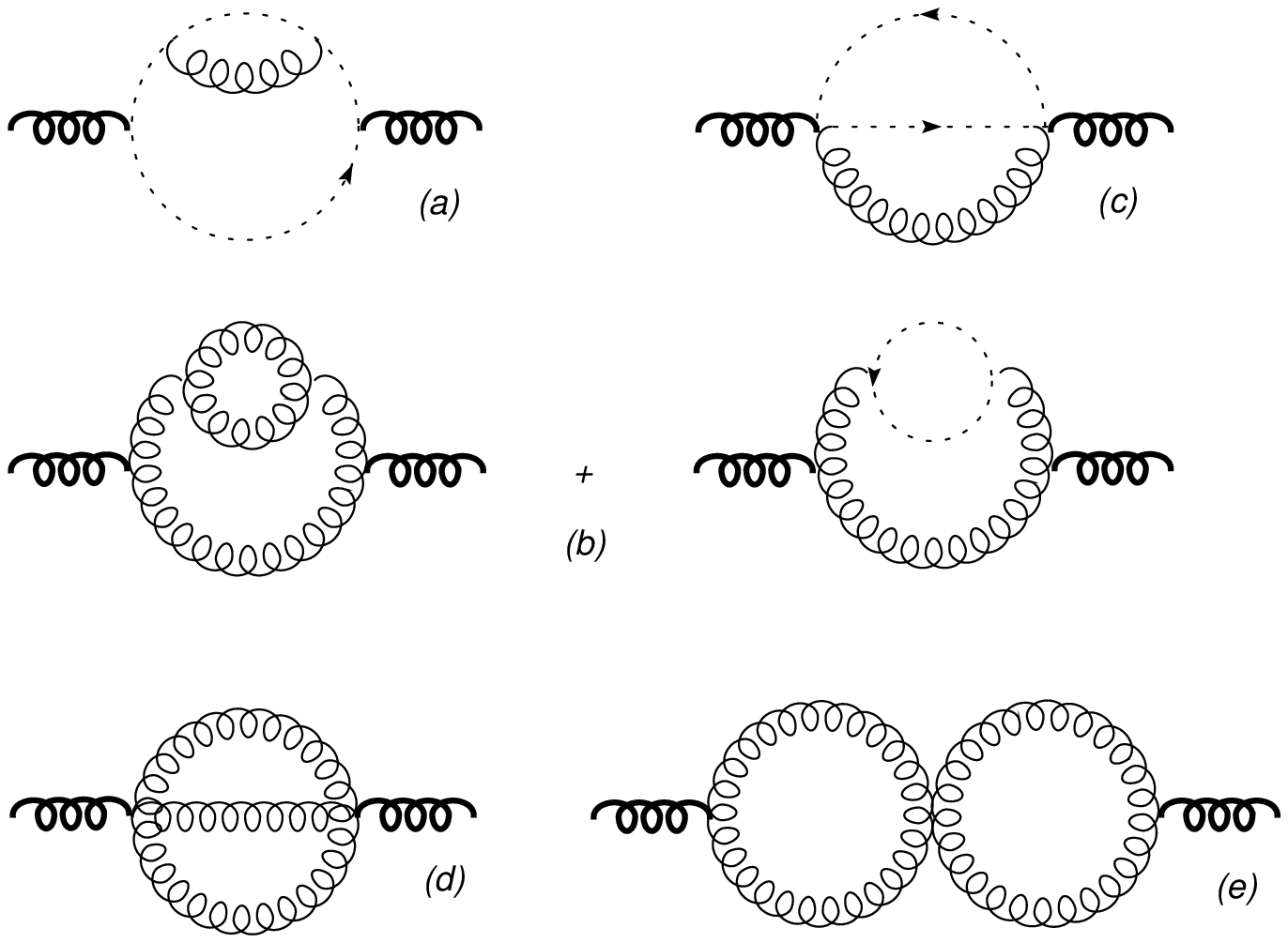}}
\caption{Diagramas de Yang-Mills a dos bucles (a)-(e).}
\end{figure}

\begin{figure}[ht]
\centerline{\epsfbox{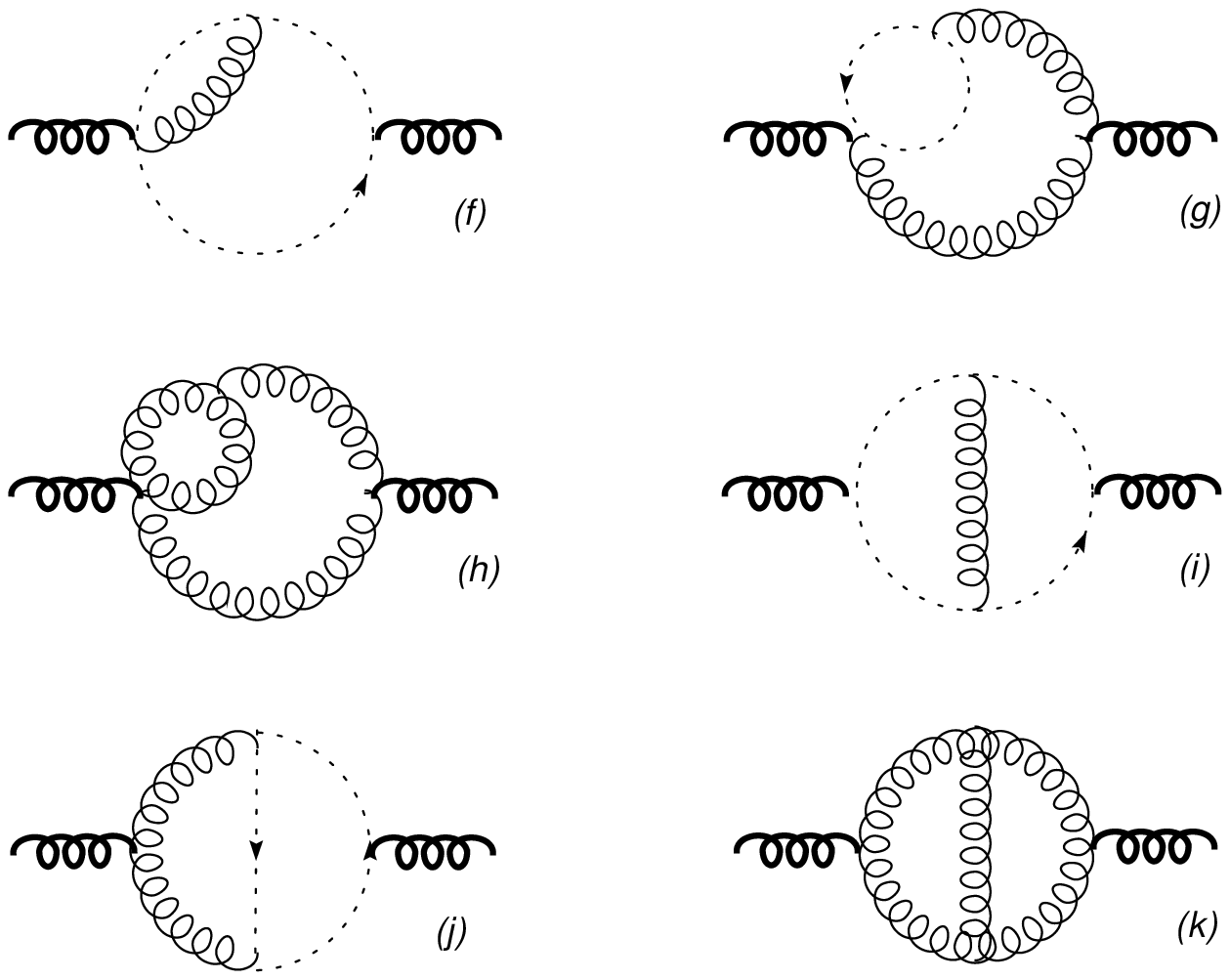}}
\caption{Diagramas de Yang-Mills a dos bucles (f)-(k).}
\end{figure}
Los diagramas $(a)$-$(h)$ tienen divergencias anidadas, mientras que $(i)$, $(j)$ y $(k)$ corresponden a expresiones con divergencias de solapamiento. En concreto, si evaluamos el diagrama $(a)$, tenemos la siguiente expresi\'on desnuda
\br
< B_{\m}^a (x) B_{\n}^b (y) >_a &=& - 2 g^4 f^{aec} f^{bcd} f^{gdf} f^{gfe} \int  d^4 u d^4 v \; \D_{xy} ( \stackrel{\leftarrow}{\pa_{\m}^x} - \pa_{\m}^x) ( \pa_{\n}^y - \stackrel{\leftarrow}{\pa_{\n}^y}) \D_{yv}  \nonumber \\
& & \times ( \pa_{\l}^v \D_{uv}) \D_{uv} ( \pa_{\l}^u \D_{xu}) \;, \nonumber \\ 
\er
que puede ser rescrita en t\'erminos de la integral $I^1$ previamente definida como\br
< B_{\m}^a (x) B_{\n}^b (y) >_a &=& - g^4 C_A^2 \d^{ab} \left[ 4 \pa_{\n} ( \D \pa_{\m} I^1 ) - \pa_{\m} \pa_{\n} ( \D I^1 ) - 4 \D \pa_{\m} \pa_{\n} I^1 \right] \;.
\er
Con los resultados mostrados para $I^1$, esto se puede renormalizar de forma inmediata y escribir como
\br
<B_{\m}^a(x) B_{\n}^b (0) >_{a\; R} &=& \frac{g^4 C_A^2 \d^{ab}}{32(4 \pi^2)^3} \left[ \pa_{\m} \pa_{\n} \Box \frac{ - \frac{1}{3} \ln^2 x^2 M^2 - \frac{8}{9} \ln x^2 M^2}{x^2} \right. \nonumber \\
& & \left. + \d_{\m \n} \Box \Box \frac{\frac{1}{3}  \ln^2 x^2 M^2 + \frac{11}{9} \ln x^2 M^2}{x^2} \right] + \textrm{(termin.~locales)} \;. \nonumber \\
\er

Tomando ahora $(i)$ como un ejemplo de integrales de solapamiento, esta contribuci\'on se puede escribir en t\'erminos de las integrales $H$ como
\br
< B_{\m}^a (x) B_{\n}^b (y) >_i = - \frac{1}{2} g^4 C_A^2 \d^{ab} &\left[ \right.& \pa_{\m}^x \pa_{\n}^y H[1, \pa_{\l} \; ; \; \pa_{\l},1] - 2 \pa_{\m}^x H[1, \pa_{\l} \; ; \; \pa_{\l} \pa_{\n} , 1]  \nonumber \\
& & \left. - 2 \pa_{\n}^y H[1, \pa_{\l} \pa_{\m} \; ; \; \pa_{\l},1] + 4 H[ 1, \pa_{\m} \pa_{\l} \; ; \; \pa_{\n} \pa_{\l} , 1] \; \right] \;. \nonumber \\
\er
Una vez que tenemos esto, con la lista de expresiones $H$ renormalizadas, llegamos inmediatamente al siguiente resultado renormalizado
\br
< B_{\m}^a (x) B_{\n}^b (0) >_{i \; R} &=& \frac{g^4 C_A^2 \d^{a b}}{32 (4 \pi^2)^3} \left[ \pa_{\m} \pa_{\n} \Box \frac{ - \frac{1}{12} \ln^2 x^2 M^2 - \frac{17}{36} \ln x^2 M^2 }{x^2} \right.  \nonumber \\
& &+ \left. \d_{\m \n} \Box \Box \frac{ \frac{1}{12} \ln^2 x^2 M^2 + \frac{29}{36} \ln x^2 M^2}{x^2} \right] +~\textrm{(termin.~locales)} \;. \nonumber \\
\er

Procediendo de forma similar en el resto de los diagramas, obtenemos la renormalizaci\'on de todas las contribuciones. Sumando todos los resultados, obtenemos el valor renormalizado de la funci\'on a dos puntos del campo background como
\br
< B_{\m}^a (x) B_{\n}^b (0) >_R &=& - \frac{g^4 C_A^2 \d^{ab}}{2(4 \pi^2)^3} ( \pa_{\m} \pa_{\n} - \d_{\m \n} \Box ) \Box \frac{ \ln x^2 M^2}{x^2} +~\textrm{(termin.~locales)} \;. \nonumber \\
\er

\subsubsection*{Ecuaci\'on del grupo de renormalizaci\'on}
Evaluando la ecuaci\'on del grupo de renormalizaci\'on a un bucle para la funci\'on a dos puntos del campo cu\'antico, obtenemos el valor del coeficiente que se ocupa de las variaciones del par\'ametro gauge. Dicho valor es
\br
\g_{\xi} &=& - \frac{5 C_A}{24 \pi^2} g^2 + \cdots 
\er
Entonces, empleando $\g_{\xi}$, la acci\'on efectiva background a un bucle en un gauge gen\'erico y el resultado para la autoenerg\'ia renormalizada del campo $B_{\m}^a$ a uno y dos bucles, obtenemos a partir de la ecuaci\'on del grupo de renormalizaci\'on background el valor de la funci\'on beta como 
\br
\b (g) &=& \b_1 g^3 + \b_2 g^5 + {\cal{O}}(g^7) \nonumber \\
\b_1 &=& - \frac{11 C_A}{48 \pi^2} \nonumber \\
\b_2 &=& - \frac{17 C^2_A}{24 (4 \pi^2)^2} \;.
\er 

\section*{Super Yang-Mills}
Pasamos ahora a estudiar la versi\'on supersim\'etrica del modelo anterior, Super Yang-Mills. Al igual que con SuperQED, aplicamos los convenios de \cite{Gates:1983nr}. En este caso, la divisi\'on del campo gauge en parte cu\'antica y parte background es no lineal  $e^{g V_{(split)}} = e^{\boldsymbol{\Omega}} e^{g V} e^{\bar{\boldsymbol{\Omega}}}$, con $V$ el campo cu\'antico gauge y $\bOmega$ el prepotencial background. Por lo tanto, las derivadas covariantes gauge se escriben en una representaci\'on quiral cu\'antica y vectorial background, por lo que la acci\'on dividida tiene la forma de 
\br
S &=& - \frac{1}{2 g^2} tr \int d^4 x d^4 \th \; ( e^{-g V} \bnabla^{\a} e^{g V}) \bar{\bnabla}^2 ( e^{- g V} \bnabla_{\a} e^{g V} ) \;,
\er
donde $\bnabla_{\a}$ es la derivada covariante background. Esto implica que el la parte cuadr\'atica en $V$ de la acci\'on con el gauge fijado (de la cual se deriva el propagador cu\'antico) depende de los campos background como
\br
& & - \frac{1}{2} tr \int d^4 x d^4 \th V \left[ {\boldsymbol{\Box}} - i \bW^{\a} \bnabla_{\a} - i \bar{\bW}^{\dot{\a}} \bar{\bnabla}_{\dot{\a}} \right] V \nonumber ~~~,~~~ \bfBox = \frac{1}{2} \bnabla^{\a \dot{\a}} \bnabla_{\a \dot{\a}} \;,
\er
donde denotamos el operador cin\'etico como $\hat{\dal} = \bfBox - i \bW^{\a} \bnabla_{\a} - i \bar{\bW}^{\dot{\a}} \bar{\bnabla}_{\dot{\a}}$. Otra consecuencia del m\'etodo de campo de background en el superespacio es la aparici\'on de los fantasmas de Nielsen-Kallosh, que corresponden a la normalizaci\'on de la funci\'on que se utiliza para realizar el promedio sobre los par\'ametros gauge en el procedimiento funcional est\'andar de cuantizaci\'on. Hay que destacar que, como estos fantasmas entran cuadr\'aticamente en la acci\'on y s\'olo interaccionan con el campo background $B$ ($\bOmega = \bar{\bOmega} = \frac{1}{2} B$), s\'olo contribuyen a primer orden en teor\'ia de perturbaciones.

Debemos notar tambi\'en que para realizar los c\'alculos a dos bucles empleamos supergr\'aficos covariantes. B\'asicamente, realizamos los c\'alculos sin extraer de la derivada covariante la conexi\'on espinorial, mediante el \'algebra de las derivadas covariantes. Por lo tanto, tenemos menos gr\'aficos y estos son m\'as convergentes.

\subsubsection*{Un bucle}
A la autoenerg\'ia background a un bucle s\'olo contribuyen los fantasmas (tanto los de Fadeev-Popov como los de Nielsen-Kallosh), siendo la expresi\'on desnuda
\br
\G^{(1)} &=& - \frac{3 C_A}{2} \int d^4 x d^4 y d^4 \th \; B^a(x, \th) \left[ \bar{D}^2 D^2 B^a (y, \th) \right] \D^2_{xy}  \nonumber \\
& & - \frac{3 C_A}{2} \int d^4 x d^4 y d^4 \th \; B^a(x, \th) B^a (y, \th) \D_{xy} \Box \D_{xy}   \nonumber \\
& & + \frac{i  3 C_A}{2} \int d^4 x d^4 y d^4 \th \; B^a(x, \th) \left[ \bar{D}^{\dot{\a}} D^{\a} B^a (y, \th) \right] \D_{xy} \pa_{\a \dot{\a}}^y \D_{xy} \;,
\er
que se renormaliza de acuerdo con las reglas de RDR como
\br
\G^{(1)} &=& \frac{ 3 C_A}{16 (4 \pi^2)^2} \int d^4 x d^4 y d^4 \th \; B^a(x,\th) \left[ D^{\a} \bar{D}^2 D_{\a} B^a (y, \th) \right] \Box \frac{ \ln (x-y)^2 M^2}{(x-y)^2} \;. \nonumber \\ 
\er

En cuanto a la funci\'on a dos puntos de los campos gauge cu\'anticos, las contribuciones corresponden a los siguientes diagramas:
\begin{figure}[ht]
\centerline{\epsfbox{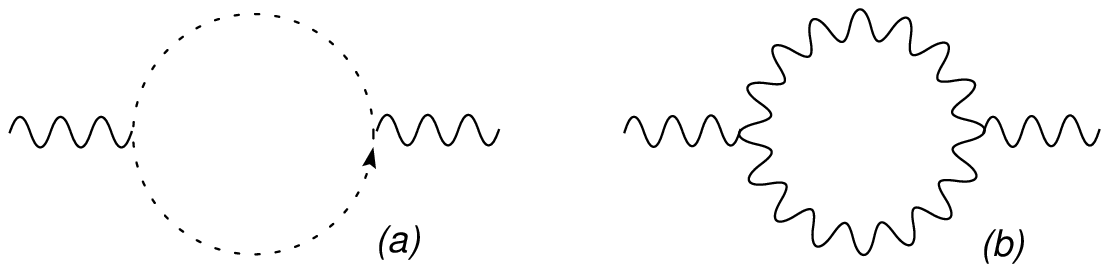}}
\caption{Contribuciones a la funci\'on  de dos puntos cu\'antica a un bucle.}
\end{figure}

La contribuci\'on final renormalizada en este caso es
\br
\G^{(1)}_V &=& - \frac{3 g^2 C_A}{16 (4 \pi^2)^2} \int d^4 x d^4 y d^4 \th \; V^a(x, \th) \Box \Pi_{1/2} V^{a} (y, \th) \Box \frac{ \ln (x-y)^2 M^2}{(x-y)^2} \;. 
\er

\subsubsection*{Acci\'on en un gauge gen\'erico}
Al igual que en caso de Yang-Mills, tenemos que evaluar la acci\'on efectiva background en un gauge gen\'erico para tener en cuenta el t\'ermino de variaci\'on del par\'ametro gauge $\xi$ (redefinido a partir del usual como $\xi + 1 = \frac{1}{\a}$) en la ecuaci\'on del grupo de renormalizaci\'on. Obtendremos el t\'ermino lineal en $\xi$ correspondiente a la contribuci\'on a segundo orden en campos background de la expansi\'on de la acci\'on efectiva. Para ello, a partir de un c\'alculo funcional, escribimos dicha acci\'on como
\br
\G_{eff} &=& - \frac{1}{2} tr \ln \left[ \hat{\dal} + \xi \left( \bnabla^2 \bar{\bnabla}^2 + \bar{\bnabla}^2 \bnabla^2 \right) \right] + tr \ln \left[ \Box_{-} + \xi \bnabla^2 \bar{\bnabla}^2 \right] \nonumber \\
&=&  - \frac{1}{2} tr \ln \hat{\dal} + tr \ln \Box_{-} + \G_{\xi} \;,
\er
donde 
\br
\Box_{+} &=& \Box - i \bW^{\a} \bnabla_{\a} - \frac{i}{2} ( \bnabla^{\a} \bW_{\a} ) \nonumber \\
\Box_{-} &=& \Box - i \bar{\bW}^{\dot{\a}} \bar{\bnabla}_{\dot{\a}} - \frac{i}{2} ( \bar{\bnabla}^{\dot{\a}} \bar{\bW}_{\dot{\a}}) \;.
\er

Considerando la expresiones inversas de los operadores, y qued\'andonos a segundo orden en campos background, obtenemos el valor renormalizado como
\br
\G_{\xi} &=& - \frac{\xi}{16 (4 \pi^2)^2} tr \int d^4 x d^4 y d^2 \th \; \bW^{\a}(x, \th) \bW_{\a} (y, \th) \Box \frac{ \ln (x-y)^2 M^2_{IR}}{(x-y)^2} + {\cal{O}}(\xi^2 ; B^3) \;. \nonumber \\ 
\er 

\subsubsection*{Dos bucles}
Para realizar la renormalizaci\'on a dos bucles con supergr\'aficos covariantes, simplemente tenemos que considerar el siguiente diagrama de vac\'io:
\begin{figure}[h]
\centerline{\epsfbox{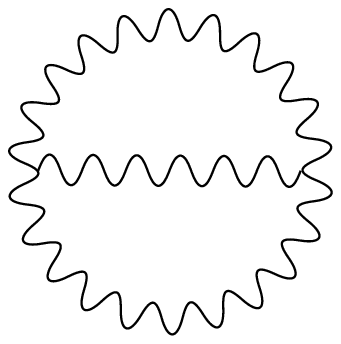}}
\caption{Contribuci\'on a dos bucles a la acci\'on efectiva background.}
\end{figure}

En dicho diagrama, los propagadores gauge $\hat{\dal}^{-1}$ dependen de los campos background, por lo que hemos de expandirlos y quedarnos con el segundo orden en $B$. De esta expansi\'on tenemos dos tipos de contribuciones. Unas tienen campos $\bW_{\a}$ expl\'icitos, mientras que otras se construyen en base a las conexiones espacio-temporales background $\bG_{\a \dot{\a}}$. As\'i, tenemos los siguients resultados renormalizados para las contribuciones con campos $\bW_{\a}$
\br
\sum^{2}_{i=1}\G^{(2)}_i |_R &=& - \frac{3 i g^2 C_A^2}{2} tr \int d^4 x d^4 y d^4 \th \; \left[ \bW^{\a}(x,\th) \pa_{\a \dot{\a}}^y \bar{\bW}^{\dot{\a}} + \bar{\bW}^{\dot{\a}}(x,\th) \pa_{\a \dot{\a}}^y \bW^{\a}(y,\th) \right]  \nonumber \\
& & \times \left[ \D I^0 \right]_R (x-y) \nonumber \\ 
& & + \frac{3 i g^2 C_A^2}{16 (4 \pi^2)^3} tr \int d^4 x d^4 y d^2 \th \; \bW^{\a} (x, \th) \bar{\bW}^{\dot{\a}}(y, \th) \pa_{\a \dot{\a}}^x \frac{\ln (x-y)^2 M^2}{(x-y)^2} + {\cal{O}}(B^3) \;, \nonumber \\ 
\er  
la cual, mediante una identidad de Bianchi se puede escribir en una forma invariante gauge como
\br
\sum^{2}_{i=1}\G^{(2)}_i |_R &=& 3 g^2 C_A^2 tr \int d^4 x d^4 y d^2 \th \; \bW^{\a} (x, \th) \bW_{\a} (y,\th) \Box \left[ \D I^0 \right]_R (x-y)  \nonumber \\
& & - \frac{3 g^2 C_A^2}{16(4 \pi^2)^3} tr \int d^4 x d^4 y d^2 \th \; \bW^{\a}(x,\th) \bW_{\a}(y, \th) \Box \frac{ \ln (x-y)^2 M^2}{(x-y)^2} + {\cal{O}}(B^3) \;. \nonumber \\
\er

Por otro lado, la suma de los diagramas con conexiones espacio-temporales es
\br
\sum^{5}_{i=3} \G^{(2)}_i |_R &=& - 3 g^2 C_A^2 tr \int d^4 x d^4 y d^4 \th \; \bG^{\a \dot{\a}} (x, \th) \bG^{\b \dot{\b}}(y,\th) \left( \pa_{\a \dot{\a}}^x \pa_{\b \dot{\b}}^x - (2 C_{\a \b} C_{\dot{\a} \dot{\b}}) \Box \right) \nonumber \\
& & \times \left[ \frac{1}{4} [ \D I^0 ]_R (x-y) - \frac{1}{32(4 \pi^2)^3} \frac{\ln (x-y)^2 M^2}{(x-y)^2} \right] + {\cal{O}}(B^3) \;,
\er
que, debido a ser una expresi\'on transversa se puede escribir en t\'erminos de $\bW_{\a}$ como
\br
& & tr \int d^4 x d^4 y d^4 \th \; \bG^{\a \dot{\a}} (x, \th) \bG^{\b \dot{\b}} (y, \th) \left( \pa_{\a \dot{\a}}^x \pa_{b \dot{\b}}^x - 2 C_{\a \b} C_{\dot{\a} \dot{\b}} \Box \right) f(x-y) \nonumber \\
& & = - 3 tr \int d^4 x d^4 y d^4 \th \; \left[ D^{\a} B(x, \th) \right] \left[ \bar{D}^2 D_{\a} B(y, \th) \right] \Box f(x-y) + {\cal{O}}(B^3) \nonumber \\
& & = 3 tr \int d^4 x d^4 y d^2 \th \; \bW^{\a} (x, \th) \bW_{\a} (y, \th) \Box f(x-y) + {\cal{O}}(B^3) \;.
\er

Por lo tanto, la expresi\'on total renormalizada a dos bucles y segundo orden en los campos background es
\br
\frac{1}{2}\sum^{5}_{i=1} \G^{(2)}_i &=& tr \int d^4 x d^4 y d^2 \th \bW^{\a} (x, \th) \bW_{\a} (y, \th) \G^{(2)}(x-y) \;,
\er
con
\br
\G^{(2)}(x) &=& \frac{3 g^2 C_A^2}{64(4 \pi^2)^3} \Box \frac{\frac{1}{4} \ln^2 x^2 M^2_{IR} + \frac{1}{2} \ln x^2 M^2_{IR} ( 1 - \ln x^2 M^2) + \ln x^2 M^2}{x^2}  \nonumber \\
& & +~\textrm{(termin.~locales)} \;.
\er

\subsubsection*{Ecuaci\'on del grupo de renormalizaci\'on}
A la hora de evaluar la ecuaci\'on del grupo de renormalizaci\'on, los pasos que hemos de dar son id\'enticos al caso de Yang-Mills. En primer lugar, con la ecuaci\'on correspondiente a la funci\'on a dos puntos de los campos cu\'anticos a un bucle, obtenemos el valor del coeficiente de variaci\'on del par\'ametro gauge, $\g_{\xi} \pa / \pa \xi$. En concreto, tenemos el siguiente resultado
\br
\g_\xi &=& - \frac{3 C_A}{4 (4 \pi^2)}g^2 + {\cal{O}}(g^4) \;.
\er 

Entonces, con $\g_{\xi}$, la acci\'on efectiva en una gauge gen\'erico y la contribuci\'on a uno y dos bucles a la autoenerg\'ia background, de la ecuaci\'on del grupo de renormalizaci\'on satisfecha por los campos $B$ 
\br
\left. \left[ M \frac{\pa}{\pa M} + \b(g) \frac{\pa}{\pa g} + \g_{\xi} (g) \frac{\pa}{\pa \xi} \right] \G (x) \right|_{\xi = 0} = 0 \;
\er
obtenemos la expansi\'on de la funci\'on beta como
\br
\b(g_{SYM}) = - (3/2) [ C_A/ (8\pi^2)] g^3_{SYM} - (3/2) [ C_A / (8 \pi^2) ]^2 g^5_{SYM} + {\cal{O}}(g^7_{SYM}) \;, 
\er
donde hemos utilizado, igual que en el caso de SuperQED, la constante de acoplamiento usual, que difiere en un factor $\sqrt{2}$ de la que se emplea en \cite{Gates:1983nr}. Notar que, al igual que en el resto de las teor\'ias consideradas, no tenemos dimensi\'on an\'omala ($\g_B$), ya que con la normalizaci\'on que tenemos del campo background y la relaci\'on entre la renormalizaci\'on de dicho campo y la constate de acoplamiento, $\g_B$ se anula.

Este c\'alculo nos permite dar una nueva visi\'on sobre un punto controvertido: el origen de las correcciones m\'as all\'a de un bucle a la funci\'on beta de Super Yang-Mills. En un principio, Novikov, Shifman, Vainshtein and Zakharov (NSVZ) en \cite{Novikov:1983uc} obtuvieron la ``funci\'on beta exacta'' ($\b_{NSVZ}$) empleando un c\'alculo de instantones, siendo posteriormente reobtenido el coeficiente a dos bucles de esta funci\'on  mediante c\'alculos perturbativos (empleando reducci\'on dimensional) \cite{Abbott:1984pz,Grisaru:1985tc,Shifman:1986zi}. Aunque algunos c\'alculos parec\'ian indicar un origen IR a las correcciones de orden superior de $\b_{NSVZ}$ \cite{Novikov:1983uc,Novikov:1985rd,Shifman:1986zi}, esto fue cuestionado en \cite{Arkani-Hamed:1997ut,Arkani-Hamed:1997mj}, donde empleando un formalismo wilsoniano (por lo tanto, en principio s\'olo dependiente del comportamiento UV de la teor\'ia) y diferenciando entre una constante de acoplamiento holom\'orfica y una can\'onica, se obtuvo un flujo NSVZ para la \'ultima.

Nuestro c\'alculo tiene la virtud de evitar uno de los puntos conflictivos de la aplicaci\'on de m\'etodos dimensionales a Super Yang-Mills: la regularizaci\'on de tanto las divergencias UV como IR con el mismo par\'ametro infinitesimal, que implica que se mezclen ambas contribuciones en los resultados renormalizados. Nosotros en cambio tenemos las divergencias claramente diferenciadas al tener asociadas dos escalas independientes. Lo que hemos encontrado es que la escala correspondiente a la renormalizaci\'on a un bucle es la que genera la contribuci\'on a dos bucles de $\b_{NSVZ}$. No existe una escala a dos bucles UV (lo cual coincide con la conclusi\'on obtenida en \cite{Grisaru:1985tc}, conforme a la cual, en un esquema de regularizaci\'on en cuatro dimensiones no hay divergencias superficiales), aunque esto no implica que el coeficiente a dos bucles de la funci\'on beta sea nulo. Como se ve en este caso, la escala UV a un bucle sobrevive a dos bucles al tener en cuenta los efectos IR.

\end{document}